\documentclass[aps,
               pra,
               twocolumn,
               showpacs,
               amsmath,
               amssymb,
               superscriptaddress,
               reprint,
               10pt
              ]{revtex4-1}

\usepackage[colorlinks=true,breaklinks=true,allcolors=blue]{hyperref}
\usepackage{url}
\usepackage{txfonts}

\usepackage{amsmath, amssymb}

\usepackage{pstricks}
\usepackage{graphicx}
\usepackage{tikz}
\usepackage[utf8x]{inputenc}
\usepackage{color}
\usepackage{hyperref}

\usepackage{pstool}
\usepackage{pgf}
\usepackage{import}

\usetikzlibrary{snakes}
\usetikzlibrary{arrows}

\usepackage{tabularx}
\usepackage{multirow}
\usepackage{array}
\newcolumntype{L}[1]{>{\raggedright\let\newline\\\arraybackslash\hspace{0pt}}m{#1}}
\newcolumntype{C}[1]{>{\centering\let\newline\\\arraybackslash\hspace{0pt}}m{#1}}
\newcolumntype{R}[1]{>{\raggedleft\let\newline\\\arraybackslash\hspace{0pt}}m{#1}}

\usepackage{balance}
\usepackage{lastpage}

\newcommand{\dF}[1]{\mathrm{d} #1 \,}

\newcommand{\df}[1]{\mathrm{d}^{d+1}\! #1 \,}

\newcommand{\ddf}[1]{\mathrm{d}^{d} #1 \,}

\renewcommand{\vec}[1]{\mathbf{#1}}

\DeclareMathOperator{\Trace}{Tr}
\newcommand{\Tr}[1]{\Trace\left[{#1}\right]}

\newcommand{\eq}[1]{(\ref{eq:#1})}
\newcommand{\Eq}[1]{Eq.\,\eqref{eq:#1}}

\newcommand{\Fig}[1]{Fig.~\ref{fig:#1}}
\newcommand{\fig}[1]{\ref{fig:#1}}

\newcommand{\Sect}[1]{Sect.~\ref{sec:#1}}
\newcommand{\sect}[1]{\ref{sec:#1}}

\newcommand{\app}[1]{\ref{app:#1}}
\newcommand{\App}[1]{App.~\ref{app:#1}}
\newcommand{\Tab}[1]{Table~\ref{tab:#1}}

\newcommand{\1}{\uparrow}

\newcommand{\footnoteremember}[2]{%
\footnote{#2}%
\newcounter{#1}%
\setcounter{#1}{\value{footnote}}%
}
\newcommand{\footnoterecall}[1]{%
\footnotemark[\value{#1}]%
}

\makeatletter
\let\cat@comma@active\@empty
\makeatother

\makeatletter
\let\cat@comma@active\@empty
\makeatother

\begin{document}

\title{Kinetic theory of non-thermal fixed points in a Bose gas}

\author{Isara Chantesana}
\affiliation{Kirchhoff-Institut f\"ur Physik, 
             Ruprecht-Karls-Universit\"at Heidelberg,
             Im Neuenheimer Feld 227, 
             69120 Heidelberg, Germany}
\affiliation{Institut f\"ur Theoretische Physik,
             Ruprecht-Karls-Universit\"at Heidelberg,
             Philosophenweg~16,
             69120~Heidelberg, Germany}

\author{Asier Pi{\~n}eiro Orioli}
\affiliation{Institut f\"ur Theoretische Physik,
             Ruprecht-Karls-Universit\"at Heidelberg,
             Philosophenweg~16,
             69120~Heidelberg, Germany}

\author{Thomas~Gasenzer}
\email{t.gasenzer@uni-heidelberg.de}
\affiliation{Kirchhoff-Institut f\"ur Physik, 
             Ruprecht-Karls-Universit\"at Heidelberg,
             Im Neuenheimer Feld 227, 
             69120 Heidelberg, Germany}

\date{\today}

\begin{abstract}
We outline a kinetic theory of non-thermal fixed points for the example of a dilute Bose gas, partially reviewing results obtained earlier, thereby extending, complementing, generalizing and straightening them out.
We study universal dynamics after a cooling quench, focusing on situations where the time evolution represents a pure rescaling of spatial correlations, with time defining the scale parameter. 
The non-equilibrium initial condition set by the quench induces a redistribution of particles in momentum space.
Depending on conservation laws, this can take the form of a wave-turbulent flux or of a more general self-similar evolution, signaling the critically slowed approach to a non-thermal fixed point.
We identify such fixed points using a non-perturbative kinetic theory of collective scattering between highly occupied long-wavelength modes.
In contrast, a wave-turbulent flux, possible in the perturbative Boltzmann regime, builds up in a critically accelerated self-similar manner. 
A key result is the simple analytical universal scaling form of the non-perturbative many-body scattering matrix, for which we lay out the concrete conditions under which it applies.
We derive the scaling exponents for the time evolution as well as for the power-law tail of the momentum distribution function, for a general dynamical critical exponent $z$ and an anomalous scaling dimension $\eta$.
The approach of the non-thermal fixed point is, in particular, found to involve a rescaling of momenta in time $t$ by $t^{\,\beta}$, with $\beta=1/z$, within our kinetic approach independent of $\eta$.
We confirm our analytical predictions by numerically evaluating the kinetic scattering integral as well as the non-perturbative many-body coupling function. 
As a side result we obtain a possible finite-size interpretation of wave-turbulent scaling recently measured by Navon et al. 
\end{abstract}

\pacs{%
03.65.Db 	
03.75.Kk, 	
05.70.Jk, 	
47.27.E-, 	
47.27.T- 	
}

\maketitle

\section{Introduction}
\label{sec:Intro}
A general characterisation of the relaxation dynamics of quantum many-body systems quenched far out of equilibrium remains a largely open problem.  
In particular, it is interesting to ask to what extent analogues of the universal descriptions arising from the equilibrium theory of critical fluctuations \cite{Goldenfeld1992a,Cardy1996a} exist for nonequilibrium systems.  

The standard classification scheme of dynamical critical phenomena applies to the linear response of classical systems driven, in a stochastic way, out of equilibrium \cite{Hohenberg1977a,Janssen1979a}, as well as to non-linear critical relaxation \cite{Racz1975a.PLA.53.433,Fisher1976a.PhysRevB.13.5039,Bausch1976a,Bausch1979a}.
Building on the theory of boundary critical phenomena \cite{Diehl1986a}, relaxation after a quench to an equilibrium critical point has been studied \cite{Janssen1989a,Janssen1992a} as well as the phenomenon of ageing \cite{Calabrese2002a.PhysRevE.65.066120,Calabrese2005a.JPA38.05.R133,Gambassi2006a.JPAConfSer.40.2006.13}.
Quenches deeper into the ordered phase induce phase-ordering kinetics and coarsening  \cite{Bray1991a.PhysRevLett.67.2670, 
Bray1994a.AdvPhys.43.357,Bray2000PhRvL..84.1503B}.
Closely related dynamical scaling phenomena which we are particularly interested in here, are  \mbox{(wave-)}turbulence \cite{Frisch2004a,Zakharov1992a,Nazarenko2011a}, as well as superfluid or quantum turbulence \cite{Tsubota2008a, Vinen2006a}.
The study of universal phenomena far from equilibrium has recently intensified, considering different types of quantum quenches
\cite{Braun2014a.arXiv1403.7199B,Lamacraft2007.PhysRevLett.98.160404,Rossini2009a.PhysRevLett.102.127204,DallaTorre2013.PhysRevLett.110.090404,Gambassi2011a.EPL95.6,Sciolla2013a.PhysRevB.88.201110,Smacchia2015a.PhysRevB.91.205136,Smacchia2015a.PhysRevB.91.205136,Maraga2015a.PhysRevE.92.042151,Maraga2016b.PhysRevB.94.245122,Chiocchetta2015a.PhysRevB.91.220302,Chiocchetta2016a.PhysRevB.94.134311,Chiocchetta:2016waa.PhysRevB.94.174301,Chiocchetta2016b.161202419C.PhysRevLett.118.135701,Marino2016a.PhysRevLett.116.070407,Marino2016PhRvB..94h5150M,Damle1996a.PhysRevA.54.5037,Mukerjee2007a.PhysRevB.76.104519,Williamson2016a.PhysRevLett.116.025301,Hofmann2014PhRvL.113i5702H,Williamson2016a.PhysRevA.94.023608,Bourges2016a.arXiv161108922B.PhysRevA.95.023616}, many of them in the context of quenches in ultracold Bose gases.

If an initially equilibrated system is quenched in a way that it eventually reequilibrates closer to or on the other side of a symmetry-breaking phase transition one would expect the universal characteristics of the (quantum) critical point to have an influence also on the non-equilibrium dynamics of that system.
However, seen from a more general perspective, the time evolution of the quenched system can approach a non-thermal fixed point, i.e., show universal behaviour away from equilibrium, in general independent of the equilibrium fixed point(s).
Such fixed points have been discussed and experimentally observed without \cite{Berges:2008wm,Berges:2008sr,Scheppach:2009wu,Berges:2010ez,Orioli:2015dxa,Berges:2015kfa,Prufer:2018hto} and with \cite{Nowak:2010tm,Nowak:2011sk,Schole:2012kt,Karl2017b.NJP19.093014,Erne:2018gmz} reference to ordering patterns and kinetics, and topological defects, paving the way to a unifying description of universal dynamics.
This concept builds on a scaling analysis of non-perturbative dynamic equations for field correlation functions in the spirit of a renormalization-group approach to far-from-equilibrium dynamics \cite{Eyink1994a,Gurarie1995a,Gasenzer:2008zz,Berges:2008sr,Berges:2012ty,Mathey2014a.PhysRevA.92.023635,Marino2016a.PhysRevLett.116.070407,Marino2016PhRvB..94h5150M,Gasenzer:2010rq}.
Near a non-thermal fixed point, correlation functions show a time evolution which takes the form of a rescaling in space and time  \cite{Berges:2014bba,Orioli:2015dxa}.
In consequence, the relaxation is critically slowed, i.e., correlations evolve as a power law rather than exponentially in time.

\begin{figure*}[t]
\centering
\includegraphics[width=0.36\textwidth]{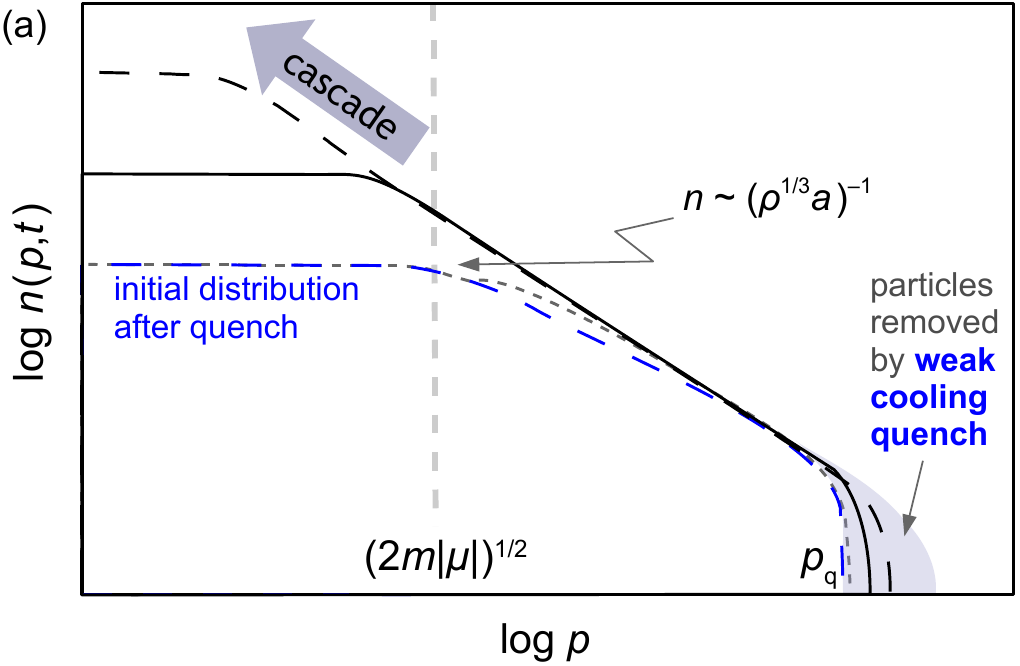}
\hspace*{0.005\textwidth}
\includegraphics[width=0.36\textwidth]{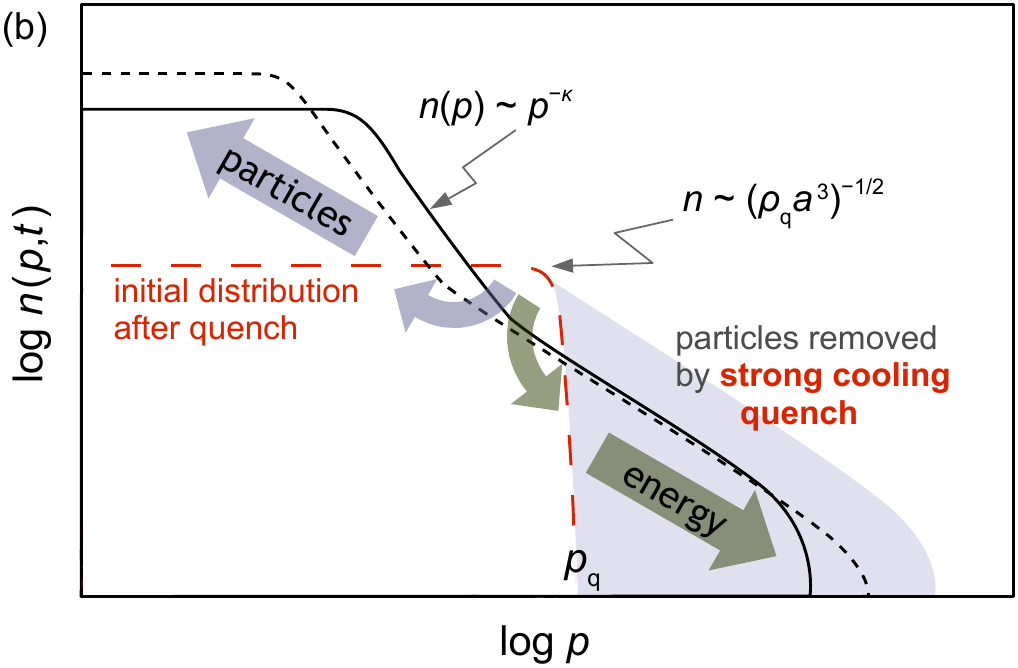}
\hspace*{0.005\textwidth}
\includegraphics[width=0.2438\textwidth]{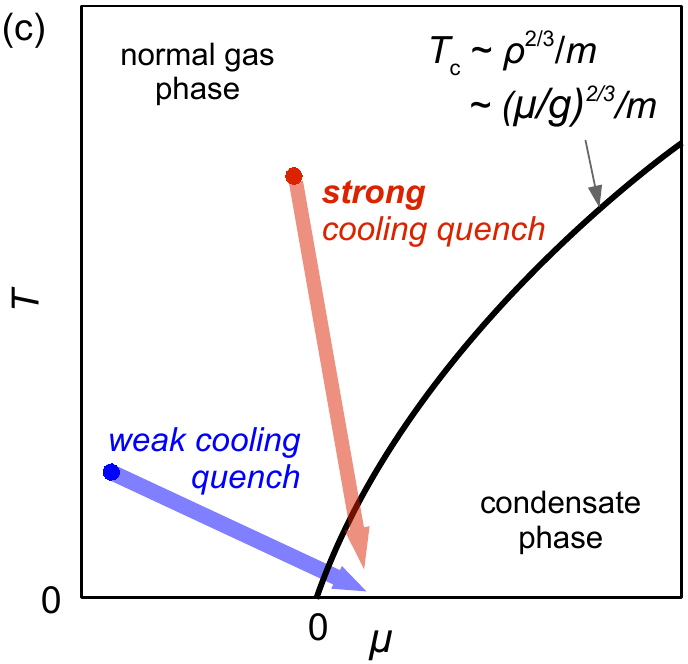}
\caption{Universal dynamics of a dilute Bose gas, as induced by cooling quenches of different strength.
(a) \emph{Weak cooling quench}. Sketch of the time evolution of the single-particle radial number distribution $n(p,t)$ as a function of momentum $p$ at three different times $t$ (grey short-dashed, solid and long-dashed lines). 
Starting from the initial distribution $n(p,t_{0})$ (blue dashed line) after a weak quench, removing part of the thermal tail (grey shaded area), an inverse wave-turbulent cascade develops, transporting particles towards lower momenta. See text for more details.
(b) \emph{Strong cooling quench}.
The initial distribution $n(p,t_{0})$ (red dashed line) is created by a strong cooling quench which removes the thermal tail (grey shaded area).
In the aftermath, a strong bi-directional redistribution of particles in momentum space (arrows) occurs.
This eventually builds up a quasicondensate in the infrared while refilling the thermal tail at large momenta. 
In (a) and (b), the particle transport towards zero as well as large momenta is characterized by self-similar scaling evolution in space and time, $n(p,t)=t^{\alpha}f(t^{\beta}p)$, with characteristic scaling exponents $\alpha$, $\beta$, in general different for the two directions. 
The infrared transport moves particles to low-$p$ modes (grey arrow) while their energy is deposited by the scattering of much fewer particles to higher momenta (green arrow), conserving total energy and particle number.
Note the double-logarithmic scale.
(c) Schematic sketch of the cooling quenches in the equilibrium phase diagram of the dilute Bose gas, leading to the same final $\mu=g\rho$, for illustrating the initial and long-time states of the system, before the quench and after re-equilibration, respectively, cf.~\App{CoolingQuenches}.
We emphasize that the dynamics ensuing the quenches can not be described \emph{within} this equilibrium plane.
}
\label{fig:NTFP}
\end{figure*}
Here we outline a kinetic theory of non-thermal fixed points, for the example of a dilute Bose gas, partially reviewing results obtained earlier \cite{Berges:2008wm,Berges:2008sr,Scheppach:2009wu,Berges:2010ez,Orioli:2015dxa,Berges:2015kfa},  thereby extending, complementing, generalizing and straightening them out.
We discuss the distinction between the approach to a non-thermal fixed point and a wave-turbulent evolution on the basis of conservation laws, using ideas presented in \cite{Svistunov1991a}, and we derive the scaling properties for these different scenarios. 
We find simple analytical forms for the universal many-body scattering-matrix elements entering the kinetic equations at the non-thermal fixed point. 
To confirm our predictions we perform numerical evaluations.
As a side result we obtain a possible finite-size interpretation of wave-turbulent momentum scaling recently measured  in a turbulent Bose gas \cite{Navon2016a.Nature.539.72}. 

We have written the paper in a way which allows the reader to consider only that part which is of interest to her or him. 
The remainder of this introductory section gives an overview of the entire theory, the key questions and situations considered and the key results provided by the theory.
Readers interested in the details of the derivations are referred to the following sections and the appendices, an overview of which is given at the end of this introductory section.

\subsection{Typical scenario of how a non-thermal fixed point can be approached in a Bose gas}
Bose condensation dynamics after a cooling quench represents a typical scenario in which the approach of a non-thermal fixed point can be observed.
In a Bose gas cooled from above the critical temperature, condensate formation is generally understood  to proceed through different stages \cite{Svistunov1991a,Kagan1992a,Kagan1994a,Kagan1995a,Semikoz1995a.PhysRevLett.74.3093,Semikoz1997a,Berloff2002a,Kozik2004a.PhysRevLett.92.035301,Kozik2005a.PhysRevLett.94.025301,Kozik2005a.PhysRevB.72.172505,Kozik2009a}. 

If the condensation dynamics occurs sufficiently far out of equilibrium the system can thereby approach a non-thermal fixed point and  show universal scaling in space and time \cite{Nowak:2012gd,Berges:2012us,Orioli:2015dxa,Davis:2016hwt}.
When this is the case, the evolution of the occupation number $n_{a}(\mathbf{p},t)=\langle\Phi_{a}^{\dag}(\mathbf{p},t)\Phi_{a}(\mathbf{p},t)\rangle$, at a non-thermal fixed point, corresponds to a rescaling in time and space like 
\begin{equation}
n_{a}(\mathbf{p},t) = (t/t_\mathrm{ref})^{\alpha}f_{\mathrm{S},a}([t/t_\mathrm{ref}]^{\beta}\mathbf{p})\,.
 \label{eq:intro:NTFPscaling}
\end{equation}
Here, we consider the more general case of an $N$-component Bose field $\Phi_{a}(\mathbf{p},t)$, $a=1,\dots,N$, and
$f_{\mathrm{S},a}(\mathbf{p})=n_{a}(\mathbf{p},t_\mathrm{ref})$ is a  universal scaling function which depends on a single $d$-dimensional momentum only and, within a scaling regime, assumes a form exhibiting power-law momentum dependence,
\begin{equation}
f_{\mathrm{S},a}(\mathbf{p})=(|\mathbf{p}|/p_{\lambda})^{-\kappa}\,,\quad p_{\Lambda}\lesssim|\mathbf{p}|\lesssim p_{\lambda}\,,
 \label{eq:intro:statNTFPscaling}
\end{equation}
with universal exponent $\kappa$.
In \eq{intro:NTFPscaling}, $t_\mathrm{ref}$ is some reference time within the scaling regime. 
The scaling exponents $\alpha$ and $\beta$ characterise the universality class of the fixed point \cite{Orioli:2015dxa}.
They are typically related by a conservation law such as the total particle density $\int\mathrm{d}\mathbf{p}\,n_{a}(\mathbf{p},t)=$ const., which implies $\alpha=\beta\,d$. 
Transport to lower $|\mathbf{p}|$ occurs when $\beta>0$ while, for $\beta<0$, the distribution shifts to larger momenta.
Generically, the scaling regime is reached in a certain scaling limit, such as for asymptotic time, $t\to\infty$, and infinite volume, $|\mathbf{p}|\to0$.
In the scaling limit, all correlation functions of the system are expected to exhibit scaling.

Cooling, i.e., removal of hot particles induces transport of most of the remaining particles towards lower energies.
In an isolated system where particle number and energy are conserved this is possible because the collisions between particles induce their energy to be transported to higher wave numbers $\mathbf{p}$.
To understand this, note that the energy of a particle scales, generally, as the momentum to the power of a positive number, $\omega(\mathbf{p})\sim p^{z}$, with $p=|\mathbf{p}|$, with dynamical exponent $z>0$. 
Therefore, the energy $\omega(\mathbf{p})\,n(\mathbf{p})$ of the system is always concentrated at a higher wave number than the particle number $n(\mathbf{p})$. 
Hence, while many particles assume lower energies, a few particles are scattered to modes where the cooling quench has removed particles from the distribution, i.e., decreased the occupation number of momentum modes. 
As a result, the dynamics following a cooling quench generically induces a bidirectional transport, moving particles towards the infrared (IR) and energy to the ultraviolet (UV). 
This is illustrated in \Fig{NTFP}(a,b) and has been discussed for various examples in Refs.~\cite{Svistunov1991a,Kagan1992a,Kagan1994a,Gurarie1995a,Nowak:2012gd,Berges:2012us,Orioli:2015dxa}.

As long as the coherences between particles in distinct eigenstates of the appropriate single-particle Hamiltonian are negligible the transport is well described by a quantum Boltzmann kinetic equation for the occupation numbers of these single-particle modes.  
This equation describes elastic inter-particle collisions which can induce particle transport towards lower energies if energy and momentum are exchanged in the collision. 
As a result, the occupation numbers of low-energy modes increase, leading to the possibility of condensate formation.  
Once the occupation numbers are sufficiently large, phase correlations between the modes develop. 
The system is then more easily described in terms of a classical field, even if it still exhibits large phase fluctuations or topological structures and undergoes turbulent dynamics.

These phenomena can occur sequentially and depend on the specifications of the system, including its dimensionality, density, and strength of interactions.  
During this stage, the system is sometimes called a nonequilibrium {quasicondensate}, in analogy to the phase-fluctuating equilibrium states of low-dimensional Bose systems~\cite{Popov1972a,Popov1983a}.  
The relaxation of the quasicondensate establishes phase coherence across the sample, leading eventually to a Bose-Einstein condensate.

\subsection{Cooling quenches}
\label{sec:intro:CoolingQuenches}
To set the stage for the presentation of the kinetic theory of non-thermal fixed points and to provide concrete examples of how to induce the corresponding far-from-equilibrium universal scaling dynamics, we discuss a few important aspects of condensate formation following a cooling quench.
We distinguish, in particular \emph{weak} and \emph{strong quenches} for obtaining different types of scaling behavior, namely wave turbulence or self-similar evolution at a non-thermal fixed point.
A more detailed discussion of the distinction between these types of quenches is given in \App{CoolingQuenches}.

Throughout this work, we focus on initially homogeneous, non-degenerate Bose gases in a closed volume. 
In a cooling quench one typically removes particles with momentum $p$ whose energy $\omega(p)$ is above a certain energy scale, $\omega(p)\gtrsim \omega_{\mathrm{q}}\equiv\omega(p_\mathrm{q})$.
This generically leads to a non-equilibrated particle distribution $n(p)$ over the momenta $p=|\vec p|$ as indicated by the colored long-dashed lines in Fig.~\ref{fig:NTFP}(a, b).

Transport and ordering stages in condensate formation after a quench were analysed in a series of papers by Svi\-stunov, Kagan, and Shlyapnikov \cite{Svistunov1991a,Kagan1992a,Kagan1994a,Kagan1995a}, by Semikoz and Tkachev \cite{Semikoz1995a.PhysRevLett.74.3093,Semikoz1997a}, and by Berloff, Kozik and Svistunov \cite{Berloff2002a,Kozik2004a.PhysRevLett.92.035301,Kozik2005a.PhysRevLett.94.025301,Kozik2005a.PhysRevB.72.172505,Kozik2009a} in the context of weak wave turbulence as well as self-similar transport.
In the re-equilibration process following the quench, energy and number conservation imply a bi-directional redistribution of particles. 
This transport process is sketched in \Fig{NTFP}(a,b) where the particle distribution $n(p,t)$ is depicted, at different times, on a double-logarithmic scale.
The initial distribution after the quench is indicated by the blue (red) long-dashed line, while the post-quench evolution is indicated by three (two) curves (grey short-dashed, solid, long-dashed).

If the state resulting from the quench is sufficiently far away from equilibrium, the ensuing time evolution can exhibit universality.
This implies that the distribution can take a simple power-law form as indicated in \Fig{NTFP}(a,b), and that the evolution becomes a self-similar rescaling in time and momentum as in \Eq{intro:NTFPscaling}, within a certain region of momenta.

A central aspect of the phenomena discussed in this work is that the character of the evolution depends very much on the `strength' of the cooling quench which determines how far the system can get out of equilibrium.
A \emph{weak} cooling quench is depicted in \Fig{NTFP}(a) and generically leads to the build-up of a weak wave-turbulent cascade.
In contrast, the approach to a non-thermal fixed point rather requires a \emph{strong} cooling quench, leading to the dynamics shown in~\Fig{NTFP}(b).

\subsubsection{Weak cooling quench}
\label{sec:intro:WeakCoolingQuench}
Consider a three-dimensional thermal dilute Bose gas 
just above the condensation temperature $T_\mathrm{c}$, see the phase-diagram sketch in \Fig{NTFP}(c).
Removing, from such a system, a few of the high-energy particles, the subsequent particle transport in momentum space towards lower energies is described by the perturbative quantum Boltzmann equation in the classical-wave limit (wave-Boltzmann equation) as long as mode occupation numbers are much larger than 1 and their wave length is shorter than the coherence length set by the chemical potential $\mu$~\cite{Svistunov1991a,Kagan1992a,Kagan1994a}.
At lower energies, however, phase correlations between momentum modes become significant, and a (non-perturbative) description beyond the wave-Boltzmann equation is needed \cite{Svistunov1991a,Kagan1992a,Kagan1994a,Kozik2004a.PhysRevLett.92.035301,Kozik2005a.PhysRevLett.94.025301,Kozik2005a.PhysRevB.72.172505,Kozik2009a}.  

For higher energies, where the wave-Boltzmann approach is still viable, Svistunov discussed different transport scenarios based on weak wave turbulence \cite{Zakharov1992a,Nazarenko2011a}.
Taking into account that the scattering matrix elements in the perturbative wave-Boltzmann equation for such a dilute gas are independent of the mode energies, he concluded that the initial kinetic transport stage of the condensation process in momentum space evolves as a weakly non-local particle wave towards lower momenta.  
Within this cascade, particles are transported locally, from momentum shell to momentum shell, from the scale $\omega(p_\mathrm{q})$ of the energy concentration in the initial state (see \Fig{NTFP}(a)) to the low-energy regime $\omega\lesssim\mu$ where coherence formation sets in and the description in terms of the perturbative wave-Boltzmann equation ceases to be valid.
The flux-wave and weak-wave-turbulence stage of condensate formation following a weak cooling quench was later confirmed numerically by Semikoz and Tkachev \cite{Semikoz1995a.PhysRevLett.74.3093,Semikoz1997a}.

\subsubsection{Strong cooling quench}
\label{sec:intro:StrongCoolingQuench}
Distinctly different universal dynamical processes are possible in a dilute Bose gas which is excited by a \emph{strong}  cooling quench at the initial time $t_{0}$.
An extreme version of such a quench would be to first tune adiabatically to a chemical potential $0<-\mu\ll k_\mathrm{B}T_\mathrm{c}$ and then remove all particles with energies higher than $\omega(p_\mathrm{q})\sim|\mu|$ (shaded area in \Fig{NTFP}(b)).

Such an extreme initial condition, in experiment, can alternatively be prepared by means of an instability, building up strong overoccupation such that the majority of particles and energy is around the momentum scale set by $\mu$ and interaction energy dominates over kinetic energy (see \App{CoolingQuenches}).
This can be achieved by starting with a system with unstable internal modes \cite{Stamper-Kurn2013a.RevModPhys.85.1191,Berges:2008wm,Karl:2013mn,Karl:2013kua,Zache:2017dnz,Prufer:2018hto}, or with an inhomogeneous condensate \cite{Nowak:2010tm,Nowak:2011sk,Schole:2012kt,Nowak:2012gd,Karl2017b.NJP19.093014,Deng:2018xsk,Navon2016a.Nature.539.72}, and can be technically advantageous.

In \Fig{NTFP}(c) we have sketched both, a weak and a strong cooling quench which are leading to the same final condensate density.
Note that, for the strong quench, one needs to prepare the pre-quench state with a chemical potential already much closer to the critical value.
To obtain the same post-quench density and thus post-quench chemical potential $\mu$ as after the weak quench, the pre-quench distribution needs a higher temperature and density.

The post-quench state induces a transport toward the IR which essentially starts at the coherence-length scale set by $\mu$ and immediately requires a description beyond the perturbative wave-Boltzmann equation.
The evolution during this period evolves universally in the sense that it becomes largely independent of the precise initial conditions set by the quench as well as of the precise values of the parameters of the theory, indicating the approach to a \emph{non-thermal fixed point} of the time evolution \cite{Orioli:2015dxa}.
At this fixed point the distribution ideally shows a universal form, cf.~\Eq{intro:NTFPscaling}, \Fig{NTFP}(b), as well as \App{CQuenchesUnivDyn}.

The details of the evolution from the post-quench distribution to the stage of universal scaling \cite{Schmied:2018upn.PhysRevLett.122.170404} are beyond the scope of the present work.
We will mostly assume strong cooling quenches as leading to the evolution near a non-thermal fixed point.
In contrast to this, weak quenches are implicitly assumed when discussing wave turbulent cascades and comparing them to the self-similar rescaling which characterises the universal dynamics near non-thermal fixed points.

\subsection{Kinetic theory of non-thermal fixed points: Key results}
We briefly summarize the key results obtained in this work.
We begin with a systematic discussion of global conservation laws and show, along the lines of \cite{Svistunov1991a}, that the exponent of $n(p)\sim p^{-\kappa}$ of the universal scaling function (\Eq{intro:statNTFPscaling} and \Fig{NTFP}b) determines which type of universal scaling evolution is possible:
Depending  on the spatial dimension $d$ and the dynamical exponent $z$, for $\kappa<d$ particle number and energy are concentrated both in the IR, for $\kappa>d+z$ both are concentrated in the UV, while in between these limits, particles are in the IR while energy is in the UV.
As a result, only the latter case allows for a bidirectional transport of particle and energy concentrations as described above.
In contrast, if both, particles and energy are at the same end of the spectrum, global conservation laws strongly constrain the  dynamics such that only single-side wave-front evolutions leading to the buildup of wave-turbulent cascades are possible, cf.~\cite{Svistunov1991a}.  

We focus on fixed-point solutions, obeying $n(p,t)=t^{\,\alpha}f(t^{\,\beta}p)$ as well as spatial scaling $n(p)\sim p^{-\kappa}$ within a regime of momenta $p$.
We anti\-cipate that the fixed point, strictly speaking, is reached in the scaling limit, i.e., at asymptotic times.
However, at any finite time, scaling according to the fixed-point solution can be realised already to a very good approximation.
To obtain these solutions, one performs a scaling analysis of wave-Boltzmann-type kinetic equations using different (non-)perturbative approximations \cite{Berges:2010ez,Orioli:2015dxa}.

While the bidirectional transport at a non-thermal fixed point is a general phenomenon, the specific situation we focus on here is that of an $N$-component Bose gas with U$(N)$ symmetric interactions.
One can analytically describe this by means of a kinetic equation derived within a Schwinger-Keldysh functional path integral approach \cite{Berges:2010ez,Orioli:2015dxa}.

Our analytic computation of the non-perturbative many-body $T$-matrix entering the kinetic equation forms  the main basis of the present paper.
This $T$-matrix involves a coupling function $g_\mathrm{eff}(p)$ obtained within a next-to-leading-order large-$N$ approximation of the dynamic equations for Greens functions \cite{Scheppach:2009wu}.
This builds on and goes beyond, e.g., Refs. \cite{Berges:2008wm,Berges:2008sr,Scheppach:2009wu,Berges:2010ez,Orioli:2015dxa,Berges:2015kfa}, see also closely related work in Refs.~\cite{Karl2017b.NJP19.093014,Deng:2018xsk,Berges:2015ixa,Schachner:2016frd,Berges:2017ldx,Walz:2017ffj}.
In physical terms, the non-perturbative $T$-matrix reflects the fact that in the scaling regime, at energies below $\mu$, the dynamics is dominated by phase excitations, which in the large-$N$ limit, are mainly the relative phases between the fields \cite{Mikheev2018a.arXiv180710228M}. 

We analytically and numerically evaluate the many-body coupling.
For $p\gtrsim p_\Xi$, with healing-length scale $p_{\Xi}=(2mg\rho)^{1/2}$ defined in terms of mass $m$, bare coupling $g$ and density $\rho$, the coupling reduces to $g$, implying standard perturbative wave-Boltzmann transport and, typically, weak wave turbulence.
In contrast, it describes collective scattering for momenta $p\lesssim p_\Xi$.
In this region, the coupling function is strongly modified due to the high IR occupation numbers.
Remarkably, we find that it assumes a universal form, scaling as $g_\mathrm{eff}(p)\sim p^{2}$, independent of the microscopic interaction strength and largely independent of the precise form of the low-energy momentum distribution $n(p)$.
We emphasise that, although this coupling function leads to a momentum- and frequency-dependent scattering $T$-matrix exhibiting scaling in space and time, the kinetic equation it enters still assumes the usual wave-Boltzmann form describing $2$-to-$2$-scattering.

Assuming a scaling solution of the kinetic equation, power counting allows one to determine the relevant scaling exponents.
In the case of weak-wave-turbulent evolution ($p>p_\Xi$) we recover the scaling exponents $\alpha$, $\beta$, and $\kappa$ obtained by Svistunov.
In the IR regime ($p<p_\Xi$), however, we find a different set of exponents, among which the ones obtained in Refs.~\cite{Orioli:2015dxa,Walz:2017ffj} are recovered.
In our analysis of global conservation laws leading to the distinction between self-similar evolution and wave turbulence we combine the general arguments of Ref.~\cite{Svistunov1991a} with the non-perturbative description of non-thermal fixed points as laid out in Refs.~\cite{Berges:2008wm,Berges:2008sr,Scheppach:2009wu,Berges:2010ez,Orioli:2015dxa,Berges:2015kfa}.
We recapture and use these arguments to obtain a systematic picture of the differences and common aspects of the various kinds of universal dynamics discussed in the literature: non-thermal fixed points and wave-turbulent cascades, critically accelerated wave fronts and critically slowed scaling, perturbative and non-perturbative processes.

At a non-thermal fixed point describing self-similar transport of particles to the IR in a non-perturbative regime (Fig.~\fig{NTFP}b), we find, in $d$ spatial dimensions and for a range of dynamical exponents $z$, the temporal scaling exponents  $\alpha=\beta d$, $\beta=1/z$.
They are fixed by $z$ and the global conservation of particles.
As was shown in Ref.~\cite{Mikheev2018a.arXiv180710228M}, this solution, obtained in the large-$N$ limit, applies in particular to the case of free relative-phase excitations between the $N$ components of the gas with $\omega(p)=p^{2}/2m$, and thus $z=2$.
The spatial exponent then results as $\kappa=d+(3z-4)/2+\eta)=d+1+\eta$ which allows bi-directional self-similar transport according the above criterion.
This result includes an anomalous dimension $\eta$ which needs to be determined by solving the equation for the spectral function.
Note that the scaling solutions derived are not expected to capture any effects caused by non-linear excitations such as vortices, solitons, or domain walls \cite{Nowak:2010tm,Nowak:2011sk,Gasenzer:2011by,Schole:2012kt,Nowak:2012gd,Schmidt:2012kw,Gasenzer:2013era,Karl:2013mn,Karl:2013kua,Moore:2015adu,Karl2017b.NJP19.093014,Deng:2018xsk}. 

The rescaling of $n(p,t)$ is critically slowed,  $p_{\Lambda}\sim t^{-\beta}$ as $t\to\infty$ and, since $\beta>0$, implies transport of particles towards the IR.
This is in contrast to the weak-wave-turbulent flux of particles, which occurs in the perturbative wave-Boltzmann regime. There, the exponent $\beta=1/(z-8/3)=-3/2$ is negative for $z=2$, and $\kappa=d-2/3$ is smaller than $d$ such that energy \emph{and} particles are concentrated at the UV end of the distribution.
 Hence, the transport towards the IR can only happen in the form of an accelerating wave front, weakly violating the locality of the transport.
This corresponds to rescaling in momentum as $p_{\Lambda}(t)\sim (t_{*}-t)^{-\beta}$, see~Ref.~\cite{Svistunov1991a}.

The main part of our paper, which provides the details of how the above summarized results are obtained, is organized as follows.
In \Sect{2PIEA} we introduce the general formalism of scaling and scaling forms for the occupancy spectrum and discuss the important consequences arising from global conservation laws.
In \Sect{KineticDescrUniDyn} we introduce the non-perturbative kinetic equation.
This encompasses the central analytic result of a universal coupling function in the collective-scattering regime.
On this basis we derive, in \Sect{ScalingAnalysisKinEqs}, the scaling exponents for self-similar dynamics, which we cross-check, in \Sect{SummaryNumerical}, on the basis of numerical evaluations of the perturbative and non-perturbative scattering integrals.
We draw our conclusions in \Sect{Conclusions}.
An extended Appendix provides further technical details.

\section{Universal dynamics in a Bose gas after a quench}
\label{sec:2PIEA}
%

%
\subsection{Model and observables}
\label{sec:ModelObservables}
In this paper we focus on the universal dynamics of a dilute, i.e., weakly interacting homogeneous Bose gas in $d=3$ spatial dimensions. 
In this subsection, we briefly summarize its field theoretical description and a few basic properties needed in the following.
The model is defined by the Gross-Pitaevskii Hamiltonian for $N$ boson fields of mass $m$, which is symmetric under U($N$) rotations of the field vector $(\Phi_a)$, $a=1,\dots,N$,
\begin{equation}
 H =   \int  \mathrm{d}^dx \, \left[ -\Phi_{a}^\dag\frac{\nabla^2}{2m}\Phi_{a}  
     +  \frac{g}{2} \, \Phi_{a}^\dag\Phi_{b}^\dag \Phi_{b} \Phi_{a} \right] \,.
\label{eq:GPHamiltonian}
\end{equation}
The time- and space-dependent fields $\Phi_{a} \equiv \Phi_{a}(\mathbf{x},t)$ satisfy Bose equal-time commutation relations, $[\Phi_{a}(\vec x,t),\Phi_{b}^{\dagger}(\vec x',t)]=\delta_{ab}\delta(\vec x-\vec x')$, $[\Phi_{a}(\vec x,t),\Phi_{b}(\vec x',t)]=0$
(We use units where $\hbar=k_{B}=1$), and summations over double indices are implied.
A single real coupling $g=4\pi a/m$ quantifies the contact interactions, defined in terms of the $s$-wave scattering length $a$. 
For simplicity of notation, we will suppress the field indices in the following.
The single-particle momentum distribution 
\begin{equation}
  \label{eq:nSP}
  n(\mathbf{p},t)=\langle\Phi^{\dagger}(\mathbf{p},t)\Phi(\mathbf{p},t)\rangle
  \end{equation}
counts the directly measurable number of particles with momentum $\mathbf{p}$.
In the absence of a condensate and for sufficiently weak interactions and occupation numbers $n$, the single-particle excitations described by $\Phi$ are eigenmodes of $H$ with free particle dispersion 
\begin{equation}
  \label{eq:freedispersion}
  \varepsilon_{\mathbf{p}}=\mathbf{p}^{2}/(2m)\,.
  \end{equation}
For comparisons with cases of a different momentum scaling $\sim\mathbf{p}^{z}$ of the dispersion, we will also consider a dilute Bose gas with a macroscopic condensate fraction
\footnote{The U($N$) symmetric model, at low energies and a total density set by a chemical potential, in fact has $N-1$ Goldstone modes with free dispersion \eq{freedispersion} and one density mode, characterized by strongly suppressed density fluctuations and enhanced phase fluctuations, with linear, Bogoliubov-sound dispersion \eq{BogDispersion}, cf.~Ref.~\cite{Mikheev2018a.arXiv180710228M}.}.
The condensate density, $\rho_{0}=|\phi_{0}|^{2}$, is given in terms of the spatially constant field expectation value $\phi_{0}\equiv\langle\Phi(x)\rangle$ which, after a suitable shift of the energy zero, is also independent of time.
(We use 4-vector notation, $x=(x_{0},\vec x)$).
In this case, the Hamiltonian can be expanded to second order in the fluctuation fields $\tilde\Phi(x)$ around the condensate, $\Phi(x)=\phi_{0}+\tilde\Phi(x)$, where, by definition, $\langle\tilde\Phi(x)\rangle\equiv0$. 
A Bogoliubov canonical transformation to bosonic quasiparticle operators $\Phi_{Q}$, defined in momentum space by $\tilde\Phi(\vec p,t)=u_{\vec p}\Phi_{Q}(\vec p,t)-v_{\vec p}\Phi^{\dagger}_{Q}(-\vec p,t)$, with $u_{\vec p}^{2}-v_{\vec p}^{2}=1$, diagonalises the resulting quadratic Hamiltonian,
\begin{equation}
 H =  \sum_{\vec p}  \omega_{\vec p}\left(\Phi^{\dagger}_{Q}(\vec p,t)\Phi_{Q}(\vec p,t)+1/2\right) \,.
\label{eq:BogHamiltonian}
\end{equation}
The Bogoliubov dispersion and mode functions read
\begin{align}
 \label{eq:BogDispersion}
 \omega_{\vec p} = \big[& \varepsilon_{\vec p}\big(\varepsilon_{\vec p}+2 g\rho_{0}\big) \big]^{1/2}\,, \\
 \label{eq:BogModeFunctions}
u_{\vec p} = \left( \frac{\varepsilon_{\vec p}+g\rho_{0} + \omega_{\vec p}}{2\omega_{\vec p}} \right)^{1/2}&\,,\quad
v_{\vec p} = \left( \frac{\varepsilon_{\vec p}+g\rho_{0} - \omega_{\vec p}}{2\omega_{\vec p}} \right)^{1/2}\,.
\end{align}
For momenta much larger than the healing-length momentum scale, $|\mathbf{p}|\gg p_{\xi}$,
\begin{align}
 \label{eq:healingmomentum}
  p_{\xi}&=\sqrt{2mg\rho_{0}}=\sqrt{8\pi a\rho_{0}}\,, 
\end{align}
the Bogoliubov dispersion resembles that of the free fundamental bosons, $\omega_{\vec p}\simeq \varepsilon_{\vec p}+g\rho_{0}$, and thus $u_{\vec p}\simeq 1$, $v_{\vec p}\simeq0$.
In the opposite limit, $|\mathbf{p}|\ll p_{\xi}$, 
the quasiparticles are sound waves,
\begin{align}
 \label{eq:sounddispersion}
 \omega_{\vec p} &\simeq c_\mathrm{s}|\vec p|\,, \\
 \label{eq:SoundModeFunctions}
u_{\vec p}^{2} &\simeq v_{\vec p}^{2} \simeq g\rho_{0}/(2\omega_{\vec p})\simeq mc_\mathrm{s}/(2|\vec p|)\,,
\end{align}
with speed of sound $c_\mathrm{s}=\sqrt{g\rho_{0}/m}=p_{\xi}/(\sqrt{2}m)$.
The occupation number of sound-wave field modes with wave-number $\mathbf{p}$ is measured by
\begin{equation}
  \label{eq:nQP}
  n_{Q}(\mathbf{p},t)=\langle\Phi_{Q}^{\dagger}(\mathbf{p},t)\Phi_{Q}(\mathbf{p},t)\rangle\,.
  \end{equation}
According to the Bogoliubov transformation, particle and quasiparticle mode occupation numbers are related by   
\begin{align}
 \label{eq:nvsnQ}
 n(\vec p,t) 
 &= (u_{p}^{2}+v_{p}^{2})\, n_{Q}(\vec p,t) + v_{p}^{2} \,.
\end{align}

In thermal equilibrium, the particle and quasiparticle distributions are given by grand-canonical and canonical Bose-Einstein distributions, respectively.
In general, $n_\mathrm{BE}(\mathbf{p})=\{\exp[(\omega(\mathbf{p})-\mu)/T]-1\}^{-1}$ for excitations with dispersion $\omega_{\mathbf{p}}$ is set by the temperature $T$ and the chemical potential $\mu$. 
We point out that, in the sound-wave limit, $0<|\mathbf{p}|\ll p_{\xi}$, and for large quasiparticle occupations, $n_{Q}(\vec p,t)\gg1$,  relation \eq{nvsnQ} together with \Eq{SoundModeFunctions} means that 
\begin{align}
 \label{eq:nvsnQscaling}
 n(\vec p,t)\simeq n_{Q}(\vec p,t)\,g\rho_{0}/\omega_{\vec p} \,\qquad(0<|\mathbf{p}|\ll p_{\xi}).
\end{align}
Hence, in the Rayleigh-Jeans regime of the equilibrium Bose-Einstein distribution, $-\mu\ll\omega(\vec p)\ll T$, where the occupancies in an ideal gas are $n(\vec p,t)\sim T/\varepsilon_{\vec p}\sim T/p^{2}$ and those of Bogoliubov sound are $n_{Q}(\vec p,t)\simeq T/\omega_{\vec p}\sim T/p$, the extra factor $1/p$ from the mode functions, $u_{p}^{2}+v_{p}^{2}\sim1/p$, ensures the same power-law dependence on $p$ of the left-hand side of \Eq{nvsnQ} for free and interacting Bosons. 
The factor thus adjusts the quasiparticle number distribution to the modified density of states in the sound-wave limit.

\subsection{Momentum scaling and universal scaling functions}
\label{sec:UniversalStates}
%
\subsubsection{Momentum scaling}
\label{sec:Scaling}
We are predominantly interested in the question to what extent non-equilibrium states exist for which the momentum distributions \eq{nSP} and \eq{nQP} assume a universal form. 
Let us, for the first, disregard the time evolution.
Simple examples of universal momentum distributions are those which, at least in a limited regime of scales, show power-law scaling, 
\begin{equation}
n(s\mathbf{p}) = s^{-\zeta}n(\mathbf{p})\,.
 \label{eq:npPowerLaw}
\end{equation}
Here $s$ is a positive, real scaling factor, and $\zeta$ is a universal scaling exponent which we assume, in the following, to be a real number.
See \Tab{OverviewScalingExponents} for an index of all scaling exponents appearing in this work.

Also the dispersion relation can, at least in certain regions, satisfy scaling 
\begin{align}
  \omega(s{\vec p})=s^{z}\omega({\vec p})\,.
  \label{eq:dynscaling}
\end{align}
This is, e.g., everywhere the case for the free dispersion \eq{freedispersion}  and,  in the free-particle and sound-wave limits, for the Bogoliubov dispersion \eq{BogDispersion}.

In the following we will account for the scaling of $\omega(\vec p)$, as far as possible, by means of an arbitrary dynamical exponent $z$.
We anticipate in this way that self-energy corrections can lead to a modified scaling of the quasiparticle dispersion and that, in a treatment beyond kinetic scattering of free modes, a more general scaling between frequency and momentum is expected.
In \App{2PIKineticTheory} we summarize the general field-theoretic approach we use to analyze our model with respect to universal dynamics.
See, in particular, \App{SpectralFcts} for the definition of the scaling of the spectral properties for a general $z$.

Given a scaling \eq{npPowerLaw} of $n(\vec p)$, the quasiparticle distribution, in the scaling region, would satisfy
\begin{equation}
n_{Q}(s\mathbf{p}) = s^{-\kappa}n_{Q}(\mathbf{p})\,,
 \label{eq:nQpPowerLaw}
\end{equation}
with
\begin{equation}
 \kappa = \zeta + z-2\,.
 \label{eq:nQpPowerLawExponent}
\end{equation}
Here, the exponent $\kappa-\zeta=z-2$ governing the relative scaling of the two distributions accounts for the $z$-dependent density of states which is encoded in the Bogoliubov coefficients $u_{p}$ and $v_{p}$. 
Note that $z-2$ interpolates between the Bogoliubov quasiparticle case, $z=1$, for which the relative scaling is given by \Eq{nvsnQscaling} and free particles, $z=2$, for which $n_{Q}\equiv n$, cf.~also the discussion in \App{ScalingHypoth}.

\subsubsection{Bulk integrals}
\label{sec:Bulk integrals}
The momentum integral over the single-particle distribution $n(\vec p)$ yields the density of non-condensed atoms $\rho_\mathrm{nc}$ and thus the observable total particle density is given by
\begin{equation}
\rho_\mathrm{tot}=\rho_{0}+\rho_\mathrm{nc}=\rho_{0}+\int \frac{\mathrm d^{d}p}{(2\pi)^{d}}\, n(\vec p) \,.
 \label{eq:ParticleDensity}
\end{equation}
Hence, the integral must be finite, and if $n(\vec p)$ shows power-law scaling \eq{npPowerLaw} in a certain range of momenta $p=|\vec p|$, this range can not extend over all possible $p$ from $0$ to $\infty$.
This is because the radial, i.e., $p$-integral over $p^{d-1-\zeta}$, which includes a volume factor, has a power-law divergence either in the ultraviolet (UV), or in the infrared (IR), or is logarithmically divergent in both limits.

This means that in any physically meaningful situation, in the continuum and thermodynamic limits, the distribution $n(\vec p)$ must take the form of a more general scaling function which ensures convergence of the integral \eq{ParticleDensity}.
Alternatively, the finite size of a generic physical system and its definition on a discrete grid would provide IR and UV cutoffs, respectively.
We are, however, interested in universal dynamics which, within first approximation, is not affected by such boundary conditions. To this end, we demand the scaling region to be sufficiently far away from the boundaries of the system and will study the intrinsic conditions under which scaling dynamics can occur.

The integral over the occupancies of the quasiparticle eigenmodes of the Hamiltonian defines the density
\begin{equation}
\rho_{Q}=\int \frac{\mathrm d^{d}p}{(2\pi)^{d}}\, n_{Q}(\vec p) \,
 \label{eq:QParticleDensity}
\end{equation}
which is in general different from the particle density \eq{ParticleDensity}.
In situations where the interactions between quasiparticles are dominated by elastic $2$-to-$2$ scattering, their total number and thus the density $\rho_{Q}$ are conserved in time.
In this paper we will restrict ourselves to situations where quasiparticle number is conserved within the scaling region.

With \Eq{nvsnQ}, the particle density $\rho_\mathrm{nc}$ can be expressed in terms of the quasiparticle density $\rho_{Q}$.
In the sound-wave regime, the relation is a pure power law, cf.~\Eq{nvsnQscaling}.
Assuming contributions from outside the scaling region with a fixed $z$ to be negligible, the relation between the particle density and the quasiparticle spectrum hence is
\begin{equation}
\rho_\mathrm{nc}=\int \frac{\mathrm d^{d}p}{(2\pi)^{d}}\, B p^{z-2}n_{Q}(\vec p,t) \,,
 \label{eq:ParticleDensityfromnQ}
\end{equation}
with some constant $B$ which, for $z=1$, is $B=g\rho_{0}/c_{s}$, cf.~\Eq{nvsnQscaling}, while for $z=2$ one has $B=1$, and quasiparticles and particles are identical.

Besides the density of (quasi)particles, also the energy density, 
\begin{equation}
\varepsilon=\int \frac{\mathrm d^{d}p}{(2\pi)^{d}}\, \omega(\vec p) n_{Q}(\vec p) \,,
 \label{eq:EnergyDensity}
\end{equation}
is a physical observable and therefore must be finite.

\subsubsection{Scaling function}
\label{sec:ScalingFunction}
Where not otherwise stated, we assume the momentum distributions to be isotropic in the following, $n_{Q}(\vec p)\equiv n_{Q}(p)$.
Assume, for the first, that $n_{Q}(p)\sim p^{-\kappa}$ is a pure power law in the radial momentum direction, satisfying \Eq{nQpPowerLaw} for all momenta. 
Furthermore, presume a power-law form for $\omega(p)$, \Eq{dynscaling},  with $z\not=0$, such that the integrand in \Eq{EnergyDensity} is a pure power law.

The exponent $\kappa$ then determines whether the IR or the UV regime dominates quasiparticle and energy densities.
If $\kappa>d$, the integral \eq{QParticleDensity} is dominated by quasiparticles with IR momenta, while for $\kappa< d$ UV momenta dominate.
Similarly, $\kappa> d+z$ leads to an IR dominance of the integral \eq{EnergyDensity} for the energy density whereas, for $\kappa<d+z$, energy is concentrated in the high-momentum modes.
In summary, the exponent $\kappa$ determines where quasipar\-ticle and energy densities are concentrated,
\begin{align}
   \kappa&<d\,,&&\mbox{quasiparticles and energy: UV}\,; 
 \label{eq:UVDominance}
 \\
   d\leq \kappa&\leq d+z\,,&&\mbox{quasiparticles: IR; energy: UV}; 
 \label{eq:SelfSimilarWindow}
 \\
   \kappa&> d+z\,,&&\mbox{quasiparticles and energy: IR}\,. 
 \label{eq:IRDominance}
\end{align}

According to the above, a power-law momentum distribution requires regularization in the IR or in the UV limit.
Note that a regularization on one side only is sufficient  when both, quasiparticles and energy are concentrated at the same end of the spectrum, i.e.~for $\kappa$ satisfying \eq{IRDominance} or \eq{UVDominance}. 
Different functional forms for the distribution $n_{Q}$ are possible for providing such a regularization and for describing the crossover between the power-law region and the regularized region.
Determining the precise form of this functional form requires solving the dynamic equations.

For $\kappa$ obeying \eq{UVDominance} or \eq{IRDominance}, a simple parametrization of such a regularized quasiparticle distribution is given by
\begin{equation}
  n_{Q}(p)=f(p/p_{\Lambda};f_{1}) \,
 \label{eq:nQbyScalingFunction}
\end{equation}
in terms of a scaling function $f(x)$ of the form
\begin{equation}
f(x;f_{1}) = 2f_{1}\left[x^{\bar\kappa}+x^{\kappa}\right]^{-1}\,.
 \label{eq:scalingf}
\end{equation}
$f(x)$ interpolates, in $x$, between two different power laws, with exponents $\bar\kappa\not=\kappa$.
Around $x=1$ the scaling crosses over from one power law to the other. 
In the distribution function $n_{Q}$, this crossover thus occurs at the momentum scale $p=p_{\Lambda}$.
While the shape of the scaling function will be a universal characteristics, its `position' will be fixed by the non-universal scale $p_{\Lambda}$ and amplitude $f_{1}=f(1;f_{1})$.

Note that in the regularized region, the function does not need to exhibit a power law, but may have, e.g., an exponential form.
Furthermore, a sharp IR (UV) cutoff, i.e., $f(x)\sim\Theta(x-1)\,x^{-\kappa}$ ($f(x)\sim\Theta(1-x)\,x^{-\kappa}$), can be  realized by choosing $\bar\kappa\to-\infty$ ($\to\infty$).

The simultaneous IR and UV convergence of both integrals, Eqs.~\eq{QParticleDensity} and \eq{EnergyDensity}, requires 
  $\bar\kappa< d$ and  $\kappa >d+z$, or vice versa. 
Within the interval \eq{SelfSimilarWindow} either the quasiparticle or the energy density diverges, such that an extended scaling function, with an additional regulator, is required.
A straightforward extension of the scaling function \eq{scalingf} involves two crossover scales, $p_{\lambda}>p_{\Lambda}$.
To make the expression more transparent, we introduce a third scale $p_{0}$.
Hence, we write
\begin{equation}
  n_{Q}(p)=f(p/p_{0};\,p_{\Lambda}/p_{0},p_{\lambda}/p_{0},f_{1}) \,,
 \label{eq:nQbyScalingFunctionf2}
\end{equation}
with the scaling function
\begin{equation}
f(x;\,y,z,f_{1})
= f_{1} \,
\left[y^{\kappa}(x/y)^{\kappa_{\Lambda}}+x^{\kappa}+z^{\kappa}(x/z)^{\kappa_{\lambda}}\right]^{-1}\,,
 \label{eq:scalingf2}
\end{equation}
such that, for $\kappa_{\Lambda}<\kappa<\kappa_{\lambda}$, the amplitude $f_{1}$ fixes $f$ at $x=1$ if the crossover scales are taken to the IR and UV limits, $f_{1}=f(1;y\to0,z\to\infty,f_{1})$, see the sketch in \Fig{SelfSimilar}.
$\kappa_{\Lambda}<d$ ensures convergence of the integral for the quasiparticle density in the IR, while $\kappa_{\lambda}>d+z$ renders the energy integral finite in the UV.

As $p_{0}$ above only sets the unit, we can simplify the parametrization such that the scaling function has only two arguments,
\begin{align}
  n_{Q}(p)&=f_{\Lambda}(p/p_{\Lambda};\,p_{\lambda}/p_{\Lambda},f_{0}[p_{0}/p_{\Lambda}]^{\kappa}) \,,
 \label{eq:nQbyScalingFunctiongL}
  \\
  f_{\Lambda}(x;\,y,f_{1})&=f_{1}\,
  \left[x^{\kappa_{\Lambda}}+x^{\kappa}+x^{\kappa_{\lambda}}y^{\kappa-\kappa_{\lambda}}\right]^{-1}\,,
 \label{eq:scalingfgL}
\end{align}
with $f_{0}=f_{\Lambda}(p_{0};y\to0,z\to\infty,f_{0}[p_{0}/p_{\Lambda}]^{\kappa})$, or, equivalently,
\begin{align}
  n_{Q}(p)&=f_{\lambda}(p/p_{\lambda};\,p_{\Lambda}/p_{\lambda},f_{0}[p_{0}/p_{\lambda}]^{\kappa}) \,,
 \label{eq:nQbyScalingFunctiongl}
  \\
  f_{\lambda}(x;\,y,f_{1})&=f_{1}\,
  \left[x^{\kappa_{\Lambda}}y^{\kappa-\kappa_{\Lambda}}+x^{\kappa}+x^{\kappa_{\lambda}}\right]^{-1}\,,
 \label{eq:scalingfgl}
\end{align}
with $f_{0}=f_{\Lambda}(p_{0};y\to0,z\to\infty,f_{0}[p_{0}/p_{\lambda}]^{\kappa})$.
In the parametrisations \eq{nQbyScalingFunctiongL} and \eq{nQbyScalingFunctiongl}, all momenta are expressed in units of the IR scale $p_{\Lambda}$ and the UV scale $p_{\lambda}$, respectively.
In the special cases that $\kappa_{\lambda}=\kappa$ or that the UV scale is sent to $p_{\lambda}\to\infty$, the scaling function \eq{scalingfgL}, up to constant factors, reduces to the function \eq{scalingf}.
The same applies to the function \eq{scalingfgl} if $\kappa_{\Lambda}=\kappa$ or $p_{\Lambda}\to0$.

In general, the precise form of the scaling function requires solving the dynamic equations.
Consequently, in realistic situations, it can, e.g., exhibit regions with different momentum power laws as sketched in \Fig{NTFP} which can relax the condition \eq{SelfSimilarWindow}, allowing, without cutoffs, $\kappa>d+z$ in the IR and $\kappa<d$ in the UV. 
Such scaling functions have been discussed in Refs.~\cite{Nowak:2012gd,Berges:2012us,Nowak:2011sk,Schole:2012kt,Orioli:2015dxa} in the context of non-thermal fixed points and the formation of topological defects.

\subsection{Universal dynamics}
\label{sec:UniversalDynamics}
%
\subsubsection{Global conservation laws}
\label{sec:GlobalConservLaws}
Our aim is to describe possible forms of universal dynamics realized in the model \eq{GPHamiltonian}.
We assume that, at a given instant in time, the quasiparticle number distribution $n_{Q}(p,t)$ is parametrized by a suitably regularized scaling function corresponding to finite total quasiparticle and energy densities.
This scaling function could be of the type \eq{scalingf2} which disposes of the essential properties discussed in the previous subsection, i.e., power-law behavior \eq{nQpPowerLaw} within a  region of momenta, $p_{\Lambda}\ll p\ll p_{\lambda}$, and convergence of the integrals \eq{ParticleDensity} and \eq{EnergyDensity} for quasiparticle and energy density, respectively.

The question then is, how such a distribution can evolve in time in a universal manner, i.e., in a way that it keeps its parametrization in terms of the initial scaling function, varying only the non-universal scales $p_{\Lambda}$ and $p_{\lambda}$, and the amplitude $f_{1}$.
The considerations of the previous subsection already provide an intuition of what types of dynamics are possible, depending on the scaling exponent $\kappa$.

In most cases, one or more global conservation laws constrain the dynamics, and these laws play an important role for the dynamical scaling phenomena possible in the system.
For a closed system and if quasiparticle-number changing processes are absent, the total quasiparticle density is conserved in time,
\begin{equation}
\rho_{Q}=\int \frac{\mathrm d^{d}p}{(2\pi)^{d}}\, n_{Q}(\vec p,t)\equiv\mathrm{const.} \,
 \label{eq:NConserv}
\end{equation}
If, furthermore, neither internal excitations nor interactions with an energy reservoir are possible, also the energy density is a constant of motion,
\begin{equation}
\varepsilon=\int \frac{\mathrm d^{d}p}{(2\pi)^{d}}\, \omega(\vec p) n_{Q}(\vec p,t)\equiv\mathrm{const.} \,
 \label{eq:EConserv}
\end{equation}
In addition to the above also real particle number $\rho_\mathrm{nc}$, \Eq{ParticleDensity}, is a viable conserved quantity.
In the present work we will eventually only consider quasiparticle number conserving processes. 
A generalization to dynamics which, for $z\not=2$,  explicitly accounts also for particle number conservation  will be done elsewhere.
In the following, we will use the terms ``quasiparticles'' for the respective quasiparticle eigenmodes of the Hamiltonian, and ``particles'' to refer to the distribution $n_\mathrm{nc}(p)\sim p^{z-2}n_{Q}(p)$, where both are identical for $z=2$.

As was pointed out in Ref.~\cite{Svistunov1991a} and as we will discuss in detail, the conservation laws \eq{NConserv} and \eq{EConserv} limit the possibilities of how the cutoff scales $p_{\Lambda}$ and $p_{\lambda}$, and the amplitude $f_{1}$ can vary in time.

For example,  if $\kappa>d$, quasiparticles are concentrated in the IR.
In this case, shifting the infrared cutoff $p_{\Lambda}$ implies a violation of the conservation law \eq{NConserv} unless the amplitude $f_{1}$ is adjusted appropriately.
Similarly, for $\kappa<d+z$, the bulk of energy sits in the UV, and $p_{\lambda}$ can in general only be varied together with $f_{1}$.
If these conditions are simultaneously fulfilled, $d\leq \kappa\leq d+z$, both, IR and UV cutoffs are needed,  cf.~Eqs.~\eq{SelfSimilarWindow} and \eq{scalingf2}, such that a change of $f_{1}$ requires also a shift of both of these cutoff scales.

On the contrary, if $\kappa>d+z$, both, quasiparticles and energy are concentrated at the IR cutoff scale.
In this case, an additional UV cutoff $p_{\lambda}\gg p_{\Lambda}$, which is expected to limit a realistic physical distribution, can  be shifted without significantly `renormalizing' the entire function since neither conservation law is strongly affected by the shift.
The same applies to shifting the IR cutoff $p_{\Lambda}$ if $\kappa<d$.

We remark again that for bimodal distributions such as the one sketched in \Fig{NTFP}, energy and particle number are concentrated in either the low- or the high-momentum power law. 
Therefore, one can have the case that, e.g., $\kappa>d+z$ in the low-momentum regime, yet the energy of the whole system is concentrated in the UV part of the high-momentum power law. 
Hence, the dynamics anticipated for the case of $d\leq\kappa\leq d+z$ would be allowed in the low-momentum regime, which is indeed what happens for strong cooling quench dynamics~\cite{Orioli:2015dxa}.

In the next two subsections we discuss in more detail how these constraints allow to classify the kinds of universal dynamics possible in the system.

\subsubsection{Non-thermal fixed points}
\label{sec:NTFPIntro}
\emph{Dynamical scaling hypothesis}.---
Similar to universal RG `evolution', an isolated system can show universal evolution in real time, for instance when it is quenched out of equilibrium and subsequently re-equilibrates.
This evolution is captured, in the simplest case, by a scaling hypothesis for the time-dependent, angle-averaged quasiparticle  distribution,
\begin{equation}
n_{Q}(p,t) = (t/t_{0})^{\alpha}f([t/t_{0}]^{\beta}p)\,.
 \label{eq:NTFPscaling}
\end{equation}
Here, $f$ is a universal scaling function in momentum space, and $t_{0}$ is an arbitrary reference time within the temporal scaling regime, where $n_{Q}(p,t_{0})=f(p)$.
The universal exponents $\alpha$ and $\beta$ define the non-equilibrium RG fixed point which the system approaches in time.
In contrast to the attractive thermal `fixed point' of the evolution where both, $\alpha$ and $\beta$ are by definition zero, non-vanishing exponents indicate the existence of a {non-thermal fixed point} \cite{Berges:2008wm,Berges:2008sr,Scheppach:2009wu,Berges:2014xea,Berges:2014bba,Orioli:2015dxa}. 

Here, we \emph{define} a closed system to be at a \emph{non-thermal fixed point} whenever  correlation functions, e.g., the quasi-particle occupation $n_{Q}$, obey a dynamical scaling behaviour of the form \eq{NTFPscaling}, characterised by universal scaling exponents and a non-thermal universal scaling function $f$, in general in a suitably defined scaling limit such as $t\to\infty$.

Determining the universal scaling function $f(p)$ in general requires solving the dynamic equations. 
Instead of this, we will work with a minimal ansatz for $f$ as the one given in \Eq{scalingf}, 
which interpolates between two momentum power laws. 
\Eq{NTFPscaling} will then be satisfied if the parameters have the power-law time dependence $f_{\Lambda}(t)$ $\equiv n_{Q}(p_{\Lambda}(t),t)\sim t^{\alpha}$, $p_{\Lambda}(t)\sim t^{-\beta}$, giving the scaling evolution $n_{Q}(p,t)=f(p/p_{\Lambda}(t);\,f_{\Lambda}(t))$, with $f$ defined in \Eq{scalingf}.
Choosing both exponents, $\alpha$ and $\beta$, to be, e.g.,  positive real numbers, time evolution shifts the distribution $n_{Q}(p,t)$ self-similarly to smaller momenta and larger values of $n_{Q}(p_{\Lambda}(t),t)\equiv f_{1}(t)$.\\[-1ex]

\emph{Constraints from conservation laws}.---
As discussed in the previous section, global conservation laws strongly constrain the form of the correlations in the system. 
They also constrain the dynamics and thus play an important role for the possible scaling phenomena.
With regard to the scaling hypothesis \eq{NTFPscaling}, they imply scaling relations between the exponents $\alpha$ and $\beta$. 
For example, if the dynamics conserves the total quasiparticle density, \Eq{NConserv},
the relation 
\begin{equation}
  \alpha=\beta d\,
 \label{eq:alphabetaNConserv}
\end{equation}
must be fulfilled.
Analogously, the conservation of the energy density, \Eq{EConserv}, requires  
\begin{equation}
  \alpha=\beta (d+z)\,.
 \label{eq:alphabetaEConserv}
\end{equation}
Here we always presuppose, that the respective integrals converge.
Given one of the above relations, the remaining exponent can be determined by a scaling analysis of the dynamic equations as we discuss in more detail in Sects.~\sect{KineticDescrUniDyn} and \sect{ScalingAnalysisKinEqs}. 

\begin{figure}[t]
\centering
\includegraphics[width=0.4\textwidth]{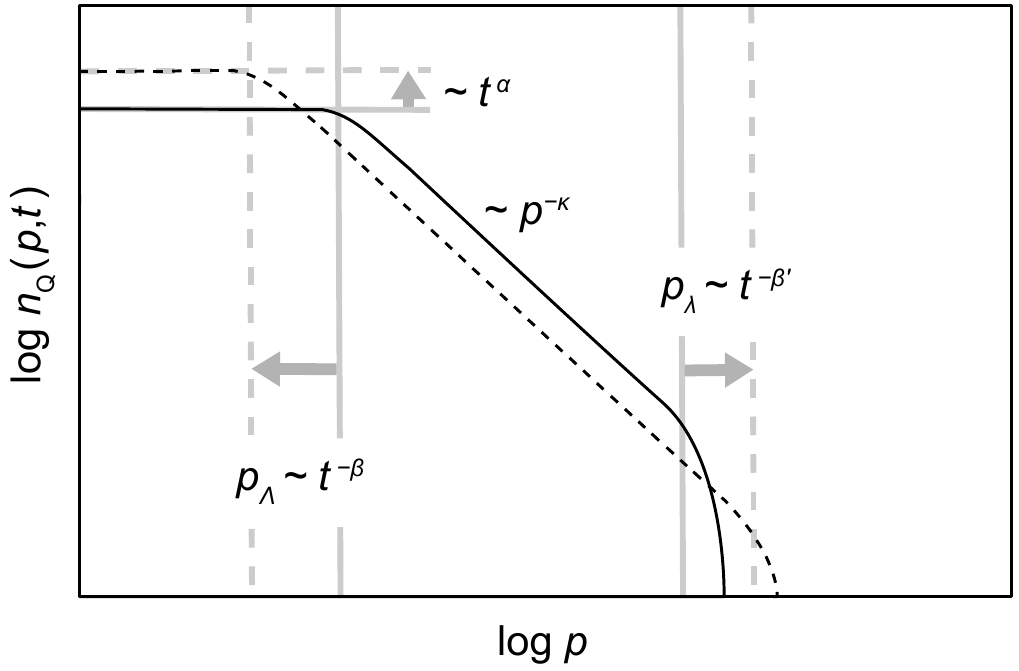}
\caption{Sketch of the self-similar evolution of the scaling form \eq{NTFPscaling0f3} for $n_{Q}(p,t)$ according to \Eq{SelfSimilarScaling}.
Note the double-logarithmic scale.
The IR cutoff scale $p_{\Lambda}$ and the UV scale $p_{\lambda}$, as well as the amplitude $f_{1}$ rescale with time $t$ such that the total quasiparticle density remains invariant.
The sketch shows the case of an inverse particle transport following a strong cooling quench.
See \Tab{ScalingExponents} for our predictions for the scaling exponents (first row, NTFP).
}
\label{fig:SelfSimilar}
\end{figure}
\emph{Scaling evolution of the closed system}.---
As discussed above, we consider the case that quasiparticle number and energy are, in each scattering process, simultaneously conserved in time.
For non-zero exponents $\alpha$ and $\beta$, however, the scaling relations \eq{alphabetaNConserv} and \eq{alphabetaEConserv} can not both be satisfied for $z\not=0$.
This means that either $\alpha=\beta=0$ or that the scaling hypothesis \eq{NTFPscaling} has to be extended.

Suppose that the scaling function has a single regulator such as in \Eq{scalingf}.
As discussed in \Sect{ScalingFunction}, quasiparticles and energy are concentrated at the same end of the momentum scaling region, within which $n_{Q}(p,t)\sim p^{-\kappa}$, if $\kappa$ is outside the interval \eq{SelfSimilarWindow}. 
In this case, $\alpha=\beta=0$ is required, and a scaling evolution is only possible at the opposite end of the scaling region.
This evolution leads to a wave-turbulent cascade which we discuss in more detail in \Sect{Turbulence} below. 

On the contrary, if $\kappa$ is within the interval \eq{SelfSimilarWindow}, or if energy and particles are concentrated in different scaling regimes (see end of \Sect{GlobalConservLaws}), particles and energy are concentrated at opposite ends of the scaling region.
In this case, a more general scaling hypothesis is needed which allows for different rescalings of the IR and the UV parts of the scaling function, see \Fig{SelfSimilar}.
We consider the example \eq{nQbyScalingFunctionf2} in terms of the scaling function \eq{scalingf2}, and suppose that the non-universal parameters follow the scaling evolution 
\begin{equation}
f_{1}(t) \sim \tau^{\alpha_{0}}\,,\quad
p_{\Lambda}(t) \sim \tau^{-\beta}\,,\quad
p_{\lambda}(t) \sim \tau^{-\beta'}\,,
 \label{eq:SelfSimilarScaling}
\end{equation}
with the dimensionless scaling parameter    
\begin{equation}
 \tau = t/t_{0}\,.
 \label{eq:tauSelfSim}
\end{equation}
Here, $t_{0}$ denotes a time which could mark the beginning of the scaling evolution.
This ansatz satisfies the extended scaling hypothesis
\begin{equation}
n_{Q}(p,t) 
= \tau^{\,\alpha_{0}+(\beta+\beta')\kappa}
   f(\tau^{\,\beta+\beta'}p;\tau^{\,\beta'}p_{\Lambda},\tau^{\,\beta}p_{\lambda})\,,
 \label{eq:NTFPscaling0f3}
\end{equation}
see the sketch in \Fig{SelfSimilar}.
It is useful to express the momenta $p$ and $p_{\lambda}$ in \Eq{NTFPscaling0f3} alternatively in terms of the IR scale $p_{\Lambda}$ and to rewrite the scaling hypothesis in terms of a scaling function of the type \eq{scalingfgL},
\begin{align}
n_{Q}(p,t) 
&= \tau^{\,\alpha}
   f_{\Lambda}(\tau^{\,\beta}p/p_{\Lambda};\,\tau^{\,\beta-\beta'}p_{\lambda}/p_{\Lambda})\,,
 \label{eq:NTFPscaling0gL}
\end{align}
suppressing the third argument of $f_{\Lambda}$.
Here we introduced the exponent $\alpha$, defined as
\begin{equation}
 \alpha = \alpha_{0}+\beta\kappa\,,
 \label{eq:alphaalpha0}
\end{equation}
such that the scaling hypothesis \eq{NTFPscaling0gL} is equivalent to \Eq{NTFPscaling} in the regime $p\ll p_{\lambda}$, i.e., in the limit $p_{\lambda}\to\infty$.
Alternatively, we can rewrite \Eq{NTFPscaling0f3}, by expressing $p$ and $p_{\Lambda}$ in terms of $p_{\lambda}$, as
\begin{align}
n_{Q}(p,t) 
&= \tau^{\,\alpha'}
   f_{\lambda}(\tau^{\,\beta'}p/p_{\lambda};\tau^{\,\beta'-\beta}p_{\Lambda}/p_{\lambda})\,,
 \label{eq:NTFPscaling0gl} 
\end{align}
with
\begin{equation}
 \alpha' = \alpha_{0}+\beta'\kappa\,.
 \label{eq:alphasalpha0}
\end{equation}
Also \Eq{NTFPscaling0gl} is equivalent to the simpler scaling hypothesis \eq{NTFPscaling}, with the replacements $\alpha\leftrightarrow\alpha'$, $\beta\leftrightarrow\beta'$, $p_{\Lambda}\leftrightarrow p_{\lambda}$, in the limit $p\gg p_{\Lambda}$ or $p_{\Lambda}\to0$.

The scaling hypotheses \eq{NTFPscaling0gL}, in the limit $p_{\lambda}\to\infty$, and \eq{NTFPscaling0gl}, in the limit $p_{\Lambda}\to0$, can now be used, in the same way as before, to obtain the scaling relations \eq{alphabetaNConserv} between $\alpha$ and $\beta$ and \eq{alphabetaEConserv} between $\alpha'$ and $\beta'$, respectively.
In summary, and eliminating $\alpha'$ by means of Eqs.~\eq{alphaalpha0} and \eq{alphasalpha0}, i.e., $\alpha'=\alpha+(\beta'-\beta)\kappa$, energy and (quasi)particle densities are both time independent if 
\begin{align}
 \label{eq:alphabetaENConserv}
  \alpha&=\beta d\,,\\
  \beta'(d+z-\kappa)&=\beta(d-\kappa)\,.
 \label{eq:betabetasENConserv}
\end{align}
Recall that these relations apply to the range of exponents \eq{SelfSimilarWindow} where particles are concentrated at small momenta and energy at large momenta.
This implies $\beta\,\beta' \leq 0$, i.e., the IR and UV scales $p_{\Lambda}$, $p_{\lambda}$ rescale in opposite directions. 
These relations hold in the limit of a large scaling region, i.e., for $p_{\lambda}\gg p_{\Lambda}$.
Thereby, particle conservation only affects the infrared shift with $\beta$, \Eq{alphabetaENConserv}, while energy conservation gives the condition \eq{betabetasENConserv} for $\beta'$ in the UV.
The scalings \eq{SelfSimilarScaling} represent the leading power-law behavior in $t$ while further non-leading terms account for the exact conservation of the energy and particle densities.

In the following, we will refer to the rescaling of a distribution as sketched in \Fig{SelfSimilar}, which implies a \emph{bi-directional}, non-local transport of particles and energy as \emph{self-similar}.
With this we distinguish it from the build-up of wave-turbulent \emph{cascades} discussed in the next section which is, in leading order, also self-similar but uni-directional and local.
In particular, the amplitude $f_{1}$ does not scale with time and the power-law part of the distribution remains approximately stationary.

\subsubsection{Wave-turbulent transport}
\label{sec:Turbulence}
A special case of non-thermal scaling solutions includes those which, in a certain regime of momenta become stationary in time.
Such solutions are studied in wave-turbulence theory, usually within a Boltzmann kinetic approach~\cite{Zakharov1992a,Nazarenko2011a}, see \App{WWTSolutions} for a summary.
In analogy to the self-similar case in \Sect{NTFPIntro} above, we study below the constraints set on such wave-turbulent solutions by global conservation laws in a closed system. Explicit solutions will be discussed in \Sect{WWTSolutions}.
\\[-1ex]

\begin{figure}[t]
\centering
\includegraphics[width=0.4\textwidth]{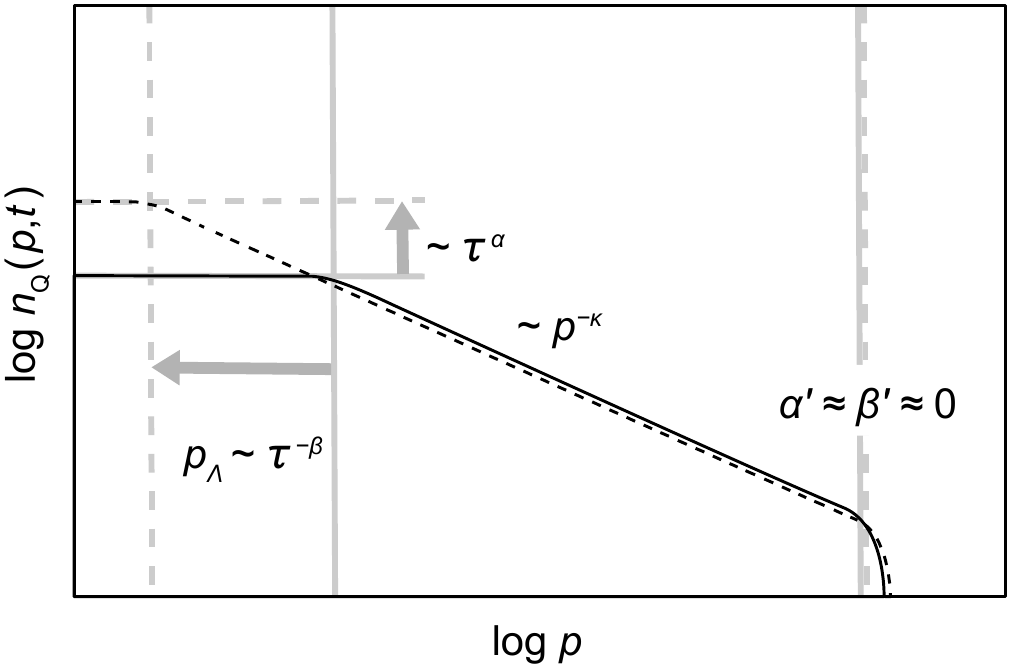}
\caption{Sketch of the build-up of the inverse quasiparticle cascade, defined as a scaling evolution of the form \eq{WaveFrontScalingForm} for $n_{Q}(p,t)$ according to \Eq{WaveFrontScaling}, with $\alpha=\beta\kappa$ and, to leading order, $\alpha'=\beta'=0$. 
Note the double-logarithmic scale.
The IR cutoff scale $p_{\Lambda}$ shifts without significantly changing the total quasiparticle density. 
Only a small non-leading-order rescaling of  the UV scale $p_{\lambda}$ is required to satisfy number conservation..
The sketch shows the case of an inverse particle transport following a weak cooling quench.
See \Tab{ScalingExponents} for our predictions for the scaling exponents.
In the case of a weak-wave-turbulence quasiparticle cascade, $p_{\Lambda}$ shifts in an accelerated way, with $\tau$ replaced by $\tau^{*}$, see \Eq{scalingParam1}.
}
\label{fig:WWTCascade}
\end{figure}
\emph{Build-up of wave-turbulence in a closed system}.---
Let us consider the build-up of wave turbulence in a closed system from an initially non-equilibrated quasiparticle distribution. Suppose that this distribution has the form \eq{scalingf2}, with a power law $n_{Q}(p)\sim p^{-\kappa}$ in the region $p_{\Lambda}\ll p\ll p_{\lambda}$ between the IR and UV cutoff scales.
Again, taking into account the integrals for (quasi)particle and energy densities, Eqs.~\eq{ParticleDensity}, \eq{EnergyDensity}, the value of the exponent $\kappa$ tells us at which end of this region energy and particle number are concentrated.

If the power law is sufficiently flat, $\kappa<d$, both, particles and energy are in the UV, and both, Eqs.~\eq{alphabetaNConserv} and \eq{alphabetaEConserv} need to be fulfilled, presupposing that the IR cutoff is sufficiently small, $p_{\Lambda}\ll p_{\lambda}$.
This is only possible for $\alpha'=\beta'=0$.
As a consequence, the amplitude $f_{1}$ and the UV cutoff scale $p_{\lambda}$ are, to a first approximation, constant in time, cf.~\Fig{WWTCascade}.

Nonetheless, a wave-turbulent, quasi-local flux can build up and thereby satisfy the global conservations laws while $p_{\Lambda}$ decreases.
As before, global conservation laws require that this process in leading order confirms the scaling hypothesis \eq{NTFPscaling0gl}, with $\alpha'=\beta'=0$, and thus $\alpha_{0}=0$, cf.~\Eq{alphasalpha0},
\begin{align}
n_{Q}(p,t) 
&= f_{\lambda}(p/p_{\lambda};\tau^{-\beta}p_{\Lambda}/p_{\lambda})\,,
 \label{eq:WaveFrontScalingForml} 
\end{align}
with scaling function of the type \eq{scalingfgl}, of which we suppress the third argument.
In turn, \Eq{alphaalpha0} implies that $\alpha=\beta\kappa$ in an equivalent scaling hypothesis of the form \eq{NTFPscaling0gL},
\begin{align}
n_{Q}(p,t) 
&= \tau^{\,\alpha}
   f_{\Lambda}(\tau^{\,\beta}p/p_{\Lambda};\,\tau^{\,\beta}p_{\lambda}/p_{\Lambda})\,.
\label{eq:WaveFrontScalingForm}
\end{align}

To determine $\beta$ requires analyzing the kinetic equation and the scaling properties of the interactions.
We will do this in Sects.~\sect{KineticDescrUniDyn} and \sect{ScalingAnalysisKinEqs}.
Depending on these scaling properties, $\beta$ can be positive or negative.
If $\beta>0$, the wave-turbulent flux builds up analogously to the self-similar scaling evolution.
The non-universal scales evolve according to
\begin{equation}
f_{\Lambda}(t) \sim  \tau^{\alpha}\,,\quad
p_{\Lambda}(t) \sim \tau^{-\beta}\,,\quad
p_{\lambda}(t) \sim \mbox{const.}\,,
 \label{eq:WaveFrontScaling}
\end{equation}
as in \Eq{SelfSimilarScaling}, with $\tau=t/t_{0}$, but  with $\alpha'=\beta'=0$, see \Fig{WWTCascade}. 

On the contrary, if $\beta<0$, building up a wave-turbulent cascade towards the IR, which requires $p_{\Lambda}$ to decrease in time, is not possible with $\tau=t/t_{0}$. 
$\tau$ rather needs to shrink with time.
The scaling analysis of the dynamic equations will show that a scaling parameter algebraic in time is required, and the simplest such parameter is given by, cf.~Refs.~\cite{Svistunov1991a,Berloff2002a},
\begin{align}
\tau=\frac{t^{*}-t}{t^{*}-t_{0}}\equiv \tau^{*}\,.
  \label{eq:scalingParam1}
\end{align}
As a result, the cascade builds up behind a wave front which, at time $t^{*}$, reaches zero momentum. 
Before this happens, however, the solution ceases to be valid as spreading information over infinite distances in a finite time is impossible.

The wave front can be imagined to look like the evolution shown in \Fig{WWTCascade}, however, with the cutoff scales evolving according to \eq{WaveFrontScaling}, \eq{scalingParam1}.
The value of $t^{*}$ is determined by the given initial distribution at time $t_{0}$.
The scaling evolution \eq{WaveFrontScaling} is valid for $\tau^{*}<1$, and $t^{*}-t$ much smaller than the overall evolution time $t^{*}-t_{0}$.
In this limit, the distribution behind the wave front becomes nearly stationary when taking into account the global conservation of particle and energy density.
Note that, as a result of these conservation laws, the scalings \eq{WaveFrontScaling} can only represent the leading behavior, while subleading terms form corrections which are the more important the further away $t$ is from $t^{*}$.

For $\kappa>d+z$, both, quasiparticles and energy are concentrated in the IR, and a direct cascade can build up according to the scaling form \eq{NTFPscaling0f3}, with $\alpha'=\beta'\kappa$, $\alpha=\beta=0$.
If $\beta'<0$ the evolution takes the form of a critically slowed wave front while for $\beta'>0$ an accelerating scaling evolution occurs.
At $t=t^{*}$, the wave front reaches infinite momentum, before which the solution, however, is expected to break down as 
arbitrarily high momenta are usually not captured.

\begin{table}[t]
\centering 
\caption{Scaling relations.
The table summarizes the relations between the scaling exponents as obtained,  in \Sect{UniversalDynamics}, from the constraints set by global conservation laws.
Depending on the relative size of the momentum-scaling exponent $\kappa$, the dimension $d$, and the dynamical exponent $z$, one expects either the build-up of an inverse cascade, a bi-directional self-similar evolution, or the build-up of a direct cascade of quasiparticles.
} 
\label{tab:ScalingRelConservationLaws} 
\begin{tabular}{  C{2.5cm}  C{1.2cm} C{0.7cm} C{1.2cm} C{1.5cm} C{0.7cm}  }
\hline
\hline 
{} & $\alpha$ & $\beta$ & $\alpha'$ & $\beta'$ & $\tau$  \\
\hline
inverse cascade&  \multirow{2}{*}{$ \beta \kappa$} & $<0$ & \multirow{2}{*}{0} & \multirow{2}{*}{0}& ${\tau_1}$ \\
$\kappa < d$   &  {} & $>0$ & {} & {} & $t/t_0$ \\
\hline 
bi-directional self-similar&  \multirow{2}{*}{$\beta d$} & {} & \multirow{2}{*}{$\beta'(d+z)$} & 
\multirow{2}{*}{$\beta\displaystyle\frac{d-\kappa}{d+z-\kappa}$}    & {} \\
$d<\kappa < d+z$  &  {} & {} & {} & {} & {} \\
\hline
direct cascade&  \multirow{2}{*}{0} & \multirow{2}{*}{0} & \multirow{2}{*}{$\beta' \kappa$} & $<0$& $t/t_0$ \\
$ d +z < \kappa$   &  {} & {} & {} & $>0$ & $\tau_1$ \\
\hline
\end {tabular}\par
\end{table}
%
We will show in \Sect{ScalingAnalysisKinEqs} that, for the cases of free particles and Bogoliubov sound in the perturbative wave-Boltzmann regime, following a weak cooling quench, an inverse cascade builds up behind a wave front described by the scaling evolution \eq{WaveFrontScaling}, with scaling parameter $\tau=\tau^{*}$, \Eq{scalingParam1}, see also Refs.~\cite{Svistunov1991a,Semikoz1995a.PhysRevLett.74.3093,Semikoz1997a}.
One may say that this wave-front scaling evolution is critically accelerating.

We emphasize that, in physically realistic situations, this scaling evolution breaks down at a finite length scale, $1/p_{\Lambda}<\infty$, i.e., before $t=t^{*}$ is reached, when the processes underlying the kinetics of the system change in a fundamental way. 
In Ref.~\cite{Nowak:2012gd} it was shown that in such cases, the change of the non-condensate distribution comes to a halt and a condensate, i.e., a macroscopic zero-mode population starts to rise up.

In the following and in accordance with the (wave-)tur\-bu\-lence literature, we will refer to  the rescaling of a distribution as sketched in \Fig{WWTCascade}, which implies a \emph{uni-directional}, quasi-local transport of particles or energy within the inertial range, to the IR or the UV, as a \emph{cascade}.
With this we distinguish it from the bi-directional self-similar non-local transport discussed in the previous section.
Recall also that one can have mixed forms with, e.g., both, a self-similar evolution in the IR and a wave turbulent cascade in the UV, see \Fig{NTFP}(b).

\subsection{Summary}
\label{sec:SummaryScalingRel}
Global conservation laws of bulk, i.e., momentum-integrated quantities play an important role in distinguishing between the different types of scaling dynamics possible in an isolated system.
A wave-turbulent, quasi-local transport of either quasiparticles or energy, which, in the inertial range, does not change $n_{Q}$ in time, is possible only if both these quantities are concentrated at the same end of the inertial range.
Depending on the relative size of $\kappa$, cf.~Eqs.~\eq{UVDominance}, \eq{IRDominance}, one expects either an inverse cascade or a direct cascade.
In contrast, if the scaling function has a single power-law $\sim p^{-\kappa}$ and $\kappa$ is inside the interval \eq{SelfSimilarWindow}, a self-similar evolution is expected.
Exemptions to this can appear when the function has more than one power law, e.g., one in the IR and one in the UV as sketched in \Fig{NTFP}b. 
In such a case, self-similar evolution is also possible outside the interval \eq{SelfSimilarWindow}.
Scaling relations between the exponents for these different cases are summarized in \Tab{ScalingRelConservationLaws}.
Note that at the boundaries, $\kappa=d$ and $\kappa=d+z$, a more careful analysis would be in order.

\section{Kinetic description of universal time evolution}
\label{sec:KineticDescrUniDyn}
%
\subsection{Kinetic equation}
\label{sec:KineticEq}
The main goal of the present work is to find scaling solutions of Boltzmann-type kinetic equations for the occupation number distributions \eq{nSP} or \eq{nQP} and compute the respective scaling exponents.
These equations can be written as
\begin{align}
  \partial_{t}n_{Q}({\vec p},t)
  &= I[n_{Q}](\vec p,t),
  \label{eq:QKinEq}
\end{align}
where $I[n_{Q}]$ is a scattering integral to be specified below.

We assume the scaling functions to be regularized such that the total (quasi)particle density and energy are finite. 
This is, e.g., satisfied by the form \eq{scalingf2}, with $\kappa_{\Lambda}=0$, $\kappa_{\lambda}\gg1$. 
We then derive the scaling exponents $\beta$ or $\beta'$, and $\kappa$, from which one obtains the remaining exponents by means of conservation-law constraints as summarized in \Tab{ScalingRelConservationLaws}.
Which solutions are possible will depend crucially on the interaction properties of the Bose fields encoded in the scattering integral.

In the following sections we introduce the scattering integral $I$ and derive universal exponents of scaling solutions of the kinetic equation \eq{QKinEq}.
We begin with determining the scaling properties of the scattering integral and subsequently discuss the different possible solutions following from these.

\subsection{Scaling properties of the scattering integral}
\label{sec:ScattInt}
%
\subsubsection{Classical-wave limit}
\label{sec:ClassWaveLimit}
The time evolution of the momentum distribution $n_{\mathbf{p}}\equiv n_{Q}(\mathbf{p},t)$ of Bose-field excitations is described, in the kinetic approximation, by a generalized 
Quantum Boltzmann Equation (QBE).
For details of the derivation within the quantum field theoretic approach we choose here see \App{DerivKinEq}.
The QBE takes the form \eq{QKinEq}, with scattering integral 
\begin{align}
   &I[n_{Q}](\vec p,t)
  = \int_{\vec k\vec q\vec r}|T_{\vec p\vec k\vec q\vec r}|^{2}\delta(\vec p+\vec k -\vec q - \vec r)
  \nonumber\\
  &\quad\times\
  \delta(\omega_{\vec p}+\omega_{\vec k}-\omega_{\vec q}-\omega_{\vec r})
  \nonumber\\
  &\quad \times\
  [(n_{\vec p}+1)(n_{\vec k}+1)n_{\vec q}n_{\vec r}
  -\
  n_{\vec p}n_{\vec k}(n_{\vec q}+1)(n_{\vec r}+1)],
  \label{eq:KinScattInt}
\end{align}
where we use the short-hand notation $\int_{\vec k}\equiv\int \ddf{k}(2\pi)^{-d}$ and $T_{\vec p\vec k\vec q\vec r}$ is the scattering $T$-matrix discussed in more detail below.
The scattering integral $I[n_{Q}](\mathbf{p},t)$ describes the redistribution of the occupations $n_{\vec p}$ of momentum modes $\vec p$ with eigenfrequency $\omega_{\vec p}$ due to elastic $2\to2$ collisions. 
If a Bose condensate is present, the occupation numbers describe, in general, quasiparticle excitations, and the scattering matrix as well as the mode eigenfrequencies are modified as we discuss in more detail in the following. 
Note that in this work we will only consider transport entirely within the ranges where a fixed scaling exponent $z$ applies. 
As a consequence, particle-number changing processes are suppressed.
Collective scattering effects beyond two-to-two exchange of occupation numbers will be captured by the $T$-matrix. 

The QBE scattering integral \eq{KinScattInt} has two classical limits:
If $n_{\vec p}\ll1$, the scattering integral reduces to the usual Boltzmann integral for classical particles with its integrand proportional to $n_{\vec q}n_{\vec r} -  n_{\vec p}n_{\vec k}$.
In the opposite, classical-wave limit of large bosonic mode occupations, $n_{\vec p}\gg1$, one obtains the wave-Boltzmann scattering integral,
\begin{align}
   I[n_{Q}](\vec p,t)
  &= \int_{\vec k\vec q\vec r}|T_{\vec p\vec k\vec q\vec r}|^{2}\delta(\vec p+\vec k -\vec q - \vec r)
  \nonumber\\
  &\quad\quad\times\
  \delta(\omega_{\vec p}+\omega_{\vec k}-\omega_{\vec q}-\omega_{\vec r})
  \nonumber\\
  &\quad\quad \times\
  [(n_{\vec p}+n_{\vec k})n_{\vec q}n_{\vec r}
  -n_{\vec p}n_{\vec k}(n_{\vec q}+n_{\vec r})]\,,
  \label{eq:KinScattIntCWL}
\end{align}
as the terms of third order in the distribution function $n_{\mathbf{p}}$ dominate over the classical-particle, second-order Boltzmann terms.
As we are interested, in this work, in the kinetics of the near-degenerate Bose gas, with $n_\mathbf{p}\gg1$, we will restrict our discussion to the integral \eq{KinScattIntCWL} of the wave-Boltzmann equation (WBE).

Since we are assuming isotropic distributions $n_{Q}(\vec p,t)=n_{Q}(p,t)\equiv n_{p}$, it is convenient to write the WBE in the form
\begin{align}
  \partial_{t}n_{Q}(p,t)
  &= I[n_{Q}](p,t)\,,
  \label{eq:WBKinEq}
  \\
  I[n_{Q}](p,t)
  &= \int_{kqr}W_{pkqr}\delta(\omega_{p}+\omega_{k}-\omega_{q}-\omega_{r})
  \nonumber\\
  &\quad\quad \times\
  [(n_{p}+n_{k})n_{q}n_{r}
  -n_{p}n_{k}(n_{q}+n_{r})],
  \label{eq:WBKinScattInt}
\end{align}
with the angle-averaged transition matrix squared ($d=2,3$)
\begin{align}
   W_{pkqr}
  =& 2^{1-d}\pi^{-1} \int \dF{\Omega_{\mathbf{p}}}\dF{\Omega_{\mathbf{k}}}\dF{\Omega_{\mathbf{q}}}\dF{\Omega_{\mathbf{r}}}
  k^{d-1}q^{d-1}r^{d-1}
  \nonumber\\
  &\times\ |T_{\vec p\vec k\vec q\vec r}|^{2}\delta(\vec p+\vec k -\vec q - \vec r)\,.
  \label{eq:WitoTq}
\end{align}
%

\subsubsection{Scaling behaviour}
\label{sec:KinIntScaling}
The scattering integral in the classical-wave limit, \Eq{WBKinScattInt}, is a homogeneous function of momentum and time, i.e., scales in these variables if the same holds for the quasiparticle distribution $n_{p}=n_{Q}(p,t)$,
\begin{equation}
n_{Q}(p,t) = s^{\alpha/\beta}n_{Q}(sp,s^{-1/\beta}t)\,,
 \label{eq:NTFPscaling0nQ}
\end{equation}
cf.~\Eq{NTFPscaling}, and for the modulus of the $T$-matrix,  
\begin{align}
  |T({\vec p,\vec k,\vec q,\vec r};t)|
  =s^{-m}|T({s\vec p, s\vec k, s\vec q, s\vec r};s^{-1/\beta}t)|\,,
  \label{eq:Tscaling0}
\end{align}
with scaling dimension $m$ of the $T$-matrix.
Choosing the scaling parameter $s$ to be
\begin{align}
 s=(t/t_{0})^{\beta}
  \label{eq:scalingParam0}
\end{align}
shows that \Eq{NTFPscaling0nQ} is equivalent to the scaling form \eq{NTFPscaling} of $n_{Q}$, with $n_{Q}(p,t_{0})=f(p)$.
In the following we will keep this notation in terms of a more general scaling parameter $s$, keeping in mind that also a different scaling such as with $\tau^{*}$, \Eq{scalingParam1}, is possible. 

The scaling \eq{Tscaling0} of the $T$-matrix squared implies that
\begin{align}
  W({p,k,q,r},t)
  =s^{-2(d+ m)+3}W({s p, s k, s q, s r};s^{-1/\beta}t)\,. 
  \label{eq:Wscaling0}
\end{align}
From the above, one obtains the spatio-temporal scaling of the scattering integral,
\begin{align}
   I[n_{Q}](p,t)
  &= s^{-\mu}I[n_{Q}](sp,s^{-1/\beta}t) 
  \nonumber\\
  &= (t/t_{0})^{-\beta\mu}I[n_{Q}]([t/t_{0}]^{\beta} p),
  \label{eq:IScalingForm0}
\end{align}
where the second line follows by inserting the scale parameter \eq{scalingParam0}.
The exponent $\mu$ is obtained from Eqs.~\eq{WBKinScattInt} and \eq{Wscaling0} as
\begin{align}
  \mu = 2(d+ m)-z-3\alpha/\beta\,.
  \label{eq:muExponent}
\end{align}

While the exponent $m$ is required for deriving the scaling exponent $\beta$ of the time evolution
\footnote{And of $\alpha$, in principle, which however follows from $\beta$ by means of conservation laws, cf.~\Sect{UniversalDynamics}.}, 
the exponent $\kappa$ characterizing the spatial scaling of $n_{Q}(p,t_{0})$, \Eq{nQpPowerLaw}, depends on the purely spatial momentum scaling of the $T$-matrix and thus of the scattering integral, at a fixed instance in time, e.g., at $t=t_{0}$.
Suppose that the distribution function $n_{Q}$ scales, within a region of momenta, according to \Eq{nQpPowerLaw}, being regularized by an IR cutoff $p_{\Lambda}$ (or a UV cutoff $p_{\lambda}$) and that in the limit $p\gg p_{\Lambda}$ (or $p\ll p_{\lambda}$) the scattering integral is finite. 
The $T$-matrix then is expected to scale in the momenta, at a fixed time $t_{0}$, according to
\begin{align}
  |T({\vec p,\vec k,\vec q,\vec r};t_{0})|
  =s^{-m_{\kappa}}|T({s\vec p, s\vec k, s\vec q, s\vec r};t_{0})|\,. 
  \label{eq:Tscaling}
\end{align}
The exponents $m_{\kappa}$ and $m$ are, in general, different.
{Note that, in the following we will write scaling exponents such as $m_{\kappa}$ with a subscript $\kappa$ to emphasise that they characterize spatial scaling at a fixed time $t_{0}$ while their counterparts without subscript apply to scaling in momentum and time.}

As we will exemplarily show below, the scaling \eq{Tscaling} applies only within a limited regime between the IR and UV cutoffs, $p_{\Lambda}\ll p\ll p_{\lambda}$.

Provided that the integrand is sufficiently local in momentum space
\footnote{Locality of the integral in momentum space is to be understood on a logarithmic scale. It ensures that the scaling remains largely unaffected by the cutoffs, cf., e.g.~Ref.~\cite{Zakharov1992a}.}, 
such that $I[n_{Q}](p)$ is independent of the respective cutoff, the momentum scaling in the respective region reads
\begin{align}
   I[n_{Q}](p,t_{0})
  &= s^{-\mu_{\kappa}}I[n_{Q}](sp,t_{0}) \,.
  \label{eq:IScalingForm0t0}
\end{align}
The scaling exponent $\mu_{\kappa}$ is obtained from Eqs.~\eq{WBKinScattInt}, \eq{Wscaling0}, and \eq{Tscaling}, with $n_{p}\sim p^{-\kappa}$, as
\begin{align}
  \mu_{\kappa} = 2(d+ m_{\kappa})-z-3\kappa\,.
  \label{eq:mukappaExponent}
\end{align}
Note that the exponents $\mu_{\kappa}$ and $\mu$, \Eq{muExponent}, are, in general different, distinguishing pure spatial from space-time scaling.

We furthermore remark that, to prove the locality of the scattering integral, it is possible to make estimates as in  Ref.~\cite{Zakharov1992a}.
Numerically, it is difficult to prove locality explicitly from the calculation of the scattering integrand.
However, as we will show in \Sect{NumericalFreeNonperturbative}, cf.~\Fig{IpSelfSimCollFree}, the scattering integral, in the scaling momentum region, shows independence of the particular value of the IR cutoff and thus proves to be local to a sufficiently large extent within the scaling region.

\subsection{Properties of the scattering $T$-matrix}
\label{sec:ScattTM}
In general, the scaling hypothesis for the $T$-matrix, \Eq{Tscaling0}, is not justified throughout the entire space of possible momentum arguments.
Scaling rather holds, with different exponents, within separate limited scaling regions, for the following reasons.
The scaling exponent $\kappa$ will turn out to be a positive real number such that the momentum occupation numbers $n_{Q}({p})\sim p^{-\kappa}$ grow large in the IR regime of small $p$. 
If the $T$-matrix stays finite in the same regime, as is the case in the perturbative approximation, the WBE can eventually fail.
As we will argue in the following, this problem can, however, be remedied by taking into account in a systematic way non-perturbative collective-scattering effects in the $T$-matrix, beyond the Boltzmann perturbative order of approximation.

\subsubsection{Perturbative region: two-body scattering}
\label{sec:ScalingScattIntPerturb}
For the non-condensed, weakly interacting cold Bose gas away from unitarity 
the $T$-matrix is known to be well approximated by 
\begin{align}
  |T_{\vec p\vec k\vec q\vec r}|^{2}
  =\ &(2\pi)^{4}g^{2} \,,
  \label{eq:Titogbare}
\end{align}
where, in $d=3$ dimensions, $g=4\pi a/m$ is proportional to the $s$-wave scattering length $a$.
\Eq{Titogbare} applies up to an ultraviolet cutoff scale $p_{u}$ and falls off to zero beyond this scale, which ensures the unitarity of the scattering amplitude \cite{Newton1982a}.
$p_{u}\sim1/a$ scales with the inverse of the scattering length $a$ and is typically much larger than the highest significantly occupied momentum mode.
As we discuss in more detail in Appendix \app{KineticScatteringInt}, \Eq{Titogbare} represents the leading perturbative approximation of the full momentum-dependent many-body coupling function.

The matrix elements \eq{Titogbare} are independent of the momenta and thus, $|T_{\vec p\vec k\vec q\vec r}|$ scales according to Eqs.~\eq{Tscaling0}, \eq{Tscaling}, with the scaling exponents
\begin{align}
   m_{\kappa}=m=0 \quad(\mbox{free particles; perturbative})\,. 
  \label{eq:mWWT}
\end{align}

If a condensate with density $\rho_{0}\leq\rho$ is present, the quasiparticle excitations below the healing-length scale $p_{\xi}=\sqrt{2g\rho_{0}m}$ take the form of sound waves on the background of the bulk condensate, cf.~\Sect{ModelObservables}.
The elastic scattering of these sound waves is captured, in leading-order perturbative approximation, by a wave-Boltzmann equation with $T$-matrix  
\footnote{Taking also into account three-wave scattering of Bogoliubov modes involving the condensate is beyond the scope of this article and will be done elsewhere. 
This implies that exchange of particles with the condensate mode is not captured here yet.}
\begin{align}
  &|T_{\vec p\vec k\vec q\vec r}|^{2}
  =\ (2\pi)^{4}\frac{(mc_{s})^{4}}{pkqr} \frac{3g^{2}}{2}  \,,
  \label{eq:TitogbareQP}
\end{align}
where the speed of sound $c_{s}$ is defined in terms of the healing-length momentum scale $mc_{s}=p_{\xi}/\sqrt{2}=\sqrt{g\rho_{0}m}$. 
See \App{KinScattIntBog} for a derivation of the corresponding wave-Boltzmann scattering integral.
According to \Eq{TitogbareQP}, the scaling exponents defined in Eqs.~\eq{Tscaling0}, \eq{Tscaling} are related by
\begin{align}
   m_{\kappa}=m=-2 \quad(\mbox{Bogoliubov sound; perturbative})\,. 
  \label{eq:mWsWT}
\end{align}
We emphasize that the above perturbative expressions are, in general, of limited applicability for solutions showing scaling in the far infrared.
The validity of Eqs.~\eq{Titogbare}, \eq{TitogbareQP},  in the limit $p\to0$ depends crucially on the occupancies $n_{p}$ of the momentum modes.
In the following we discuss highly-occupied free-particle and quasiparticle distributions and the respective modification of the scaling properties of the $T$-matrix. 

\subsubsection{Collective scattering: many-body $T$-matrix}
\label{sec:ScalingScattIntResummed}
Consider as an example the case $d=3$.
With the $T$-matrices \eq{Titogbare} and \eq{TitogbareQP} inserted, contributions to the scattering rates in the integral $I[n_{Q}]({p})$, which are of higher order than $\zeta^{2}n_{Q}^{2}$ and thus beyond the approximation \eq{KinScattIntCWL} are expected to become important  where $\zeta^{2}n_{Q}^{2}\gg1$, with $\zeta=a\rho^{1/3}$ being the diluteness parameter. 
To find universal scaling solutions in such a highly occupied region of strong collective scattering, an approach beyond the Boltzmann, leading-order perturbative approximation is required. 

Here, we use an $s$-channel loop resummation derived within a quantum-field-theoretic approach based on the two-particle irreducible (2PI) effective action or $\Phi$-functional, see \App{2PIKineticTheory}.
The scheme is equivalent to a large-$N$ approximation at next-to-leading order.
It gives an effective momentum-dependent coupling constant $g_\mathrm{eff}(p)$ replacing the bare coupling $g$ in \Eq{Titogbare} and changing the scaling exponent $m$ of the $T$-matrix within the IR region of large occupancies, see Appendices \app{DerivKinEq} and \app{geff} for technical details of the derivation.
This builds on and goes beyond, e.g., Refs.~\cite{Berges:2008wm,Scheppach:2009wu,Orioli:2015dxa,Berges:2015kfa}.
As is furthermore shown in Refs.~\cite{Mikheev2018a.arXiv180710228M,Schmied:2018upn.PhysRevLett.122.170404} the momentum dependence of the effective coupling is recovered in an equivalent momentum dependence of the bare coupling obtained for the phase angles of the Bose fields within a low-energy effective-theory approach.
\\

\emph{Many-body $T$-matrix for scattering of free particles}.---
The resummation scheme provides us with an expression for the $T$-matrix elements which, for free particles ($z=2$), reads
\begin{align}
  |T_{\vec p\vec k\vec q\vec r}|^{2}
  =\ &(2\pi)^{4}g^{2}_\mathrm{eff}(\varepsilon_\mathbf{p}-\varepsilon_\mathbf{r},\vec p-\vec r)
       \,.
  \label{eq:Titogeff}
\end{align}
Here, $\vec p-\vec r$ and $\varepsilon_\mathbf{p}-\varepsilon_\mathbf{r}$ are the momentum and energy transfer in a scattering process, respectively.
$g_\mathrm{eff}$ is an effective coupling function derived in the 2PI approach, as described in detail in \App{2PIKineticTheory}.
Making a scaling ansatz for $n(\vec p,t)$, \Eq{nQbyScalingFunctiongL}, with scaling function of the type \eq{scalingfgL}, the dependence of $g_\mathrm{eff}$ on the energy and momentum arguments can be calculated explicitly, as shown in  \App{EffCouplingFctFree}.
As a result of the ansatz in terms of the scaling function \eq{scalingfgL}, the coupling depends on the IR cutoff scale $p_{\Lambda}$.
%
\begin{figure}[t]
    \centering
    \includegraphics[width=0.4\textwidth]{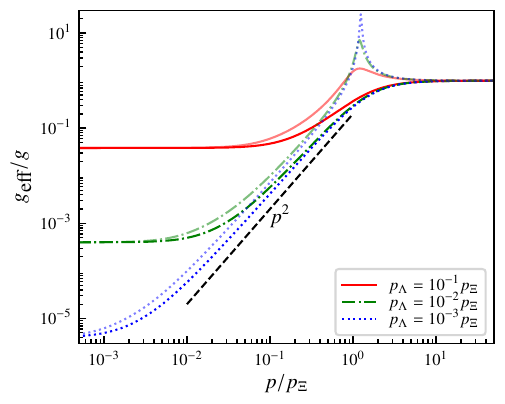}
    \caption{Effective coupling $g_\mathrm{eff}(p_{0},p)$ in $d=3$ dimensions as a function of the spatial momentum  $p=|\vec p|$, on a double-logarithmic scale. 
    The figure shows cuts in the $p_{0}$--$p$-plane, with $p_{0} = 0.5\varepsilon_{\vec p}$ (dark lines) and $p_{0} = 1.5\varepsilon_{\vec p}$ (transparent bright lines). 
    Different colors (line styles) refer to different infrared cutoff scales $p_{\Lambda}$ as listed in the legend.
    $p_{\Lambda}$ is set by the scaling form of the occupation number distribution entering the non-perturbative coupling function, see, e.g., \Fig{SelfSimilar}.
    All momenta are measured in units of the `healing'-length wave number $p_{\Xi}=(2g\rho_\mathrm{nc} m)^{1/2}$ which is  set by the \emph{non-condensed} particle density $\rho_\mathrm{nc}$.
    Note that $p_{\Xi}$ sets the scale separating the perturbative region at large $p$ from the non-perturbative collective-scattering region within which the coupling assumes the form given in \Eq{geffFreeUniversal}.
    See   \Fig{EffCouplingFree} for the dependence in the full $p_{0}$--$p$-plane.
    }
    \label{fig:EffCouplingfreeCuts-b}
\end{figure}
%
%
\begin{figure}[t]
    \centering
    \includegraphics[width=0.4\textwidth]{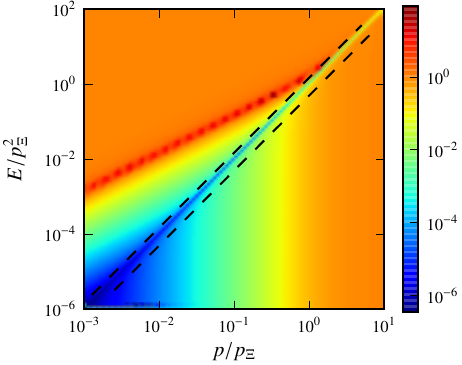}
    \caption{Contour plot of the effective coupling function $g_\mathrm{eff}(E,p)/g$ of free particles, for $d=3$, as a function of $E = 2mp^0$ and $p=|\mathbf{p}|$, for $p_\Lambda = 10^{-3}p_{\Xi}$. 
    This function enters the $T$-matrix as defined in \Eq{Titogeff}. 
    Cuts through this function, for $E = 1.5\,p^{2}$ and $E = 0.5\,p^{2}$ (black dashed lines), are shown in \Fig{EffCouplingfreeCuts-b}.
    The quasiparticle distribution $n_{Q}$ was chosen to scale with $\kappa = 3.5$.
    The coupling function in the collective-scattering region \eq{CollScattRegime}, is well described by the universal form \eq{geffFreeUniversal} which does not depend on $\kappa$.
    The scattering resonance marked by the dark red `ridge' is centered around the Bogoliubov-type energy-momentum transfer $E=[2p_{\Xi}^{2}p^{2}+p^{4}]^{1/2}$.
    }
    \label{fig:EffCouplingFree}
\end{figure}
%

\Fig{EffCouplingfreeCuts-b} shows the resulting momentum-dependent effective coupling function $g_{\mathrm{eff}}(p_{0},\vec p)$ along exemplary cuts $p_{0} = 0.5\varepsilon_{\vec p}$ and $p_{0} = 1.5\varepsilon_{\vec p}$ in frequency-momentum space, for different infrared cutoffs $p_{\Lambda}$.
The function $g_{\mathrm{eff}}$ is plotted as a function of $p/p_{\Xi}$, where $p_{\Xi}$ is defined as
\begin{align}
   p_{\Xi}=\sqrt{2g\rho_\mathrm{nc} m}\,.
  \label{eq:pmu}
\end{align}
$p_{\Xi}$ is the analogue of the healing-length scale $p_{\xi}$, cf.~\Eq{healingmomentum}, which, instead of the condensate density $\rho_{0}$, is set by the \emph{non-condensed} particle density $\rho_\mathrm{nc}=\rho_\mathrm{tot}-\rho_{0}$, recall \Eq{ParticleDensity}.

At large momenta, the effective coupling is constant and agrees with the perturbative result, $g_{\mathrm{eff}}=g$. 
However, below $p_{\Xi}$, the coupling deviates from the perturbative result. 
Within
\begin{align}
   p_{\Lambda}\ll p\ll p_\mathrm{np}=p_{\Xi}\, ,
  \label{eq:CollScattRegime}
\end{align}
the effective coupling assumes the universal scaling form
\begin{equation}
 \label{eq:geffFreeUniversal}
 g_{\mathrm{eff}}(p_{0},\vec p) 
 \simeq \frac{\left|\varepsilon_{\vec p}^{2}-p_{0}^{2}\right|}{2\rho_\mathrm{nc}\, \varepsilon_{\vec p}}\,,
 \qquad(\kappa>3)
\end{equation}
independent of both, the microscopic interaction constant $g$, and the particular value of the scaling exponent $\kappa$ of $n_{Q}$, see  \App{EffCouplingFctFree}.
Below the IR cutoff, $p<p_{\Lambda}$, the coupling settles in to become constant again.

Note, in particular, that this low-momentum cutoff, for the case of universal scaling evolution at a non-thermal fixed point, \eq{NTFPscaling0nQ}, implicitly leads to a time dependence of $g_{\mathrm{eff}}(p_{0},\vec p;t)$ as $p_{\Lambda}(t)\sim t^{-\beta}$ decreases in time, cf.~\Eq{SelfSimilarScaling}.

Before we discuss our results further, we briefly remark that the effective coupling $g_{\mathrm{eff}}$ has a slightly more complicated structure than what has been discussed so far. 
The areas of constant $g_{\mathrm{eff}}$ shown in \Fig{EffCouplingFree} (blue to orange) for $E = 2mp_{0}>0$ reflect the  dependence on $p_{0}$ and $p$ of the form \eq{geffFreeUniversal}. 
The two dashed black diagonal lines in \Fig{EffCouplingFree} correspond to the exemplary cuts $E=1.5\,p^{2}$ and $E = 0.5\,p^{2}$ along which $g_{\mathrm{eff}}(E(p),\vec p)$ is depicted in \Fig{EffCouplingfreeCuts-b}.
The coupling features, along the former cut, a peak at $p\simeq1.5 p_{\Xi}$. 
This peak can be seen at different frequencies and momenta, shown as a red band in \Fig{EffCouplingFree}. 
It appears, in fact, at the energy transfer 
\begin{equation}
E(p)=[2p_{\Xi}^{2}p^{2}+p^{4}]^{1/2} 
\end{equation}
taking Bogoliubov form for a momentum transfer $p=|\vec{p}|\lesssim p_{\Xi}$, and thus has the form of a many-body scattering resonance indicating the exchange of  sound-like quasiparticles for $p\ll p_{\Xi}$, see also \App{EffCouplingFctFree}. 
According to our numerical evaluation, the peak does not seem to play a visible role for the wave-turbulent scattering integrals discussed later.

\Fig{EffCouplingFree} furthermore indicates that the coupling for exchange of $E(p)=p^{2}$ is suppressed and scales differently, seen as a narrow line along the diagonal, $E=p^{2}$.
This does not appear to affect the scaling of the solutions of the kinetic equations either. 
One finds in this case  $g_\mathrm{eff}(E(p),\vec p)$ $\sim p$, see the discussion of \Fig{EffCouplingfreeCuts-a} in \App{EffCouplingFctFree}, and \cite{Walz:2017ffj}. 

\emph{Universality of the effective coupling}.---
The simple universal form \eq{geffFreeUniversal} of the effective coupling in the collective-scattering regime is one of the central results of the present work.
It results from the $s$-channel resummation and only requires the quasiparticle occupancies $n_{Q}(p)\sim p^{-\kappa}$ to show a sufficiently steep power law, $\kappa>3=d$, and to be regularized in the IR below the momentum scale $p_{\Lambda}$ such that $n(\vec p)$ gives a finite particle number and energy, cf.~\App{EffCouplingFctFree} for details of the derivation.

Note that, while expression \eq{geffFreeUniversal} is independent of g, the dependence of the full effective coupling on the bare coupling $g$ appears in the fact that the  universal increase of $g_\mathrm{eff}(p_{0},p)$ with its increasing arguments $p_{0}$ and $p$ reaches $g$ at the scales  $p\simeq p_{\Xi}$ and $p_{0}\simeq\varepsilon(p_{\Xi})$ which, in turn, depend on $g$.

Note furthermore that the effective coupling is universal in a similar sense as in the unitary limit of $g\to\infty$, as at a Feshbach resonance. 
There the quantum corrections to the self energy which have been neglected here lead to a UV scaling $|g_\mathrm{eff}|^{2}\sim p^{-2}$ as $p\to\infty$.

We strongly emphasize that the result \eq{geffFreeUniversal} is independent of $\kappa$ if $\kappa>3$, see \App{Universalitygeffkappagtthree} for details.
In \Sect{ScalingAnalysisKinEqs} we will show that, at an IR non-thermal fixed point, one has $\kappa>d$ if $z>4/3$, which is fulfilled for the free particles considered here.
Hence, universality is ensured in the case of $d=3$ dimensions.

The scaling properties of $g_\mathrm{eff}$ allow us to confirm the scaling hypothesis \eq{Tscaling} for the $T$-matrix. 
Within the perturbative regime, the coupling function scales as
\begin{align}
  g_\mathrm{eff}(p_{0},\vec p) 
  = s^{-\gamma_{\kappa}}g_\mathrm{eff}(s^{z}p_{0},s\vec p)
       \,,
  \label{eq:geffscaling}
\end{align}
with $z=2$ and
\begin{align}
  \gamma_{\kappa}=0\qquad\mbox{(perturbative regime)}\,,
  \label{eq:gammapert}
\end{align}
while, in the collective-scattering regime, 
\begin{align}
  \gamma_{\kappa}=2\qquad\mbox{(collective-scattering regime)}\,.
  \label{eq:gammanonpert}
\end{align}
Together with \Eq{Titogeff} this gives the scaling exponent 
\begin{align}
   m_{\kappa}=2  \quad(\mbox{free particles; collective})\, 
  \label{eq:mSWT}
\end{align}
of the $T$-matrix in the collective-scattering regime \eq{CollScattRegime}.
We will discuss the scaling properties for general $d$ and $z$ in the collective-scattering regime in more detail in \Sect{TempScalinggeff}.

We emphasize that, according to the above non-perturbative results, the breakdown of the perturbative wave-Boltzmann scattering integral which appears due to the rise of occupation numbers in the IR, is counteracted by a strong power-law fall-off of the scattering $T$-matrix.
At very low scales, below the IR cutoff $p_{\Lambda}$, the effective coupling saturates again to a much smaller constant.
This saturation occurs due to the growth of occupation numbers being regularized to ensure convergence of physical quantities such as particle and energy densities.
As a consequence, this effectively reinstates in this lowest-$p$ regime the same scaling as in the perturbative region albeit at a constant value generally different from the perturbative approximation. 

We finally remark that, as can be shown within a Luttinger-Liquid-type low-energy effective theory \cite{Mikheev2018a.arXiv180710228M}, the power-law suppression of the coupling in the IR limit can be understood, in a multi-component system, to be due to the fact that relative phases and densities in a multicomponent system fluctuate similarly while only the total density fluctuations are suppressed by the repulsive interactions.

\emph{Many-body $T$-matrix for Bogoliubov sound waves}.---
If the strongly occupied IR modes have a Bogoliubov-like dispersion scaling with $z=1$, the $T$-matrix elements are calculated analogously, as we show in \App{EffCouplingFctBog}.

\subsection{Summary}
\label{sec:SummaryKinetic}
The kinetic approach outlined in this section forms the basis of the scaling analysis presented in the following sections.
Using a non-perturbative large-$N$ approximation of Kadanoff-Baym type equations of motion for single-particle Greens functions we derived kinetic equations of the quantum-Boltzmann form \eq{KinScattIntCWL}, with a scattering $T$-matrix which depends on the distribution function itself.
The analytic form \eq{geffFreeUniversal} of the universal effective coupling function defining the momentum-frequency dependence \eq{Titogeff} of the $T$-matrix is a central result of the present work. 
It decreases the effective coupling below the bare coupling $g$ where it shows a universal power-law form, see \Fig{EffCouplingFree}. 

\section{Scaling analysis of the kinetic equations}
\label{sec:ScalingAnalysisKinEqs}
We are now in the position to derive the scaling properties of the solutions to the kinetic equations \eq{WBKinEq}, \eq{WBKinScattInt}.
Before we proceed with this final step we discuss the generalization of the scaling properties of the effective coupling function, and thus of the $T$-matrix and the scattering integral to the time-dependent case, on the basis of a scaling analysis of the self-energy entering the  field equations in the Schwinger-Keldysh formalism.
This will allow us to predict the scaling of the many-body $T$ matrix for general dimensions $d$ and dynamical exponent $z$, as well as a possible anomalous scaling dimension $\eta$ to be introduced in the following.

\subsection{Field theoretic description of the scattering integral}
\label{sec:FieldTheoryScattInt}
In order to perform the generalized scaling analysis we first need to summarize how the effective coupling $g_{\mathrm{eff}}(p)$ is obtained.
We follow for this a field theoretic approach, from which we derive the kinetic equation and the scattering integral, see Apps.~\app{2PIKineticTheory} and \app{DerivKinEq} for details.
We determine $g_{\mathrm{eff}}(p)$, which defines the $T$-matrix elements \eq{Titogeff} by means of an $s$-channel loop resummation. 
As a result, it can be written as
\begin{align}
    g_{\mathrm{eff}}(p_0,\vec p;t)
    &= \frac{g}{\left|1+g\,\Pi^R(p_0,\vec p;t)\right|},
\label{eq:geff}
\end{align}
see \Fig{2PI} for a diagrammatic representation.
The retarded one-loop self-energy $\Pi^R(p_0,\vec p;t)$ is defined by
\begin{align}
  &\Pi^R(p_0,\vec p) 
  = \frac{i}{2} \int \frac{\mathrm{d}q_0}{2\pi} \frac{1}{q_0+i\epsilon} \int \frac{\mathrm{d}^{d}k\,\mathrm{d}k_0}{(2\pi)^{d+1}}  
  \nonumber \\
  &\qquad \times \big[ 
    \rho_{ab}(p_0-q_0+k_0,\vec p + \vec k)\,
    F_{ba}(k_0,\vec k;t) 
    \nonumber \\
  &\qquad \quad -\ 
    F_{ab}(p_0-q_0+k_0,\vec p + \vec k;t)\, 
    \rho_{ba}(k_0,\vec k)
    \big]\,,
\label{eq:PiRdefAppG}
\end{align}
cf.~Eqs.~\eq{PiA}, \eq{PiFrhopStatClass}, in terms of the statistical and spectral functions
\begin{align}
  \label{eq:mainT:Fmatrix}
  F_{ab}(x,y)&=\langle\{\Phi_{ a}(x),\Phi_{\bar b}(y)\}\rangle/2\,,
  \\
  \rho_{ab}(x,y)&=i\langle[\Phi_{ a}(x),\Phi_{\bar b}(y)]\rangle\,,
  \label{eq:mainT:rhomatrix}
\end{align}
with indices $a,b\in\{1,2\}$, with $\bar a=3-a$ and $\Phi_{1}(x)\equiv\Phi(x)$, $\Phi_{2}(x)\equiv\Phi^{\dagger}(x)$, and their momentum-space analogues in \Eq{PiRdefAppG} follow from Fourier transformation with respect to the four-vector $x-y$ while $t=(x_{0}+y_{0})/2$.
Within the kinetic approximation considered in this work, $\rho(p_0,\vec p)$ remains constant in time $t$.

The scaling properties of $F$ and $\rho$ are defined by 
\begin{align}
\label{eq:mainT:ScalingF}
  F_{ab}(p_{0},\mathbf{p};t)
  &= s^{2-\eta+\alpha/\beta}\,F_{ab}(s^{z}p_{0},s\mathbf{p};s^{-1/\beta}t)\,,
   \\
\label{eq:mainT:ScalingFstat}
  F_{ab}(p_{0},\mathbf{p};t_{0})
  &= s^{2-\eta+\kappa}\,F_{ab}(s^{z}p_0,s\mathbf{p};t_{0})\,,
  \\
\label{eq:mainT:Scalingrho}
   \rho_{ab}(p_0,\mathbf{p})
   &= s^{2-\eta}\,\rho_{ab}(s^{z}p_0,s\mathbf{p})\,.
\end{align}
Besides the already known exponents,  $\eta$ appears as an anomalous scaling dimension of the spectral function 
\footnoteremember{fn4}{We point out that we choose a notation different from that in earlier work, e.g.~Refs.~\cite{Berges:2008wm,Scheppach:2009wu,Orioli:2015dxa}, effectively replacing $\kappa\to\kappa-\eta$, to take into account that in equilibrium, both, $\rho$ and $F$ scale with the same anomalous dimension $\eta$ while out of equilibrium the scaling dimension of $F$ gets modified by $\kappa\not=z$, cf.~\App{RelEqCrPh}.
Note furthermore that the scaling \eq{mainT:ScalingF} along the center-time axis, $x_{0}=y_{0}=t$, is different in character from scaling known in the context of initial-slip and ageing dynamics \cite{Janssen1989a,Janssen1992a,Calabrese2002a.PhysRevE.65.066120,Calabrese2005a.JPA38.05.R133,Gambassi2006a.JPAConfSer.40.2006.13}.
See \App{RelEqCrPh}.}.

For $z=1$ and $z=2$, $\rho(p_0,\vec p)$ is given in Eqs.~\eq{rhofree} and \eq{rhoBog}, respectively, encoding the dispersion relation as well as the density of states.
For the free quasiparticles considered here, $\eta$ modifies the scaling of the density of states,
\begin{align}
 \rho_{ab}(p_0,\mathbf{p}) &\sim \frac{i}{p^{2-\eta-z}}\,
 \big[ B_{ab}\delta(p_{0}-\omega_{\vec p})-B'_{ab}\delta(p_{0}+\omega_{\vec p})\big]\,,
  \label{eq:mainText:rhoGenDispersion}
\end{align}
where $B$ and $B'$ are constant matrices, cf.~\App{ScalingHypoth}.

The particle and quasiparticle distributions, obtained by frequency integrations over $F$, 
\begin{align}
 \label{eq:nDefF}
 n(\vec p,t) 
 &= \int_{0}^{\infty}\frac{\mathrm{d}\omega}{2\pi}\mathrm{Tr}F(\omega,\vec{p};t)-1/2 \,,
 \\
 \label{eq:nQDefF}
 n_{Q}(\vec p,t) 
 &= \int_{0}^{\infty}\frac{\mathrm{d}\omega}{2\pi}\mathrm{Tr}[\sigma^{3}F(\omega,\vec{p};t)]-1/2 \,,
\end{align}
scale, according to Eqs.~\eq{mainT:ScalingF}, \eq{mainT:ScalingFstat}, as
\begin{align}
  n(\vec p,t)&=s^{\alpha/\beta-\eta+2-z}\,n(s\vec p,s^{-1/\beta}t)\,,
  \label{eq:app:nptScaling}
  \\
  n(\vec p,t_{0})&=s^{\kappa-\eta+2-z}\,n(s\vec p,t_{0})=s^{\zeta}\,n(s\vec p,t_{0})\,,
  \label{eq:app:npt0Scaling}
  \\
  n_{Q}(\vec p,t)&=s^{\alpha/\beta}\,n_{Q}(s\vec p,s^{-1/\beta}t)\,,
  \label{eq:app:nQptScaling}
  \\
  n_{Q}(\vec p,t_{0})&=s^{\kappa}\,n_{Q}(s\vec p,t_{0})\,,
  \label{eq:app:nQpt0Scaling}
\end{align}
which for $\eta=0$ is consistent with Eqs.~\eq{npPowerLaw}, \eq{nQpPowerLaw}, and \eq{NTFPscaling0nQ}.

In the perturbative regime, the contribution from the self-energy $\Pi^{R}$ in the denominator of the coupling function \eq{geff} can be neglected as compared to the $1$, and $g_\mathrm{eff}\equiv g$.
In the collective-scattering regime, the scaling behaviour of $g_\mathrm{eff}$,
\begin{align}
  g_\mathrm{eff}(p_0,\vec p;t) 
  &= s^{-\gamma}g_\mathrm{eff}(s^{z}p_0,s\vec p;s^{-1/\beta}t)
       \,,
  \label{eq:geffScalingHypothesis0}
  \\
  g_\mathrm{eff}(p_0,\vec p;t_{0}) 
  &= s^{-\gamma_{\kappa}}g_\mathrm{eff}(s^{z}p_0,s\vec p;t_{0})
       \,,
  \label{eq:geffScalingHypothesis}
\end{align}
can be determined by inserting the scaling of $F$ and $\rho$ into \Eq{PiRdefAppG}.
Straightforward power counting gives
\begin{align}
  \Pi^R &(p_{0},\vec p;t)=s^{\alpha/\beta-d+z+2(2-z-\eta)}\,\Pi^R (s^{z}p_{0},s\vec p;s^{-1/\beta}t)\,.
  \label{eq:PiRScalingpt}
\end{align}
The scaling analysis at a fixed time $t_{0}$ requires more care. 
If the integral \eq{PiRdefAppG} is infrared divergent  it requires the quasiparticle distribution $n_{Q}$ entering $F$ to be regularized as discussed earlier and will then be sensitive only to the infrared part of the distribution. 
In this case it leads to a universal coupling and with this to the scaling of the solution of the kinetic equation at an IR non-thermal fixed point.
On the contrary, if the integral is IR-finite it becomes sensitive to the UV part of $n_{Q}$ and in general will not give rise to fixed-point scaling.

In the IR-dominated case, the integral $\Pi^{R}$ depends also on the cutoff scale $p_{\Lambda}$ such that simple power counting gives
\begin{align}
  \Pi^R &(p_{0},\vec p;t_{0};p_{\Lambda})=s^{\kappa-d-z+2(2-\eta)}\,\Pi^R (s^{z}p_{0},s\vec p;t_{0};sp_{\Lambda})\,,
  \label{eq:PiRScalingppL}
\end{align}
and it remains, at first, unclear, to what extent the scaling is provided by the rescaling of $p_{\Lambda}$ and how much by that of $p$.

\subsubsection{Universal scaling at an IR dominated non-thermal fixed point}
Inspection of the integral \eq{PiRdefAppG} allows to infer the contribution from the rescaling of $p_{\Lambda}$ from the power of the divergence of the integral.
The terms in the integrand in square brackets give rise to divergencies where the arguments of $\rho$ \emph{or} $F$ vanish. 
However, for $p\gg p_{\Lambda}$, the arguments of $F$ and $\rho$ in each product can not be zero simultaneously.
Among the different divergent terms, for $\kappa>0$, those arising from $F$ dominate the integral, compare~Eqs.~\eq{mainT:ScalingFstat} and \eq{mainT:Scalingrho}. 
As a result, the divergences originating from $F$ determine the scaling in $p_{\Lambda}$, 
\begin{align}
  \Pi^R &(p_{0},\vec p;t_{0};p_{\Lambda})=s^{\kappa-d-z+2-\eta}\,\Pi^R (p_{0},\vec p;t_{0};sp_{\Lambda})\,.
  \label{eq:PiRscaling-in-pL}
\end{align}
This scaling behaviour is consistent with Eqs.~\eq{PiRFree_fepsilonk} and \eq{app:Pif_free_withf} in \App{EffCouplingFctFree} where we exemplarily discuss the case $d=3$, $z=2$, $\eta=0$ 
\footnote{$\Pi^{R}$ according to \Eq{app:Pif_free_withf} scales as $p_{\Lambda}^{2-\kappa}$, and the integral over $y$ in \eq{app:Pif_free_withf} scales, by Eqs.~\eq{sol_pi_free} and \eq{pitilde_xgge}, as $\sim x^{-1}\sim p_{\Lambda}/p$, giving in total $\Pi^{R}\sim p_{\Lambda}^{3-\kappa}$, as in \eq{PiRscaling-in-pL} for $d=3$, $\eta=0$, $z=2$.}.
Factoring  \eq{PiRscaling-in-pL} out of \eq{PiRScalingppL} leaves
\begin{align}
  \Pi^R &(p_{0},\vec p;t_{0})=s^{2-\eta}\,\Pi^R (s^{z}p_{0},s\vec p;t_{0})\,,
  \label{eq:PiRScalingp}
\end{align}
as for the spectral function \eq{mainT:Scalingrho} itself.
We further note that, according to \Eq{PiRscaling-in-pL}, if 
\begin{align}
  \kappa>d+z-2+\eta\,,\quad(\mbox{IR dominance})
    \label{eq:NTFPcondition}
\end{align}
the value of $\Pi^{R}$ increases, for fixed $p$ and $p_{0}$, when $p_{\Lambda}$ is lowered, decreasing $g_\mathrm{eff}$ accordingly, cf.~\Eq{geff}.
We also see that in this case the total density $\rho_\mathrm{nc}=\int_{\vec p}n(\vec p,t_{0})$, whose scaling is determined by \eq{app:npt0Scaling}, is dominated by the IR cutoff, i.e., $\rho_\mathrm{nc}\sim p_{\Lambda}^{d}(\Lambda/p_{\Lambda})^{\kappa-z+2-\eta}$, where the fixed scale parameter $\Lambda$ adjusts the `engineering' dimension to that of the density $\rho_\mathrm{nc}$. 

Hence, assuming $\rho_\mathrm{nc}=const.$, the dependence of $\Pi^{R}$ on $p_{\Lambda}$ can be exchanged for a dependence on the invariant density, leaving a universal momentum-energy dependence $g\Pi^{R}(p)\sim p^{-2+\eta}g\rho_\mathrm{nc}\sim p^{-2+\eta}p_{\Xi}^{2}$. 

As a result, for $p\ll p_{\Xi}$, the $1$ in the denominator of the coupling \eq{geff} can be neglected compared with $g\Pi^{R}$, leaving a universal $g_\mathrm{eff}$ of the type found earlier, cf.~\Eq{geffFreeUniversal}.
We will later show that, for our IR non-thermal fixed points, the condition \eq{NTFPcondition} is fulfilled, such that \eq{PiRScalingp} defines the universal IR momentum scaling of the effective coupling function \eq{geff}.

\subsubsection{Thermal case and wave turbulence}
On the other hand, weak wave turbulence and thermal states in general have exponents $\kappa<d+z-2+\eta$.
From the scaling \eq{PiRscaling-in-pL} of $\Pi^{R}$ in $p_{\Lambda}$ it follows that for $\kappa< d+z-2+\eta$ the integral \eq{PiRdefAppG} is no longer IR divergent and thus sensitive also to the UV end, while the dominant contribution in $p_{\Lambda}$ still arises from the $F$ functions.
Moreover, the overall scaling \eq{PiRScalingppL} implies that for $\kappa>d+z-2(2-\eta)$ the integral is UV-finite 
\footnoteremember{fn2}{The integral \eq{sol_pi_free} in the case $z=2$, $d=3$, $\eta=0$ is UV divergent already for $\kappa<2$. However, for $\kappa>d+z-2(2-\eta)$, the leading divergence is an imaginary constant, the last term in \Eq{pikappa-one-x-pl} which drops out when inserting $\tilde\pi_{\kappa}$ into $\Pi^{R}$.}.  
For $0<\kappa< d+z-2(2-\eta)$ \footnoterecall{fn2}, the integral \eq{PiRdefAppG} is UV divergent, so the cutoff $p_{\lambda}$ is needed, which for $\kappa> d+z-2(2-\eta)$ is the case only for the normalisation.

In these cases, in the UV, both, $F$, and $\rho$ are equally important in \eq{PiRdefAppG}, giving a momentum scaling, cf.~\eq{PiRScalingp},
\begin{align}
  \Pi^R &(p_{0},\vec p;t_{0})=s^{\kappa-d-z+2(2-\eta)}\,\Pi^R (s^{z}p_{0},s\vec p;t_{0})\,
  \label{eq:PiRScalingpUVdiv}
\end{align}
for fixed IR and UV cutoff scales, i.e., $g\Pi^{R}(p)\sim p^{-\kappa+d+z-4+2\eta}$, consistent with \Eq{gPiR-scaling-regime-kltd} for $d=3$, $z=2$, and $\eta=0$.

Moreover, as the total density is dominated, for $\kappa< d+z-2+\eta$, by the UV end of the spectrum $n(\vec p)$, i.e., $\rho_\mathrm{nc}\sim p_{\lambda}^{d}(\Lambda/p_{\lambda})^{\kappa-z+2-\eta}$, substituting $p_{\Lambda}$ for $p_{\Xi}$ leads to an overall scaling $g\Pi^{R}(p)\sim p^{-2+\eta}(p_{\lambda}/p)^{\kappa-d-z+2-\eta}p_{\Xi}^{2}$, cf.~the explicit example in \Eq{gPiR-scaling-regime-kltd}.

We point out that in Refs.~\cite{Berges:2008wm,Berges:2008sr,Scheppach:2009wu,Berges:2010ez}, the scaling \eq{PiRScalingpUVdiv} was employed to derive stationary wave-turbulent scaling exponents, while rather Eq.~\eq{PiRScalingp} applies.
The temporal scaling analysis in Ref.~\cite{Orioli:2015dxa}, though, is not affected by the IR divergence of the $\Pi^{R}$ integral and, for $\eta=0$, gave the result  \eq{PiRScalingpt}.

\subsection{General scaling of the effective coupling $g_\mathrm{eff}(p,t)$}
\label{sec:TempScalinggeff}
The scaling properties presented in the previous section are used, in the following, to derive general scaling relations between the exponents $\gamma$, $\gamma_{\kappa}$, $m$, $m_{\kappa}$, $\alpha/\beta$, $\kappa$, $z$, and $\eta$, for spatial dimension $d$.
We start with the scaling in time and momentum.
\Eq{PiRScalingpt} implies a scaling of the effective coupling according to \Eq{geffScalingHypothesis0}, with
\begin{align}
  \gamma=\alpha/\beta-d-z+4-2\eta\,.
  \label{eq:gammaalphabetadzFree}
\end{align}
Comparison of Eqs.~\eq{Tscaling0}, \eq{Titogeff}, \eq{TitogeffQP}, and \eq{geffScalingHypothesis0} yields 
\begin{align}
  m=\gamma+2z-4+2\eta\,
  \label{eq:mitogamma}
\end{align}
between $\gamma$ and the scaling exponent $m$ of the $T$-matrix. 
Combining Eqs.~\eq{gammaalphabetadzFree} and \eq{mitogamma} eliminates $\gamma$ and $\eta$, 
\begin{align}
  m=\alpha/\beta-d+z
  \qquad\mbox{(collective scattering)}\,.
  \label{eq:malphabetadzFree}
\end{align}
Note that, both, the distribution function $n_{Q}(p,t)$ and the $T$-matrix  show a spatio-temporal scaling independent of $\eta$, cf.~\Eq{app:nQptScaling} for $n_{Q}$.
Hence, the anomalous dimension is not affected by the scaling properties of the kinetic equation and, in turn, can not depend on them.
This result is consistent with the fact that $\eta$ accounts for the scaling of the spectral properties which remain unchanged during the kinetic evolution within the leading-order gradient approximation \cite{Branschadel:2008sk} of the scattering integral considered here.

As we have anticipated in \Sect{UniversalDynamics} and will discuss further in the forthcoming sections, depending on the type of scaling evolution, the quotient $\alpha/\beta$ is either fixed by the global conservation of quasiparticle number, $\alpha/\beta=d$, cf. \eq{alphabetaENConserv}, or by the momentum scaling law, $\alpha/\beta=\kappa$, cf.~\Tab{ScalingRelConservationLaws}. 
In the first case, i.e., for (bi-directional) self-similar evolution, one finds 
\begin{align}
  m=z\quad\mbox{(self-similar evol., collect. scatt.)}\,.
  \label{eq:mselfsimz}
\end{align}
In the case of wave-turbulent cascades, the exponent reads 
\begin{align}
  m=\kappa-d+z\quad\mbox{(cascade, collective scatt.)}\,.
  \label{eq:mWTz}
\end{align}

In analogy to the above, the momentum scaling at fixed time $t_{0}$ defined by \Eq{geffScalingHypothesis} is governed by \Eq{PiRScalingp},
\begin{align}
  \gamma_{\kappa}=2-\eta\,,\qquad(\mbox{collective scatt., universal})
  \label{eq:gammakappageneral}
\end{align}
provided that \eq{NTFPcondition} is satisfied, which we will later confirm to be the case for $\eta<2$.
Comparison of Eqs.~\eq{Tscaling}, \eq{Titogeff}, \eq{TitogeffQP}, and \eq{geffScalingHypothesis} gives the general relation between the $T$-matrix and $g_\mathrm{eff}$ exponents at fixed time $t_{0}$,
\begin{align}
  m_{\kappa}=\gamma_{\kappa}+2z-4+2\eta\,.
  \label{eq:mkappaitogammakappa}
\end{align}
Hence, inserting Eq.~\eq{gammakappageneral}  yields
\begin{align}
  m_{\kappa}=2(z-1)+\eta\qquad\mbox{(collective scatt., universal)}\,,
  \label{eq:mkappazBog}
\end{align}
which for $\eta=0$ is consistent with our earlier results \eq{mSWT}, \eq{mSsWT}.
Note that, as is the case for the universal scaling forms \eq{geffFreeUniversal} and \eq{app:geffBogUniversal}, the exponents $\gamma_{\kappa}$ and $m_{\kappa}$ are independent of $\kappa$ in the collective-scattering regime, cf.~Eqs.~\eq{gammakappageneral} and \eq{mkappazBog}.

The above relations generalize our results obtained in \Sect{ScalingScattIntResummed} for the momentum scaling exponents $\gamma_{\kappa}$ and $m_{\kappa}$ in $d=3$ and for Bogoliubov sound ($z=1$) and free particles ($z=2$), to arbitrary dimensions and dynamical exponents $z$, including a possible anomalous dimension $\eta$.

Finally, if $\kappa< d+z-2+\eta$, i.e., outside the universal scaling regime \eq{NTFPcondition}, the UV-sensitive coupling scales as \eq{PiRScalingpUVdiv}, 
\begin{align}
  \gamma_{\kappa}=\kappa-d-z+2(2-\eta)\,,\quad(\mbox{collective, non-univ.})
  \label{eq:gammakappageneral-nonuniv}
\end{align}
and thus 
\begin{align}
  m_{\kappa}=\kappa-d+z\qquad\mbox{(collective scatt., non-univ.)}\,
  \label{eq:mkappazBog-nonuniv}
\end{align}
%

\subsection{Scaling analysis of the kinetic equation}
\label{sec:ScalingSols}
We are now in the position to use the scaling properties of the scattering integral derived in the previous subsections to determine the remaining exponents of the scaling solutions of the wave-Boltzmann equations.
Recall that we need to distinguish, depending on the value of the exponent $\kappa$ of $n_{Q}\sim p^{-\kappa}$, between a self-similar rescaling of $n_{Q}$, with bi-directional transport of particles and energy in momentum space, and the build-up of wave-turbulence cascades behind a wave front, cf.~\Tab{ScalingRelConservationLaws} and Figs.~\fig{SelfSimilar}, \fig{WWTCascade}.
In the following we will analyse the kinetic equation separately for these cases, starting with the self-similar evolution of the quasiparticle spectrum.
It will turn out that this type of scaling evolution only applies in the non-perturbative regime while wave-turbulent cascades exist predominantly in the perturbative case.

Our results for the different types of scaling dynamics will be summarized in a compact way in \Tab{ScalingExponents} and \Sect{SummaryNumerical}.

\subsubsection{Self-similar scaling evolution}
\label{sec:SelfSimKinetics}
%
\emph{Uni-directional self-similar transport}.---
We begin with the scaling form \eq{NTFPscaling} for $n_{Q}$, with one non-universal momentum scale, disregarding for a moment a possible second cutoff, and insert it into the left-hand side of the wave kinetic equation \eq{WBKinEq}.
Using \Eq{IScalingForm0}, one finds
\begin{align}
  &\left.(t/t_{0})^{\alpha-1}(\alpha + \beta\, q \,\partial_{ q})f(q)\right|_{q=(t/t_{0})^{\beta} p}
  \nonumber\\
  &\qquad\qquad=\   t_{0}(t/t_{0})^{-\beta\mu}I[f]([t/t_{0}]^{\beta} p)\,.
  \label{eq:NTFPScalingKinEq0}
\end{align}
For the kinetic equation \eq{NTFPScalingKinEq0} 
to hold at any momentum $p$ and time $t$ within the scaling regime, in particular at $t=t_{0}$, the time-independent fixed-point equation
\begin{align}
  (\alpha + \beta\,p\,\partial_{ p})f(p)
  &=   t_{0}I[f]( p)
  \label{eq:FPEq}
\end{align}
for the scaling function $f(p)=n_{Q}(p,t_{0})$ needs to hold.
Moreover,  both sides of \Eq{NTFPScalingKinEq0}, at a given momentum $q$, scale in the same way in time only if the scaling relation
\begin{align}
  \alpha =1-\beta\mu\,
  \label{eq:IScalingRelation}
\end{align}
is satisfied.
Inserting the scaling exponent \eq{muExponent} of the collision integral this relation implies
\begin{align}
  \alpha = \beta[d+ m-z/2]-1/2\,.
  \label{eq:alphaofdmz}
\end{align}
Combining \Eq{alphaofdmz} with the scaling relations \eq{alphabetaNConserv} and \eq{alphabetaEConserv} which arise from the global conservation of quasiparticle and energy density, respectively, yields the exponents $\alpha$ and $\beta$ describing the scaling evolution near the respective non-thermal fixed point,
\begin{align}
 \beta&=(2m-z)^{-1} &&(\mbox{number conservation})\,,
 \label{eq:betaSolutionN}
 \\
 \beta&=(2 m-3z)^{-1} &&(\mbox{energy conservation})\,.
 \label{eq:betaSolutionE}
\end{align}
If the scaling function $f$ in the scaling form \eq{NTFPscaling} is parametrized by \Eq{scalingf}, the exponents $\alpha$, \Eq{alphaofdmz}, and $\beta$, \Eq{betaSolutionN} or \eq{betaSolutionE}, describe the rescaling of the parameters $f_{\Lambda}(t)=n_{Q}(p_{\Lambda}(t),t)\sim t^{\alpha}$, $p_{\Lambda}(t)\sim t^{-\beta}$ in time.

With the kinetic equation \eq{NTFPScalingKinEq0} at hand we can now also infer the momentum exponent $\kappa$ characterizing the scaling function \eq{scalingf}.
Let us assume that either $\bar\kappa=0$ provides an IR regularization and $f(p\to\infty)\sim p^{-\kappa}$, or that $\bar\kappa\to\infty$ regularizes the scaling function in the UV and $f(p\to0)\sim p^{-\kappa}$.

In both cases, the scaling function must obey, in the region where $f(p)\sim p^{-\kappa}$, the fixed-point equation \eq{FPEq}, i.e.
\begin{align}
  (\alpha - \beta\kappa)f(p) &= t_{0}I[f](p)\,.
  \label{eq:FPEqkappa}
\end{align}
If both sides of \Eq{FPEqkappa} are non-zero, they must scale in the same way in $p$. 
This means that the momentum exponent $\kappa_{S}$ of the self-similarly evolving scaling form  is $\kappa_{S}=-\mu_{\kappa}$, cf.~Eqs.~\eq{IScalingForm0t0} and \eq{mukappaExponent}, i.e.,
\begin{align}
  \kappa_{S} = d+m_{\kappa}-z/2\,.
  \label{eq:kappaSelfSim}
\end{align}
This result is equivalent to the scaling given in Eqs.~(2.7) and (2.11) of Ref.~\cite{Svistunov1991a}, as can be seen with the relations between the relevant scaling exponents
\footnoteremember{fn3}{The exponent $\alpha$ in Ref.~\cite{Svistunov1991a} translates to $d/z-1$ in our work, $\gamma$ to $2(d+m_{\kappa})/z-3$. Hence, $\varepsilon^{\gamma/2+1}\sim p^{z(\gamma/2+1)}\sim p^{d+m_{\kappa}-z/2}$.}.

As we will see below, the self-similar scaling evolution only occurs in the infrared collective-scattering regime. 
Hence, we can combine the result \eq{kappaSelfSim} with Eqs.~\eq{mselfsimz}, \eq{mkappazBog}, and \eq{alphaofdmz} to obtain
\begin{align}
  \alpha - \beta(\kappa_{S}-z+2-\eta)&= -1/2 \,.
  \label{eq:alphambetakappa}
\end{align}
Note that, for free particles ($z=2$, $\eta=0$), \Eq{alphambetakappa} implies that the left-hand side of \Eq{FPEqkappa} is not identically zero. 
This ensures the above derivation of $\kappa_{S}$ to be justified.
In \Sect{SummaryNumerical} we will check numerically that the scaling of $f$ and of the scattering integral $I[f]$ is consistent with the exponent \eq{kappaSelfSim}.\\[-1ex]

\emph{Bi-directional self-similar evolution}.---
As in \Sect{NTFPIntro} we proceed to the case of a quasiparticle distribution $n_{Q}(p,t)$ parametrized according to \Eq{nQbyScalingFunctionf2} in terms of the scaling function \eq{scalingf2} with $\kappa_{\Lambda}<d$ and $\kappa_{\lambda}>d+z$.

If $\kappa$ is within the interval \eq{SelfSimilarWindow}, (quasi)particles and energy are concentrated at opposite ends of the scaling region. 
Recall that in this case, temporal rescaling can occur according to \Eq{NTFPscaling0f3}, describing a bi-directional transport towards the IR and the UV.
For the momentum exponent $\kappa_{S}$, \Eq{kappaSelfSim}, to be inside the interval \eq{SelfSimilarWindow}, the scaling exponent $m_{\kappa}$ of the $T$-matrix must satisfy
   $z/2\leq m_{\kappa}\leq 3z/2$.
With \Eq{mkappaitogammakappa} this gives
   $z/2\leq 4-\gamma_{\kappa}-2\eta\leq 3z/2$
which can be rewritten as
\begin{equation}
   2(4-\gamma_{\kappa}-2\eta)/3\leq z \leq 2(4-\gamma_{\kappa}-2\eta)\,.
 \label{eq:mSelfSimilarWindowz}
\end{equation}
In the collective-scattering regime, inserting \eq{gammakappageneral} this translates into
\begin{equation}
   2(2-\eta)/3\leq z \leq 2(2-\eta)\,.
 \label{eq:NTFPzregime}
\end{equation}
For $\eta=0$, this is fulfilled, e.g., for $z=2$ (free particles) but not for $z=1$ (Bogoliubov quasiparticles).

Recall that global particle and energy conservation during the evolution lead to two scaling relations between the four exponents $\alpha$, $\beta$, $\beta'$, and $\kappa$, Eqs.~\eq{alphabetaENConserv} and \eq{betabetasENConserv}.
Combining these with Eqs.~\eq{kappaSelfSim} and \eq{alphambetakappa} one finds
\begin{align}
 \alpha_{S}&= {d}\,({2m-z})^{-1}\,,
 \label{eq:alphaS}
 \\
 \beta_{S}&=(2m-z)^{-1}\,,
 \label{eq:betaS}
 \\
 \beta'_{S}&=\beta_{S}(2 m_{\kappa}-z)(2 m_{\kappa}-3z)^{-1} \,.
 \label{eq:betasS}
\end{align}
Using furthermore the relations \eq{mselfsimz} and \eq{mkappazBog} we find
\begin{align}
 \alpha_{S}&= {d}/{z}\,,
 \qquad
 \label{eq:alphaSfreecoll}
 \\
 \beta_{S}&=1/z\,,
 \qquad
 \label{eq:betaSfreecoll}
 \\
 \beta'_{S}&=\beta_{S}(3z-4+2\eta)(z-4+2\eta)^{-1} \,,
 \label{eq:betasSfreecoll}
 \\
 \kappa_{S}&=d+(3z-4)/2+\eta\,,
 \label{eq:kappaSfreecoll}
 \\
 \zeta_{S}&=d+z/2\,.
 \label{eq:zetaSfreecoll} 
\end{align}
where 
we used \Eq{app:npt0Scaling} to obtain the particle-number momentum exponent $\zeta$.
Note that \eq{NTFPzregime} requires $z\ge0$, $\eta\le2$, such that the scaling relation \eq{kappaSfreecoll} is consistent with the condition \eq{NTFPcondition} for the applicability of the universal scaling \eq{gammakappageneral} of the effective coupling. 

The above exponents belong to the main results of the present work and characterize the universal bi-directional self-similar transport in the regime where collective scattering dominates and the $T$-matrix must be determined beyond the perturbative Boltzmann-type approximation.
We remark that $\beta=1/z$ has been proposed on the grounds of numerical simulations in Ref.~\cite{Schachner:2016frd}. 
An exponent $\zeta_{S}\simeq d+1$ was seen in semi-classical simulations, during the early-time evolution after a strong cooling quench, for $d=3$ in Refs.~\cite{Nowak:2012gd,Orioli:2015dxa}, for $d=2$ in \cite{Nowak:2011sk}, and for $d=1$ in \cite{Schmidt:2012kw}.
In Ref.~\cite{Walz:2017ffj}, a numerical evaluation of the kinetic equation in $d=3$ dimensions also gave $\kappa\simeq4$.

\subsubsection{Wave-turbulent scaling evolution}
\label{sec:WWTSolutions}
Within the regime of applicability of the QBE, zeroes of the scattering integral \eq{KinScattInt} correspond to stationary solutions. 
Examples are the constant solution for which the occupation number is independent of ${\vec p}$, as well as the maximum-entropy thermal equilibrium Bose-Einstein distribution $n_\mathrm{BE}(p)$. 
For these solutions, the scattering integral vanishes due to detailed balance, and thus $n_{Q}({p},t)$ is independent of $t$.

Further non-trivial scaling solutions, with different exponents $\kappa$, can be derived with the methods of wave-turbulence theory \cite{Zakharov1992a,Nazarenko2011a} as summarized in \Sect{Turbulence}.
Each such scaling solution corresponds to a different locally conserved current in \emph{momentum} space [see Eqs.~\eq{BalEqQ} and \eq{BalEqP}].
Combining these conserved currents with the kinetic equation \eq{WBKinEq} one can determine the exponents $\kappa$ and $\beta'$ from which the remaining exponents $\alpha$ and $\beta$ are fixed by global conservation laws, see \Tab{ScalingRelConservationLaws}.

The wave-turbulent scaling mainly prevails in the perturbative regime and has been discussed since the advent of weak-wave turbulence \cite{Zakharov1992a,Nazarenko2011a}.
We present the details of the derivations in \App{WWTSolutions} and only give the resulting exponents here.

\emph{Perturbative regime}.---
The most relevant case concerns the perturbative regime. 
There, a self-similar buildup of an inverse particle cascade occurs with temporal exponent
\begin{align}
   \beta=\beta_{Q}= (z-8/3+4\eta/3)^{-1}\, ,
  \label{eq:betaQ}
\end{align}
which is negative for weak-wave-turbulent transport of both, free particles and Bogoliubov sound. 
Therefore, cf.~\Tab{ScalingRelConservationLaws}
and the discussion in \Sect{Turbulence}, the build-up of the inverse quasiparticle cascade occurs in the form of a critically accelerating wave-front evolution with scaling parameter $\tau=\tau^{*}$, cf.~\Eq{scalingParam1}.
This result, with $\beta_{Q}$ given by \Eq{betaQ}, is equivalent to Eq.~(2.22) of Ref.~\cite{Svistunov1991a} 
[cf.~endnote \footnoterecall{fn3} for the translation between exponents].
The results of semi-classical simulations \cite{Berloff2002a} corroborate these predictions.

The corresponding momentum exponent $\kappa$ for the quasi-particle cascade reads
\begin{align}
   \kappa_{Q}&=d+z-8/3+4\eta/3\quad\mbox{(perturbative)}\,.
  \label{eq:kappaQWWT}
\end{align}

\emph{Collective-scattering regime}.---
With the scattering $T$-matrix in the non-perturbative limit, one finds that a similar cascade solution as for weak-wave turbulence applies for $z<2(2-\eta)/3$, cf.~\eq{NTFPzregime}, and thus, e.g., to Bogoliubov quasiparticles ($z=1$, $\eta=0$). 
The temporal exponent results as
\begin{align}
   \beta=1/z\, .
  \label{eq:betaSWCascade}
\end{align}
Hence, if $z>0$, the scaling parameter is $\tau=t/t_{0}$ and the evolution is critically slowed at large times.
The momentum exponent $\kappa$, for the quasi-particle cascade building up in this way is obtained as 
\begin{align}
   \kappa_{Q}&=d+z-(4-2\eta)/3\quad\mbox{(collective scatt., $z>0$)}\,.
  \label{eq:kappaQPSWT}
\end{align}
For $\eta<2$ and $z>0$ these scaling relations imply that the condition \eq{NTFPcondition} for the applicability of the scaling \eq{gammakappageneral} of the effective coupling is fulfilled. 

Note that only for the rather unlikely case of $\eta>2$, $z<0$,  the exponent formally obtained in Refs.~\cite{Berges:2008wm,Berges:2008sr,Scheppach:2009wu,Berges:2010ez} applies,
\begin{align}
\begin{aligned}
   \kappa_{q}&=d+z\quad\mbox{(collective scatt., $z<0$)}\,,
\end{aligned}
  \label{eq:kappaqpSWT}
\end{align}
cf.~\footnoterecall{fn4}.
In a closed system, this $\kappa_{q}$ fulfills the condition \eq{UVDominance} for the buildup of an inverse cascade, behind an accelerated wave front since $\beta=1/z<0$ for $z<0$.

\begin{table*}[t]
\centering 
\caption{Scaling exponents.
The table summarizes the scaling exponents for non-perturbative and perturbative scaling evolution following a cooling quench, describing quasiparticle transport for general dimension $d$, dynamical exponent $z$, and anomalous dimension $\eta$.
The exponents are defined by the scaling forms for $n_{Q}$ and $n$ given in Eqs.~\eq{npPowerLaw}, \eq{nQpPowerLaw}, \eq{NTFPscaling0gL}, \eq{NTFPscaling0gl}, and depicted in Figs.~\fig{SelfSimilar} and \fig{WWTCascade}.
Only transport to lower momenta is relevant for such a quench.
The self-similar and cascade evolutions in the non-perturbative collective-scattering regime ($p\ll p_{\Xi}$) and for $z>0$ have a positive $\beta$ and slow down algebraically at large times due to their proximity to an IR non-thermal fixed point (NTFP).
In contrast, $\beta<0$ for the build-up of a strong-wave-turbulence (SWT) or weak-wave-turbulence (WWT) inverse (quasi)particle cascade, implying a critically accelerated wave-front evolution.
While the WWT cascade occurs in the perturbative regime, the SWT cascade is non-perturbative, occurring below the non-universal scale $p_\mathrm{np}\sim p_{\Xi}^{{1}/({2-\eta})}p_{\lambda}^{1/2}$.  
The rightmost column is obtained from $\zeta=\kappa+2-z-\eta$, cf.~\Eq{app:npt0Scaling}.
See \Tab{OverviewScalingExponents} for an overview and index of all exponents.
\label{tab:ScalingExponents} 
}
\begin{tabular}{  C{5.2cm}  C{1.1cm} C{1.9cm} C{1.3cm} C{2.8cm} C{2.3cm}  C{2.3cm}  }
\hline
{} & $\alpha$ & $\beta$ & $\alpha'$ & $\beta'$ & $\kappa-d$ & $\zeta-d$  \\
\hline
\hline 
\multicolumn{2}{l}{\textbf{Non-perturbative dynamics}}  &  {} &  {} &  {} &  {} & {} \\
\hline
\multicolumn{2}{l}{IR NTFP (self-similar, algebraically slowed)}  
&  {} &  {} &  {} &  {}\\
\hline
$2(2-\eta)/3\leq z \leq 2(2-\eta)$
&  \multirow{2}{*}{$\beta d$} & \multirow{2}{*}{$\displaystyle{1}/{z}$} & \multirow{2}{*}{$\beta'({d+z})$} & \multirow{2}{*}{$\displaystyle\beta\frac{3z-4+2\eta}{z-4+2\eta}$} & \multirow{2}{*}{$\displaystyle\frac{3z}{2}-2+\eta$}  & \multirow{2}{*}{$\displaystyle\frac{z}{2}$}\\
$p\ll p_{\Xi}$  &  {} & {} & {} & {} & {} & {} \\
\hline
\multicolumn{2}{l}{IR NTFP (cascade, algebraically slowed)}  
&  {} &  {} &  {} &  {} & {} \\
\hline
$0<z<2(2-\eta)/3$, $\eta<2$
&  \multirow{2}{*}{$  \beta\kappa$} & \multirow{2}{*}{$\displaystyle{1}/{z}$} & \multirow{2}{*}{0} & \multirow{2}{*}{0}& \multirow{2}{*}{${z-\displaystyle\frac{4-2\eta}{3}}$} & \multirow{2}{*}{${\displaystyle\frac{2-\eta}{3}}$} \\
$p\ll p_{\Xi}$   &  {} & {} & {} & {} & {} & {} \\
\hline
\multicolumn{2}{l}{SWT (cascade, accelerated wave front)}  
&  {} &  {} &  {} &  {} & {} \\
\hline
$z<0$, $\eta>2$
&  \multirow{2}{*}{$  \beta\kappa$} & \multirow{2}{*}{$\displaystyle{1}/{z}$} & \multirow{2}{*}{0} & \multirow{2}{*}{0}& \multirow{2}{*}{${z}$} & \multirow{2}{*}{${\displaystyle{2-\eta}}$} \\
$p\ll p_\mathrm{np}\ll p_{\lambda}$   &  {} & {} & {} & {} & {} & {} \\
\hline
\hline 
\multicolumn{2}{l}{\textbf{Perturbative dynamics}}  &  {} &  {} &  {} &  {} & {} \\
\hline
\multicolumn{2}{l}{WWT (cascade, accelerated wave front)}  
&  {} &  {} &  {} &  {} & {} \\
\hline
no constraints on $z$, $\eta$
&  \multirow{2}{*}{$ \beta \kappa$} & \multirow{2}{*}{$\displaystyle\frac{1}{z-(8-4\eta)/3}$} & \multirow{2}{*}{0} & \multirow{2}{*}{0}& \multirow{2}{*}{${z-\displaystyle\frac{8-4\eta}{3}}$} & \multirow{2}{*}{${-\displaystyle\frac{2-\eta}{3}}$} \\
$p_\mathrm{np}\ll p$   &  {} & {} & {} & {} & {} & {} \\
\hline 
\end {tabular} \par
\end{table*}

\subsection{Kinetic time}
The build-up of wave-turbulent cascades has also been discussed in terms of kinetic time $\tau_\mathrm{kin}$, a time scale on which the critically accelerated wave fronts move to a certain momentum scale $p$, cf., e.g., Ref.~\cite{Svistunov2015a.SuperfluidStatesofMatter}.
We point out that the exponents $\beta_{Q,P}$, describing quasiparticle and energy cascades, respectively, also define the momentum scaling of the kinetic time, cf.~\App{KineticTime} for details, as
\begin{align}
   \tau^{Q}_\mathrm{kin}&\sim p^{-1/\beta_{Q}}\sim p^{d-\kappa_{Q}}\,,
   \label{eq:taukinQ}
   \\
   \tau^{P}_\mathrm{kin}&\sim p^{-1/\beta_{P}'}\sim p^{d+z-\kappa_{P}}\,.
   \label{eq:taukinP}
\end{align}
Inserting, e.g., the exponent \eq{betaQ} for the perturbative scattering of modes with $z=2$, $m=2$, $\eta=0$, one finds that the kinetic time $\tau^{Q}_\mathrm{kin}\sim p^{2/3}$ decreases algebraically for $p\to0$ which illustrates the accelerating character of the wave front. 
According to Eqs.~\eq{taukinQ}, \eq{taukinP}, this is the case for $\kappa_{Q}<d$ and, for energy cascades to the UV, for $\kappa_{P}>d+z$, consistent with the constraints from Eqs.~\eq{UVDominance} and \eq{IRDominance}, respectively. 

The same expression \eq{taukinQ} holds for critically slowed self-similar evolutions at an IR non-thermal fixed point in the collective scattering regime, where the scaling is critically slowed.
The kinetic time $\tau^{S}_\mathrm{kin}\sim p^{-1/\beta_{S}}$ expresses the fact that the time it takes for the distribution $n_{Q}(p,t)$ to show momentum scaling $\sim p^{-\kappa}$ above the momentum scale $p_{\Lambda}\to0$ diverges as  $\tau^{S}_\mathrm{kin}\sim p^{-2}$, for the case $z=1$, $m=2$, $\eta=0$.

\section{Summary of scaling behaviour and numerical comparison}
\label{sec:SummaryNumerical}
In this final section we summarize the different types of universal dynamics possible in a closed system and the corresponding  scaling exponents, and we evaluate the scattering integral numerically to confirm our analytical predictions.

\subsection{Overview of universal scaling exponents}
\label{sec:OverviewTable}
\Tab{ScalingExponents} on p.~\pageref{tab:ScalingExponents} gives an overview of the exponents obtained for the universal dynamics of a closed system following a cooling quench, defined by the scaling forms for $n_{Q}$ given in Eqs.~\eq{npPowerLaw}, \eq{nQpPowerLaw}, \eq{NTFPscaling0gL}, \eq{NTFPscaling0gl}, and depicted in Figs.~\fig{SelfSimilar} and \fig{WWTCascade}.
The exponents are valid for general $d$, $z$, and $\eta$, within the regions of validity indicated, which includes the cases of free particles ($z=2$, $\eta=0$) and Bogoliubov sound ($z=1$, $\eta=0$) in $d$ dimensions, discussed explicitly in \Sect{KineticDescrUniDyn} and \App{EffCouplingFctBog}, respectively.
In order to compare to numerical results we concentrate in the following on the $d=3$ case for vanishing $\eta$.
We recall that, to obtain non-perturbative dynamics one generically needs to apply a strong cooling quench while weak cooling leads to the perturbative evolution, cf.~\Sect{intro:CoolingQuenches}.

\subsection{Scaling evolution of free-particle distributions}
\label{sec:NumericalFree}
We begin with the evolution of free particles, i.e.~the high-energy single-particle excitations of the model \eq{GPHamiltonian} and, in the low-momentum regime below the healing-length scale, its linear relative-phase excitations \cite{Mikheev2018a.arXiv180710228M}.
Their dispersion \eq{freedispersion} gives the dynamical exponent $z=2$, cf.~\eq{dynscaling}.

\subsubsection{Perturbative, wave-turbulent regime}
\label{sec:NumericalFreePerturbative}
We found that, in the \emph{perturbative regime} of low occupation numbers, i.e., for momenta $p\gg p_{\Xi}=(2mg\rho_\mathrm{nc})^{1/2}$, with non-condensate density $\rho_\mathrm{nc}$, \Eq{ParticleDensity}, the scaling evolution of a closed system takes the form of a wave-turbulent cascade towards lower wave numbers, see \Fig{WWTCascade}.
This means that the transport of particles or energy is locally conserved, i.e., the number distribution $n(p,t)$ is stationary within the inertial region $p_{\Lambda}\ll p \ll p_{\lambda}$ while the limiting scale $p_{\Lambda}$ evolves algebraically in time.
For the closed system, due to number and energy conservation, this evolution takes the form of a critically accelerating wave front, Eqs.~\eq{WaveFrontScaling}, \eq{scalingParam1}.
Such an evolution is generically induced by a weak quench, as described in \Sect{intro:WeakCoolingQuench} and sketched in \Fig{NTFP}(a).

%
\begin{figure*}[t]
    \centering
    \includegraphics[width=0.38\textwidth]{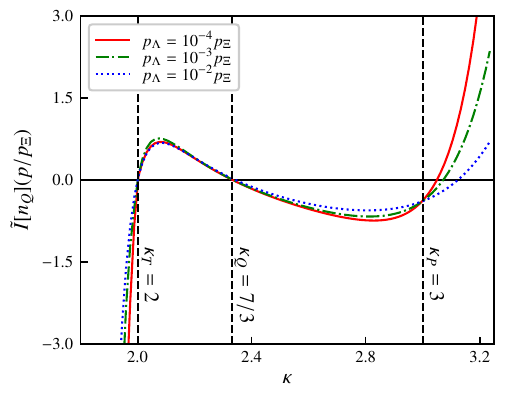}
    \hspace*{0.1\textwidth}
     \includegraphics[width=0.37\textwidth]{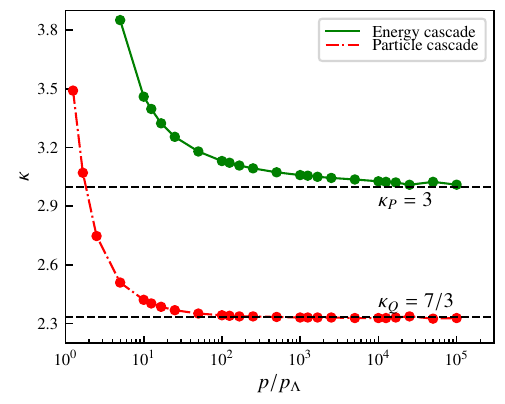} 
    \caption{\emph{Left panel:} Dependence of the  perturbative Boltzmann scattering integral $I[n_{Q}](p)$ for free particles ($z=2$, $\eta=0$; $n=n_{Q}$) in $d=3$ spatial dimensions, at the momentum $p=1.5p_{\Xi}$, 
    on the momentum scaling exponent $\kappa$ characterizing the occupation number distribution $n(p)\sim p^{-\kappa}$.
    The vertical dashed lines mark, from the left, the thermal zero at $\kappa_{T}=2$, the particle-cascade exponent $\kappa_{Q}=7/3$, and the energy-cascade exponent $\kappa_{P}=3$.
     In the figure, the rescaled integral $\tilde{I}[n_Q](p/p_\Xi) = (2 \pi)^3\,p^{3\kappa-4}_\Xi (2m \, g^2 \Lambda^{3\kappa})^{-1} \, I[n_{Q}](p)$ is shown, see \Eq{fmychoice_free} for the definition of $\Lambda$. 
    The different colors (line styles) correspond to different values of the IR cutoff $p_{\Lambda}$, as indicated in the legend.
    As the cutoff is lowered, the zeroes approach the predicted values.
    The sign of the slope $\partial I[n_{Q}]/\partial\kappa$ at the zeroes determines the direction of the cascade. 
    \emph{Right panel:}
    Scaling exponents $\kappa$ of occupation number distribution $n(p)\sim p^{-\kappa}$ for which the perturbative Boltzmann scattering integral $I[n](p)$ for free particles ($z=2$, $\eta=0$; $n=n_{Q}$) in $d=3$ spatial dimensions has a zero, for different momenta $p$ in units of the IR cutoff scale $p_{\Lambda}$.
   Lower line: particle cascade for which $\kappa$ approaches $\kappa_{Q}=7/3$ for $p_{\Lambda}\to0$. 
   Upper line: zeroes of the energy cascade, approaching $\kappa_{P}=3$.
   We emphasize that the two lines represent different physical situations and, as $\kappa_{Q}=7/3<d=3$, due to constraints from conservation laws, only the inverse quasiparticle cascade, is relevant in an isolated system, cf.~\Sect{GlobalConservLaws}. 
    }
    \label{fig:IofkappaPertFree}
%
%
    \vspace*{0.01\textwidth}
    \includegraphics[width=0.38\textwidth]{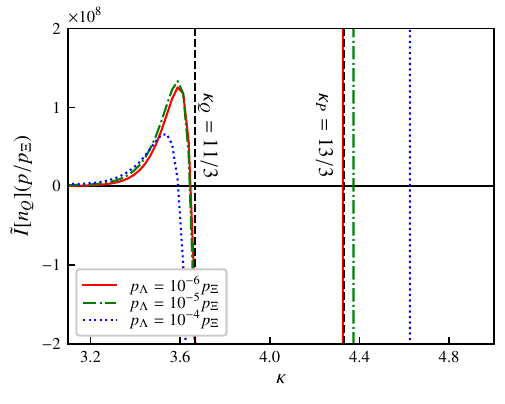} 
    \hspace*{0.1\textwidth}
    \includegraphics[width=0.37\textwidth]{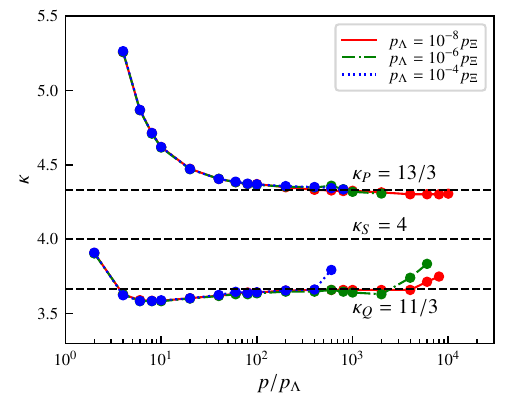}
    \caption{\emph{Left panel:}
    Dependence of the non-perturbative scattering integral $I[n](p)$ in the collective-scattering regime, for free particles ($z=2$, $\eta=0$; $n=n_{Q}$) in $d=3$ spatial dimensions, at the momentum $p=0.001p_{\Xi}$, 
    on the momentum scaling exponent $\kappa$ characterizing the occupation number distribution $n(p)\sim p^{-\kappa}$.
    The vertical dashed lines mark, from the left, the inverse particle-cascade exponent $\kappa_{Q}=11/3$, and the direct energy-cascade exponent $\kappa_{P}=13/3$.
     In the figure, the rescaled integral $\tilde{I}[n_Q](p/p_\Xi) = (2 \pi)^3\,p^{3\kappa-4}_\Xi (2m \, g^2 \Lambda^{3\kappa})^{-1} \, I[n_{Q}](p)$ is shown, see \Eq{fmychoice_free} for the definition of $\Lambda$. 
    The different colors (line styles) correspond to different values of the IR cutoff $p_{\Lambda}$, as indicated in the legend.
    As the cutoff is lowered, the zeroes approach the predicted values.
    The sign of the slope $\partial I[n]/\partial\kappa$ at the zeroes determines the direction of the cascade. 
    Note that the slope at $\kappa_{P}\simeq13/3$ is finite and positive.
    \emph{Right panel:}
    Scaling exponents $\kappa$ of occupation number distribution $n(p)\sim p^{-\kappa}$ for which the non-perturbative Boltzmann scattering integral $I[n](p)$ for free particles ($z=2$, $\eta=0$; $n=n_{Q}$) in $d=3$ spatial dimensions has a zero.
   The figure applies to the IR region of large occupation numbers where the effective many-body coupling describing collective scattering scales with $\gamma_{\kappa}=2$ and modifies the scaling properties.
   The colors (line styles) mark different choices of the IR cutoff scale $p_{\Lambda}$.
   The upper line corresponds to an energy cascade and approaches $\kappa_{P}=d+4/3$ for $p_{\Lambda}\to0$, as obtained from \Eq{kappaPWT} with $\gamma_{\kappa}=2$, while the lower line approaches the particle-cascade exponent $\kappa_{Q}=d+2/3$, cf.~\Eq{kappaQWT}.
   The exponent $\kappa_{S}=4$, cf.~\Eq{alphabetasSkappaSFreecoll}, characterizing the scaling function of the self-similar evolution which is the only one relevant in the isolated system after a quench is marked by the middle dashed line.
    }
    \label{fig:IofkappaCollFree}
\end{figure*}
%
In this regime, the effective many-body coupling equals the bare coupling, $g_\mathrm{eff}(p)\equiv g$, and thus its scaling, Eqs.~\eq{geffScalingHypothesis}, \eq{geffScalingHypothesis0}, and that of the $T$-matrix, Eqs.~\eq{Tscaling}, \eq{Tscaling0}, is fixed by the exponents  $\gamma=\gamma_{\kappa}=m=m_{\kappa}=0$, see also Fig.~\fig{EffCouplingfreeCuts-b}. 
One then obtains, from \Tab{ScalingExponents},
the exponents \cite{Scheppach:2009wu}
\begin{align}
 \alpha_{Q}= 1-3d/2\,,
 \qquad
 \beta_{Q}&=-3/2\,,
 \qquad
 \kappa_{Q}=d-2/3\,.
 \label{eq:kappaQfreepert}
\end{align}
The  exponents for an energy cascade to lower wave numbers are obtained from Eqs.~\eq{kappaPWT} and \eq{betaP} and read
\begin{align}
 \alpha_{P}= -d/2\,,
 \qquad
 \beta_{P}&=-1/2\,,
 \qquad
 \kappa_{P}=d\,.
 \label{eq:kappaPfreepert}
\end{align}
In both cases, $\alpha'=\beta'=0$ vanish, and $\alpha=\kappa\beta$, implying a cascading evolution which leaves $n(p,t)$ stationary in the inertial range $p\gtrsim p_{\Lambda}$.

To check the validity of the above predictions we have numerically determined the zeroes of the scattering integral \eq{KinScattIntCWL}, for fixed values of the external momentum $p$ and the cutoff scale $p_{\Lambda}$.
At these zeroes the particle distribution function at $p$ becomes stationary in time, indicating a wave-turbulent cascade. 
In evaluating the scattering integral we use the bare $T$-matrix \eq{Titogbare} and choose a scaling-function ansatz for the particle distribution $n(p)$ of the type \eq{scalingfgL}. 
Specifically, we choose the form  \eq{fmychoice_free} with IR cutoff scale $p_{\Lambda}$, and map the integral from $p$ to $p_{\lambda}\sim p_{\Lambda}^{-1}$ back onto the interval $(p_{\Lambda},p)$ 
\footnote{Note that we choose the scaling form \eq{fmychoice_free} as it is easier to integrate analytically than the form \eq{scalingf}. Sufficiently far away from the cutoff scale we do not expect this to affect our results.}.
We simplify the $3d$-dimensional integral such that two numerical integrations remain, cf.~ \App{KinScattIntFree}.
We checked that the result was numerically independent of $p_{\lambda}$.

In \Fig{IofkappaPertFree}, we show the dependence of the scattering integral $I[n](p)$, at the momentum $p=1.5p_{\Xi}$, on the momentum exponent $\kappa$ describing the scaling of the distribution $n(p)\sim p^{-\kappa}$ in the inertial range, for three different values of the infrared cutoff scale $p_{\Lambda}$.
The results indicate the way how the zeroes of $I[n](\vec p,t)$ approach the predicted values $\kappa_{Q}=7/3$ and $\kappa_{P}=3$ as the IR cutoff is lowered.
The figure also shows the thermal zero at $\kappa=2$ where the number distribution exhibits Rayleigh-Jeans scaling $n(p)= T/\varepsilon_{p}\sim p^{-2}$.

The sign of the slope $\partial I[n]/\partial\kappa$ at the zeroes in $\kappa$ determines the direction of the cascade, implying a direct cascade for $\partial I[n]/\partial\kappa>0$ and an inverse cascade otherwise \cite{Zakharov1992a}.
Since a direct cascade requires $\kappa>d+z$, \Eq{IRDominance}, only the inverse particle cascade depicted in \Fig{WWTCascade} plays a role in the perturbative dynamics of the closed system considered here, recall  \Sect{Turbulence}.
%
\begin{figure}[tb]
    \centering
    \includegraphics[width=0.39\textwidth]{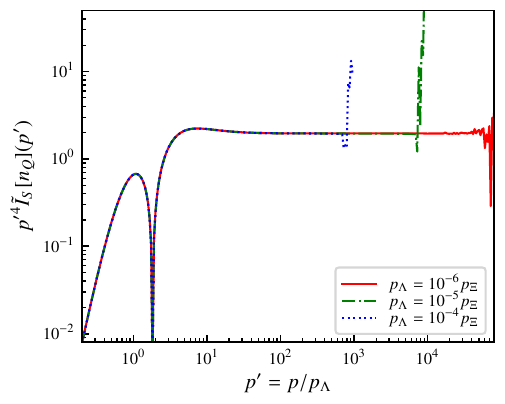}
   \caption{Momentum dependence of the scattering integral \eq{KinScattIntCWL} multiplied by ${p'}^{4}=(p/p_{\Lambda})^{4}$, for free particles ($z=2$, $\eta=0$, $n=n_{Q}$) in $d=3$ dimensions.
   The integral is evaluated in the collective-scattering regime, with effective many-body $T$-matrix, Eqs.~\eq{Titogeff}, \eq{geff}, \eq{PiRFree_fepsilonk}, for different values of the IR cutoff scale $p_{\Lambda}$ (distinguished by color and line style), and the results are scaled on top of each other by showing 
   $p'^4\,\tilde{I}_S[n_Q](p') =  (p/p_\Xi)^{4}(2 \pi)^3\,p^{3\kappa-4}_\Xi (2m \, g^2 \Lambda^{3\kappa})^{-1} \, I[n_{Q}](p)$. 
   The horizontal plateau demonstrates the power-law dependence $I[n_{Q}](p)\sim p^{-4}$ predicted by $\kappa_{S}=4$, cf.~\Eq{alphabetasSkappaSFreecoll}.
      }
   \label{fig:IpSelfSimCollFree}
\end{figure}
\Fig{IofkappaPertFree} (right panel) shows the dependence of the wave-turbulent zeroes of the scattering integral on the momentum where the integral is evaluated, relative to the infrared cutoff.
Red dots correspond to the particle cascade for which $\kappa$ approaches $\kappa_{Q}=7/3$ in the scaling limit $p\gg p_{\Lambda}$, marked by the lower dashed line and confirming the analytically predicted value. 
In the same way, the green dots, marking the zeroes of the energy cascade, confirm the value $\kappa_{P}=3$.
Note that the result only depends on the ratio $p/p_{\Lambda}$.

We note that the deviation of the momentum scaling exponent $\zeta \simeq 3.5$ from the predicted value $\zeta = 3$, observed in the experiment by Navon et al.~\cite{Navon2016a.Nature.539.72}, may be due to the finite size of the experimental apparatus. 
\Fig{IofkappaPertFree} (right) shows that the direct-cascade exponent evaluates to $\zeta = \kappa \simeq 3.5$ if the observed momentum scale is on the order of ten times the infrared cutoff scale, $p \simeq 10\,p_\Lambda$. Considering Fig.~3a of Ref.~\cite{Navon2016a.Nature.539.72}, one estimates $p_\Lambda \xi \simeq 0.5 $ and observes power-law fall-off with $\kappa = 3.5$ ($\gamma$ in their notation) in a momentum region $p\xi \lesssim 5$, consistent with our finite-size estimate.  

We furthermore find that the momentum-dependent deviation from the predicted weak-wave-turbulence exponent $\kappa_{P}=3$, for $p/p_{\Lambda}\gtrsim10^{2}$, approximately follows a logarithmic form, $\kappa_{P}(p)\simeq3+\bar{\kappa}_{P}/\ln(p/p_{0})$, with some momentum scale $p_{0}$.
This is qualitatively similar to the behaviour found in Ref.~\cite{Falkovich1991a:PhysFlB4.92.594} away from the scale $p_{0}$ marking the position of the source which drives the direct energy cascade.
As compared to the value $\bar{\kappa}_{P}=2/3$ predicted in \cite{Falkovich1991a:PhysFlB4.92.594} to quantify the logarithmic correction of the constant scaling exponent $\kappa_{P}=3$, 
our data is rather consistent with $\bar{\kappa}_{P}\simeq0.9$, and closer to the IR cutoff, i.e., for $p/p_{\Lambda}\lesssim10$,  with $\bar{\kappa}_{P}\simeq2$.

Note that according to our general arguments, a direct cascade should not appear in an isolated quenched system. 
This is in accordance with the experimental observation \cite{Navon2016a.Nature.539.72} that the distribution, after an end of the driving, comes to a halt at $p\xi \simeq 5$ after having built up, rather than continuing to spread towards higher momenta. 

\subsubsection{Non-perturbative, self-similar transport regime}
\label{sec:NumericalFreeNonperturbative}
In the \emph{collective-scattering regime} of high occupation numbers, i.e., for momenta $p\ll p_{\Xi}=(2mg\rho_\mathrm{nc})^{1/2}$,  the scaling evolution of a closed system describes a bi-directional non-local transport of particles towards the IR and the UV, leading to a self-similar rescaling of the particle distribution, which is critically slowed at large times, see the sketch in \Fig{SelfSimilar}.
Such an evolution is generically induced by a strong quench, as described in \Sect{intro:StrongCoolingQuench} and sketched in \Fig{NTFP}(b).

The occupation number scales according to \Eq{NTFPscaling0f3}.
In this regime, for momenta larger than the IR cutoff, $p\gg p_{\Lambda}$, the effective many-body coupling takes the universal scaling form \eq{geffFreeUniversal} with scaling exponent $\gamma_{\kappa}=2$, which is constant in time as long as the density of non-condensed particles $\rho_\mathrm{nc}$ remains invariant, see also Fig.~\fig{EffCouplingfreeCuts-b}.

From Eqs.~\eq{betasSfreecoll}, \eq{kappaSfreecoll}, one obtains the scaling exponents for the bi-directional self-similar transport,
\begin{align}
 \alpha_{S}&= {d}/{2}\,,
 &\beta_{S}&=1/2\,,
  &\kappa_{S}&=d+1\,,
 \nonumber\\
 \alpha'_{S}&= -d/2-1\,,
 &\beta'_{S}&=-1/2 \,,
 \label{eq:alphabetasSkappaSFreecoll}
\end{align}
As the self-similar evolution does not leave $n(p,t)$ stationary within the momentum scaling regime where $n(p,t)\sim p^{-\kappa}$, the transport is non-local, and the scattering integral does not vanish. 
\Fig{IofkappaCollFree} shows the scattering integral as a function of $\kappa$, at a fixed momentum $p=10^{-3}\,p_{\Xi}$, for different IR cutoffs $p_{\Lambda}$, and its zeroes as functions of $p$ for the wave-turbulent exponents $\kappa_{Q}=11/3$ and $\kappa_{P}=13/3$, cf.~\Tab{ScalingExponents}.
As we discussed in the previous section, these zeroes, however, do not play a role in the evolution of a quenched isolated system.
The integral evaluates to negative values in between the wave-turbulence zeroes shown in Fig.~\fig{IofkappaCollFree} (left panel), in analogy to the perturbative case in Fig.~\fig{IofkappaPertFree}.
To allow for the self-similar scaling solution in this interval of exponents $\kappa$, the scattering integral, within the momentum scaling region, needs to show the same power-law dependence as the solution $n(p,t)\sim p^{-\kappa_{S}}$, recall our discussion leading to \Eq{kappaSelfSim}.

\Fig{IpSelfSimCollFree} shows the respective momentum dependence of the scattering integral \eq{KinScattIntCWL} in $d=3$ dimensions, where $\kappa_{S}=4$, cf.~\Eq{alphabetasSkappaSFreecoll}.
As one sees, $p^{4}I(p)$ becomes $p$-independent in the scaling regime, demonstrating the power-law dependence $I[n_{Q}](p)\sim p^{-4}$ predicted by $\kappa_{S}=4$, cf.~\Eq{alphabetasSkappaSFreecoll}.

\section{Conclusions}
\label{sec:Conclusions}
We have presented a kinetic-theory description of universal dynamics at non-thermal fixed points of a near-degenerate multicomponent Bose gas based on a non-perturbatively approximated effective action functional.
Our work unites previous treatments which concentrated on wave-turbulent cascades \cite{Svistunov1991a,Berges:2008wm,Berges:2008sr,Scheppach:2009wu,Berges:2010ez} and self-similar evolutions \cite{Svistunov1991a,Orioli:2015dxa,Berges:2015kfa}, complements them with a systematic discussion of the global conservation of energy and particle number and extends them to more general cases.
These include universal dynamics in general dimensions $d$, for a general scaling of the dispersion $\omega(p)\sim p^{z}$, and including a possible anomalous scaling dimension of the density of states $\eta$.

We have focused on universal time evolution after a cooling quench which leads to a redistribution of the particle occupancies towards lower wave numbers, while energy is deposited by means of a few particles being scattered to higher momentum modes. 
We have shown in detail how the initial quench relates to the type of universal transport dynamics that can be observed, with special focus on the particle transport. 
Weak cooling quenches where only a few of the high-energy particles are removed and after which kinetic energy still dominates over interaction energy lead to the build-up of a quasi-local wave-turbulent cascade behind a wave-front. 
On the other hand, strong cooling quenches after which interaction energy dominates over kinetic energy induce transport by means of a non-local self-similar rescaling.

We described both cases using a kinetic equation for momentum mode occupation numbers with a resummed many-body effective coupling $g_\mathrm{eff}$ which results from a $1/N$ expansion to NLO \cite{Berges:2008wm,Berges:2008sr,Scheppach:2009wu,Berges:2010ez,Orioli:2015dxa,Berges:2015kfa}, and which accounts for the non-perturbative suppression of the interactions between highly occupied field modes in the low-energy collective-scattering regime. 
We have analytically and numerically evaluated the momentum-dependent effective coupling $g_\mathrm{eff}(p)$ and the corresponding many-body $T$-matrix.
The kinetic equation is still governed by two-to-two scattering. 
We have clearly identified two distinct regimes. 
For $p>p_{\Xi}$, we obtain $g_\mathrm{eff}\approx g$ as expected and thus recover the usual perturbative form of the wave-Boltzmann kinetic equation. 
This corresponds to the regime of intermediate occupation numbers $1/\zeta\gg n\gg1$, with diluteness parameter $\zeta= \rho^{1/3}a$.
For $p<p_{\Xi}$, the effective coupling is renormalized and becomes frequency and momentum dependent. 
Remarkably, we found it to be given by a simple universal form $g_\mathrm{eff}\sim p^2$, independent of the particular scaling of the resulting occupation number and otherwise depending only on the total particle density and the mass. 
This regime corresponds to highly-occupied modes $n\gg1/\zeta$. 

We have employed the kinetic equation with the effective coupling $g_\mathrm{eff}$ to search for non-trivial self-similar scaling solutions obeying $n(p,t)=t^\alpha f(t^\beta p)$, as well as for solutions with spatial scaling $n(p)\sim p^{-\kappa}$ within a given momentum range. 
By means of power counting we obtained general expressions for the exponents $\alpha$, $\beta$ and $\kappa$ in both, perturbative and non-perturbative regimes as a function of the dynamical exponent $z$, and the anomalous dimension $\eta$. 
The results for particle transport are summarized in \Tab{ScalingExponents}. 
In deriving the exponents, we explicitly distinguished between particle and quasiparticle distributions, and we analyzed the dependence on $z$ and $\eta$ in a consistent manner. 
In this way, our results generalize previous results to arbitrary $z$ and $\eta$, and correct some of the non-perturbative expressions derived in the past.

To corroborate the analytical predictions for $\kappa$, we numerically evaluated the kinetic scattering integral for the specific cases of free particle modes with quadratic dispersion ($z=2$) and Bogoliubov sound quasiparticles with linear dispersion ($z=1$). 
In all cases we found excellent agreement with the analytical results. 
Moreover, our numerical evaluation of the scattering integral shows that the scaling exponent $\kappa=3.5$ observed in the recent experiment \cite{Navon2016a.Nature.539.72} for a direct wave-turbulent energy cascade, can be interpreted to deviate from the predicted value $\kappa_P = 3$ due to finite-size effects in the trap.

From the analysis and results presented in this paper the following unifying picture of universal dynamics after cooling quenches emerges. 
For a weak cooling quench, the transport of particles occurs first within the perturbative regime and is characterized by the build up of a wave-turbulent cascade behind an accelerated wave front. 
The dynamics in this stage are determined by the exponents $\beta=1/(z-8/3)=-3/2$  for $z=2$, $\kappa=d-2/3$, and $\alpha=\beta\kappa=1-3d/2$ \cite{Svistunov1991a}.
Once the wave has reached momenta on the order of $p_{\Xi}$, corresponding to the chemical potential $\mu=g\rho_\mathrm{nc}$ for the total density $\rho_\mathrm{nc}$ of non-condensed particles, collective scattering sets in and brings the wave to a halt.
The subsequent stages in the evolution have been studied in \cite{Svistunov1991a,Kagan1992a,Kagan1994a,Kagan1995a,Semikoz1995a.PhysRevLett.74.3093,Semikoz1997a,Nowak:2012gd,Berges:2012us,Davis:2016hwt} and will lead to the build up of a condensate zero-mode.

For a strong cooling quench the transport of particles towards lower momenta takes place within the non-perturbative regime of collective scattering. 
The dynamics during this stage proceeds in an algebraically slowed manner as a self-similar shift of the infrared momentum distribution. 
We find that the evolution is fully determined by the exponents $\kappa=d+(3z-4)/2$, $\alpha=\beta d$, $\beta=1/z$. 
For $z=2$, this agrees with the result $\beta=1/2$ of \cite{Orioli:2015dxa} and generalizes it otherwise to general $z$ in a consistent way. 
The value of $\kappa=d+1$ obtained differs from previous results \cite{Berges:2008wm,Berges:2008sr,Scheppach:2009wu,Berges:2010ez} due to our careful treatment of the scaling properties of $g_\mathrm{eff}$.

The algebraic slowing down of the dynamics reflects the approach to a non-thermal fixed point and a diverging kinetic time scale. 
At the fixed point, the redistribution of the particles occurs in a bi-directional manner, with particles, and thus potential energy, transported to the infrared while kinetic energy is shifted towards higher wave numbers.
This kind of process is at the basis of the dynamics of Bose-Einstein condensation.

We expect the exponents derived in this work to be valid for a Bose gas in $d=3$ dimensions as well as in $d=2$. 
However, the development of non-linear and topological excitations together with strong phase coherence is likely to modify our results, potentially through an appropriate modification of $z$ and $\eta$. 
The cases considered here apply to quasiparticle-like excitations, e.g.~to the relative phase excitations of a multicomponent Bose gas with $z=2$  \cite{Mikheev2018a.arXiv180710228M}, which are dominant for a large number of field components. 
For such cases, our work provides a unifying approach to  universal transport phenomena characterized by self-similar scaling in space and time.

\section*{Acknowledgments}
The authors thank J.~Berges, K.~Boguslavski, S.~Czi\-schek, S.~Diehl, S.~Erne, M.~G\"arttner, M.~Karl, P.~Kunkel, D.~Linnemann, S.~Jochim, J.~Marino, A.N.~Mikheev, P.~Murthy, B.~Nowak,  M.~K.~Oberthaler, J.~M.~Pawlowski, M.~Pr\"ufer, C.-M.~Schmied,  H.~Strobel, and R.~Walz for discussions and collaboration on the topics described here. 
This work was supported by the Development and Promotion of Science and Technology Talents Project (DPST) of the Royal Thai Government, Thailand,  the Horizon-2020 framework programme of the European Union (FET-Proactive project AQuS, No. 640800),  by Deutsche Forschungsgemeinschaft (SFB 1225 ISOQUANT and Grant No.~GA677/8), by the Helmholtz Association (HA216/EMMI), and by Ruprecht-Karls-Universit\"at Heidelberg (CQD).\\

\begin{appendix}
\begin{center}
{\bf APPENDIX}  
\end{center}
\vspace{-7mm}
\numberwithin{equation}{section}
\numberwithin{figure}{section}
\setcounter{section}{0}
\setcounter{equation}{0}
\setcounter{figure}{0}

\begin{table}[t]
\centering 
\caption{Scaling exponents.
The table lists the scaling exponents appearing in this work and refers to the defining equations.
} 
\label{tab:OverviewScalingExponents} 
\begin{tabular}{  C{0.7cm}  L{5.7cm}  C{1.9cm}   }
\hline
\hline 
&\textbf{Occurence}   &  see Eqs.   \\
\hline
&\multicolumn{2}{l}{\textbf{Quasiparticle distribution in time and momentum}}\\
\hline
$\alpha$ 		&IR rescaling, e.g.~of $f_{\Lambda}$	&  \eq{NTFPscaling}, \eq{NTFPscaling0gL}, \eq{mainT:ScalingF}  \\
$\alpha'$ 		&UV rescaling, e.g.~of $f_{\lambda}$	&  \eq{NTFPscaling0gl}  \\
$\alpha_{0}$ 	&rescaling of amplitude $f_{1}$&  \eq{SelfSimilarScaling}, \eq{NTFPscaling0f3}  \\
$\alpha_{P}$ 	&$\alpha$ for wave-turbulent energy cascade		&  \eq{kappaPfreepert}  \\
$\alpha_{Q}$ 	&$\alpha$ for wave-turbulent quasiparticle cascade	&  \eq{kappaQfreepert}  \\
$\alpha_{S}$ 	&$\alpha$ for self-similar evolution at NTFP (IR)	&  \eq{alphaS}  \\
$\alpha'_{S}$ 	&$\alpha'$ for self-similar evolution at NTFP (UV)	&  \eq{alphabetasSkappaSFreecoll}  \\
\hline
$\beta$ 		&IR rescaling, e.g.~of $p_{\Lambda}$			&  \eq{NTFPscaling}, \eq{mainT:ScalingF}  \\
$\beta_{P}$ 	&$\beta$ for wave-turbulent energy cascade		&  \eq{betaP}  \\
$\beta_{Q}$ 	&$\beta$ for wave-turbulent quasiparticle cascade	&  \eq{betaQ}  \\
$\beta_{S}$ 	&$\beta$ for self-similar evolution at NTFP (IR)		&  \eq{betaS}  \\
$\beta'$ 		&UV rescaling, e.g.~of $p_{\lambda}$			&  \eq{SelfSimilarScaling}, \eq{NTFPscaling0f3}  \\
$\beta'_{P}$ 	&$\beta'$ for wave-turbulent energy cascade		&  \eq{betasP}  \\
$\beta'_{Q}$ 	&$\beta'$ for wave-turbulent quasiparticle cascade	&  \eq{betasQ}  \\
$\beta'_{S}$ 	&$\beta'$ for self-similar evolution at NTFP (UV)	&  \eq{betasS}  \\
\hline
$\kappa$ 		&momentum scaling of quasipart.~distribution		&  \eq{nQpPowerLaw}  \\
$\bar\kappa$ 	&second momentum scaling of distribution			&  \eq{scalingf}  \\
$\kappa_{\Lambda}$ &momentum scaling, $p\ll p_{\Lambda}$		&  \eq{scalingfgL}, \eq{scalingfgl}  \\
$\kappa_{\lambda}$ 	&momentum scaling, $p\gg p_{\lambda}$		&  \eq{scalingfgL}, \eq{scalingfgl}  \\
$\kappa_{p}$ 	&$\kappa$ for anomalous energy cascade		&  \eq{kappaqpSWT}  \\
$\kappa_{P}$ 	&$\kappa$ for wave-turbulent energy cascade		&  \eq{kappaPWT}, \eq{app:kappaQPWWT}  \\
$\kappa_{q}$ 	&$\kappa$ for anomalous  quasiparticle cascade	&  \eq{kappaqpSWT}  \\
$\kappa_{Q}$ 	&$\kappa$ for wave-turbulent quasiparticle cascade &  \eq{kappaQWT}, \eq{app:kappaQPWWT}  \\
$\kappa_{S}$ 	&$\kappa$ for self-similar evolution at NTFP 		&  \eq{kappaSelfSim}  \\
\hline
&\multicolumn{2}{l}{\textbf{Particle distribution in momentum}}\\
\hline
$\zeta$ 		&momentum scaling of particle distribution			&  \eq{npPowerLaw}  \\
$\zeta_{S}$ 	&$\zeta$ for self-similar evolution at NTFP		&  \eq{zetaSfreecoll}  \\
\hline
&\multicolumn{2}{l}{\textbf{Effective coupling} $g_\mathrm{eff}$}\\
\hline
$\gamma$ 	&scaling  in time and momentum				&  \eq{geffScalingHypothesis0}  \\
$\gamma_{\kappa}$ 	&scaling in momentum at fixed time			&  \eq{geffscaling}, \eq{geffScalingHypothesis}  \\
\hline
&\multicolumn{2}{l}{$T$-\textbf{matrix}}\\
\hline
$m$ 			&scaling  in time and momentum				&  \eq{Tscaling0}  \\
$m_{\kappa}$ 	&scaling in momentum at fixed time				&  \eq{Tscaling}  \\
\hline
&\multicolumn{2}{l}{\textbf{Scattering integral} $I_{Q}$}\\
\hline
$\mu$ 		&scaling  in time and momentum				&  \eq{IScalingForm0}  \\
$\mu_{\kappa}$&scaling in momentum at fixed time				&  \eq{IScalingForm0t0}  \\
\hline
&\multicolumn{2}{l}{\textbf{Other}}\\
\hline
$d$ 			&spatial dimension							&  \eq{QParticleDensity}  \\
$\eta$ 		&anomalous dimension						&  \eq{mainT:Scalingrho}  \\
\hline
$z$ 			&dispersion, dynamical exponent				&  \eq{dynscaling}, \eq{mainT:Scalingrho}  \\
\hline 
\end {tabular}\par
\end{table}
\section{Index of scaling exponents and notation used}
\label{app:notation}
In \Tab{OverviewScalingExponents} we provide an index of all scaling exponents appearing in this work, linking to equations where the exponents are defined in their context.

Choosing the ($+$\,$-$\,$-$\,$-$) convention for the metric, 
the Minkowski product of ($d+1$)-vectors
$p = (p_{0},p_{1},...,p_{d}) = (p_0,\vec p) = (\omega,\vec p)$, etc.
reads
$px = p_0x_0 - \vec p\cdot\vec x$.
Defining the $(d+1)$-dimensional Fourier transform as 
$\mathcal{F}[f(p)](x) = f(x) = (2\pi)^{-(d+1)}\int{\mathrm{d}^{d+1}p}\exp\{-ipx\}f(p)$,
the following convention is used for convolutions:
\begin{align}
  (f\ast h)(x) 
  &= \int\df{y}f(x-y)h(-y),
\label{eq:notConvolution}
  \\
  (f\ast h)(p) 
  &= \int\frac{\df{q}}{(2\pi)^{d+1}}f(p-q)h(-q).
\end{align}
The convolution theorem is then
\begin{align}
  \mathcal{F}[(f\ast h)](x) 
  &= (f\cdot h)(x) = f(x)\, h(-x),
  \\
  \mathcal{F}[(f\ast h)](p) 
  &= (f\cdot h)(p) = f(p)\, h(-p).
\end{align}
%

\section{Universal scaling dynamics after a cooling quench}
\label{app:CoolingQuenches}
A central aspect of the phenomena discussed in this work is that the character of the evolution depends very much on the `strength' of the cooling quench which determines how far the system can get out of equilibrium.
In this appendix we have a closer, more technical look at the \emph{weak} and \emph{strong} cooling quenches depicted in Figs.~\fig{NTFP}(a) and (b), respectively. While the former typically leads to weak-wave turbulence, the latter is required to induce an approach to a non-thermal fixed point.
We use estimates on the basis of ideal-gas thermodynamic quantities for critical properties as well as on the dilute-gas approximation for the $s$-wave interactions between particles. 

\subsection{Weak cooling quench}
Consider a three-dimensional thermal dilute alkali Bose gas for which the gas parameter $\zeta=a/l$, relating the interparticle distance $l=\rho^{-1/3}$, given by the density $\rho$, and the $s$-wave scattering length $a$,  is typically on the order of a few percent,  $\zeta= \rho^{1/3} a\ll1$.
We assume that, before the quench, the gas is just above the condensation temperature $T_\mathrm{c}$, see the phase-diagram sketch in \Fig{NTFP}(c).
We assume the scale set by the chemical potential $\mu$ to be below the temperature $T$ of the gas, i.e., $|\mu| \ll\hbar^{2}\rho^{2/3}/m\sim k_\mathrm{B}T_\mathrm{c}\lesssim k_\mathrm{B}T$ where we use the ideal-gas expression for the critical temperature and assume corrections to this to be small. 
In this case, the Bose-Einstein distribution shows, in the energy range $|\mu|\ll\omega(p)\ll k_\mathrm{B}T_\mathrm{c}$, Rayleigh-Jeans behaviour and is much larger than unity, $n_\mathrm{BE}(\omega(p))\simeq k_\mathrm{B}T_\mathrm{c}/\hbar\omega(p)\gg 1$.

Removing, from such a system, a few of the high-energy particles, the subsequent particle transport in momentum space towards lower energies is, for sufficiently large momenta, described by the perturbative quantum Boltzmann equation in the classical-wave limit~\cite{Svistunov1991a,Kagan1992a,Kagan1994a}.
This equation is valid for modes with energies above the scale set by the zero-temperature chemical potential $\mu$, and below the scale $\sim k_{B}T$ where mode occupancies fall below 1.
Here, $\mu=g\rho_{0}\equiv g\rho\sim a\rho/m\sim\zeta\rho^{2/3}/m\sim\zeta k_\mathrm{B} T_\mathrm{c}$ is defined in terms of the density $\rho_{0}$ of the entirely Bose-condensed gas with interaction constant $g=4\pi \hbar^{2}a/m$.
At lower energies, however, where the near-critical Rayleigh-Jeans distribution is $n_\mathrm{BE}(\omega<g\rho)> T_\mathrm{c}/(g\rho)\sim\zeta^{-1}$, phase correlations between momentum modes become significant, and a (non-perturbative) description beyond the quantum Boltzmann equation is needed \cite{Svistunov1991a,Kagan1992a,Kagan1994a,Kozik2004a.PhysRevLett.92.035301,Kozik2005a.PhysRevLett.94.025301,Kozik2005a.PhysRevB.72.172505,Kozik2009a}.  

For higher energies, where the quantum Boltzmann approach is still viable, Svistunov discussed different transport scenarios based on weak wave turbulence, in analogy to similar processes underlying Langmuir-wave turbulence in plasmas \cite{Zakharov1992a,Nazarenko2011a}.
Taking into account that the scattering matrix elements in the perturbative wave-Boltzmann equation for such a dilute gas are independent of the mode energies, he concluded that the initial kinetic transport stage of the condensation process in momentum space evolves as a weakly non-local particle wave towards lower momenta.  
Specifically, he proposed that this particle-flux wave followed the self-similar form $n(p,t)\sim p_{\Lambda}(t)^{-\alpha/\beta}f(p/p_{\Lambda}(t))$, with $p_{\Lambda}(t)\sim(t_{*}-t)^{-\beta}$, and scaling function $f$ falling off as $f(x)\propto x^{-\kappa}$ for $x\gg1$.
He found the scaling exponents $\alpha=-7/2$, $\beta=-3/2$, and $\kappa=7/3$
\footnote{Note that in Ref.~\cite{Svistunov1991a}, the formulation was given in terms of $\varepsilon_{1}(t)\sim p_{\Lambda}(t)^{2}$, i.e., $n(\varepsilon,t)\sim\varepsilon_{1}(t)^{-7/6}f(\varepsilon/\varepsilon_{1}(t))$, with $f(x)\propto x^{-\alpha}$ where $\alpha=7/6$.}.
Following the arrival of this wave at time $t=t_*\simeq t_0 +\hbar\omega(p_\mathrm{q})/\mu^2$, a quasi-stationary wave-turbulent cascade forms.
Within this cascade, particles are transported locally, from momentum shell to momentum shell, from the scale $\omega(p_\mathrm{q})$ of the energy concentration in the initial state (see \Fig{NTFP}(a)) to the low-energy regime $\omega\lesssim\mu$ where coherence formation sets in and the description in terms of the wave-Boltzmann equation ceases to be valid.

This flux-wave and weak-wave-turbulence stage of condensate formation following a cooling quench was investigated in more detail by Semikoz and Tkachev \cite{Semikoz1995a.PhysRevLett.74.3093,Semikoz1997a}, who solved the wave-Boltzmann equation numerically and found results consistent with the above scenario, albeit with slightly different power-law exponents $\alpha\simeq-2.6$ and $\kappa\simeq2.48$ for the wave-turbulence spectrum during the build-up stage.  
Dynamical classical-field simulations of the condensation process by Berloff and Svistunov \cite{Berloff2002a} corroborated the above picture.

\begin{figure}[t]
\centering
\includegraphics[width=0.4\textwidth]{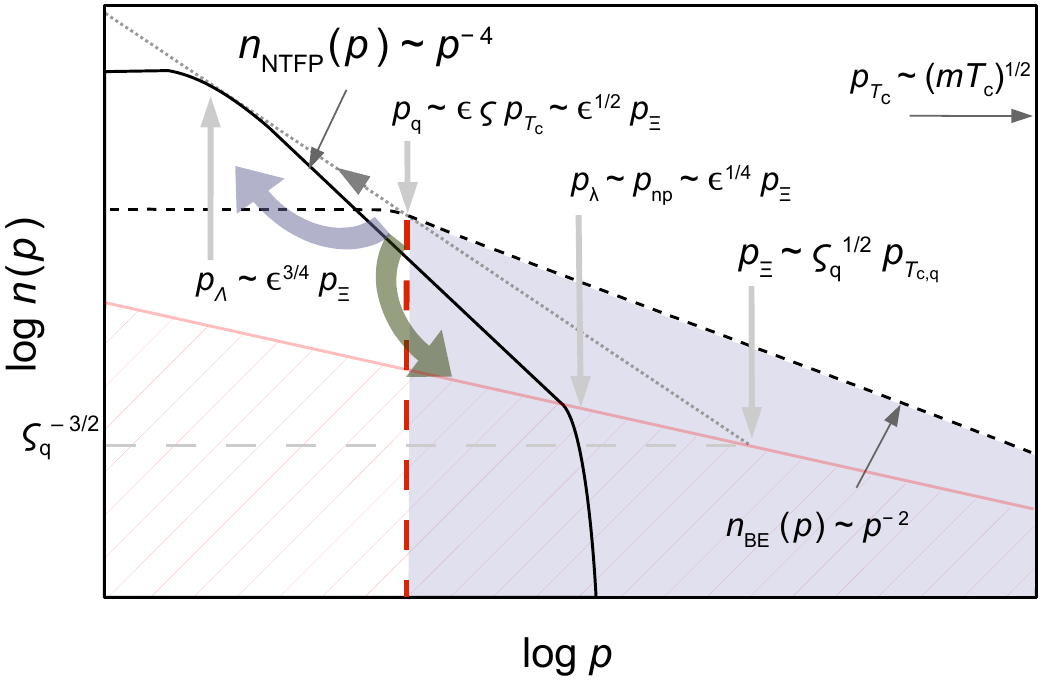}
\caption{Exemplary momentum distribution at a non-thermal fixed point ($n_\mathrm{NTFP}(p)$, solid line), vs.~pre-quench thermal Bose-Einstein distribution in the Rayleigh-Jeans regime, $|\mu|\ll k_\mathrm{B}T_\mathrm{c}$ (short-dashed line).
All particles to the right of the red long-dashed vertical line at $p=p_\mathrm{q}$ are removed in the strong cooling quench, $p_\mathrm{q}<\zeta p_{T_\mathrm{c}}$ (shaded area, see main text).
In $d=3$ dimensions, according to our kinetic theory, the NTFP distribution of modes with frequency $\omega(p)\sim p^{2}$ shows power-law scaling as $n_\mathrm{NTFP}(p)\sim p^{-4}$ and evolves self-similarly in time, see \Fig{NTFP}(b). 
The figure shows the special, transient case of such a distribution with a \emph{single} momentum power law (black solid line).
As the evolution conserves the total density $\rho=\int_\mathbf{k}n(\mathbf{p})$, the point $[p_{\Lambda}(t),n_\mathrm{NTFP}(p_{\Lambda}(t))]$ moves along the grey fine-dotted arrow $n_\mathrm{NTFP}(p_{\Lambda}(t))\sim p_{\Lambda}(t)^{-3}$.
The momentum scale where the pre-quench distribution is cut off in the UV, corresponding approximately to the pre-quench critical temperature, $p_{T_\mathrm{c}}\sim(mT_\mathrm{c})^{1/2}$, is outside the graph's frame. 
The grey vertical arrows mark the momenta corresponding to  the quench scale $p_\mathrm{q}\sim\epsilon \zeta p_{T_\mathrm{c}}$, with pre-quench diluteness parameter $\zeta$ and some positive number $\epsilon<1$,
the post-quench chemical potential scale $p_{\Xi}=(g\rho_{q})^{1/2}=\zeta_{q}^{1/2}p_{T_\mathrm{c,q}}$, with post-quench diluteness parameter $\zeta_\mathrm{q}$ and crit.~temperature scale $p_{T_\mathrm{c,q}}$, 
and the IR and UV cutoff scales $p_{\Lambda}\sim \epsilon^{3/4} p_{\Xi}$ and $p_{\lambda}\sim\epsilon^{1/4}p_{\Xi}$, at the moment when the scaling function is given by the solid line and still entirely within the non-perturbative regime. 
The regime where the perturbative wave-Boltzmann equation is applicable ($n(p)\ll (ap)^{-1}$, cf.~\Eq{perturbativeWBlimit}) and  \cite{Svistunov2015a.SuperfluidStatesofMatter}, is marked by the red dashed area. 
See Secs.~\app{StrongCoolingQ} and \app{CQuenchesUnivDyn} for details. 
}
\label{fig:NTFP2}
\end{figure}
\subsection{Strong cooling quench}
\label{app:StrongCoolingQ}
Distinctly different universal dynamical processes are possible in a dilute Bose gas which is excited by a \emph{strong}  cooling quench at the initial time $t_{0}$.
An extreme version of such a quench would be to first tune adiabatically to a chemical potential $0<-\mu\ll k_\mathrm{B}T_\mathrm{c}$ and then remove all particles with energies higher than $\omega(p_\mathrm{q})\sim|\mu|$ (grey shaded area in \Fig{NTFP2}).
This reduces the particle density $\rho$ of the Rayleigh-Jeans pre-quench distribution near $T_\mathrm{c}\sim\rho^{2/3}/m$ to a post-quench density on the order of $\rho_\mathrm{q}\sim\rho^{2/3}(m|\mu|)^{1/2}$. 
Alternatively, circumventing the requirement of adiabatic tuning, one can build up the same box-like distribution through an instability or parametric amplification.

For illustration we discuss the cooling quench in more detail. 
Suppose that, by order of magnitude, 
\begin{equation}
|\mu|
\sim\epsilon^{2}\zeta^{2}k_\mathrm{B}T_\mathrm{c}
\sim\epsilon^{2}\zeta^{2}\rho^{2/3}/m
\sim\epsilon^{2}\zeta g\rho\,,
\label{eq:DefStrongQuench}
\end{equation}
with pre-quench gas parameter $\zeta=a\rho^{1/3}$  and a real positive number $\epsilon<1$. 
Hence, the post-quench distribution is cut off at the momentum scale given by $p_\mathrm{q}^{2}/2m\sim \epsilon^{2}\zeta g\rho$, i.e., $p_\mathrm{q}\sim\epsilon\zeta p_{T_{c}}$, with $p_{T_{c}}^{2}={2m\,k_\mathrm{B}T_\mathrm{c}}$. 
The remaining density is $\rho_\mathrm{q}\sim\rho^{2/3}(m|\mu|)^{1/2}\sim\epsilon\zeta\rho$.
This implies  $p_\mathrm{q}^{2}/2m\sim\epsilon g\rho_\mathrm{q}$, and thus all of the particles end up below the coherence scale of the post-quench density $\rho_\mathrm{q}$, 
\begin{equation}
p_\mathrm{q}=\epsilon^{1/2}p_{\Xi}\,, 
\end{equation}
with $p_{\Xi}=[2mg\rho_\mathrm{q}]^{1/2}$, and thus $\epsilon$ determines how far below.

Such a strong cooling quench leads to an extreme initial condition for the re-equilibration dynamics:
The remaining post-quench distribution is strongly over-occupied, at momenta $p<p_\mathrm{q}$, as compared to the final equilibrium distribution, on the order of $n_\mathrm{q}(p<p_\mathrm{q},t_{0})\sim n_\mathrm{BE}(p_\mathrm{q})\sim (\epsilon\zeta_\mathrm{q})^{-3/2}\sim (\epsilon\zeta)^{-2}$,  with postquench gas parameter $\zeta_\mathrm{q}=a\rho_\mathrm{q}^{1/3}$.

At the cutoff, one finds the ratio of the interaction to kinetic energy to scale as 
\begin{equation}
[gn(p_\mathrm{q})p_\mathrm{q}^{3}]/(p_\mathrm{q}^{2}/m)
\sim mg\,n_\mathrm{BE}(p_\mathrm{q})p_\mathrm{q}
\sim\epsilon^{-1}\,,
\end{equation}
which needs to be smaller than one for the perturbative wave-Boltzmann equation to yield a valid description of the subsequent collisional redistribution \cite{Svistunov2015a.SuperfluidStatesofMatter}.

In general, occupation numbers fulfilling 
\begin{equation}
n(p)<(mgp)^{-1}\sim\zeta_\mathrm{q}^{-3/2}p_{\Xi}/p
\label{eq:perturbativeWBlimit}
\end{equation}
break the condition for perturbative approaches to be valid and are marked by the red-dashed area in \Fig{NTFP2}.
Hence, as the occupation number is constant below the cutoff, $n_\mathrm{BE}(p<p_\mathrm{q})\sim (\epsilon\zeta_\mathrm{q})^{-3/2}$, cf.~\Fig{NTFP2}, the perturbative (wave-Boltzmann) equation, at the time of the quench, turns out to be inapplicable for the narrow range of momenta $\epsilon p_\mathrm{q}<p<p_\mathrm{q}$, which is, however, strongly relevant as most of the particle density and energy is concentrated here.

As a consequence, for $\epsilon<1$ the subsequent dynamics requires a description beyond the wave-Boltzmann equation.
In three dimensions, $\rho_\mathrm{q}$ and the post-quench kinetic energy density $\varepsilon_\mathrm{q}$ are then concentrated at this highest momentum scale $p_\mathrm{q}$.
One finds $\varepsilon_\mathrm{q}/\rho_\mathrm{q}\sim\epsilon^{2}\zeta^{2} T_\mathrm{c}\sim \epsilon^{2}\zeta^{4/3} T_\mathrm{c,q}$, in terms of the post-quench critical temperature $T_\mathrm{c,q}\simeq\hbar^{2}\rho_\mathrm{q}^{2/3}/(m k_\mathrm{B})$.
The same applies to the post-quench interaction energy density per particle, $u_\mathrm{q}/\rho_\mathrm{q}\sim g\rho_\mathrm{q}\sim\epsilon^{2}\zeta^{2}T_\mathrm{c}$.
Hence, the post-quench mean energy per particle $(\varepsilon_\mathrm{q}+u_\mathrm{q})/\rho_\mathrm{q}$ and thus the expected final temperature $T$ are far below the post-quench critical temperature $T_\mathrm{c,q}$.
Therefore, the particles, after re-equilibration, are expected to end up Bose-condensed, and  $\mu_\mathrm{q}=g\rho_\mathrm{q}$ well approximates the chemical potential of the final condensate.
In \Fig{NTFP}(c) we have sketched both, a weak and a strong cooling quench which are leading to the same final condensate density.

\subsection{Universal scaling dynamics after a strong cooling quench}
\label{app:CQuenchesUnivDyn}
During the evolution starting from the overoccupied distribution introduced above, a much steeper momentum power law develops at low momenta as compared to the case of weak wave turbulence. 
We sketch such a power-law distribution in \Fig{NTFP2}  (solid line).
As we show in this work, cf.~\Sect{NumericalFreeNonperturbative} and \Eq{alphabetasSkappaSFreecoll}, in three dimensions, for particle modes with frequency $\omega(p)\sim p^{2}$, this distribution scales as $n_\mathrm{NTFP}(p)\sim p^{-4}$.
This power law develops below the momentum scale $p_{\Xi}=[2mg\rho_\mathrm{q}]^{1/2}=1/\Xi$ corresponding to the healing length $\Xi$ for the total post-quench density.
\Fig{NTFP2} shows the special case where the  distribution contains particles with momenta predominantly within the region $p<\epsilon^{1/4}p_{\Xi}<p_{\Xi}$ where the perturbative approximation breaks down.

The sketch further demonstrates that one only needs a sufficiently small $\epsilon<1$ to allow the overoccupied initial distribution to build up a distribution $n_\mathrm{NTFP}(p)$ of the type shown in the figure, with a single power-law fall-off $\sim p^{-4}$ in between the IR and UV limiting scales, entirely within the non-perturbative region above the red solid line.
As the total density $\rho=\int_\mathbf{k}n(\mathbf{p})$ is conserved during the transport, the point $[p_{\Lambda}(t),n_\mathrm{NTFP}(p_{\Lambda}(t))]$, moves along the grey fine-dotted arrow $n_\mathrm{NTFP}(p_{\Lambda}(t))\sim [p_{\Lambda}(t)]^{-3}$ where the IR cutoff scale evolves as $p_{\Lambda}(t)\sim t^{-\beta}$, with $\beta=1/2$.
At the same time, energy conservation causes the UV scale $p_{\lambda}$ of the distribution to grow as $p_{\lambda}(t)\sim t^{-\beta'}$, with $\beta'=-1/2$, cf.~Eqs.~\eq{SelfSimilarScaling}, \eq{betaSfreecoll}, \eq{betasSfreecoll}, with $z=2$, $\eta=0$.

The maximum momentum $p_\mathrm{np}$ up to which the perturbative wave-Boltzmann equation remains invalid is found from the constraint \eq{perturbativeWBlimit}  to scale as 
\begin{equation}
p_\mathrm{np}\sim \epsilon^{1/4}p_{\Xi},
\end{equation}
as illustrated in \Fig{NTFP2}.
As the IR scale $p_{\Lambda}$ decreases further, this scale will shrink, too, according to the limit set by the red line.
Once the UV scale $p_{\lambda}$ exceeds $p_\mathrm{np}$, the transport of energy towards higher momenta is described by the perturbative wave-Boltzmann equation, the power law in this part of the tail changes, leading to the build-up of a direct weak-wave-turbulent cascade and eventually a thermal tail as sketched in~\Fig{NTFP}(b).

The distribution evolves universally in time, in the sense that it becomes largely independent of the precise initial conditions set by the quench as well as of the precise values of the parameters of the theory, indicating the approach to a \emph{non-thermal fixed point} of the time evolution \cite{Orioli:2015dxa}.

\section{2PI effective action approach to kinetic theory}
\label{app:2PIKineticTheory}
In this appendix we summarize the most relevant elements of the description of the universal dynamics of a dilute ultra cold Bose gas on the basis of the two-particle irreducible (2PI) effective action \cite{Luttinger1960a,Baym1962a,Cornwall1974a,Berges:2001fi,Aarts:2002dj,Berges:2004yj,Gasenzer2009a} approach in a semi-classical approximation \cite{Berges:2007ym}.
We aim at a compromise between rendering the article compact and, at the same time, sufficiently self-contained.
For more details we refer to Ref.~\cite{Scheppach:2009wu}.

\subsection{Real-time 2PI effective action approach}
\label{subsec:2PI}
%
\subsubsection{Effective action and dynamical correlators}
\label{subsec:2PIEA}
The 2PI effective action $\Gamma[\phi,G]$  \cite{Luttinger1960a, Baym1962a, Cornwall1974a}  is defined as the double Legendre transform of the generating functional for connected correlation functions (cumulants).
Applied to real-time dynamics, the latter is defined, up to a factor $i$, as the logarithm of the Schwinger-Keldysh functional integral $Z$ \cite{Keldysh1964a,KadanoffBaym1995a,Rammer2007a}.
The 2PI action is a functional of the field expectation value $\phi(x)=\langle\Phi(x)\rangle$ as well as of the full Greens function $G(x,y)=\langle\mathcal{T}_\mathcal{C}\Phi^{\dagger}(x)\Phi(y)\rangle$, time-ordered on the Schwinger-Keldysh closed time path (CTP) $\mathcal{C}$ (here, we use four-vector notation $x=(x_{0},\mathbf{x})$).
As in classical Hamiltonian mechanics, one derives, from this action, the equation of motion for the field $\phi$  by means of the variational principle, $\delta\Gamma/\delta \phi=0$, and the Dyson-type, Kadanoff-Baym dynamic equations for $G(x,y)$ from $\delta\Gamma/\delta G=0$. 
The 2PI approach ensures that the dynamic equations conserve energy and particle number irrespective of the approximation chosen for the effective action $\Gamma$ \cite{Arrizabalaga:2005tf,Gasenzer:2005ze}.
For the model \eq{GPHamiltonian}, the 2PI effective action $\Gamma[\phi,G]$ is typically written as a sum of loop integrals involving the bare coupling $g$, the field $\phi(x)$, and the Greens function $G(x,y)$.

To be more specific, we define the correlator matrix 
\begin{equation}
\label{eq:Gmatrix}
  G_{ab}(x,y)=\langle\mathcal{T}\Phi_{a}(x)\Phi_{\bar b}(y)\rangle=G_{ba}^{\dagger}(y,x)
\end{equation}
where the indices are $a,b\in\{1,2\}$, with $\bar a=3-a$ and $\Phi_{1}(x)\equiv\Phi(x)$, $\Phi_{2}(x)\equiv\Phi^{\dagger}(x)$.
Hence, $G_{11}(x,y)=\langle\mathcal{T}\Phi(x)\Phi^{\dagger}(y)\rangle$, etc.
Wherever indices are suppressed in the following the full matrix is meant. 
The time ordering in $G$ is conveniently treated by  decomposing it into  
\begin{equation}
\label{eq:GitoFrho}
  G(x,y) = F(x,y) -\frac{i}{2}\mathrm{sgn}_\mathcal{C}(x_0-y_0)\,\rho(x,y),
\end{equation}
where the signum function $\mathrm{sgn}_\mathcal{C}(x_0-y_0)$ evaluates to $1$ ($-1$) for $x_{0}$ later (earlier) on the CTP than $y_{0}$.
Here, the statistical and spectral parts,  
\begin{align}
  \label{eq:Fmatrix}
  F_{ab}(x,y)&=\langle\{\Phi_{ a}(x),\Phi_{\bar b}(y)\}\rangle/2\,,
  \\
  \rho_{ab}(x,y)&=i\langle[\Phi_{ a}(x),\Phi_{\bar b}(y)]\rangle\,,
  \label{eq:rhomatrix}
\end{align}
are defined in terms of the anticommutator and commutator of the bosonic fields, respectively.\\

\subsubsection{Model and dynamic equations for correlation functions}
We consider the evolution of a dilute interacting bosonic quantum gas described by the complex-valued field operators $\Phi(t,\mathbf{x})$, in $d$ spatial dimensions, obeying the commutation relations $[\Phi(t,\mathbf{x}),\Phi^\dagger(t,\mathbf{y})]=\delta(\mathbf{x}-\mathbf{y})$, $[\Phi(t,\mathbf{x}),\Phi(t,\mathbf{y})]=0$.
The action functional of the system reads (in units where $\hbar=1$)
\begin{align}
  S[\Phi]
  =& \frac{1}{2}\int_{x y} \Phi^{\dagger}_a(x)\, iD_{ab}^{-1}(x,y)\,\Phi_b(y)
  \nonumber\\
  & 
    -\frac{g}{8} \int_{x} \Phi^{\dagger}_a(x)\Phi_a(x)\Phi^{\dagger}_b(x)\Phi_b(x) ,
\label{eq:Sclassphi4}
\end{align}
where we use the notation $\int_x \equiv \int \mathrm{d} x_0 \int \mathrm{d}^d x$ with $(x_0,\mathrm{x}) = (t,\mathrm{x})$, and $g$ denotes the interaction strength.
Note that, as compared to \cite{Scheppach:2009wu}, we here use the standard notation with field components $\Phi_{1}\equiv\Phi$, $\Phi_{2}\equiv\Phi^{\dagger}$.
The free inverse propagator
\begin{align}
\label{eq:G0inv}
  &iD^{-1}_{ab}(x,y)
   =\delta(x-y)\left[i\sigma^3_{{a}{b}}\partial_{x_0}
   -\delta_{ab}H_\mathrm{1B}(x)
     \right],
\end{align}
with $\sigma^3$ the Pauli 3-matrix, involves the single-particle Hamiltonian $H_\mathrm{1B}(x)=-\sum_{j=1}^{d}\partial^2_j/2m+V(x)$, and we choose, in the following, the external potential $V(x)$ to vanish.

We use the two-particle irreducible effective action for the above model \cite{Gasenzer:2005ze,Scheppach:2009wu} to obtain coupled evolution equations for the field expectation value $\phi_{a}=\langle\Phi_{a}(x)\rangle$ and for the time-ordered two-point correlation function \eq{Gmatrix}.
For this, it is advantageous to decompose $G$ according to \Eq{GitoFrho} into the statistical (Keldysh) function $F$, \eq{Fmatrix}, and the spectral function $\rho$, \eq{rhomatrix}.
The resulting integro-differential dynamic equations for $F$ and $\rho$, assuming Gaussian initial conditions, have the form of Schwinger-Dyson (or Kadanoff-Baym) equations,
\begin{widetext}
\begin{align}
 &   \Big[i\sigma^3_{ab}\partial_{x_0}
    -g\,F_{ab}(x,x)\Big]\,\phi_b(x) 
    -\Big(H_\mathrm{1B}(x)
    +\frac{g}{2}\big[\phi_c(x)\phi^*_c(x)
    +F_{cc}(x,x)\big]\Big)\,
    \phi_a(x) 
   = \int_{t_0}^{x_0} \!\mathrm{d}y\,
   \Sigma^\rho_{ab}(x,y;\phi\equiv 0)\,\phi_b(y)\, ,
\label{eq:EOMphi}
  \\
  &\left[i\sigma^3_{ac}\partial_{x_0} - M_{ac}(x) \right]
    F_{cb}(x,y)
  =  \int_{t_0}^{x_0} \! \mathrm{d} z\,
    \Sigma^{\rho}_{ac}(x,z;\phi) F_{cb}(z,y)
  - \int_{t_0}^{y_0} \! \mathrm{d}z\,
    \Sigma^{F}_{ac}(x,z;\phi) \rho_{cb}(z,y)\, ,
\label{eq:EOMF}
  \\[1.5ex]
  &
  \left[i\sigma^3_{ac}\partial_{x_0}
    -M_{ac}(x) \right] \rho_{cb} (x,y)  
  = \int_{y_0}^{x_0} \! \mathrm{d}z\,
  \Sigma^{\rho}_{ac} (x,z;\phi)
  \rho_{cb}(z,y)\,,
\label{eq:EOMrho}
\end{align}
where $\int_{t}^{t'}\mathrm{d}z = \int_t^{t'}\mathrm{d}z_0\int \mathrm{d}^dz$.
The ``mass'' matrix $M$ contains the free Hamiltonian and mean-field shifts,
\begin{equation}
  M_{ab}(x)
   = \delta_{ab}
  \Big[H_\mathrm{1B}(x)
   + \frac{g}{2}\Big(\phi_c(x)\phi^*_c(x)+F_{cc}(x,x)\Big)\Big]
   + g\Big(\phi_a(x)\phi^*_b(x)+F_{ab}(x,x)\Big).
\label{eq:MM}
\end{equation}
\end{widetext}

The self energy $\Sigma$  is obtained as the derivative of the 2PI part $\Gamma_{2}$ of $\Gamma$ which consists of two-loop and higher-order graphs built of field expectation values $\phi$, full correlators $G$ and bare vertices, see \Fig{2PI} and, for more details, Ref.~\cite{Gasenzer2009a},
\begin{equation}
\label{eq:SigmafromGamma2}
  \Sigma_{ab}(x,y;\phi,G)=2i\frac{\delta\Gamma_2[\phi,G]}{\delta G_{ba}(y,x)}.
\end{equation}
In analogy to $F$ and $\rho$, the two-point function $\Sigma$ is decomposed into a local mean-field part $\Sigma^{(0)}_{ab}(x)$ adding to the mass matrix, and nonlocal `statistical' and `spectral' parts,
\begin{align}
\label{eq:Sigma0Frho}
  \Sigma_{ab}(x,y)
  &=\Sigma^{(0)}_{ab}(x)\delta (x-y)
  \nonumber\\
  &+\ \Sigma^F_{ab}(x,y)-\frac{i}{2}\mathrm{sgn}(x_0-y_0)\Sigma^\rho_{ab}(x,y).
\end{align}
The non-local parts appear as kernels in the memory integrals on the right-hand sides of the integro-differential dynamic equations \eq{EOMphi}--\eq{EOMrho}.

\subsection{Resummed self-energy and effective coupling}
\label{app:NLO1N}
\label{app:lambdaeff}
\begin{figure}[t]
\begin{center} 
\includegraphics[width=0.35 \textwidth]{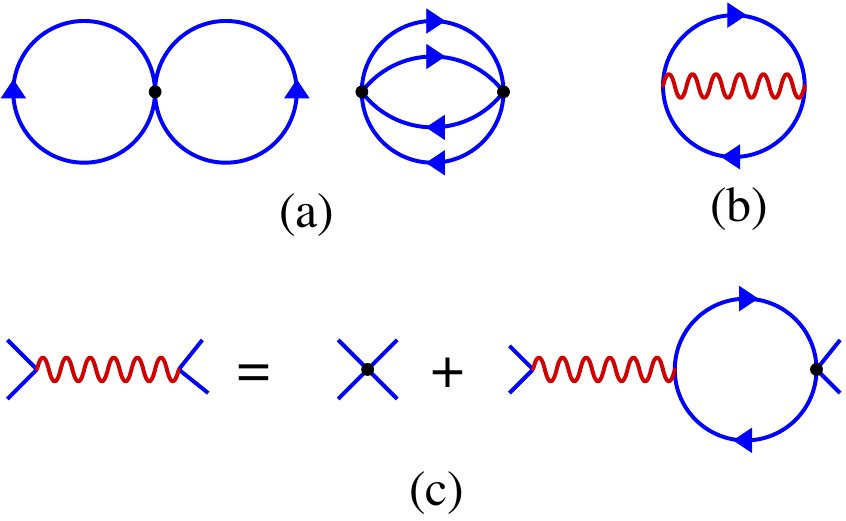}
\caption{Diagrams which are contributing to $\Gamma_2[G]$ and are relevant in the present work.
(a) The two lowest-order diagrams of the loop expansion which lead to the Quantum-Boltzmann equation and thus to the coupling $g_\mathrm{eff}=g$ in the perturbative region. 
Black dots represent the bare vertex $\sim g\delta(x-y)$, solid lines the propagator $G(x,y)$.
(b) Diagram representing the resummation approximation which, in the IR, replaces the diagrams in (a) and gives rise to the modified scaling of the $T$-matrix.
(c) The wiggly line is the effective coupling function which is represented as a sum of bubble-chain diagrams.
Summation of the geometric series gives the expression in \Eq{app:geff}. 
\label{fig:2PI}}
\end{center}
\end{figure}

The dynamic equations \eq{EOMphi}--\eq{EOMrho} are closed within any approximation of the self energy $\Sigma$.
In practice this requires choosing a specific approximation of $\Gamma_{2}$.
The leading-order truncation of $\Gamma_{2}$ in an expansion in $g$ includes the two-loop diagrams, i.e., 
the first diagram 
shown in \Fig{2PI}(a) and leads to the self-consistent Hartree-Fock-Bogoliubov mean-field dynamic equations \cite{Gasenzer:2005ze}.
To describe kinetic transport processes in momentum space such as wave turbulence requires approximations beyond the mean-field order which account for collisional redistribution processes.

The next step is to include the $3$-loop, `basket-ball' diagram (second diagram in \Fig{2PI}(a))
which constitutes the next-lowest perturbative order accounting for elastic `two-to-two' scattering processes.
This leads to dynamic equations which, in the kinetic-theory limit, reduce to the quantum Boltzmann Equation (QBE), \Eq{QKinEq}.
When going beyond the perturbative order of the QBE, we employ an expansion of $\Gamma_2$ in powers of the inverse number of field degrees of freedom $\cal N$ \cite{Berges:2001fi,Aarts:2002dj,Berges:2004yj}.
We expand up to next-to-leading order (NLO), i.e., include the contribution 
corresponding to an $s$-channel resummation of all bubble chains \cite{Gasenzer:2008zz}.
This $1/{\cal N}$ expansion at NLO is equivalent to replacing the vertex in the one-loop leading-order term in  \Fig{2PI}(a) by a bubble-resummed vertex \cite{Aarts:2002dj,Gasenzer:2005ze,Berges:2007ym,Gasenzer:2008zz,Scheppach:2009wu}, see \Fig{2PI}(b,c).

As we focus on the semi-classical region of large occupation numbers where  $\mathrm{Tr}(F^{2}) \gg \mathrm{Tr}(\rho^{2})$ and the self-energies derived from the 2PI effective action shown in  \Fig{2PI}(b), in position space, reduce to \cite{Berges:2007ym}
\begin{align}
  \Sigma^F_{ab}(x)  
  &  =    -g F_{ab}(x)I^F(-x)
  \nonumber\\
  \Sigma^\rho_{ab}(x) 
  & =   -g \big[ \rho_{ab}(x) I^F(-x) - F_{ab}(x)I^\rho(-x) \big],
\label{eq:Sigmarhox}
\end{align}
where we shorten our notation to $F_{ab}(x-y)\equiv F_{ab}(x,y)$, assuming translation invariance and suppressing the dependence on the evolution time $t=(x_{0}+y_{0})/2$.
Note that, in the following derivation of the effective coupling a leading-order gradient expansion in $t$ of the integrals appearing in the self-energy is implied \cite{Branschadel:2008sk}.
In \Eq{Sigmarhox}, we use
\begin{align}
  I^F(x)
  & = 
  g \{ \Pi^F(x) - [\Pi^F\ast(\theta^-\cdot I^\rho)](x)-(\Pi^R \ast I^F)(-x) \},
  \nonumber\\
  I^\rho (x)
  & =   
  g\{ \Pi^\rho(x) - [\Pi^\rho\ast(\theta^-\cdot I^\rho)](x)-(\Pi^R \ast I^\rho)(-x) \},
\label{eq:Irhorecursive}
\end{align}
with the retarded and advanced functions
\begin{align}
  \Pi^R(x) 
  & =  -(\theta\cdot\Pi^{\rho})(x),
  \quad
  \Pi^A(x) 
   =   (\theta^-\cdot\Pi^{\rho})(x).
\label{eq:PiA}
\end{align}
In the classical limit, the functions $\Pi^{F,\rho}$ are
\begin{align}
  \Pi^F (  x)
  & =  \mathrm{Tr}[F\cdot F](x)/2, 
  \nonumber\\
  \Pi^\rho (  x)
   &=  \mathrm{Tr}[\rho \cdot F - F \cdot \rho](x)/2, 
\label{eq:PiFrhoStatClass2}
\end{align}
which, in momentum space, read
\begin{align}
  \Pi^F(  p) 
  & =  \mathrm{Tr}[F\ast  F](p)/2, 
    \nonumber\\
  \Pi^\rho(  p) 
  & =  \mathrm{Tr}[\rho\ast F - F \ast \rho](p)/2. 
\label{eq:PiFrhopStatClass}
\end{align}
The trace of the product/convolution is defined as $\mathrm{Tr}[F\cdot \rho](x)=F_{ab}(x)\rho_{ba}(-x)$.
Finite integration limits in time in the above convolutions are taken into account by the theta function $\theta(x)\equiv\theta(x_{0})$, with $\theta^{-}(x)\equiv\theta(-x)$.

The recursive equations \eq{Irhorecursive} for $I^F$ and $I^\rho$, after a gradient expansion in time, can be solved explicitly.
Using the convolution theorem one obtains, in momentum space,
\begin{align}
  gI^F(p) 
  & =  (\Pi^{F} \cdot g_{\mathrm{eff}}^{2})(p),
  \nonumber\\
  gI^\rho(p) 
  & = (\Pi^{\rho} \cdot g_{\mathrm{eff}}^{2})(p),
\label{eq:IF}
\end{align}
where 
\begin{align}
    g_\mathrm{eff}^{2}(p)
    &=g^{2}\frac{1-(\theta^- \ast I^\rho)(p)}{1+ g\Pi^R(p)} 
    = \frac{g^{2}}{(1 + g\Pi^A (p))(1+g\Pi^R (p))} 
    \nonumber\\
    &= \frac{g^{2}}{\left|1+g\Pi^R(p)\right|^2}.
\label{eq:lambdaeff}
\end{align}
Here we used the symmetry relation $\Pi^A(p) = \Pi^{R \ast}(p)=\Pi^R(-p)$, showing that the effective coupling $g_\mathrm{eff}^{2}(p)$ is real and symmetric. 
The second equality in \Eq{lambdaeff} follows from 
\begin{equation}
  (1+g\Pi^{A}(  p)) \left( 1-(\theta^- \ast I^\rho)( p) \right) = 1
\label{eq:corelambda}
\end{equation}
which in $x$-space reads:
\begin{equation}
  g\{\Pi^R (-x) - [(\theta^-\cdot I^\rho) \ast \Pi^R ](  x) \} - (\theta^-\cdot I^\rho)(x)   = 0.
\end{equation}
This identity is proven by substituting the expression \eq{Irhorecursive} for $I^\rho(x)$  into the third term giving (after some cancellation)
\begin{equation}
  \{ [ (\theta^- \cdot I^\rho) \ast \Pi^\rho ] \cdot \theta\}(x)+[ (\Pi^R \ast I^\rho)\cdot \theta  ](x) - [\Pi^R \ast (\theta^-\cdot I^\rho)](x) 
   = 0
\end{equation}
which can be verified by rewriting all terms in the integral form and combining the integrals.  

Summarizing, the momentum-space self-energies read, in the semi-classical limit,
\begin{align}
  \Sigma^F_{ab}(  p)  
    &=    -[F_{ab} \ast (\Pi^F \cdot g_{\mathrm{eff}}^{2})](p)
  \label{eq:SigmaF}
  \\
  \Sigma^\rho_{ab}(  p) 
   &=-[\rho_{ab} \ast (\Pi^F\cdot g_{\mathrm{eff}}^{2})- F_{ab} \ast (\Pi^\rho \cdot g_{\mathrm{eff}}^{2})](p) ,
\label{eq:Sigmarho}
\end{align}
with the statistical and spectral parts of the loop-integral given in Eqs.~\eq{PiFrhopStatClass}.

\subsection{Particle and quasiparticle correlation functions}
\label{app:KineticQP}
Deriving, from the Kadanoff-Baym equations (KBE)  \eq{EOMphi}--\eq{EOMrho}, a Boltzmann-type kinetic equation \eq{QKinEq} requires the definition of a set of quasiparticles.
The spectral properties of these quasiparticles are encoded in the spectral function $\rho$ while $F$ accounts for the occupation of the respective quasiparticle modes.

We assume our system to be translationally invariant, such that the correlation functions depend on a single momentum $\mathbf{p}$ only.
From the Kadanoff-Baym dynamic equations for $F(x_{0},\mathbf{x};y_{0},\mathbf{y})$ and $\rho(x_{0},\mathbf{x};y_{0},\mathbf{y})$, by means of a gradient expansion in evolution time, kinetic equations for their Fourier transforms with respect to the relative space-time dependence, 
\begin{align}
  F(\omega,\mathbf{p};t)
  =\int{\mathrm{d}\tau\,\mathrm{d}\mathbf{r}}\,F(\tau,\mathbf{r};t,0)\exp[i(\omega\tau-\mathbf{p}\mathbf{r})]\,, 
\end{align}
 can be derived, where $\tau=x_{0}-y_{0}$, $\mathbf r=\mathbf{x}-\mathbf{y}$, and $t=(x_{0}+y_{0})/2$, see, e.g., Ref.~\cite{Branschadel:2008sk}.

\subsubsection{Spectral functions $\rho(\omega,\mathbf{p})$}
\label{app:SpectralFcts}
The Boltzmann-type equation determining the time evolution of the quasiparticle occupancies is obtained by formally solving the obtained kinetic equation for  $\rho(\omega,\mathbf{p};t)$ and inserting the result into the kinetic equation for the statistical function $F(\omega,\mathbf{p};t)$.
We note that, to leading order of the gradient expansion, the spectral function  is found to be independent of time, $\rho(\omega,\mathbf{p};t)\equiv\rho(\omega,\mathbf{p})$, cf., e.g., Ref.~\cite{Branschadel:2008sk}.
Hence, at leading order, the spectral function is approximated to be that of free quasiparticles.
Between the scattering processes, these quasiparticles evolve freely.

\emph{Free particles}.--- The relevant information contained in the spectral function is the dispersion relation $\omega(\mathbf{p})$ defining the quasiparticle frequencies in terms of the mode momentum $\mathbf{p}$.
In the case of free bosons, the matrix elements of the spectral function read 
\begin{align}
\label{eq:rhofree}
  &\rho_{11}(\omega,\mathbf{p})= \rho^*_{22}(-\omega,-\mathbf{p}) = -\rho_{22}(-\omega,-\mathbf{p})
  = 2\pi i\,\delta(\omega-\varepsilon_{\mathbf{p}})\,,
  \nonumber\\
  &\rho_{12}(\omega,\mathbf{p}) = \rho_{21}^*(\omega,\mathbf{p}) = 0\,,
\end{align}
with free dispersion $\varepsilon_{\vec p}=p^{2}/2m$, cf.~\Eq{freedispersion}.

\emph{Bogoliubov sound waves}.--- In a dilute Bose gas with a macroscopic condensate fraction, linear excitations are described in terms of Bogoliubov quasiparticles as summarized in \Sect{ModelObservables}.
The matrix elements of the spectral function of these quasiparticles, in the basis of the fundamental Bose fluctuation field $\tilde\Phi$, cf.~\Sect{ModelObservables}, read 
\begin{align}
 \rho_{\mathrm{B},11}(p) &= 2 \pi i\,\big[ u^2_{\vec p}\delta(p_{0}-\omega_{\vec p})-v^2_{\vec p}\delta(p_{0}+\omega_{\vec p})\big]\,,
  \nonumber\\
  \rho_{\mathrm{B},22}(p) &= 2 \pi i\, \big[v^2_{\vec p}\delta(p_{0}-\omega_{\vec p})-u^2_{\vec p}\delta(p_{0}+\omega_{\vec p})\big]\,,
  \nonumber\\
 \rho_{\mathrm{B},12}(p) &= 2 \pi i\, u_{\vec p} v_{\vec p} \big[\delta(p_{0}-\omega_{\vec p})-\delta(p_{0}+\omega_{\vec p})\big]
  \nonumber\\
 &= \rho^{*}_{\mathrm{B},21}(-p)\,,
  \label{eq:rhoBog}
\end{align}
with Bogoliubov dispersion \eq{BogDispersion} and mode functions \eq{BogModeFunctions}. 
In the  sound-wave limit, $|\mathbf{p}|\ll p_{\xi}$, cf.~Eqs.~\eq{healingmomentum}--\eq{SoundModeFunctions}, the spectral function simplifies to
\begin{align}
 \rho_{\mathrm{B}}(p) &= \frac{i\pi g\rho_{0}}{\omega_{\vec p}}
 (1+\sigma^{1})\,
 \big[ \delta(p_{0}-\omega_{\vec p})-\delta(p_{0}+\omega_{\vec p})\big]\,.
  \label{eq:rhoBogSound}
\end{align}
%

\subsubsection{Relations between $F$ and $\rho$}
\label{app:KineticFDT}
In thermal equilibrium, $F$ and $\rho$ are related by a fluctuation-dissipation relation which, for the example of an ideal uniform Bose-Einstein gas, takes the Callan-Welton form 
\begin{align}
\label{eq:CWFDT}
F_{ab}(\omega,\mathbf{p})= -i\, [n_\mathrm{BE}(\omega)+1/2]\,\rho_{ab}(\omega,\mathbf{p})\,,
\end{align}
where the Bose-Einstein function 
\begin{align}
\label{eq:BEDistribution}
n_\mathrm{BE}(\omega)=\{\exp[\beta(\omega-\mu)]-1\}^{-1}\,
\end{align}
depends explicitly on frequency and chemical potential only.

Based on this, mode occupation numbers for particles, \Eq{nSP} and quasiparticles, \Eq{nQP}, are obtained from the statistical correlator $F(\omega,\vec{p};t)$ as defined in Eqs.~\eq{nDefF}, \eq{nQDefF},
where we included a time dependence for use in dynamics and emphasize that the integrals run over positive frequencies only.
Inserting \eq{CWFDT}, with spectral function \eq{rhofree}, into \eq{nDefF} gives the thermal mode occupation for free particles, $n(\mathbf{p})=n_\mathrm{BE}(\varepsilon_{\mathbf{p}})$.
Using, instead, the spectral function \eq{rhoBog} for Bogoliubov quasiparticles, one obtains the single-particle spectrum $n(\mathbf{p})=(u_{p}^{2}+v_{p}^{2})\,n_\mathrm{BE}(\omega_{\mathbf{p}})+v_{p}^{2}$, cf.~\Eq{nvsnQ}.

To obtain a kinetic description of non-equilibrium dynamics of stable (quasi)particles one assumes, based on the fluctuation-dissipation relation, that the time-evolving statistical function $F(p;t)$ is related to $\rho(p)$ by 
\begin{align}
  F_{ab}(p;t) = -if_{\mathrm{neq}}(p_{0};t)\rho_{ab}(p)\,.
  \label{eq:CallanWelton}
\end{align}
This takes the form of a standard equilibrium Callan-Welton fluctuation-dissipation relation where $f_{\mathrm{neq}}(p_{0};t)$ is a time-dependent quasiparticle frequency spectrum.
As the quasiparticle eigenfrequencies entering the kinetic description are assumed to be real and thus quasiparticles to be stable, an explicit $\vec p$-dependence of $f_{\mathrm{neq}}$ can be neglected.
Furthermore, restricting ourselves as before to the classical-wave limit, $|f_{\mathrm{neq}}|\gg1$, we have neglected, in \eq{CallanWelton}, the quantum ground-state fluctuation term $-i\rho_{ab}/2$ adding to $F_{ab}$, as compared to \Eq{CWFDT} which still includes it. 

Since the definitions \eq{Fmatrix} and \eq{rhomatrix} imply the symmetries $F_{ab}(p;t)=F_{\bar b\bar a}(-p;t)$ and $\rho_{ab}(p)=-\rho_{\bar b\bar a}(-p)$, the real-valued quasiparticle frequency distribution must obey 
\begin{align}
  f_{\mathrm{neq}}(-p_{0};t) = -f_{\mathrm{neq}}(p_{0};t)\,.
  \label{eq:fantisymmetry}
\end{align}
%

\subsubsection{Scaling hypotheses for $F$, $\rho$, $f_{\mathrm{neq}}$, $n$, and $n_{Q}$}
\label{app:ScalingHypoth}
We emphasize that in the case of universal scaling dynamics considered here, $f_{\mathrm{neq}}$ is not an equilibrium distribution function.
We will rather assume that $f_{\mathrm{neq}}$ shows, for $p_{0}>0$, universal power-law behaviour such that the quasiparticle occupancies \eq{nDefF}, \eq{nQDefF} show scaling according to Eqs.~\eq{app:nptScaling}--\eq{app:nQpt0Scaling}, while for $p_{0}<0$ the values of $f_{\mathrm{neq}}$ follow from \Eq{fantisymmetry}.

To obtain the scaling forms one introduces, within the respective scaling regimes, the following scaling hypothesis for the spectral correlator \eq{rhomatrix},
\begin{align}
\label{eq:Scalingrho}
   \rho_{ab}(\omega,\mathbf{p})
   &= s^{2-\eta}\rho_{ab}(s^{z}\omega,s\mathbf{p})\,,
\end{align}
where $s$ is a positive scaling factor and we included a possible anomalous dimension $\eta$.
For the statistical correlator \eq{Fmatrix}, there are two scaling hypotheses, one in space and time and one for fixed time,
\begin{align}
\label{eq:ScalingF}
  F_{ab}(\omega,\mathbf{p};t)
  &= s^{2-\eta+\alpha/\beta}F_{ab}(s^{z}\omega,s\mathbf{p};s^{-1/\beta}t)\,,
   \\
\label{eq:ScalingFstat}
  F_{ab}(\omega,\mathbf{p};t_{0})
  &= s^{2-\eta+\kappa}F_{ab}(s^{z}\omega,s\mathbf{p};t_{0})\,.
\end{align}
From the relation \eq{CallanWelton}, one obtains the scaling of $f_{\mathrm{neq}}$,
\begin{align}
\label{eq:Scalingfanomalous}
   f_{\mathrm{neq}}(\omega;t)
   &= s^{\alpha/\beta}f_{\mathrm{neq}}(s^{z}\omega;s^{-1/\beta}t)\,,
   \\
\label{eq:Scalingfanomaloust0}
   f_{\mathrm{neq}}(\omega;t_{0})
   &= s^{\kappa}f_{\mathrm{neq}}(s^{z}\omega;t_{0})\,.
\end{align}
Here we furthermore point to the important implication that, in a kinetic approximation, with the dispersion scaling as $\omega(p)\sim p^{z}$, the transformation coefficients $u_{p}$ and $v_{p}$ take up the anomalous scaling defined by \eq{Scalingrho}.
For example, a spectral function matrix which, in the basis of the fundamental Bose fields $\Phi$, takes the form \eq{rhoBog}
with mode functions scaling, for $|p|\to0$ as
\begin{align}
 \label{eq:SoundModeFunctionsGeneta}
u_{\vec p}^{2} &\sim v_{\vec p}^{2} \sim u_{\vec p}v_{\vec p} \sim p^{-2+\eta+z}\,,
\end{align}
satisfies \Eq{Scalingrho}. 
This is consistent with both the free ($z=2$, $\eta=0$), \Eq{rhofree}, and the sound-wave dispersion ($z=1$, $\eta=0$), \Eq{rhoBogSound}.
For $n(\vec p,t)$, \Eq{nDefF}, the scaling form \eq{ScalingF} can be used to deduce the IR scaling \eq{app:nptScaling}:
$n(\vec p,t)\sim\int d\omega \Tr{F(\omega,\mathbf{p};t)}\sim \int d\omega\, f_\mathrm{neq}(\omega;t)(u_\mathbf{p}^{2}+v_\mathbf{p}^{2})[\delta(\omega-\omega_\mathbf{p})-\delta(\omega+\omega_\mathbf{p})]\sim f_\mathrm{neq}(\omega_\mathbf{p};t)\,p^{-2+\eta+z}$.
Hence, using \eq{Scalingfanomalous}, one obtains \eq{app:nptScaling}.
Analogously, Eqs.~\eq{ScalingFstat} and \eq{Scalingfanomaloust0} imply the momentum scaling \eq{app:npt0Scaling}
at a fixed moment of time,  cf.~also \Eq{nQpPowerLawExponent}.

Note, however, that the Pauli matrix inside the trace in \Eq{nQDefF} leads to a cancellation of the leading-order terms, 
$n_{Q}(\vec p,t)\sim\int d\omega\, \mathrm{Tr}[\sigma^{3}F(\omega,\mathbf{p};t)]\sim \int d\omega\, f_\mathrm{neq}(\omega;t)(u_\mathbf{p}^{2}-v_\mathbf{p}^{2})$ $[\delta(\omega-\omega_\mathbf{p})-\delta(\omega+\omega_\mathbf{p})]\sim f_\mathrm{neq}(\omega_\mathbf{p};t)$, where we use the invariance of the commutator under the symplectic canonical transformation between the particle and quasiparticle algebras, which requires that $u_\mathbf{p}^{2}-v_\mathbf{p}^{2}=1$.
Then,  \Eq{Scalingfanomalous} is identical to the scaling \eq{app:nQptScaling} of the quasiparticle distribution, cf.~also \Eq{NTFPscaling0nQ}.
Analogously one derives the stationary scaling \eq{app:nQpt0Scaling}.

\subsubsection{Relation to near-equilibrium critical scaling}
\label{app:RelEqCrPh}
We would finally like to add a few remarks concerning the relation between the non-equilibrium scaling behaviour discussed in this paper and standard equilibrium scaling as well as near-equilibrium dynamical critical phenomena.
To make this comparison we consider the special case that the system is situated precisely at a non-thermal fixed point such that the momentum distribution $n(\vec p,t)$, which corresponds to the (time-dependent) structure factor of the fundamental Bose field, obeys IR scaling behaviour according to Eqs.~\eq{app:nptScaling} and \eq{app:npt0Scaling}, the latter for momenta above the inverse ``correlation length'' scale, $p>p_{\Lambda}\equiv1/\xi_{\Lambda}$.
From \Eq{app:npt0Scaling} one obtains the spatial scaling of the first-order correlation function in space, $g^{(1)}(\vec r,t)=\int_{\vec p}\exp(i\,\vec p\vec r)\,n(\vec p,t)$, at a fixed time,
\begin{align}
g^{(1)}(\vec r,t_{0})=s^{d-\zeta}g^{(1)}(s\vec r,t_{0})=s^{d+z-\kappa-2+\eta}g^{(1)}(s\vec r,t_{0}).
\label{eq:g1rt0Scaling}
\end{align}
In comparison, the standard definition of equilibrium scaling exponents at a critical point is given by $g^{(1)}(\vec r)=s^{d-2+\eta}g^{(1)}(s\vec r)$ where $\eta$ is the equilibrium anomalous scaling dimension of the field operator at the fixed point.
One observes that, in our formulation, $\eta_\mathrm{neq}=\eta+z-\kappa$ is a non-equilibrium variant of the anomalous dimension $\eta$.

For an equilibrium Bose gas just above the critical point, the exponent $\kappa$ is set by the Rayleigh-Jeans power-law scaling $n_{Q}(p)\sim T/\omega(p)$ of the Bose-Einstein distribution between the scales set by the chemical potential and the temperature, $-\mu\ll\omega(p)\ll T$.
Hence, $\kappa=z$, such that $\eta_\mathrm{neq}\equiv\eta$ 
\footnote{Alternatively, one could look at a $T=0$ quantum critical point where the equilibrium scaling reads $g^{(1)}(\vec r)=s^{d+z-2+\eta}g^{(1)}(s\vec r)$ and  $\eta_\mathrm{neq}=\eta-\kappa$. In this case, the equilibrium momentum exponent is $\kappa=0$, consistent with a vacuum distribution.}.

The scaling evolution in time and space at a non-thermal fixed point which is to be compared to the above equilibrium critical correlations is defined by the self-similar transport towards lower momenta in the far infrared.
This fixed point is characterised by $\kappa_{S}=d+(3z-4)/2+\eta$, cf.~\Eq{kappaSfreecoll}, such that  $\eta_\mathrm{neq}=2-z/2-d$ which is defined in terms of canonical exponents and appears being unrelated to $\eta$.
To clarify such relations further is an interesting task beyond the scope of the present work.

One can take this comparison one step further, to the scaling of the correlation length which, in equilibrium, is related to the inverse gap or mass parameter of the action.
It can be expressed as $\xi(\tau)\sim\tau^{-\nu}$ where $\tau$ is the tuning parameter, e.g., $\tau=(T-T_\mathrm{c})/T_\mathrm{c}$ near a finite-temperature phase transition, and $\nu$ is the related critical exponent which is $\nu=1/2$ in mean-field approximation.
We compare this with $\xi_{\Lambda}\sim t^{\,\beta}$ at a non-thermal fixed point  which plays the role of the inverse gap parameter of the statistical correlator, $F^{-1}(p=0)\sim p_{\Lambda}^{\kappa}$.
This suggests that $\beta=1/z$ at a non-thermal fixed point could be considered to be a non-equilibrium analogue of $\nu$ at a thermal fixed point.

At last, we point out that the dynamics at a non-thermal fixed goes beyond the well-discussed initial-slip dynamics and ageing phenomena \cite{Janssen1989a,Janssen1992a,Calabrese2002a.PhysRevE.65.066120,Calabrese2005a.JPA38.05.R133,Gambassi2006a.JPAConfSer.40.2006.13}, which relate to the surface critical dimension \cite{Diehl1986a} of the relevant operator in time and manifest as scaling, both in the relative-time direction $t-t'$ of two-point correlators $C(\vec r;t,t')$, and in $t/t'$.
The difference to non-thermal fixed points is that the spectral function $\rho$ and the statistical correlator $F$, in ageing, are assumed to show the same scaling \cite{Gambassi2006a.JPAConfSer.40.2006.13}.
It is an interesting question beyond the scope of this work how initial-slip scaling manifests in the context of a non-thermal fixed point.

\section{Derivation of the kinetic equation}
\label{app:DerivKinEq}

\subsection{Transport equations}
\label{app:NLOScattInt}
From the Kadanoff-Baym dynamic equation for $F(\omega,\vec{p};t)$, time evolution equations for the particle and quasiparticle numbers can be derived.
Here, we are interested in scaling solutions for the low-energy, strongly populated modes, with $n_{Q}(\vec p;t)\gg1$.
In this semi-classical regime, quantum fluctuations can be neglected, $\mathrm{Tr}(F^{2}) \gg \mathrm{Tr}(\rho^{2})$, and thus the $-1/2$ on the right of Eqs.~\eq{nDefF}, \eq{nQDefF} can be neglected.

Therefore, the kinetic equation for the quasiparticle number takes the form
\begin{equation}\label{eq:BoltzmannF}
 \partial_{t}n_{Q}(\vec p,t) 
 = \int_{0}^{\infty}\frac{\mathrm{d}\omega}{2\pi}\partial_t \sigma^{3}_{ab}F_{ba}(\omega,\vec{p};t) 
 = I(\vec{p},t)\,, 
\end{equation}
with the scattering integral being obtained in a leading-order gradient expansion of the Kadanoff-Baym dynamic equations, cf., e.g., Ref.~\cite{Branschadel:2008sk},
\begin{equation}
  I(\mathbf{p},t) = -i \int_{0}^{\infty} \frac{\mathrm{d}\omega}{2\pi}\,\mathrm{Tr}\big[\Sigma^\rho(p;t)F(p;t) - \Sigma^F(p;t)\rho(p)\big].
\label{eq:I2PI}
\end{equation}
Here we use again 4-vector notation $p\equiv(p_{0},\mathbf{p})=(\omega,\mathbf{p})$. 
Where not explicitly stated we will suppress the time argument $t$ in the following.
$\Sigma^{\rho}$ and $\Sigma^{F}$ are the spectral and statistical components of  the self energy, cf.~\App{NLO1N}.
Through the 2PI effective action, $\Sigma^{\rho}$ and $\Sigma^{F}$ are fully determined by the functions $F$  and $\rho$, and the resulting kinetic equation is closed.

\subsection{Kinetic scattering integral}
\label{app:KineticScatteringInt}
Next, we derive the kinetic scattering integral  within the non-perturbative $s$-channel or next-to-leading order $1/\mathcal{N}$ approximation.
Substituting  the self energies \eq{SigmaF} and \eq{Sigmarho} into \Eq{I2PI} gives 
\begin{align}
  I(\vec p) 
   =  -i\,\frac{(2\pi)^3}{2}\int_{kqr}&\!\mathrm{d}p_{0}\,
   \delta(p+k-q-r)\,\theta(p_{0})\,g_\mathrm{eff}^{2}(p+k)
    \nonumber\\
   \times\ \big[ 
     &F_{ab}(p)F_{ba}(-k)F_{cd}(q)\rho_{dc}(-r) 
     \nonumber\\
   -\ &F_{ab}(p)F_{ba}(-k)\rho_{cd}(q)F_{dc}(-r) 
     \nonumber\\
    -\ &F_{ab}(p)\rho_{ba}(-k)F_{cd}(q)F_{dc}(-r) 
     \nonumber\\
    +\ &\rho_{ab}(p)F_{ba}(-k)F_{cd}(q)F_{dc}(-r)
     \big]\,,
     \label{eq:Ip2PIFFFrho}
\end{align}
where $\int_{k}\equiv\int\!\mathrm{d}^{d+1}k/(2\pi)^{d+1}$, etc. 
Inserting the relation \eq{CallanWelton} we rewrite the scattering integral \eq{Ip2PIFFFrho} as
\begin{align}
  I(\vec p) 
      =\  \frac{(2\pi)^3}{2}&\int_{kqr}\!\mathrm{d}p_{0}\,
    \delta(p+k-q-r)\,\theta(p_{0})\,g_\mathrm{eff}^{2}(p-r)
    \nonumber\\
   &\times\  
     \rho_{ab}(p)\rho_{ba}(r)\,\rho_{cd}(q)\rho_{dc}(k) 
     \nonumber\\
  &\quad\  \,
   \times\ \big\{ \big[f(p_{0})+f(k_{0})\big]\,f(q_{0})f(r_{0})
     \nonumber\\
  & \quad\  \, 
    \phantom{\big\{\big[}-\ f(p_{0})f(k_{0})\,\big[f(q_{0})+f(r_{0})\big]
     \big\}\,,
     \label{eq:ScattIntwithrhos}
\end{align}
where we have interchanged the integration variables $r\leftrightarrow-k$, and suppressed the time arguments. 
For ease of notation we also shorten, in this appendix, $f_\mathrm{neq}(p_{0};t)\to f(p_{0})$. 
It is now useful to rewrite the integrals to range over positive frequencies only,
\begin{align}
  I(\vec p) 
      =\  \frac{(2\pi)^3}{2}&\int_{0}^{\infty}\!\mathrm{d}p_{0}\,
     \mathrm{d}k\,\mathrm{d}q\,  \mathrm{d}r\,\, \delta(\vec p+\vec k-\vec q-\vec r)
    \nonumber\\
   &\times\ \sum_{s,s'\in\{-1,1\}}
   \big[\mathcal{I}_{1sss'}+\mathcal{I}_{1s(-s)s'}
    \big]\,,
     \label{eq:ScattIntwithrhosPosFreq}
\end{align}
where $\int_{0}^{\infty}\mathrm{d}k=(2\pi)^{-d-1}\int_{0}^{\infty}\mathrm{d}k_{0}\int\mathrm{d}^{d}k$, and
\begin{align}
    \mathcal{I}_{s\sigma\sigma's'}\ 
    =&\ \delta(sp_{0}+\sigma k_{0}-\sigma'q_{0}-s'r_{0})
     \nonumber\\
   &\times\   g_\mathrm{eff}^{2}(sp_{0}-s'r_{0},\vec p-\vec r)\, 
     \nonumber\\
   &\times\   \rho_{ab}(sp_{0},\vec p)\rho_{ba}(s'r_{0},\vec r)\,\rho_{cd}(\sigma'q_{0},\vec q)\rho_{dc}(\sigma k_{0},\vec k) 
     \nonumber\\
  &\ \  \,
   \times\ \big\{ \sigma's'\big[sf(p_{0})+\sigma f(k_{0})\big]\,f(q_{0})f(r_{0})
     \nonumber\\
  & \ \  \, 
    \phantom{\big\{\big[}-\ s\sigma f(p_{0})f(k_{0})\,\big[\sigma'f(q_{0})+s'f(r_{0})\big]
     \big\}\,,
     \label{eq:Issisipsp}
\end{align}
%

\subsubsection{Kinetic scattering integral for free particles}
\label{app:KinScattIntFree}
Inserting the free spectral function \eq{rhofree}, only the first integrand in \Eq{ScattIntwithrhosPosFreq} contributes, with $s=1$, $\mathcal{I}_{1ss1}$, which can be written as
\begin{align}
    \mathcal{I}_{1ss1}^{(0)}
    = &\ (2\pi)^{4}\delta(p_{0}+s k_{0}-sq_{0}-r_{0})
     \nonumber\\
   &\times\  \delta(p_{0}-\varepsilon_{\vec p})\delta(k_{0}-\varepsilon_{\vec k})
   \delta(q_{0}-\varepsilon_{\vec q})\delta(r_{0}-\varepsilon_{\vec r})
     \nonumber\\
   &\times\   s\,g_\mathrm{eff}^{2}(p_{0}-r_{0},\vec p-\vec r)\, 
     \nonumber\\
  &\ \  \,
   \times\ \big\{ \big[f(p_{0})+s f(k_{0})\big]\,f(q_{0})f(r_{0})
     \nonumber\\
  & \ \  \, 
    \phantom{\big\{\big[}-\  f(p_{0})f(k_{0})\,\big[sf(q_{0})+f(r_{0})\big]
     \big\}\,.
     \label{eq:IsspspsFree}
\end{align}
Integrating over the frequencies $p_{0}$, $k_{0}$, $q_{0}$ and $r_{0}$ we obtain
\begin{align}
  I(\vec p) 
      =\  &(2\pi)^4 \, \int_{\vec k\vec q\vec r}\, \delta(\vec p+\vec k-\vec q-\vec r)
    \nonumber\\
   &\ \ \,\times\ 
   \,g_\mathrm{eff}^{2}(\varepsilon_{\vec p}-\varepsilon_{\vec r},\vec p-\vec r)\,
    \,\delta(\varepsilon_{\vec p}+\varepsilon_{\vec k}-\varepsilon_{\vec q}-\varepsilon_{\vec r})
     \nonumber\\
  &\ \  \,
   \times\ \big\{ \big[f(\varepsilon_{\vec p})+ f(\varepsilon_{\vec k})\big]\,f(\varepsilon_{\vec q})f(\varepsilon_{\vec r})
     \nonumber\\
  & \ \  \, 
    \phantom{\big\{\big[}-\  f(\varepsilon_{\vec p})f(\varepsilon_{\vec k})\,\big[f(\varepsilon_{\vec q})+f(\varepsilon_{\vec r})\big]
     \big\}\,,
     \label{eq:IsspspsFreeFreqIntegrated}
\end{align}
Hence, \Eq{BoltzmannF} takes the form of the quantum Boltzmann equation \eq{QKinEq} for $n_{Q}(\mathbf{p},t)\equiv n(\mathbf{p},t)=f_\mathrm{neq}(\varepsilon_{\mathbf{p}},t)\gg1$, with the scattering integral in the classical-wave approximation, \Eq{KinScattIntCWL}.
Comparing \eq{IsspspsFreeFreqIntegrated} to \eq{KinScattIntCWL} one obtains the expression for the $T$-matrix,
\begin{align}
  |T_{\vec p\vec k\vec q\vec r}|^{2}
  =\ &(2\pi)^{4}g^{2}_\mathrm{eff}(\varepsilon_\mathbf{p}-\varepsilon_\mathbf{r},\vec p-\vec r)
       \,,
\end{align}
given in \Eq{Titogeff}, which depends on the particle distribution $n(\vec p,t)$ themselves.
Making a scaling ansatz for $n(\vec p,t)$, we calculate, in  \App{EffCouplingFctFree}, the dependence of these matrix elements on the energy and momentum arguments explicitly.

To evaluate the scattering integral, we eliminate the $\vec q$ integration by means of the momentum-conservation delta distribution, define  $\vec r = \vec p - \vec r'$ and $\vec q = \vec k + \vec r'$,
choose the $\vec k$-orientation such that $\vec r'$ is parallel to the $z$-component of $\vec k$.
Then, $\theta_{\vec k}=\measuredangle(\vec r', \vec k)$, and we can replace the integral over $\mathrm{d}\cos \theta_{\vec k}$ by an integral over $\mathrm{d}q$, giving 
\begin{align}
I(\vec p) &= (2\pi)^{-1} \int_{\vec r'} \, \frac{1}{r'}  g_\mathrm{eff}^{2}(\varepsilon_{\vec p}-\varepsilon_{\vec p - \vec r'},\vec r') \,\nonumber \\
   & \qquad \times \int^{\infty}_{0}\mathrm{d}k \, k \int^{|k + r'|}_{| k - r'|} \mathrm{d}q \, q \, \delta(\varepsilon_{\vec p}+\varepsilon_{\vec k}-\varepsilon_{\vec q}-\varepsilon_{\vec p - \vec r'}) \nonumber \\
   & \qquad \times  \, \big\{ f(\varepsilon_{\vec p})f(\varepsilon_{\vec p - \vec r'}) \, \big[f(\varepsilon_{\vec q})- f(\varepsilon_{\vec k})\big] \nonumber \\
   & \qquad \qquad +  \big[f(\varepsilon_{\vec p - \vec r'})-f(\varepsilon_{\vec p})\big] f(\varepsilon_{\vec q})f(\varepsilon_{\vec k})\, \big\} \,.
\end{align}
We insert $f(p_0)$, \Eq{fmychoice_free}, and the dispersion $\varepsilon(\vec p) = p^2/2m$, change $\vec r'$ back to $\vec r = \vec p - \vec r' $, and define $\vec q = \vec p - \vec r$ before replacing the integration over $\mathrm{d}\theta_{\vec r}$ by one over $\mathrm{d}q$. This gives
\begin{align}
&I(\vec p) = \frac{2m }{(2\pi)^3} \frac{\Lambda^{3\kappa}}{p}  \int^{\infty}_{0}\mathrm{d}r \, r \int^{p+r}_{|p-r|}\mathrm{d}q  \, g_\mathrm{eff}^{2}(\varepsilon_{\vec p}-\varepsilon_{\vec r},\vec q) \nonumber \\
& \qquad \times \Bigg\{  \frac{ I^{\mathrm{free}}_1(\vec p, \vec q)}{(p^2+p^2_\Lambda)^{\kappa/2}(r^2+p^2_\Lambda)^{\kappa/2}} \, \nonumber \\
& \qquad +  \Bigg[ \frac{ 1}{(r^2+p^2_\Lambda)^{\kappa/2}}-\frac{ 1}{(p^2+p^2_\Lambda)^{\kappa/2}}\Bigg]  I^{\mathrm{free}}_2(\vec p, \vec q) \Bigg\}
\,,
  \nonumber
\end{align}
\begin{align}
&I^{\mathrm{free}}_1(\vec p, \vec r') = \int^{\infty}_{0}\mathrm{d}k \, k \int^{| k +  r'|}_{| k -  r'|} \mathrm{d}q \, q \, \delta(p^2+k^2-q^2-|\vec p - \vec r'|^2)  \nonumber \\
& \qquad\qquad\quad  
  \times \,  \Bigg[\frac{ 1}{(q^2+p^2_\Lambda)^{\kappa/2}}- \frac{ 1}{(k^2+p^2_\Lambda)^{\kappa/2}}\Bigg] \,, 
\nonumber\\
&I^{\mathrm{free}}_2(\vec p, \vec r') = \int^{\infty}_{0}\mathrm{d}k \, k \int^{| k +  r'|}_{| k -  r'|} \mathrm{d}q \, q \, \delta(p^2+k^2-q^2-|\vec p - \vec r'|^2)  \nonumber \\
& \qquad\qquad\quad  
  \times \, \frac{ 1}{(q^2+p^2_\Lambda)^{\kappa/2}(k^2+p^2_\Lambda)^{\kappa/2}} \,.
\label{eq:SM:IchnumFree_2}
\end{align}
It is convenient to introduce new variables $u$ and $v$,
\begin{align}
 u = (k^2+q^2)/\sqrt{2} \,, \qquad v = (k^2-q^2)/\sqrt{2} \,,
 \label{eq:uv_set1}
\end{align}
such that the delta distributions depend on a single variable $v$. 
The integral $I^{\mathrm{free}}_1$ becomes
\begin{align}
 I^{\mathrm{free}}_1(\vec p,\vec r') &= \frac{1}{4} \int^{\infty}_{-\infty} \mathrm{d}v  \, \delta(p^2 - |\vec p - \vec r'|^2 + \sqrt{2}v) \nonumber \\
 & \quad \times \int^\infty_{u(v)}\mathrm{d}u \, 
 \Bigg\{ \left[(u-v)/\sqrt{2}+p^2_\Lambda\right]^{-\kappa/2} \nonumber \\
 & \quad \quad \qquad\ \ \ - 
          \left[(u+v)/\sqrt{2}+p^2_\Lambda\right]^{-\kappa/2} \Bigg\} \,,
 \label{eq:I1freeResult}
\end{align}
where $u(v) = r'^{-2}({v^2+r'^4/2})/{\sqrt{2}}$. 
Here, the integration domain is bound by a parabola in the variables $u,v$: 
For the upper bound, $q = k+r'$, one has
$(- v- r'^2/\sqrt{2})^2 = ( v + r'^2/\sqrt{2})^2 = \sqrt{2}r'^2({u+v})$
and for the lower bound, $q = |k-r'|$, that
$( v - r'^2/\sqrt{2})^2 = \sqrt{2}r'^2({u+v})$. 
By expanding and rearranging the terms, one finds the parabolic form of $u(v)$. 
The integration over d$u$ can be done directly before that over d$v$ is evaluated with the delta distribution, giving
\begin{align}
 I^{\mathrm{free}}_1(\vec p,\vec q) &= \frac{2^{\kappa/2}}{4(\kappa-2)} \Bigg[ \Bigg( \frac{(p^2-r^2+q^2)^2 }{2q^2} + 2p^2_\Lambda\Bigg)^{1-\kappa/2} \nonumber \\
  &\qquad  - \Bigg( \frac{(p^2-r^2-q^2)^2 }{2q^2} + 2p^2_\Lambda\Bigg)^{1-\kappa/2} \Bigg] \,,
\end{align}
where we have replaced $\vec r'$ by $\vec p - \vec r$ and defined $\vec q = \vec p - \vec r$. 
$q = |\vec q|$ is used instead of the polar angle 
between $\vec r$ and $\vec p$.  

The same variable transformation and the same strategy is applied to $I^{\mathrm{free}}_2$,
\begin{align}
I^{\mathrm{free}}_2(\vec p, \vec r') 
&= \frac{\sqrt{2}}{4} \int^\infty_{-\infty} \mathrm{d}v \, \delta(p^2 - |\vec p - \vec r'|^2 + \sqrt{2}v) 
\nonumber \\
&  \times 
\, \int^\infty_{\sqrt{2}u(v)} \mathrm{d}u\, \Big[ (u + p^2_\Lambda/\sqrt{2})^2-v^2 \Big]^{-\kappa/2} \,.
\end{align}
where  $A = \sqrt{2}u(v)+2p^2_\Lambda$ is positive. 
The integral over $u$ can be written in terms of a hypergeometric function \cite{bateman1955higher},
\begin{align}
 \int^{\infty}_A & \mathrm{d}x (x^2 -y^2)^{-\alpha} 
   = \frac{A^{1-2\alpha}}{2}\int^{\infty}_{1}\mathrm{d}x'x'^{-1/2}\left(x'-\frac{y^2}{A^2}\right)^{-\alpha} \nonumber \\
 &= \frac{A^{1-2\alpha}}{2\alpha-1}{}_2F_1(\alpha,\alpha-{1}/{2};\alpha+{1}/{2};[{y}/{A}]^{2}) \,.
 \label{eq:hypergeo_int1}
\end{align}
With this and integrating over d$v$, $I^{\mathrm{free}}_2$ becomes
\begin{align}
I^{\mathrm{free}}_2(\vec p, \vec q) &
= {}_2F_1\Big(\mbox{$\frac{\kappa}{2}$},\mbox{$\frac{\kappa-1}{2}$};\mbox{$\frac{\kappa+1}{2}$};
\Big[\mbox{$\frac{2(p^2-r^2)q^2}{(p^2-r^2)^2+q^4+4p^2_\Lambda q^2}$}\Big]^2 \Big)
\nonumber \\
&\times\ \frac{2^{\kappa-1}}{4(\kappa-1)} \left[ \frac{(p^2-r^2)^2+q^4}{2q^2}+2p^2_\Lambda \right]^{1-\kappa} 
  \,.
 \label{eq:I2freeResult}
\end{align}
Inserting \eq{I1freeResult} and \eq{I2freeResult} into  \eq{SM:IchnumFree_2} and using \eq{app:geff}, \eq{PiR_free} and \eq{sol_pi_free} for $g_\mathrm{eff}$ allows the numerical evaluation of the scattering integral \eq{SM:IchnumFree_2} which gave the results in \Sect{NumericalFree}.

\subsubsection{Kinetic scattering integral for Bogoliubov sound waves}
\label{app:KinScattIntBog}
Alternatively, we consider the case of a macroscopic zero-mode population with Bogoliubov quasiparticle excitations.
In the regime where the dispersion is linear, \Eq{sounddispersion}, the Bogoliubov excitations are sound waves.
Inserting the spectral function \eq{rhoBog} into the integrand \eq{Issisipsp} gives
\begin{align}
    \mathcal{I}_{s\sigma\sigma's'}&
    = \frac{(2 \pi g\rho_{0})^{4}}{\omega_{\vec p}\omega_{\vec k}\omega_{\vec q}\omega_{\vec r}}
     \delta(sp_{0}+\sigma k_{0}-\sigma'q_{0}-s'r_{0})
     \nonumber\\
   &\times\   g_\mathrm{eff}^{2}(sp_{0}-s'r_{0},\vec p-\vec r)\, 
     \nonumber\\
   &\times\   \delta(p_{0}-\omega_{\vec p})\delta(k_{0}-\omega_{\vec k})
   \delta(q_{0}-\omega_{\vec q})\delta(r_{0}-\omega_{\vec r}) 
     \nonumber\\
  &\ \  \,
   \times\ \big\{ \big[\sigma f(p_{0})+s f(k_{0})\big]\,f(q_{0})f(r_{0})
     \nonumber\\
  & \ \  \, 
    \phantom{\big\{\big[}-\ f(p_{0})f(k_{0})\,\big[s'f(q_{0})+\sigma'f(r_{0})\big]
     \big\}\,.
     \label{eq:IssisipspBog1}
\end{align}
Only a subset of these integrands carries a non-vanishing contribution to the scattering integral \eq{ScattIntwithrhosPosFreq}.
To identify this we integrate out the frequencies and the momentum $\vec q$, and shift $\vec r=\vec p-\vec r'$. 
We then rewrite the integral over $\vec k$ into an integral over $k$ and one over the angle $\measuredangle(\vec r', \vec k)$ which we replace by an integral over $q=|\vec r'+\vec k|$.
Omitting the prime of the transfer momentum $\vec r'$, and taking into account that $\omega_{\vec p}\equiv\omega_{p}$ is isotropic, we get
\begin{align}
  I(\vec p) 
      =\  &\frac{(2\pi)^{1-3d}}{2}\int\!\frac{\mathrm{d}^{d}r}{r}\, 
        \int_{0}^{\infty}k\,\mathrm{d}k\,\int_{|r-k|}^{|r+k|}q\,\mathrm{d}q\,
    \nonumber\\
   &\times\ \sum_{s,s'\in\{-1,1\}}
   \big[\mathcal{\tilde I}_{1sss'}+\mathcal{\tilde I}_{1s(-s)s'}
   \big],
     \label{eq:ScattIntqrIntegratedOut}
\end{align}
where
\begin{align}
    \mathcal{\tilde I}_{s\sigma\sigma's'}&
    = \frac{(2 \pi g\rho_0)^{4}}{\omega_{p}\omega_{k}\omega_{q}\omega_{|\vec p-\vec r|}}
     g_\mathrm{eff}^{2}(s\omega_{p}-s'\omega_{|\vec p-\vec r|},\vec r)\,
      \nonumber\\
   &\times\   \delta(s\omega_{p}+\sigma \omega_{k}-\sigma'\omega_{q}-s'\omega_{|\vec p-\vec r|})
     \nonumber\\
  &\ \  \,
   \times\ \big\{ \big[\sigma f_{p}+s f_{k}\big]\,f_{q}f_{|\vec p-\vec r|}
     \nonumber\\
  & \ \  \, 
    \phantom{\big\{\big[}-\ f_{p}f_{k}\,\big[s'f_{q}+\sigma'f_{|\vec p-\vec r|}\big]
     \big\}\,,
     \label{eq:IssisipspBog2}
\end{align}
and $f_{p}\equiv f(\omega_{\vec p})$.
In order to reduce the set of integrands we analyze the energy conservation delta function for the case of the Bogoliubov dispersion \eq{BogDispersion}: $\delta(s\omega_{p}+\sigma \omega_{k}-\sigma'\omega_{q}-s'\omega_{|\vec p-\vec r|})=\mathrm{const.}\times\delta(|\vec p-\vec r|- ss'[p+s\sigma (k-\sigma\sigma'q)])$.
Since $|k-q|\leq r$ and $k+q\geq r$ we find that the argument of the delta distribution can vanish only if $\sigma\sigma'=1$ and $ss'=1$ or if $\sigma\sigma'=-1$ and $ss'=-1$ and $s=-\sigma$.
This leaves us with the following terms:
\begin{align}
  &I(\vec p) 
      =\  \frac{(2\pi)^{1-3d}}{2}\int\!\frac{\mathrm{d}^{d}r}{r}\, 
        \int_{0}^{\infty}k\mathrm{d}k\,\int_{|r-k|}^{|r+k|}q\mathrm{d}q\,
    \nonumber\\
   &\times\ 
   \big[\mathcal{\tilde I}_{1111}+\mathcal{\tilde I}_{1(-1)(-1)1}
      +\ \mathcal{\tilde I}_{1(-1)1(-1)}\big].
     \label{eq:ScattIntqrIntegratedOutECons}
\end{align}
where $\mathcal{\tilde I}_{s\sigma\sigma's'}\equiv\mathcal{\tilde I}_{s\sigma\sigma's'}(\vec p,k,q,\vec r)$.

Turning back to the full integral \eq{ScattIntwithrhosPosFreq} and keeping only the nonvanishing contribution results in
\begin{align}
  I(\vec p) 
    =&\  (2\pi)^{4}  \int_{\vec k\vec q\vec r}  \frac{(g\rho_0)^{4}}{\omega_{\vec p}\omega_{\vec k}\omega_{\vec q}\omega_{\vec{r}}}\,  \nonumber \\
   &  \times\ \Big[ g_\mathrm{eff}^{2}(\omega_{\vec p}-\omega_{\vec r},\vec p-\vec r) 
         + \frac{1}{2}g_\mathrm{eff}^{2}(\omega_{\vec p}-\omega_{\vec k},\vec p+\vec k) \Big] \, \nonumber \\
   &   \times\  \delta(\vec p+\vec k-\vec q-\vec r) \, \delta(\omega_{\vec p}+\omega_{\vec k}-\omega_{\vec q}-\omega_{\vec r}) \,  \nonumber \\
   &   \times\ \big\{ \big[f(\omega_{\vec p})+ f(\omega_{\vec k})\big]\,f(\omega_{\vec q})f(\omega_{\vec r})
      \nonumber \\
    &  \qquad-\  f(\omega_{\vec p})f(\omega_{\vec k})\,\big[f(\omega_{\vec q})+f(\omega_{\vec r})\big]
     \big\} \,.
     \label{eq:IsspspsBog3}
\end{align}
Exchanging, finally, in each summand the spatial integration variables suitably,
one recovers the Boltzmann equation  \eq{QKinEq} for $n_{Q}(t,\mathbf{p})=f(t,\varepsilon(\mathbf{p}))\gg1$, with the scattering integral \eq{KinScattIntCWL} in the classical-wave approximation, and the $T$-matrix elements,
\begin{align}
  &|T_{\vec p\vec k\vec q\vec r}|^{2}
  =\ (2\pi)^{4}\frac{(g\rho_{0})^{4}}{\omega_{p}\omega_{k}\omega_{q}\omega_{r}}
    \nonumber\\
    &\times\
    \big[g_\mathrm{eff}^{2}(\omega_{p}-\omega_{r},\vec p-\vec r)+ \frac{1}{2}
    g_\mathrm{eff}^{2}(\omega_{p}-\omega_{k},\vec p + \vec k)\big]
       \,.
  \label{eq:app:TitogeffQP}
\end{align}
The $T$-matrix elements depend on the quasiparticle distribution $n_{Q}(\vec p,t)$.
We evaluate these matrix elements in \App{EffCouplingFctBog}.

For the details of the evaluation of the scattering integral in the Bogoliubov case, which proceeds in the same way as in the free case, we refer to \cite{Chantesana2017a} and here quote the resulting expression:
\begin{align}
& I(\vec p) 
 = \frac{(g\rho_0)^4}{(2\pi) c^5_s}\frac{\Lambda^{3\kappa}}{p} \Bigg(\int_{\vec r'}\frac{1}{|\vec p -\vec r'| r'} g_\mathrm{eff}^{2}(\omega_{\vec p}-\omega_{\vec p - \vec r'},\vec r')\, \nonumber \\
& \qquad \times \Bigg\{ \frac{I^{\mathrm{bog}}_1(\vec p,\vec r')}{(p+p_\Lambda)^\kappa (|\vec p -\vec r'|+p_\Lambda)^\kappa } \nonumber \\
& \qquad \  + \Bigg[ \frac{1}{(|\vec p - \vec r'|+p_\Lambda)^\kappa} - \frac{1}{(p+p_\Lambda)^\kappa} \Bigg] I^{\mathrm{bog}}_2(\vec p,\vec r') \Bigg\} 
\nonumber\\
  &\quad+\ \int_{k'}\frac{1}{2k'|\vec k' -\vec p| }\, g_\mathrm{eff}^{2}(\omega_{\vec p}-\omega_{\vec k' - \vec p},\vec k') \, \nonumber \\
  & \qquad \times \Bigg\{ -\frac{I^{\mathrm{bog}}_3(\vec p,\vec k')}{(p+p_\Lambda)^\kappa (|\vec k' -\vec p|+p_\Lambda)^\kappa }   \nonumber \\
  & \qquad\ + \Bigg[ \frac{1}{(\vec p +p_\Lambda)^\kappa} + \frac{1}{(|\vec k' -\vec p|+p_\Lambda)^\kappa} \Bigg] I^{\mathrm{bog}}_4(\vec p,\vec k') \Bigg\}\Bigg) \,, 
  \nonumber 
\end{align}
\begin{align}
&I^{\mathrm{bog}}_1(\vec p, \vec q) 
= \frac{{1 }}{\kappa-1}\Big[ \Big(\mbox{$\frac{q+ p - r}{2}$} + p_\Lambda \Big)^{1-\kappa} 
- \Big( \mbox{$\frac{q- p + r}{2}$} + p_\Lambda  \Big)^{1-\kappa} \Big] \,,
\nonumber \\
 &I^{\mathrm{bog}}_2(\vec p, \vec q) 
 = \frac{ (2q+4p_\Lambda)^{2\kappa-1}}{2\kappa -1}\,  
 {}_2F_1\Big(\kappa , \kappa - \mbox{$\frac{1}{2}$},\kappa + \mbox{$\frac{1}{2}$}; \Big(\mbox{$\frac{p-r}{q+2p_\Lambda}$}\Big)^2\Big) \,,  
\nonumber \\
&I^{\mathrm{bog}}_3(\vec p,\vec q) 
= \frac{2}{\kappa-1}\, \Big[ \Big(\mbox{$\frac{p+k - q}{2}$} + p_\Lambda\Big)^{1-\kappa} 
- \Big(\mbox{$\frac{p+k + q}{2}$} + p_\Lambda\Big)^{1-\kappa} \Big] \,,
\nonumber \\
&I^{\mathrm{bog}}_4(\vec p, \vec q) 
= q\,\Big(\mbox{$\frac{p+k}{2}$} + p_\Lambda\Big)^{-2\kappa}
{}_2F_1\Big( \kappa, \mbox{$\frac{1}{2}$} ;\mbox{$\frac{3}{2}$}; \Big(\mbox{$\frac{q}{p + k+2p_\Lambda}$} \Big)^2\Big) \,.
  \label{eq:IchnumBog_2}
\end{align}
%

\section{Effective many-body coupling function $g_\mathrm{eff}(p)$}
\label{app:geff}
In this appendix we discuss in detail  the structure of the effective, momentum dependent coupling $g_{\mathrm{eff}}(p)$ in both, the perturbative and collective-scattering regimes.
We assume the momentum distribution to assume a general scaling form as defined in \Eq{scalingf}, and will discuss the whole regime of possible exponents $\kappa$, including for non-thermal fixed points, weak-wave turbulence, and thermal distributions.  
The appendix complements the discussion of the effective coupling in \Sect{ScattTM} in the main text, providing, amongst other things, details of the derivation of Eqs.~\eq{geffFreeUniversal} and \eq{app:geffBogUniversal}. 

$g_{\mathrm{eff}}(p)$, which defines the $T$-matrix elements, Eqs.~\eq{Titogeff} and \eq{TitogeffQP}, is the result of the $s$-channel loop resummation and can be written as
\begin{align}
    g_{\mathrm{eff}}(p)
    &= \frac{g}{\left|1+g\,\Pi^R(p)\right|},
\label{eq:app:geff}
\end{align}
see \App{2PIKineticTheory} for details of the field theoretical formalism, as well as \Fig{2PI}.
Here, $\Pi^R(p)= -(\theta\ast\Pi^{\rho})(p)$, cf.~\Eq{PiA}, is the retarded loop integral.
Within the kinetic approximation introduced in \App{KineticQP}, $\Pi^\rho(p)$ can be written in terms of the quasiparticle frequency spectrum $f(t,p_{0})$ and the spectral function $\rho(p)$, cf.~Eqs.~\eq{PiFrhopStatClass} and \eq{CallanWelton}, as
\begin{align}
\label{eq:PirhoKinetic}
 &\Pi^\rho(  p) = \frac{1}{2}(\rho_{ab}*F_{ba}-F_{ab}*\rho_{ba})(p) \nonumber \\
 & = \frac{1}{2} \int_{k}\Big[ \rho_{ab}(p-k)F_{ba}(-k) -F_{ab}(p-k) \rho_{ba}(-k) \Big] \nonumber \\
 & = -\frac{i}{2} \int_{k}\Big[f(-k_0)-f(p_0-k_0)\Big]
 \rho_{ab}(-k)\rho_{ba}(p-k)\,,
\end{align}
where $\int_{k}=(2\pi)^{-d-1}\int d^{d+1} k$ and, here and in the following, we suppress the time argument $t$ of $f(p_{0};t)$.

In the following, we will derive the coupling function $g_{\mathrm{eff}}(p)$ within the kinetic approximation, for given spectral functions of both, free and Bogoliubov quasiparticles in $d=3$ dimensions, and assuming a particular scaling form for the quasiparticle distribution function.

\subsection{Universal $g_\mathrm{eff}(p)$ for free particles}
\label{app:EffCouplingFctFree}
\subsubsection{2PI loop resummation}
We first discuss the case of free particles.
The retarded one-loop function $\Pi^{R}$ includes the spectral function which contains the eigenfrequencies of the (quasi)particle modes, as well as their occupation defined by the distribution function $f\equiv f_\mathrm{neq}$, \Eq{CallanWelton}.
Inserting the spectral function \eq{rhofree} into \Eq{PirhoKinetic}, one obtains
\begin{align}
\Pi^\rho&(  p^0, \vec p) 
= {-2\pi i} \int \frac{d^d k}{(2\pi)^d}f(\varepsilon_{\vec k})\Big[
\delta(p_{0}-\varepsilon_{\vec k}+\varepsilon_{\vec p-\vec k})
\nonumber\\
&\qquad\qquad\qquad\qquad-\  
\delta(p_{0}+\varepsilon_{\vec k}-\varepsilon_{\vec p-\vec k})\Big].
\label{eq:Pirhoitof}
\end{align}
Inserting this into $\Pi^R$, \Eq{PiA}, we obtain, in $d=2,3$ dimensions,
\begin{align}
\Pi^R &(E,\vec{p}) 
   =  \int^{\infty}_{-\infty} \frac{dq^0}{2\pi} \frac{i}{q^0+i\epsilon}\Pi^{\rho}(q^0-p^0,-\vec{p}) 
   \nonumber \\
   &=    \int \frac{d^d k}{(2\pi)^d}f(\varepsilon_{\vec k})\left[
   \frac{1}{p_{0}-\varepsilon_{\vec k}+\varepsilon_{\vec p-\vec k}+i\epsilon}\right.
   \nonumber\\
   &\qquad\qquad\quad\qquad -\  \left.
   \frac{1}{p_{0}+\varepsilon_{\vec k}-\varepsilon_{\vec p-\vec k}+i\epsilon}\right]  \,.
   \label{eq:PiRFullIntegral}
   \nonumber \\
   &=   \frac{2mS_{d-2}}{(2\pi)^{d}} \int^\infty_0 dk \,k^{d-1} \int^1_{-1}\frac{d\cos\theta}{\sin^{3-d}\theta} f(\varepsilon_{\vec k}) 
   \nonumber \\
   &\quad \times \left[
   \frac{1}{E-k^2+|\vec p - \vec k|^2+i\epsilon} 
  - \frac{1}{E+k^2-|\vec p - \vec k|^2+i\epsilon}\right] \,,
\end{align}
where we defined $E = 2mp^0$,  $\theta$ is the angle between $\vec p$ and $\vec k$, and
\begin{align}
  S_{d-1}=\frac{2\pi^{d/2}}{\Gamma(d/2)}
\end{align}
is the surface area of the unit sphere in $d$ dimensions.
Integrating out the angular part of the spatial convolution, one obtains, in $d=3$ dimensions, the energy and momentum dependence 
\begin{align}
&\Pi^R (E,\vec{p})
  \label{eq:PiRFree_fepsilonk}
   =   \frac{1}{(2\pi)^{2}} \frac{m}{p}\int^\infty_0 dk \,k^{2}\, \int^1_{-1}dz\, f(\varepsilon_{\vec k}) 
   \nonumber \\
   &\times \left[
   \left({\frac{E+p^2}{2p}-kz+i\epsilon}\right)^{-1} - \left({\frac{E-p^2}{2p}+kz+i\epsilon}\right)^{-1}\right] 
   \nonumber \\
   &=    \frac{1}{(2\pi)^{2}} \frac{m}{p}
   \left[ 
          \widetilde{\Pi}_{f} \left(\mbox{$\frac{E+p^2}{2pp_{\Lambda}}$}\right) 
       -  \widetilde{\Pi}_{f} \left(\mbox{$\frac{E-p^2}{2pp_{\Lambda}}$}\right) \right] \,,
\end{align}
where we defined $E = 2mp^0$, $p=|\vec p|$. 
We factor out the infrared cutoff  $p_{\Lambda}$ such that the remaining momentum integral in the one-loop function can be written as
\begin{align}
  \widetilde{\Pi}_{f}(x)
  &= p_{\Lambda}^{2}\int^\infty_{0} dy\,{y}\, f(\varepsilon_{yp_{\Lambda}})
  \ln\left(\frac{x+y+i\epsilon}{x-y+i\epsilon}\right) \,,
\label{eq:Pif_free}
\end{align}
with $\varepsilon_{k}=k^{2}/2m$.
To proceed, make an ansatz for the particle distribution $f(\varepsilon_{\vec k})$.
We assume it to take a scaling form,  
\begin{equation}
 f(p_{0}) = \theta(\varepsilon_{p_{\lambda}}-p_{0})\,\text{sgn}(p_{0})\left(\frac{\varepsilon_{\Lambda}}{p_{0} + \varepsilon_{p_\Lambda}}\right)^{\kappa/2}\,,
 \label{eq:fmychoice_free}
\end{equation}
where $p_{\Lambda}$ takes the role of an infrared cutoff, $p_{\lambda}>p_{\Lambda}$ is a sharp UV cutoff, and $\varepsilon_{\Lambda}=\Lambda^{2}/2m$ defines a further scale $\Lambda$ parametrizing the amplitude.
This ansatz interpolates in a smooth way between a constant in the infrared limit and a power-law fall-off $f(\varepsilon_{p})\sim p^{-\kappa}$, with crossover scale $p_{\Lambda}$, and thus has the same momentum scaling as the form in \Eq{scalingf}.
For simplifying the following derivations, it has a different crossover behaviour at $p_{0}\simeq\varepsilon_{p_{\Lambda}}$.
While the precise form at and below the crossover scale can be different we are, here, primarily interested in  the analytical structure of $\Pi^{R}$ and seeing its (in)dependence (of) on the power-law exponent $\kappa$.
The signum function is introduced to account for the anti-symmetry of $f$ in $p_{0}$, cf.~\Eq{fantisymmetry}.

We insert the ansatz \eq{fmychoice_free} for the quasiparticle distribution $f(\varepsilon_{\vec k})$ into \Eq{Pif_free}, and obtain
\begin{align}
  \widetilde{\Pi}_{f}(x)
  &= \frac{\Lambda^{\kappa}}{p_{\Lambda}^{\kappa-2}}
  \int^{y_{\lambda}}_{0} \frac{dy\,y}{\left({1+y^2}\right)^{\kappa/2}}
  \ln\left(\frac{x+y+i\epsilon}{x-y+i\epsilon}\right)\,. 
\label{eq:app:Pif_free_withf}
\end{align}
The scale $\Lambda$ is fixed by the normalization of the particle distribution to the density $\rho_\mathrm{nc}$ of non-condensed particles,
\begin{align}
 \rho_\mathrm{nc} 
 & = \frac{S_{d-1}\Lambda^{\kappa}p_\Lambda^{d-\kappa}}{2(2\pi)^{d}}
 \int^{y_{\lambda}^{2}}_0 \frac{du\, u^{d/2-1}}{ (1+u)^{\kappa/2}} 
 \nonumber \\
 &= \frac{\Lambda^{\kappa}p_\lambda^{d}p_\Lambda^{-\kappa}}{2^{d}\pi^{d/2}\,\Gamma({d}/{2}+1)}
       {}_2F_1\left(\mbox{$\frac{\kappa}{2},\frac{d}{2};\frac{d}{2}+1;-\left[\frac{p_{\lambda}}{p_{\Lambda}}\right]^{2}$}\right)\,,
 \label{eq:nitofscalingfree}
\end{align}
and thus the normalisation factor reads, for $d=3$,
\begin{align}
 \frac{\Lambda^{\kappa}}{p_{\Lambda}^{\kappa-2}} 
 &= {(2\pi)^{2}}
       { C_{\kappa}(y_{\lambda})}
       \, \frac{ \rho_\mathrm{nc}}{p_\Lambda}       \,,
 \label{eq:Lambdatokappa_free_UV}
\end{align}
with       
\begin{align}
  C^{-1}_{\kappa}&(y) = \frac{2y^{3}}{3}\,{}_2F_1\left(\frac{\kappa}{2},\frac{3}{2};\frac{5}{2};-y^{2}\right)
  \nonumber \\
  &\approx \frac{1}{3-\kappa}\left[ 
     2y^{3-\kappa} 
  -  \frac{\sqrt{\pi}\,\Gamma(\frac{\kappa-1}{2})}{\Gamma(\frac{\kappa}{2})} \right]  \quad (\mbox{for}\ \ y\gg1)\,.
 \label{eq:Ckappay_free}
\end{align}
We see that, while the integral \eq{app:Pif_free_withf} is UV convergent for $\kappa>2=z$, the integral \eq{nitofscalingfree} defining the normalization is UV divergent for $\kappa\leq d$.
This implies that, as $\kappa$ approaches $d$ from above, the density $\rho_\mathrm{nc}$ becomes dominated by the occupation number at the UV end of the scaling form and thus the normalization $\Lambda$, for a given total density $\rho_\mathrm{nc}$ becomes sensitive to the UV cutoff $p_{\lambda}$.
To take this effect into account is crucial for understanding the crossover behaviour of the effective coupling as $\kappa$ is varied from $\kappa\gg d$ to $\kappa=z=2$.

In the main text we derive $\kappa=d+1$ at a non-thermal fixed point, for $z=2$, $\eta=0$, cf.~\Eq{kappaSfreecoll}, such that $\kappa>d$ is fulfilled, which  turns out to be also the case for exponents $\kappa_{Q}=d+2/3$, $\kappa_{P}=d+4/3$ characterising strong-wave-turbulent cascades, cf.~\Eq{kappaQPSWT}.
For these solutions, fulfilling $\kappa>d$, the effective coupling will become insensitive to the UV cutoff $p_{\lambda}$ and thus represent a universal quantity characteristic for critical scaling phenomena.

Inserting  \eq{Lambdatokappa_free_UV} into \eq{app:Pif_free_withf}, and this into \eq{PiRFree_fepsilonk}, we obtain 
\begin{align}
  \Pi^R (E,p) 
  &= 
    \frac{p_{\Xi}^{2}}{2gpp_{\Lambda}}
  \left[ 
         \tilde{\pi}_\kappa \left(\mbox{$\frac{E+p^2}{2pp_{\Lambda}}$}\right) 
       - \tilde{\pi}_\kappa \left(\mbox{$\frac{E-p^2}{2pp_{\Lambda}}$}\right) \right] \,,
 \label{eq:PiR_free}
\end{align}
where $p_{\Xi}=\sqrt{8\pi a\rho_\mathrm{nc}}$ is the momentum scale corresponding to the `healing' length set by the density $\rho_\mathrm{nc}$ of \emph{non-condensate} particles, see the discussion following \Eq{pmu}.
$ \tilde{\pi}_\kappa$ contains the integral over $y$ which can be expressed in terms of Gaussian hypergeometric functions as \cite{bateman1955higher},
\begin{align}
  \tilde{\pi}_\kappa(x) 
  &= -\,C_{\kappa}(y_{\lambda})
  \int^{y_{\lambda}}_{0} dy\,\frac{y}{(1+y^2)^{\kappa/2}}
  \ln\left(\frac{x+y+i\epsilon}{x-y+i\epsilon}\right) 
  \nonumber \\
  &= \frac{C_{\kappa}(y_{\lambda})}{\kappa-2}\Bigg\{
  \frac{\sqrt{\pi}}{x}  \frac{\Gamma(\frac{\kappa-1}{2})}{\Gamma(\frac{\kappa}{2})\,}
  {}_2F_1\left(1,\frac{1}{2};\frac{\kappa}{2};1+\left[{(1 \pm i \epsilon)x}\right]^{-2} \right)
  \nonumber \\
  & \qquad + x\,y^{1-\kappa}_\lambda \frac{\Gamma(\frac{\kappa -1}{2})}{\Gamma(\frac{\kappa+1}{2})}\,
  {}_2F_1\Big( 1, \frac{\kappa-1}{2};\frac{\kappa+1}{2}; \big[(1\pm i\epsilon)\frac{x}{y_\lambda}\big]^2 \Big) 
  \nonumber \\
  & \qquad -\left(1+y^2_\lambda\right)^{1-\kappa/2}
  \ln\left( \frac{1+y_\lambda/x \pm i\epsilon}{1-y_\lambda/x \pm i\epsilon}\right)   
  \Bigg\}\,.
  \label{eq:sol_pi_free}
\end{align}
Here the $+(-)$ sign of the infinitesimal imaginary shift applies in the case $x>0$ ($x<0$). 
In the second line we sent the UV cutoff to infinity, resulting in the hypergeometric function, and subtracted the integral from $y_{\lambda}$ to $\infty$ which gives the third and fourth lines.

If $x\gg1$, i.e., for momenta/energies sufficiently far above the infrared cutoff, we can simplify $\tilde{\pi}_\kappa(x)$ using the analytic continuation of the hypergeometric function  \cite{Chantesana2017a} and from this derive a simple approximate scaling form of the effective coupling $g_\mathrm{eff}(p)$.
Assuming that $\kappa$ is not an integer 
\footnoteremember{fn1}{The case of an integer $\kappa$ requires further discussion due to the non-simple pole structure of the integral representation of the hypergeometric function in the form of a Mellin-Barnes integral. As our numerical results presented in \Sect{SummaryNumerical} demonstrate, there is no discontinuity in the transition from non-integer to integer values of $\kappa$. We therefore use the expressions for non-integer $\kappa$ and take the limit to an integer value.}, 
and that $1\ll x\ll y_{\lambda}$, one finds the leading behaviour 
\begin{align}
  \tilde{\pi}_\kappa(x) 
  \simeq C_{\kappa}(y_{\lambda})\,&
   \Bigg[ \frac{\sqrt{\pi}\ \ \Gamma(\frac{\kappa-1}{2})}{(\kappa-3)\,\Gamma(\frac{\kappa}{2})}\frac{1}{x} 
   -\frac{2}{\kappa-1}\,{x}\,y_{\lambda}^{1-\kappa}
   \nonumber\\
   &-\frac{i{\pi}}{\kappa-2}\left({|x|}^{2-\kappa}-{y_{\lambda}}^{2-\kappa}\right)\Bigg]\,,
  \label{eq:pikappa-one-x-pl}
\end{align}
where the $y_{\lambda}$-dependent terms arise from the subtraction. 
The last, $x$-independent term drops out when inserting $\tilde\pi_{\kappa}$ into \eq{PiR_free}, such that $\Pi^{R}$ is UV-divergent for $\kappa<1$ while $\tilde\pi_{\kappa}$ diverges already for $\kappa<2$.

Far below the IR cutoff, for $x\ll 1\ll y_{\lambda}$, one obtains
\begin{align}
  \tilde{\pi}_\kappa(x) 
  \simeq C_{\kappa}(y_{\lambda})\,x\,&
   \Bigg\{ \frac{\sqrt{\pi}\ \ \Gamma(\frac{\kappa-1}{2})}{\Gamma(\frac{\kappa}{2})}
   -\frac{2}{\kappa-1}\,y^{1-\kappa}\Bigg\}+\mbox{const.}\,.
  \label{eq:pikappa-x-one-pl}
\end{align}
Using the approximation \eq{Ckappay_free} of $C_{\kappa}(y_{\lambda})$ for large $y_{\lambda}=p_{\lambda}/p_{\Lambda}\gg1$, we need to distinguish the cases $\kappa>3$ and $\kappa\leq3$.

\subsubsection{Scaling of $g_\mathrm{eff}(p)$ for $\kappa>3$}
For $\kappa>3$, the first term in \eq{Ckappay_free} can be neglected, as well as the $y_{\lambda}$-dependent terms in Eqs.~\eq{pikappa-one-x-pl} and \eq{pikappa-x-one-pl} and obtain the leading behaviour as
\begin{align}
 \tilde{\pi}_\kappa(x)
 &\simeq \frac{1}{x}-i\sqrt{\pi}\,  
 \frac{\Gamma\big(\frac{\kappa-2}{2}\big)}{\Gamma\big(\frac{\kappa-3}{2}\big)}
 |x|^{2-\kappa} \,,
\label{eq:pitilde_xgge}
\end{align}
noting the exact limit $\lim_{x\to\infty}[x\,\tilde{\pi}_\kappa(x)]=1$.
In the opposite limit $x\ll1$, i.e., far below the IR cutoff, one finds, for $\kappa\gtrsim3$,
\begin{align}
 \tilde{\pi}_\kappa(x)
 &\simeq (\kappa-3)x\mp i\sqrt{\pi}\,  
 \frac{\Gamma\big(\frac{\kappa-2}{2}\big)}{\Gamma\big(\frac{\kappa-3}{2}\big)}
 \left(1-\frac{\kappa-2}{2}x^{2}\right)
   +\mathcal{O}(x^{3})  \,,
\label{eq:pitilde_xlle}
\end{align}
where the $-(+)$ sign of the imaginary part applies for $x>0$ ($x<0$).
This
suggests that for $\kappa\searrow3$, the function $\tilde{\pi}_\kappa(x)$, evaluated at a finite $x$, vanishes.
This is, however, an effect of having sent the UV cutoff to $\infty$ and is counterbalanced by the neglected $y_{\lambda}$-dependent terms.

From \Eq{pitilde_xgge} one finds that, for $\kappa\gtrsim4$, the real part dominates above the infrared cutoff such that
\begin{align}
g \Pi^R(E,p) 
   &\simeq   \frac{2p_\Xi^2 p^2}{p^4-E^2} 
   - i\sqrt{\pi}\,\frac{p_\Xi^{2}}{2pp_\Lambda}
   \frac{\Gamma\Big(\frac{\kappa-2}{2}\Big)}{\Gamma\Big(\frac{\kappa-3}{2}\Big)} \nonumber \\
  &\phantom{aaaa} \times\ \left( \left| \frac{E+p^2}{2p\,p_\Lambda} \right|^{2-\kappa} 
  - \left| \frac{E-p^2}{2p\,p_\Lambda} \right|^{2-\kappa}  \right) \, ,
  \label{eq:gPiR-scaling-regime}
\end{align}
As a result, for $|E \pm p^2|\gg2p_\Lambda p$, the loop integral scales as $\Pi^R(s^2 E, s p) = s^{-2}\Pi^R(E,p)$.
Inserting this into \Eq{app:geff}, we find that, in the momentum region $p_\Lambda\ll |E \pm p^2|/p\ll p_{\Xi}$ the effective coupling assumes the universal scaling form \eq{geffFreeUniversal} quoted in the main text,
\begin{equation}
 \label{eq:app:geffFreeUniversal}
 g_{\mathrm{eff}}(p_{0},p) \simeq \frac{\left|\varepsilon_{p}^{2}-p_{0}^{2}\right|}{2\rho_\mathrm{nc}\, \varepsilon_{p}}\,,
\end{equation}
independent of both, the microscopic interaction constant $g$, and the scaling exponent $\kappa$ of $f$.
Moreover, it is scaling as \Eq{geffscaling}.
Together with Eqs.~\eq{Titogeff}, \eq{geffScalingHypothesis} this gives the scaling exponents \eq{gammanonpert}, \eq{mSWT} of the many-body $T$-matrix, $\gamma_{\kappa}=m_{\kappa}=2$.

At larger energy and momentum scales, above the healing-length scale, $|E \pm p^2|/p\gg p_{\Xi}$, the one-loop function \eq{gPiR-scaling-regime} falls below $1$, and the effective coupling \eq{app:geff} saturates at the microscopic interaction constant, $g_{\mathrm{eff}}\simeq g$, recovering $\gamma_{\kappa}=0$ and the perturbative Boltzmann $T$-matrix \Eq{Titogbare} with scaling exponent $m_{\kappa}=0$.
We emphasize that, while the transition scale from the microscopic coupling $g$ to the universal scaling form \eq{app:geffFreeUniversal} is set by $p_{\Xi}$ and thus by the microscopic coupling $g$, the particular value of the \emph{universal} coupling $g_{\mathrm{eff}}$ is independent of $g$ but only depends on $\rho_\mathrm{nc}$.

\subsubsection{Universality of $g_\mathrm{eff}(p)$ for $\kappa>3$}
\label{app:Universalitygeffkappagtthree}
The independence of $\kappa$ for steep scalings, $\kappa>3$, stems from the fact that the integrals entering the effective coupling are all dominated by the IR end of the momentum range such that the coupling becomes universal in the sense that it is independent of the details in the UV, and in particular does not depend on a possible UV cutoff $p_{\lambda}$.
Moreover, the dependence on the remaining IR scale $p_{\Lambda}$ is replaced, using the constraint that the integral over the distribution must give the total density of particles $\rho_\mathrm{nc}$, by the scale $p_{\Xi}$, \Eq{pmu}, which depends on $g$ and $\rho_\mathrm{nc}$.
As a result, also the dependence on the microscopic coupling $g$ disappears, which underlines the universality.

The above results belong to the most central ones of the present work. 
They show that, for a given coupling constant $g$ (i.e., scattering length $a$), non-perturbative universal scaling to occur at a certain energy-momentum transfer $(E,p)$ requires a sufficiently strong non-condensate density 
\begin{align}
 \rho_\mathrm{nc}\gg \frac{1}{2\pi a}\frac{(E+p^{2})^{2}}{p^{2}}.
\end{align}
Alternatively, for a given $\rho_\mathrm{nc}$ and $(E,p)$, non-perturbative corrections become important for a sufficiently large coupling $g$.
Note that the effective coupling becomes universal in a similar sense as in the unitary limit of $g\to\infty$, as at a Feshbach resonance, where the quantum corrections to $\Pi^{R}$ which have been neglected here lead to a UV scaling $|g_\mathrm{eff}|^{2}\sim p^{-2}$ as $p\to\infty$.

Far below the IR cutoff, for $x\ll1$, the one-loop function, according to \eq{pitilde_xlle}, approaches, for $\kappa\gg3$,
\begin{align}
g \Pi^R(E,p) 
   &\simeq   \frac{p_\Xi^2 }{2p_{\Lambda}^2} 
   \left(\kappa-3
   \right)\, .
\end{align}
Hence, in the IR limit, for  $|E \pm p^2|/p\ll p_\Lambda\ll p_{\Xi}$, the effective coupling saturates at the constant value
\begin{equation}
 \label{eq:geffFreeUniversalIRlimit}
 g_{\mathrm{eff}}(p_{0},p) 
 \simeq
 \frac{2\varepsilon_{p_{\Lambda}}}{\rho_\mathrm{nc}}(\kappa-3)^{-1}
 \,.
\end{equation}
%

%
\begin{figure}[t]
    \centering
    \includegraphics[width=0.4\textwidth]{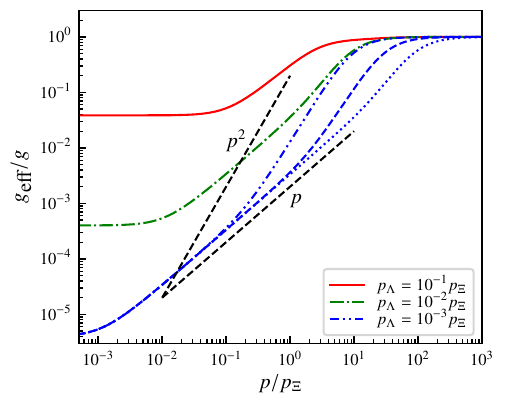}
    \caption{Effective coupling $g_\mathrm{eff}(E,p)$ as a function of momentum $p$. 
    Shown are cuts in the $p$--$E$-plane, with $E = 0.99p^2$ (red solid, green dot-dashed and blue double-dot-dashed lines), $E = 0.999p^2$ (blue dashed), $E = 0.9999p^2$ (blue dotted).
    Different colors refer to different infrared cutoff scales $p_{\Lambda}$ as listed in the legend.
    }
    \label{fig:EffCouplingfreeCuts-a}
\end{figure}
%

%
\begin{figure}[t]
    \centering
    \includegraphics[width=0.4\textwidth]{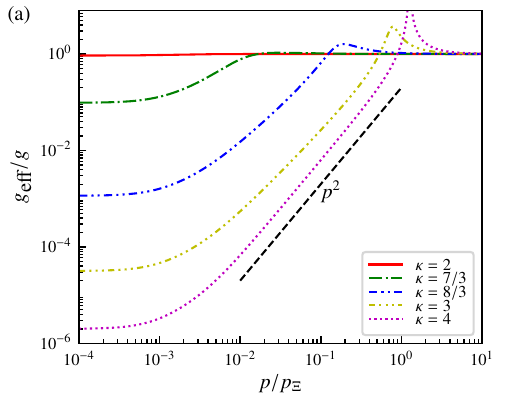}\\
    \includegraphics[width=0.23\textwidth,height=0.134\textheight]{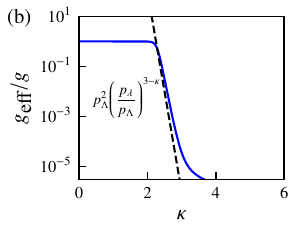}
    \includegraphics[width=0.23\textwidth]{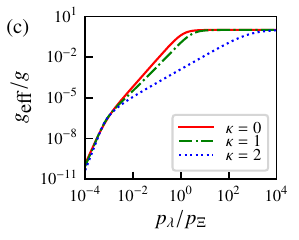}\\
    \includegraphics[width=0.23\textwidth]{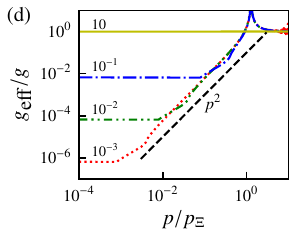}
    \includegraphics[width=0.23\textwidth]{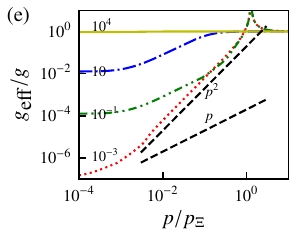}
    \caption{Effective coupling $g_\mathrm{eff}(E,p)$ for $E = 1.5p^2$, $p_{\Lambda}=10^{-3}\,p_{\Xi}$.
    The panels show the coupling as a function of (a) $p$, for different $2\leq\kappa\leq4$ (see legend), $p_{\lambda}\to\infty$;
    (b) of $\kappa$, for $p=10^{-2}p_{\Lambda}$ and $p_{\lambda}=10^{5}p_{\Xi}$ (dashed line: approximate scaling \eq{geffFreeUniversalIRlimit-kltd});
    (c) of $p_{\lambda}$, for $p=10^{-2}p_{\Lambda}$, $\kappa=0$, $1$, and $2$ (see legend);
    (d) of $p$, for $\kappa=0$ and different $p_{\lambda}$ as labeled;
    (e) of $p$, for $\kappa=2$ and different $p_{\lambda}$ as labeled.
    }
    \label{fig:EffCouplingfreeCuts-c}
\end{figure}
%

\Fig{EffCouplingFree} shows the effective coupling constant in the $E$--$p$ plane, on a double-logarithmic scale.
Cuts through this graph, for $E=0.5p^2$ and $E = 1.5p^2$, are shown in 
\Fig{EffCouplingfreeCuts-b}, for three different values of the infrared cutoff $p_{\Lambda}$.
These figures demonstrate the scaling of $g_{\mathrm{eff}}(E,p)\sim p^{2}$ within the regime $p_{\Lambda}\ll p\ll p_{\Xi}$ and the saturation to $g_{\mathrm{eff}}(E,p)=g$ for $p\gg p_{\Xi}$.

Depending on the momentum $p=|\vec{p}|$, a maximum appears, see \Fig{EffCouplingFree}, at the Bogoliubov-type energies  
\begin{align}
 E_{0}(p) &= \pm\left[  2 {p_{\Xi}^2}p^{2}+p^4\right]^{1/2}\,.
 \label{eq:hypercurve_PiR}
\end{align}
On this line, the real part of $1+g\Pi^{R}$ vanishes, cf.~\Eq{gPiR-scaling-regime} and the denominator of the effective coupling is dominated by the imaginary part,  
\begin{align}
  g_{\mathrm{eff}}(E_{0}(p),p) = 1/|\mathrm{Im}\Pi^{R}(E_{0}(p),p)|\,,
  \label{eq:Peakingeff}
\end{align}
meaning that the coupling shows a sort of many-body scattering resonance at the Bogoliubov-type energy-momentum transfer \eq{hypercurve_PiR}.
According to our numerical evaluation, however, these maxima do not appear to influence the scattering integral.

Note that the on-shell effective coupling, $g_\mathrm{eff}(\varepsilon_\mathrm{p},\vec p)\sim p$, scales linearly in $p$, seen as a narrow bright line in \Fig{EffCouplingFree}.
This can be seen in \Fig{EffCouplingfreeCuts-a} which shows cuts in the $p$--$E$-plane at $E = (1-\epsilon)p^2$, with $\epsilon=10^{-2}$, $10^{-3}$, and $10^{-4}$. 
However, at any finite deviation $\varepsilon\not=0$, this scaling gives way to the off-shell quadratic scaling $g_\mathrm{eff}(\varepsilon_\mathrm{p},\vec p)\sim p^{2}$ in the scaling limit $p_{\Lambda}\to0$, below the transition scale, $p<p_{\mathrm{on-shell},\Xi}(\epsilon)\sim \epsilon^{-1/2}p_{\Xi}$.

\subsubsection{Scaling of $g_\mathrm{eff}(p)$ for $\kappa\leq3$}
In contrast to the case $\kappa>3$ discussed so far, the integral \eq{nitofscalingfree} is UV divergent for $\kappa\leq 3$, and the physical cutoff needs to be taken into account. 
Also the integral \eq{app:Pif_free_withf} becomes UV divergent for $\kappa\searrow2$.
With the sharp cutoff at $p=p_{\lambda}$ inserted, the subtraction in the third and fourth lines of \eq{sol_pi_free} becomes relevant.
Analogously to above, neglecting the second term in \eq{Ckappay_free}, one finds,  for $1\ll x\ll y_{\lambda}$, the leading behaviour 
\begin{align}
 \tilde{\pi}_\kappa(x)
 &\simeq \frac{i\pi}{2}\frac{\kappa-3}{\kappa-2}\,
  y_{\lambda}^{\kappa-3}|x|^{2-\kappa} \,.
\label{eq:pitilde_xgge-kltd}
\end{align}
Far below the IR cutoff, $x\ll 1\ll y_{\lambda}$, one finds 
\begin{align}
  \tilde{\pi}_\kappa(x) 
  \simeq \frac{3-\kappa}{2}\,
   \sqrt{\pi}\,\frac{\Gamma(\frac{\kappa-1}{2})}{\Gamma(\frac{\kappa}{2})}xy_{\lambda}^{\kappa-3}\,.
   \label{eq:pitilde_xlle-kltd}
\end{align}
Inserting \Eq{pitilde_xgge-kltd} into \eq{PiR_free} gives, for $\kappa<3$,  
\begin{align}
g \Pi^R(E,p) 
   &\simeq   \frac{i\pi}{4}\frac{\kappa-3}{\kappa-2}\,\frac{p_{\Xi}^{2}}{(pp_{\lambda})^{3-\kappa}}
   \left(|E+p^{2}|^{2-\kappa}-|E-p^{2}|^{2-\kappa}\right)\,.
  \label{eq:gPiR-scaling-regime-kltd}
\end{align}
Inserting this into \Eq{app:geff}, we find that for 
\begin{align}
  p_\Lambda\ll \frac{|E \pm p^2|}{p}\ll p_\mathrm{np}=
  \left(p_{\Xi}^{2}p_{\lambda}^{\kappa-3}\right)^{{1}/({\kappa-1})}
  \label{eq:pnp-kltd}
\end{align}
the effective coupling assumes the scaling form
\begin{align}
 \label{eq:app:geffFreeUniversal-kltd}
 g_{\mathrm{eff}}(p_{0},p) 
 \simeq & \frac{2}{\pi}\frac{\kappa-2}{\kappa-3}\,
 \frac{pp_{\lambda}}{m\rho_\mathrm{nc}}
 \left(\left|\frac{E+p^{2}}{pp_{\lambda}}\right|^{2-\kappa}-\left|\frac{E-p^{2}}{pp_{\lambda}}\right|^{2-\kappa}\right)^{-1}
  \nonumber\\
 &(\kappa<3)\,.
\end{align}
The effective coupling \eq{app:geffFreeUniversal-kltd} depends on $p_{\lambda}$ and thus is no longer universal.
Furthermore, the IR cutoff must be sufficiently small, $p_{\Lambda}\ll p_\mathrm{np}$ for the IR suppression to occur.

This effectively lowers the momentum scale $p_\mathrm{np}$ of the onset of the non-perturbative part of $g_\mathrm{eff}$ (at $p\lesssim p_\mathrm{np}$) from \eq{CollScattRegime} to
\begin{align}
  p_\mathrm{np}=p_{\Xi}
  \left(p_{\Xi}/p_{\lambda}\right)^{{(3-\kappa)}/({\kappa-1})}\,. \qquad (\kappa< 3)
  \label{eq:main:pnp-kltd}
\end{align}

Far below the IR cutoff, for $|E \pm p^2|/p\ll p_{\Lambda}$, the one-loop function, according to \eq{pitilde_xlle-kltd}, approaches 
\begin{align}
g \Pi^R(E,p) 
   &\simeq   \frac{3-\kappa}{4}\,
   \sqrt{\pi}\,\frac{\Gamma(\frac{\kappa-1}{2})}{\Gamma(\frac{\kappa}{2})}
   \frac{p_\Xi^2 }{p_{\Lambda}^2} \left(\frac{p_{\lambda}}{p_{\Lambda}}\right)^{\kappa-3}
   \, ,
   \label{eq:gPiR-pto0-kltd}
\end{align}
resulting in the coupling
\begin{equation}
 \label{eq:geffFreeUniversalIRlimit-kltd}
 g_{\mathrm{eff}}(p_{0},p) 
 \simeq \frac{4}{3-\kappa}\,
   \,\frac{\Gamma(\frac{\kappa}{2})}{\Gamma(\frac{\kappa-1}{2})\sqrt{\pi}}
 \frac{\varepsilon_{p_{\Lambda}}}{\rho_\mathrm{nc}}\left(\frac{p_{\lambda}}{p_{\Lambda}}\right)^{3-\kappa}
 \,.
\end{equation}
The simple form \eq{app:geffFreeUniversal-kltd}, valid for $\kappa<3$ is, e.g., relevant for the equilibrium Bose-Einstein distribution in $d=3$ dimensions just above the critical point, with $0<-\mu\ll T\simeq T_{c}$.
It shows Rayleigh-Jeans scaling with $\kappa=z=2$ within the regime $p_{\Lambda}\ll p\ll p_{\lambda}$, where
the IR cutoff is given by the chemical potential, $p_{\Lambda}\simeq(2m|\mu|)^{1/2}$, the UV cutoff by the temperature, $p_{\lambda}\simeq(2mk_\mathrm{B}T_{c})^{1/2}$, and the density is approximately $\rho_{nc}\simeq(2mk_\mathrm{B}T_{c})^{3/2}$.
From \Eq{pmu} it follows that $p_{\Xi}\simeq\zeta^{1/2}p_{\lambda}$ with gas parameter $\zeta= \rho^{1/3} a\ll1$.
Inserting this into \eq{main:pnp-kltd}, the condition $p_{\Lambda}\ll p_\mathrm{np}$ requires that the chemical potential be lowered to the point that $p_{\Lambda}\lesssim \zeta^{1/2}p_{\Xi}=\zeta p_{\lambda}$.
Taking into account typical gas parameters on the order of $\zeta\approx10^{-2}$, one thus needs to cool the gas to $|\mu|/(k_{B}T_{c})\lesssim \zeta^{2}\approx10^{-4}$ to see the non-perturbative rescaling of the coupling described above.

\subsubsection{Numerical evaluation of $g_\mathrm{eff}(p)$ for $0\leq\kappa\leq4$}
\Fig{EffCouplingfreeCuts-c}(a) shows the momentum dependence of the effective coupling $g_\mathrm{eff}(E,p)$, along $E = 1.5p^2$, for $p_{\Lambda}=10^{-3}\,p_{\Xi}$, for different $2\leq\kappa\leq4$.
As can be inferred from \Eq{gPiR-pto0-kltd}, the non-perturbative IR suppression of $g_\mathrm{eff}$ below $p_\mathrm{np}$, cf.~\eq{pnp-kltd}, where the coupling scales approximately as $~p^{\kappa-1}$ (dotted lines), disappears, depending on the UV cutoff $p_{\lambda}$, when
\begin{align}
  \kappa\lesssim3-2\frac{\ln(p_{\Xi}/\sqrt2 p_{\Lambda})}{\ln(p_{\lambda}/p_{\Lambda})}
  \simeq 2.2\,.
   \label{eq:kappamin-of-plpL}
\end{align}
The dependence of $g_\mathrm{eff}(1.5p^{2},p)$ on $\kappa$, for $p=10^{-2}p_{\Lambda}$, $p_{\Lambda}=10^{-3}p_{\Xi}$, and $p_{\lambda}=10^{5}p_{\Xi}$ is shown in \Fig{EffCouplingfreeCuts-c}(b).
The dashed line indicates the approximation \eq{geffFreeUniversalIRlimit-kltd}, while the dotted line marks the behaviour \eq{geffFreeUniversalIRlimit} for $\kappa\gtrsim3$.
The perturbative limit $g_\mathrm{eff}/g=1$ is reached at the value of $\kappa$ estimated in \eq{kappamin-of-plpL}.
These results depend strongly on the UV cutoff $p_{\lambda}$ as shown in \Fig{EffCouplingfreeCuts-c}(c) for the same $p$ and $p_{\Lambda}$ for three different $\kappa$.
Note that $\kappa=0$ is relevant for the overpopulation initial condition marked as a dashed line in \Fig{NTFP}(b) while $\kappa=2=z$ represents the case of a thermal Rayleigh-Jeans distribution, cf.~\Fig{NTFP}(a).

In Figs.~\fig{EffCouplingfreeCuts-c}(d) and (e) we show respective examples of the momentum dependence of $g_\mathrm{eff}(1.5p^{2},p)$ for $p_{\Lambda}=10^{-3}p_{\Xi}$, for $\kappa=0$ (panel d) and $\kappa=2$ (panel e), comparing different UV cutoffs $p_{\lambda}$.
As \Fig{EffCouplingfreeCuts-c}(b) indicates, the case $\kappa=0$ is outside the regime of exponents where \Eq{pnp-kltd} determines the scaling region.
We find, however, that for $p_{\lambda}\lesssim p_{\Xi}$, the coupling is suppressed for $p<p_{\Xi}$, saturating to a constant value $g_\mathrm{eff}(1.5p^{2},p)\sim p_{\lambda}^{2}/(2m\rho_\mathrm{nc})$ (dotted line in \Fig{EffCouplingfreeCuts-c}(c)) at $p\lesssim p_{\lambda}$.
The coupling scales approximately as $g_\mathrm{eff}(1.5p^{2},p)\sim p^{2}$ above $p\gtrsim p_{\lambda}$ (dotted line), showing also intervals of linear scaling in $p$ (dashed line).
Finally, \Fig{EffCouplingfreeCuts-c}(e) shows the same dependence for a thermal distribution, $\kappa=2=z$, for different $p_{\lambda}$ as indicated in the legend. 
As seen already in \Fig{EffCouplingfreeCuts-c}(c), the IR suppression sets in for $p_{\lambda}\lesssim p_{\Xi}^{2}/p_{\Lambda}$. 
The coupling scales as $g_\mathrm{eff}(1.5p^{2},p)\sim p$ for $p\gg p_{\Lambda}$. 
Once $p_{\lambda}\lesssim p_{\Xi}$, this scaling changes over to $g_\mathrm{eff}(1.5p^{2},p)\sim p^{2}$.

\subsection{Universal $g_\mathrm{eff}(p)$ for Bogoliubov quasiparticles}
\label{app:EffCouplingFctBog}
We finally consider the case of a macroscopic zero-mode population such that Bogoliubov quasiparticles represent the elementary excitations of the system,  corresponding to sound waves in the linear regime of the dispersion \Eq{sounddispersion}. 
For these,
the spectral function, in the basis of the fundamental fields $\tilde\Phi_{a}$, is given in \Eq{rhoBog}.  
Inserting this into \Eq{PirhoKinetic} we obtain, in $d=3$ dimensions,
\begin{align}
  \Pi^\rho&(p^0,\vec{p}) 
  = - i\left(\frac{g \rho_{0}}{2\pi}\right)^2 \int \text{d}^3k \frac{1}{\omega_k \omega_{|\vec p- \vec k|}} f(\omega_{k}) 
  \nonumber \\
  &\times \Big[ \delta(p_0+\omega_k+\omega_{|\vec p- \vec k|})-\delta(p_0+\omega_k-\omega_{|\vec p- \vec k|}) 
  \nonumber \\
  &\ -\, \delta(p_0-\omega_k-\omega_{|\vec p- \vec k|})+\delta(p_0-\omega_k+\omega_{|\vec p- \vec k|})\Big] \,,
\end{align}
where we have replaced the integral over the angle $\measuredangle(\vec p, \vec k)$ by an integral over $r = |\vec p - \vec k|$, and defined $p_0=c_{s}E$. 
Inserting this into $\Pi^R$, \Eq{PiA}, gives
\begin{align}
  \Pi^R (E,p) &
   =   \frac{1}{(2\pi)^2 \omega_p}\Big(\frac{g \rho_{0}}{c_s}\Big)^2 \int^\infty_{0} \text{d}k \int^{p+k}_{|p-k|} \text{d}r
  f(\omega_{k}) 
  \nonumber \\
  &\quad\times\ \left[ \mbox{$\frac{1}{E+k+r+i\epsilon}-\frac{1}{E+k-r+i\epsilon}
   - \frac{1}{E-k-r+i\epsilon}+\frac{1}{E-k+r+i\epsilon}$} \right]\,.
  \nonumber \\
  & =   \frac{p_{\Lambda}p_{\xi}}{(2\pi)^2\sqrt{2}}
  \frac{m}{p} 
  \left[ \widetilde\Pi_{f}'\big(\mbox{$\frac{E+p}{2p_{\Lambda}}$}\big) - \widetilde\Pi_{f}'\big(\mbox{$\frac{E-p}{2p_{\Lambda}}$}\big)\right]\,.
 \label{eq:PiRBogFullIntegral}
\end{align}
The angular part of the spatial momentum convolution can be performed, giving,
in $d=3$ dimensions,
\begin{align}
  \Pi^R &(E,p) 
  =   \frac{p_{\Lambda}}{(2\pi)^2}
  \frac{(g\rho_0)^2}{c^3_sp} 
  \left[ \widetilde\Pi_{f}'\left(\mbox{$\frac{E+p}{2p_{\Lambda}}$}\right) 
        - \widetilde\Pi_{f}'\left(\mbox{$\frac{E-p}{2p_{\Lambda}}$}\right)\right]\,,
\label{eq:PiR_main}
\end{align}
where, again, $p_{\Lambda}$ has been factored out, $g$ is the bare coupling, $\rho_{0}$ the density of condensed particles, and $c_{s}=\sqrt{g\rho_{0}/m}$ the speed of sound.
$\widetilde\Pi_{f}'$ is defined as
\begin{align}
  \widetilde\Pi_{f}' (x) 
  &= \int^\infty_0 dy\, f(\omega_{yp_{\Lambda}}) \ln \left( \frac{x+y+i \epsilon}{x-y+i \epsilon} \right). 
\label{eq:pifBog}
\end{align}

To proceed, we need to specify the quasiparticle distribution $f(\omega_{\vec k})$.
We choose again the infrared cutoff to be $p_{\Lambda}$, and assume $f$ to have the scaling form
\begin{equation}
 f(p_0) = \text{sgn}(p_0)\left(\frac{\omega_\Lambda}{|p_0|+\omega_{p_\Lambda}}\right)^\kappa.
 \label{eq:fmychoice}
\end{equation}
where the signum function accounts for the symmetry \eq{fantisymmetry}, and the scale $\Lambda$ is fixed by the normalization of $n(p,t)$ to the density $\rho_\mathrm{nc}=\rho_\mathrm{tot}-\rho_{0}$ of non-condensed particles, see \Eq{normalisationBog}.
\eq{fmychoice} scales in the same way with $\kappa$ as \eq{fmychoice_free} for the free case.
For brevity, we only consider the case  $\kappa>d-1=2$ relevant for non-thermal fixed points, where the integrals above are UV convergent and we can neglect the dependence on $p_{\lambda}$.

%
\begin{figure}[t]
    \centering
    \includegraphics[width=0.35\textwidth]{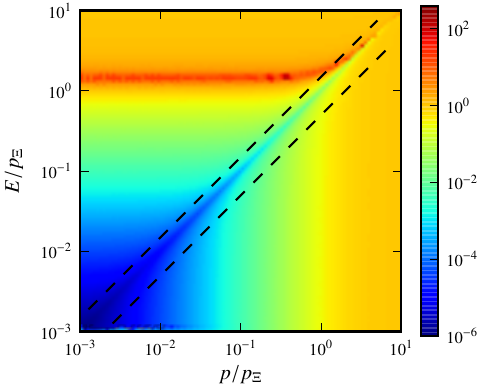}
    \caption{Contour plot of $g_\mathrm{eff}(E,p)/g$ defined in \Eq{app:geff}, for interactions of Bogoliubov quasiparticles, with $\Pi^R(E,p)$ given in \Eq{PiR_main}, for $\kappa = 3.5$ and $p_\Lambda = 10^{-3}p_{\Xi}$.
    In leading-order approximation the coupling function does not depend on $\kappa$.
    }
    \label{fig:EffCouplingBog}
\end{figure}
%
Inserting the quasiparticle distribution \eq{fmychoice} into \Eq{pifBog}  gives  \cite{bateman1955higher}, for $\kappa>d-1=2$,
\begin{align}
\widetilde\Pi_{f}' &(x) 
  = \left(\frac{\Lambda}{p_{\Lambda}}\right)^{\kappa}
  \frac{1}{(\kappa-1)^2 x} 
  \Big[ {}_2 F_1 \left(1,1;\kappa;1-[(1\pm i\epsilon)x]^{-1}\right)
  \nonumber \\
  &\qquad +{}_2 F_1 \left(1,1;\kappa;1+[(1\pm i\epsilon)x]^{-1}\right) \Big], 
\label{eq:app:Pif_Bog}
\end{align}
%
%
\begin{figure}[t]
    \centering
   \includegraphics[width=0.3\textwidth]{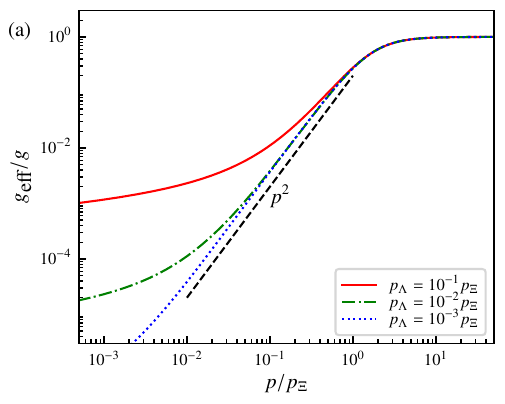}
   \includegraphics[width=0.3\textwidth]{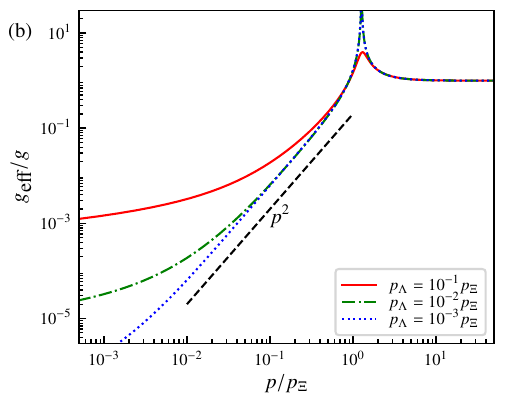}
    \caption{Effective coupling $g_\mathrm{eff}^{2}(E,p)$ as a function of momentum $p$. 
    Shown are different cuts in the $p$--$E$-plane, with (a) $E = 0.5p$ and (b) $E = 1.5p$. 
    Different colors (line styles) refer to different infrared cutoff scales $p_{\Lambda}$ as listed in the legends.
    }
    \label{fig:EffCouplingBogCut}
\end{figure}
%
where the $+ (-)$ sign of the infinitesimal imaginary shift applies for $x > 0 \, (x<0)$, see \cite{Chantesana2017a}.
In the main text we derive $\kappa=d-1/3$ (cf.~\Eq{kappaQBogcoll}) such that $\kappa>d-1$ is fulfilled.
Again, the factor $\Lambda^{\kappa}$ is fixed by the normalization of the single-particle distribution to the non-condensate density  $\rho_\mathrm{nc}=\rho_\mathrm{tot}-\rho_{0}$,
\begin{align}
 \rho_\mathrm{nc} 
 \label{eq:normalisationBog}
    &= \left(\frac{\Lambda}{p_\Lambda}\right)^\kappa \frac{g\rho_0}{c_s}\frac{p^{d-1}_\Lambda}{(2\pi)^d} S_{d-1} 
    \int^\infty_0 du u^{d-2}(u+1)^{-\kappa} \nonumber \\
    &= \left(\frac{\Lambda}{p_\Lambda}\right)^\kappa \frac{g\rho_0}{c_s}\frac{p^{d-1}_\Lambda}{(2\pi)^d} 
     \frac{2\pi^{d/2}}{\Gamma(d/2)} \frac{\Gamma(\kappa-d+1)\Gamma(d-1)}{\Gamma(\kappa)} \,.
\end{align}
For $d=3$, we then have 
\begin{align}
\label{eq:Lambdatokappa_bog}
 \left(\frac{\Lambda}{p_\Lambda}\right)^\kappa 
 &= \frac{\sqrt{2}\rho_\mathrm{nc}}{p_\xi}  \frac{2\pi^2}{p^2_\Lambda} (\kappa-1)(\kappa-2),
\end{align}
where we used $c_s = (g\rho_0/m)^{1/2} = p_\xi/(\sqrt{2}m)$, with inverse healing length $p_\xi = \sqrt{2gm\rho_0}$.
Using \Eq{Lambdatokappa_bog} to fix the scale $\Lambda$, we obtain, from Eqs.~\eq{PiR_main} and \eq{app:Pif_Bog},
\begin{align}
  \Pi^R (E,p) 
       &=   \frac{{1}}{2g}\frac{p_\Xi^2}{ p p_{\Lambda}} 
  \left[ \tilde\pi_{\kappa}'\left(\mbox{$\frac{E+p}{2p_{\Lambda}}$}\right) 
       - \tilde\pi_{\kappa}'\left(\mbox{$\frac{E-p}{2p_{\Lambda}}$}\right)\right]\,
\label{eq:PiR_Bog}
\end{align}
with
\begin{align}
  \tilde{\pi}_\kappa'(x) 
  =  \frac{1}{2x} \frac{\kappa-2}{\kappa-1}\,& 
  \Big[ {}_2 F_1 \left(1,1;\kappa;1-[(1\pm i\epsilon)x]^{-1}\right)
  \nonumber \\
  &\!\!\!\!+\ {}_2 F_1 \left(1,1;\kappa;1+[(1\pm i\epsilon)x]^{-1}\right) \Big]\,.
\label{eq:pitilde_Bog}
\end{align}

If $|x|\gg1$, i.e., sufficiently far above the infrared cutoff, and assuming that $\kappa$ is not an integer 
\footnoterecall{fn1}, the hypergeometric functions are approximated,  in leading order, by
\begin{align}
  \tilde{\pi}_\kappa'(x) 
   &\approx  \frac1x- \frac{i\pi}{2|x|^{\kappa-1}}(\kappa-2)\,, 
   \label{eq:hypergeomapproxBog}
\end{align}
while below the cutoff, $|x|\ll1$, one gets
\begin{align}
  \tilde{\pi}_\kappa'(x) 
   &\approx  - \frac{i\pi}{2}(\kappa-2)
   -\left[C(\kappa)+(\kappa-2)(\kappa-1)\ln x\right]x + \mathcal{O}(x^{2})\,, 
   \label{eq:hypergeomapproxBogIR}
\end{align}
with a $\kappa$-dependent constant $C(\kappa)$ \cite{Chantesana2017a}.
The resulting form of $\Pi^R$ depends on the relative size of $E$ and $p$. 
Here we only quote the form applying in the regions where $E>p$ and $E<|p|$ as the scattering integral will receive its dominating contributions there.
Inserting \Eq{hypergeomapproxBog} into \Eq{PiR_Bog} one finds that, for $\kappa>2$ as assumed above, the real part dominates above the infrared cutoff such that
\begin{align}
g \Pi^R&(E,p) 
  \simeq   \frac{2{p_\Xi^2}}{p^2-E^2} - i\pi(\kappa-2) \frac{{{p}_{\Xi}^{2}}p_{\Lambda}^{\kappa-2}}{4p}
  \nonumber\\
  &\quad\times\
  \left(\left| \frac{E+p}{2} \right|^{1-\kappa}-\left|\frac{E-p}{2}\right|^{1-\kappa}\right)\, .
\end{align}
Below the cutoff, the retarded loop approaches
\begin{align}
g \Pi^R&(E,p) 
  \simeq - \frac{{p_{\Xi}^{2}}}{2p_{\Lambda}^{2}}C(\kappa) + \mbox{log. corrections}\, .
\end{align}

As a result, for $p,|E \pm p|\gg p_\Lambda$ and $\kappa>2$, the loop integral scales as $\Pi^R(s^2 E, s p) = s^{-2}\Pi^R(E,p)$.
Inserting this into \Eq{app:geff}, we find that, in the momentum region $p_\Lambda\ll |E \pm p^2|/p\ll p_{\Xi}<p_{\xi}$ (with $p_{\Xi}<p_{\xi}$ ensuring that sound waves are the relevant quasiparticles) the effective coupling assumes the universal scaling form,
\begin{equation}
 \label{eq:app:geffBogUniversal}
 g_{\mathrm{eff}}(p_{0},p) \simeq  \frac{|p^{2}-(p_{0}/c_s)^{2}|}{{4m\rho_\mathrm{nc}}}\qquad(\kappa>d-1)\,,
\end{equation}
which is, again, effectively independent of the microscopic interaction constant $g$. 
In the IR limit, the coupling saturates to the same constant \eq{geffFreeUniversalIRlimit} as for the free case, but, due to the logarithmic terms in a much slower manner.

This result enters the $T$-matrix elements describing the kinetic scattering of strongly occupied IR sound modes.
In this case, direct and exchange terms are accounted for by separate terms, such that the $T$-matrix elements are parametrized, in terms of the effective coupling function for the $z=1$ case, as 
\begin{align}
  &|T_{\vec p\vec k\vec q\vec r}|^{2}
  =\ (2\pi)^{4}\frac{(g\rho_{0})^{4}}{\omega_{p}\omega_{k}\omega_{q}\omega_{r}}
    \nonumber\\
    & \, \, \times\
    \big[g_\mathrm{eff}^{2}(\omega_{p}-\omega_{r},\vec p-\vec r)+\frac{1}{2}
    g_\mathrm{eff}^{2}(\omega_{p}-\omega_{k},\vec p + \vec k)\big]
       \,,
  \label{eq:TitogeffQP}
\end{align}
where the effective coupling function \eq{app:geffBogUniversal} enters.

The function \eq{app:geffBogUniversal} scales according to \eq{geffscaling}, with $z=1$ and $\gamma_{\kappa}$, in the respective momentum regions, identical to what one finds in the case $z=2$, cf.~Eqs.~\eq{gammapert}, \eq{gammanonpert}.
Together with \Eq{TitogeffQP} this gives the scaling exponent
\begin{align}
   m_{\kappa}=0  \quad(\mbox{Bogoliubov sound; nonperturbative})\, .
  \label{eq:mSsWT}
\end{align}
of the many-body $T$-matrix.

At larger energy and momentum scales, above the healing-length scale, $|E \pm p^2|/p\gg p_{\Xi}$, the effective coupling saturates at the microscopic interaction constant, $g_{\mathrm{eff}}\simeq g$, recovering the perturbative Boltzmann $T$-matrix \Eq{TitogeffQP} with scaling exponent $m_{\kappa}=m=-2$.
We again emphasize that, while the transition scale from the microscopic coupling $g$ to the universal scaling form \eq{app:geffBogUniversal} is set by $p_{\Xi}$ and thus by the microscopic coupling $g$, the particular value of the \emph{universal} coupling $g_{\mathrm{eff}}$ in the scaling regime is independent of $g$.

\Fig{EffCouplingBog} shows the effective coupling constant in the ($p$,$E$) plane, on a double-logarithmic scale.
While the coupling is constant at large momenta and energies, it falls off as power laws in the infrared. 
To illustrate this, cuts through \Fig{EffCouplingBog}, for $E=0.5\,p$ and $E=1.5\,p$, are shown in \Fig{EffCouplingBogCut}, for three different values of the infrared cutoff $p_{\Lambda}$.
The curves again show the saturation to $g$ at momenta above the healing-length scale $p_{\Xi}$, the power-law scaling below, and a weaker scaling below the infrared cutoff scale $p_{\Lambda}$. 

Note that, similar to the case of free particles, the real part of $1+g\Pi^{R}(E,p)$  vanishes at the energy-momentum transfer $(E,p)$ defined by the now gapped sound-wave energy $E(p)=\pm[p^{2}+2p_{\Xi}^{2}]^{1/2}$, where a peak in the effective coupling appears.

\section{Wave-turbulent scaling evolution}
\label{app:WWTSolutions}
Within the regime of applicability of the QBE, zeroes of the scattering integral \eq{KinScattInt} correspond to stationary solutions. 
Examples are the constant solution for which the occupation number is independent of ${\vec p}$, as well as the maximum-entropy thermal equilibrium Bose-Einstein distribution $n_\mathrm{BE}(p)$. 
For these solutions, the scattering integral vanishes due to detailed balance, and thus $n_{Q}({p},t)$ is independent of $t$.

Consider the wave-classical limit, where the occupation number is $n_\mathrm{BE}(p)\gg1$.
The Bose-Einstein distribution in the Rayleigh-Jeans regime ($\omega(p)\ll T$) is also a zero of the scattering integral \eq{KinScattIntCWL} of the WBE.
For a scaling dispersion, \Eq{dynscaling}, $\omega({p})\sim p^{z}$, the distribution has a scaling form qualitatively like \eq{scalingf2}, with $\kappa_{\Lambda}=0$, $\kappa_{\lambda}\to\infty$, $p_{\Lambda}\sim-\mu/T$ and the UV scale $p_{\lambda}$ regularizing the Rayleigh-Jeans divergence at the temperature scale.
The exponent $\kappa$  equals $\kappa=z$ in the Rayleigh-Jeans region of momenta $-\mu\ll\omega(p)\ll T$.

Further non-trivial scaling solutions, with different exponents $\kappa$, can be derived with the methods of wave-turbulence theory \cite{Zakharov1992a,Nazarenko2011a}.

\subsection{Stationary turbulent flows}
According to Boltzmann, stationarity of the maximum-entropy state is related to detailed balance between the collision processes \cite{Huang1987a}. 
In contrast, out-of-equilibrium stationary states generally do not require detailed balance.
In particular when considering driven open systems, stationary states can exist on the basis of a balanced but directed flow through the momentum shells or energy levels.
This is possible when, e.g.,  kinetic energy is inserted into the system predominantly at one length scale while being ejected or dissipated at a different length scale.

A well-known example is turbulence in a three-dimensional incompressible fluid driven continuously at a particular length scale, e.g., by a stirrer \cite{Frisch2004a}.
Fully developed turbulence is characterized by a stationary energy distribution within an extended `inertial range' of wave numbers.
The limiting scales of the inertial range are typically set, on the low-energy side, by the size of eddies stirred into the fluid, and, at the opposite end, by viscosity, dissipating kinetic energy into heat.

Within the inertial range, on average and per unit of time, the same amount of energy is transported uni-directionally through each momentum shell, from large to small characteristic length scales, or vice versa, as is the case in Kraichnan turbulence in two dimensions \cite{Kraichnan1967a}.
This turbulent transport is in general quasi local in momentum space.

The dilute Bose gas, \Eq{GPHamiltonian}, is compressible such that also quantities other than the energy can be locally conserved in their transport through momentum space.
As the interactions are spatially isotropic, these local conservation laws can be expressed in the form of one-dimensional transport equations for either the radial quasiparticle number, 
\begin{align}
  N_{Q}(p)=(2p)^{d-1}\pi n_{Q}(p) \,,
  \label{eq:RadialNDistr}
\end{align}
(for $d=1,2,3$) or the energy distribution,
\begin{align}
  E_{Q}(p)=(2p)^{d-1}\pi \varepsilon_{Q}(p)\,.
  \label{eq:RadialEDistr}
\end{align}
Here $\varepsilon_{Q}(p)=\omega(p)n_{Q}(p)$, and as quasiparticles we again consider free particles or Bogoliubov sound waves using the same notation.
The respective transport equations are written as
\begin{align}
  \partial_{t}N_{Q}(p,t) 
  &= -\partial_{p}Q(p,t)\,,
  \label{eq:BalEqQ}\\
  \partial_{t}E_{Q}(p,t) 
  &= -\partial_{p}P(p,t)\,,
  \label{eq:BalEqP}
  \end{align}
with radial quasiparticle current $Q$ and energy current $P$.
These continuity equations relate the temporal change of a density to the momentum divergence of a current.
Hence, they describe local transport in momentum space. 
Given the in general local interactions in position space which correspond to non-local interactions in momentum space, such a local transport appears somewhat unintuitive at first sight and in general is weakly violated by sub-leading scaling terms.

\subsection{Wave-turbulent cascades}
Consider the one-dimensional transport equations \eq{BalEqQ}, \eq{BalEqP}, for the radial distributions of the quasiparticle number, \Eq{RadialNDistr}, and the energy, \Eq{RadialEDistr}, respectively.
According to \Eq{WBKinEq}, the current gradients are related to the wave-Boltzmann scattering integral \eq{WBKinScattInt} by
\begin{align}
\partial_{p}Q(p,t)&=-(2p)^{d-1}\pi\, I[n_{Q}]({p},t)\,,
\label{eq:QitoI}
\\
\partial_{p}P(p,t)&=-(2p)^{d-1}\pi\,\omega(p)\, I[n_{Q}]({p},t)\, .
\label{eq:PitoI}
\end{align}
For wave-turbulent stationary distributions $n_{Q}(p)$ or $\omega(p)n_{Q}(p)$, both gradients vanish.

The different constant local fluxes $Q$ and $P$ require, however, different values of the scaling exponent $\kappa$ of the stationary quasiparticle distribution $n_{Q}(p)$ entering the scattering integral.
These exponents  can be determined, to a first approximation, by power counting,
assuming the scattering $T$-matrix to scale according to Eqs.~\eq{Tscaling} and thus the integral $I[n_{Q}]$ as given in Eqs.~\eq{IScalingForm0t0} and \eq{mukappaExponent}.
The relative scaling of $Q$ and $I[n_{Q}]$, \Eq{QitoI}, implies that for a quasiparticle cascade, where $\partial_{p}Q=0$, the integral $\int dp\,p^{d-1}I[n_{Q}](p)$ must scale as $p^{0}$.
Counting all powers of $p$ in $I[n_{Q}](p)\sim p^{3+[2(d+m_{\kappa})-3]-z-3\kappa}$, cf.~\Eq{WBKinScattInt}, this requires  $\kappa=\kappa_{Q}$,
\begin{align}
   \kappa_{Q}&=(3d+2m_{\kappa}-z)/3
   \nonumber\\
   &=d+z+2(\gamma_{\kappa}-4+2\eta)/3\, ,
  \label{eq:kappaQWT}
\end{align}
confirming 
Eq.~(2.21) of Ref.~\cite{Svistunov1991a}, cf.~\footnoterecall{fn3}.
Analogously one infers the scaling with $\kappa=\kappa_{P}$,
\begin{align}
   \kappa_{P}=\kappa_{Q}+z/3 \, ,
  \label{eq:kappaPWT}
\end{align}
as a condition to find a stationary energy distribution $E$ and an energy cascade,
cf.~Eq.~(2.13) of \cite{Svistunov1991a}.
The above exponents can also be explicitly derived by determining the zeroes of the scattering integral by means of Zakharov integral transformations \cite{Zakharov1992a}.
We confirm,  in \Sect{SummaryNumerical}, the above power-counting results by explicitly evaluating the scattering integral.

\subsection{Build-up of the wave-turbulent flux towards the IR}
As discussed in \Sect{Turbulence}, a wave-turbulent cascade can not build up instantaneously but as a wave front, either critically slowing or accelerating in time \cite{Svistunov1991a}.
During this transient build-up, the locality of the transport, Eqs.~\eq{BalEqQ},  \eq{BalEqP},  is weakly broken due to global quasiparticle and energy conservation.
The wave front evolves according to the scaling form \eq{WaveFrontScalingForm} or \eq{WaveFrontScalingForml}.

Let us consider the case $\kappa<d$ in which quasiparticles and energy are concentrated at the UV end of the scaling region and an inverse cascade can build up towards the IR.
In this case, we find, for the quasiparticle cascade, $\kappa=\kappa_{Q}$, \Eq{kappaQWT}, that  the scaling exponent $m_{\kappa}$ of the $T$-matrix must satisfy
\begin{equation}
   m_{\kappa}< z/2\,.
 \label{eq:mInvQCascade}
\end{equation}
Note that this condition corresponds to the lower bound of the inequality~(2.8) of Ref.~\cite{Svistunov1991a}, 
cf.~\footnoterecall{fn3}.
Since $m_{\kappa}=\gamma_{\kappa}+2z-4+2\eta$, cf.~\Eq{mkappaitogammakappa}, the inequality  \eq{mInvQCascade} is fulfilled for $z<2(4-\gamma_{\kappa}-2\eta)/3$.
Amongst the cases considered here this is fulfilled, for $\eta=0$, for Bogoliubov quasiparticle transport ($z=1$) in the IR collective-scattering regime ($\gamma_{\kappa}=2$) and for free particles ($z=2$) in the perturbative regime ($\gamma_{\kappa}=0$).
Analogously, the quasiparticle energy cascade, with $\kappa=\kappa_{P}$, requires
\begin{equation}
   m_{\kappa}< 0\,,
 \label{eq:mInvPCascade}
\end{equation}
which is fulfilled for the perturbative weak-wave-turbulent transport of Bogoliubov sound waves ($m_{\kappa}=-2$).

Since, for $\kappa<d$, $\alpha'=\beta'=0$, the scaling form \eq{WaveFrontScalingForm} reads
\begin{align}
n_{Q}(p,t) 
&= \tau^{\,\alpha}
   f_{\Lambda}(\tau^{\,\beta}p/p_{\Lambda};\tau^{\,\beta}p_{\lambda}/p_{\Lambda})\,
\label{eq:WaveFrontScalingFormInvCasc}
\end{align}
and obeys the kinetic equation
\begin{align}
  &\tau^{\,\alpha-1}\left.\left(\alpha + \beta\, [x\, \partial_{x}+y\,\partial_{y}]\right)
  f_{\Lambda}(x;y)\right|_{ x=\tau^{\beta} p/p_{\Lambda};y=\tau^{\beta} p_{\lambda}/p_{\Lambda}}
  \nonumber\\
  &\qquad=\   (\partial_{t}\tau)^{-1}\tau^{-\beta\mu}
    I[f_{\Lambda}]\left(\tau^{\,\beta}p/p_{\Lambda};\tau^{\,\beta}p_{\lambda}/p_{\Lambda}\right)\,,
  \label{eq:NTFPScalingKinEq1InvCasc}
\end{align}
where
\begin{align}
   \alpha=\beta\kappa \, ,
  \label{eq:alphabetakappaQWT}
\end{align}
cf.~\Tab{ScalingRelConservationLaws},
and $\mu$ is given in \Eq{muExponent}.

The wave-turbulent transport is weakly non-local such that even in the inertial regime the scattering integral is non-zero.
Hence, for \Eq{NTFPScalingKinEq1InvCasc} to hold at different times during the scaling evolution, the exponents $\alpha$, $\beta$, and $\mu$ must obey \Eq{IScalingRelation} as in the self-similar case.
Combining \Eq{IScalingRelation} with Eqs.~\eq{alphabetakappaQWT} and \eq{muExponent}  one obtains
\begin{align}
   \beta=(\kappa+\mu)^{-1} 
   = [2(d+ m-\kappa)-z]^{-1}\, ,
  \label{eq:betakappaQWTmu}
\end{align}

\emph{Collective-scattering regime, $z>0$}.---
Here,  $\gamma_{\kappa}=2-\eta$, and the cascade solution applies for $z<2(2-\eta)/3$, cf.~\eq{NTFPzregime}, and thus, e.g., to Bogoliubov quasiparticles ($z=1$, $\eta=0$). 
The $T$-matrix exponent $m$ is then given by \Eq{mWTz}, which gives
\begin{align}
   \beta=1/z\, ,
  \label{eq:app:betaSWCascade}
\end{align}
independent of whether one considers a quasiparticle or an energy cascade, and exactly as for the self-similar evolution, cf.~\Eq{betaSfreecoll}.
Hence, if $z>0$, the scaling parameter is $\tau=t/t_{0}$ and the evolution is critically slowed at large times.

The momentum exponents $\kappa$, for the quasi-particle and energy cascades, are obtained from Eqs.~\eq{gammakappageneral}, \eq{kappaQWT}, and \eq{kappaPWT},
\begin{align}
\begin{aligned}
   \kappa_{Q}&=d+z-(4-2\eta)/3\,,\quad\mbox{(collective scatt., $z>0$)}
  \\
   \kappa_{P}&=d+(4z-4+2\eta)/3\, .\quad\mbox{(collective scatt., $z>0$)}
\end{aligned}
  \label{eq:app:kappaQPSWT}
\end{align}
Note that, for $\eta<2$ and $z>0$ these scaling relations imply that the condition \eq{NTFPcondition} for the applicability of the scaling \eq{gammakappageneral} of the effective coupling is fulfilled. 

\emph{Collective-scattering regime, $z<0$}.---
For the rather unlikely case of $\eta>2$, $z<0$, the scaling \eq{mkappazBog-nonuniv} of the $T$-matrix leads to the exponents formally obtained in Refs.~\cite{Berges:2008wm,Berges:2008sr,Scheppach:2009wu,Berges:2010ez},
\begin{align}
\begin{aligned}
   \kappa_{q}&=d+z\,,\quad\mbox{(collective scatt., $z<0$)}
   \\
   \kappa_{p}&=d+2z\, ,\quad\mbox{(collective scatt., $z<0$)}
\end{aligned}
  \label{eq:app:kappaqpSWT}
\end{align}
cf.~\footnoterecall{fn4}.
In a closed system, $\kappa_{q}$ fulfills the condition \eq{UVDominance} for the buildup of an inverse cascade, behind an accelerated wave front since $\beta=1/z<0$.

\emph{Perturbative regime}.---
In contrast, for sufficiently large momenta one has  $\gamma_{\kappa}=\gamma=0$, $\eta=0$, and thus $m=m_{\kappa}=2(z-2+\eta)$, cf.~Eqs.~\eq{mitogamma} and \eq{mkappaitogammakappa}.
Inserting the momentum exponent for the inverse quasiparticle cascade, \Eq{kappaQWT}, into \Eq{betakappaQWTmu} gives
\begin{align}
   \beta_{Q}= 3(2m-z)^{-1}=(z-8/3+4\eta/3)^{-1}\, .
  \label{eq:app:betaQ}
\end{align}
Hence, for weak-wave-turbulent transport of both, free particles and Bogoliubov sound, 
\begin{align}
   \beta_{Q}^{-1}<0\, .
  \label{eq:betaQLimitInvQCascade}
\end{align}
Therefore, cf.~\Tab{ScalingRelConservationLaws}
and the discussion in \Sect{Turbulence}, the build-up of the inverse quasiparticle cascade occurs in the form of a critically accelerating wave-front evolution with scaling parameter $\tau=\tau^{*}$, cf.~\Eq{scalingParam1}.
This result, with $\beta_{Q}$ given by \Eq{app:betaQ}, is equivalent to Eq.~(2.22) of Ref.~\cite{Svistunov1991a}, 
cf.~\footnoterecall{fn3} for the translation between exponents.
The results of semi-classical simulations \cite{Berloff2002a} corroborate these predictions.

Analogously, inserting $\kappa_{P}$, \Eq{kappaPWT}, into \Eq{betakappaQWTmu} gives
\begin{align}
   \beta_{P}= (2m/3-z)^{-1}=3(z-8+4\eta)^{-1}\, .
  \label{eq:betaP}
\end{align}
Hence, also for a weak-wave-turbulent energy cascade, $\beta_{P}$ is negative such that only a wave-front scaling with $\tau=\tau^{*}$ is possible.

The momentum exponents $\kappa$, for the quasi-particle and energy cascades, are obtained from Eqs.~\eq{gammapert} and \eq{kappaQWT},
\begin{align}
   \kappa_{Q}&=d+z-8/3+4\eta/3\,,\quad\mbox{(perturbative)}
   \nonumber\\
   \kappa_{P}&=d+4(z-2)/3+4\eta/3\, .\quad\mbox{(perturbative)}
  \label{eq:app:kappaQPWWT}
\end{align}
%

\subsection{Build-up of the wave-turbulent flux towards the UV}
Finally, in the case that $\kappa>d+z$, where particle and energy densities are concentrated in the IR, one has $\alpha=\beta=0$, such that the scaling form \eq{WaveFrontScalingForml} reads
\begin{align}
n_{Q}(p,t) 
&= \tau^{\,\alpha'}
   f_{\lambda}(\tau^{\,\beta'}p/p_{\lambda};\tau^{\,\beta'}p_{\Lambda}/p_{\lambda})\,.
 \label{eq:WaveFrontScalingFormDirCasc} 
\end{align}
and obeys the kinetic equation
\begin{align}
  &\tau^{\,\alpha'-1}\left.\left(\alpha' + \beta'\, [x\, \partial_{x}+y\,\partial_{y}]\right)
  f_{\lambda}(x;y)\right|_{ x=\tau^{\beta'} p/p_{\lambda};y=\tau^{\,\beta'} p_{\Lambda}/p_{\lambda}}
  \nonumber\\
  &\qquad=\   (\partial_{t}\tau)^{-1}\tau^{-\beta'\mu}
    I[f_{\lambda}]\left(\tau^{\,\beta'}p/p_{\lambda};\tau^{\,\beta'}p_{\Lambda}/p_{\lambda}\right)\,,
  \label{eq:NTFPScalingKinEq1DirCasc}
\end{align}
where
\begin{align}
   \alpha'=\beta'\kappa \, ,
  \label{eq:alphasbetaskappaQWT}
\end{align}
cf.~\Tab{ScalingRelConservationLaws},
and $\mu$ is given in \Eq{muExponent}.
As before, we obtain 
\begin{align}
   \beta'=(\kappa+\mu)^{-1} 
   = [2(d+ m-\kappa)-z]^{-1}\, .
  \label{eq:betaskappaQWTmu}
\end{align}
In the \emph{collective-scattering regime},  \Eq{mWTz} implies that
\begin{align}
   \beta'=1/z\, ,
  \label{eq:betasSWCascade}
\end{align}
independent of the local conservation law applying in the cascade.
Hence, if $z>0$, then $\beta'>0$, and the scaling parameter is $\tau=\tau^{*}$, i.e., the evolution is accelerated at large times.

In the \emph{perturbative regime}, inserting the exponents \eq{kappaQWT} and \eq{kappaPWT} into \Eq{betaskappaQWTmu} gives
\begin{align}
   \beta'_{Q}= 3(2m-z)^{-1}\, ,   
  \label{eq:betasQ}
  \\
   \beta'_{P}= (2m/3-z)^{-1}\, .   
  \label{eq:betasP}
\end{align}
As $\gamma_{\kappa}=\gamma=0$, one has $m=m_{\kappa}=2(z-2+\eta)$, cf.~Eqs.~\eq{mitogamma} and \eq{mkappaitogammakappa}.
The condition $\kappa_{Q}>d+z$, however, implies $m_{\kappa}> 2z$
for direct particle cascades, see \Eq{kappaQWT}, which means that ${\beta'}_{Q}^{-1}>z$ and thus ${\beta'}_{Q}$ is positive for $z>0$.
Analogously,  $\kappa_{P}>d+z$ requires 
  $ m_{\kappa}> 3z/2$ 
which implies
$  {\beta'}_{P}^{-1}>0$, 
for direct energy cascades.
Hence, direct cascades build up with $\tau=\tau^{*}$ and thus are critically accelerated at large times, cf.~\cite{Svistunov1991a}.
However, for the cases considered here, neither $m_{\kappa}> 2z$ nor $ m_{\kappa}> 3z/2$ are fulfilled.
As a result, transport of particles and energy towards the UV can not occur in the perturbative regime in the form of a direct cascade but rather involves the self-similar evolution discussed in \Sect{SelfSimKinetics}.

\subsection{Kinetic time}
\label{app:KineticTime}
The distinction between wave-turbulent cascades and self-similar evolution at a non-thermal fixed point can be also made clear using the concept of kinetic time, see, e.g., \cite{Svistunov2015a.SuperfluidStatesofMatter}.
A scaling analysis of the quantum kinetic equation \eq{QKinEq} and the wave-Boltzmann scattering integral \eq{KinScattIntCWL}, see Eqs.~\eq{IScalingForm0} and \eq{muExponent}, allows to estimate the kinetic time scale from integrating \Eq{QKinEq} to $\tau_\mathrm{kin}$ as
\begin{align}
\tau^{-1}_{\text{kin}} \sim  p^{2(d+m)-z}\,n_{Q}^{2}(p,\tau_\mathrm{kin}) \,,
\end{align}
if $\partial_t n_Q(\vec{p},t) \neq 0$ as it is the case in self-similar evolutions.
Using the scaling \eq{NTFPscaling0nQ} of $n_{Q}$ and the result \eq{betasS} for the scaling exponent $\beta$ for a self-similar particle transport to the IR, we find that the kinetic time scales as
\begin{align}
\tau^{S}_{\text{kin}} \sim  p^{-1/\beta_{S}}\,.
\end{align}
Since $\beta_{S}>0$, the kinetic time thus diverges in the scaling limit $p\to0$, reflecting critical slowing down in time in the system approaching the non-thermal fixed point.

%
\begin{figure*}[t]
    \centering
    \includegraphics[width=0.4\textwidth]{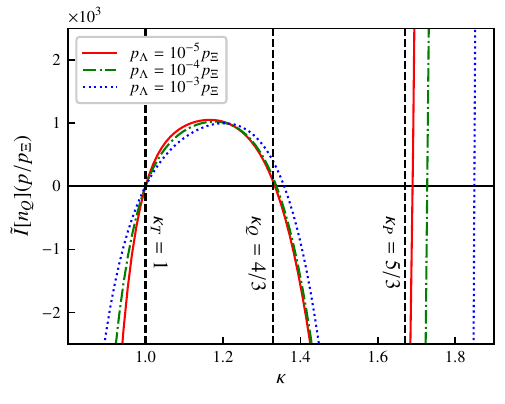}
    \hspace*{0.1\textwidth}
    \includegraphics[width=0.39\textwidth]{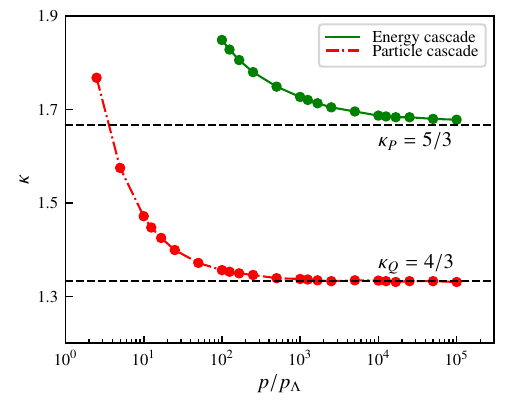}
    \caption{\emph{Left panel:}
    Dependence of the perturbative wave-Boltzmann scattering integral $I[n_{Q}](p)$ for Bogoliubov sound $(z=1, \eta=0)$ in $d=3$ spatial dimensions, at the momentum $p=0.1p_{\Xi}$, 
    on the momentum scaling exponent $\kappa$ characterizing the occupation number distribution $n_{Q}(p)\sim p^{-\kappa}$.
    The vertical dashed lines mark, from the left, the thermal zero at $\kappa_{T}=1$, the inverse quasiparticle-cascade exponent $\kappa_{Q}=4/3$, and the direct energy-cascade exponent $\kappa_{P}=5/3$.
     In the figure, $\tilde{I}[n_Q](p/p_\Xi) = 2^{3/2}\,(2\pi)^3\, p^{3\kappa-1}_\Xi [m\,g^2\,(p_\xi\,\Lambda^{\kappa})^3]^{-1} \, I[n_{Q}](p)$ is shown. 
    The different colors (line styles) correspond to different values of the IR cutoff $p_{\Lambda}$, as indicated in the legend.
    As the cutoff is lowered, the zeroes approach the predicted values. 
    The sign of the slope $\partial I[n_{Q}]/\partial\kappa$ at the zeroes determines the direction of the cascade.
    Note that the slope at $\kappa_{P}\simeq5/3$ is finite and positive.
\emph{Right panel:}
   Scaling exponents $\kappa$ of the quasiparticle distribution $n_{Q}(p)\sim p^{-\kappa}$ for which the perturbative wave-Boltzmann scattering integral $I[n_{Q}](p)$  in $d=3$ spatial dimensions has a zero, for different momenta $p$ in units of the IR cutoff scale $p_{\Lambda}$.
   As in \Fig{IofkappaPertFree}, the lower data corresponds to the particle cascade, with the zeroes approaching $\kappa_{Q}=4/3$, the upper data to the energy cascade, approaching $\kappa_{P}=5/3$ as labeled.
    }
    \label{fig:IofkappaPertBog}
\end{figure*}
%
%
\begin{figure*}[t]
    \centering
    \includegraphics[width=0.4\textwidth]{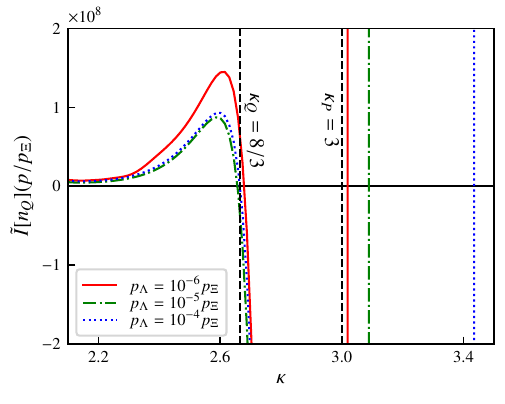}
    \hspace*{0.1\textwidth}
    \includegraphics[width=0.39\textwidth]{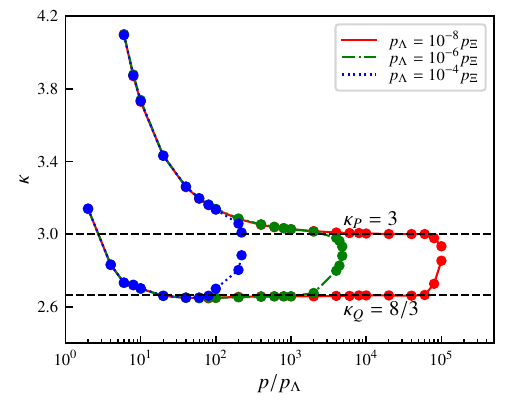}
    \caption{\emph{Left panel:}
    Dependence of the non-perturbative scattering integral $I[n_{Q}](p)$ for Bogoliubov sound $(z=1, \eta=0)$ in $d=3$ spatial dimensions, at the momentum $p=0.002\,p_{\Xi}$, 
    on the momentum scaling exponent $\kappa$ characterizing the occupation number distribution $n_{Q}(p)\sim p^{-\kappa}$.
    The vertical dashed lines mark, from the left, the quasiparticle-cascade exponent $\kappa_{Q}=8/3$, and the energy-cascade exponent $\kappa_{P}=3$.
     In the figure, $\tilde{I}[n_Q](p/p_\Xi) = 2^{3/2}\,(2\pi)^3\, p^{3\kappa-1}_\Xi [m\,g^2\,(p_\xi\,\Lambda^{\kappa})^3]^{-1} \, I[n_{Q}](p)$ is shown. 
    The different colors (line styles) correspond to different values of the IR cutoff $p_{\Lambda}$, as indicated in the legend.
    As the cutoff is lowered, the zeroes approach the predicted values. 
    The sign of the slope $\partial I[n_{Q}]/\partial\kappa$ at the zeroes determines the direction of the cascade.
\emph{Right panel:}
   Scaling exponents $\kappa$ of the quasiparticle distribution $n_{Q}(p)\sim p^{-\kappa}$ for which the non-perturbative wave-Boltzmann scattering integral $I[n_{Q}](p)$ in the IR collective-scattering region, in $d=3$, has a zero, for different momenta $p$ in units of the IR cutoff scale $p_{\Lambda}$, for three different $p_{\Lambda}$.
   As in \Fig{IofkappaPertBog}, the lower data marks the quasiparticle cascade, with the zeroes approaching $\kappa_{Q}=8/3$, while the upper data marks the energy cascade, approaching $\kappa_{P}=3$ as labeled.
   }
   \label{fig:IofkappaCollBog}
\end{figure*}
%
Analogously, using the argument given in \cite{Svistunov2015a.SuperfluidStatesofMatter}, one derives the kinetic times for the build-up of wave-turbulent quasiparticle and energy cascades, using the integrals
\begin{align}
 \partial_{t} \int^{\infty}_p  N_{Q}(p,t) 
   &= Q_0\,, \qquad
 \partial_{t} \int^{p}_0 E_{Q}(p,t) 
  = P_0\,,
\end{align}
over the transport equations \eq{BalEqQ} and \eq{BalEqP}, respectively.
These lead to the estimates of kinetic time,
\begin{align}
  \tau^{Q}_{\text{kin}} \sim p^{d-\kappa_Q} \,,
  \qquad
  \tau^{P}_{\text{kin}} \sim p^{d+z-\kappa_P} \,.
\end{align}
which, together with Eqs.~\eq{kappaQWT} and \eq{kappaPWT}, prove the relations \eq{taukinQ} and \eq{taukinP}, respectively.
Here, the kinetic time $\tau^{Q}_{\text{kin}}$ decreases, for $\kappa_{Q}<d$, with decreasing momentum and, while $\tau^{P}_{\text{kin}}$, for $\kappa>d+z$, decreases with increasing momentum.

\section{Non-perturbative scaling evolution of Bogoliubov sound}
Finally, we summarize the results for Bogoliubov quasiparticles in the sound-wave regime of momenta $p\ll p_{\xi}=(2mg\rho_0)^{1/2}$, with condensate density $\rho_{0}$.
Their eigenfrequency, \Eq{sounddispersion}, exhibits the dynamical exponent $z=1$.

In the \emph{perturbative regime} of low occupation numbers, i.e., for momenta $p\gg p_{\Xi}=(2mg\rho_\mathrm{nc})^{1/2}$, with non-condensate density $\rho_\mathrm{nc}$, \Eq{ParticleDensity}, the scaling evolution of a quasiparticle distribution takes the form of a wave-turbulent cascade towards lower wave numbers.
Within the cascade the transport of quasiparticles or energy is also locally conserved, i.e., the number distribution $n_{Q}(p,t)$ is stationary within the inertial region $p_{\Lambda}\ll p \ll p_{\lambda}$ while the limiting scale $p_{\Lambda}$ evolves algebraically in time.

As before, we consider the elastic two-to-two scattering processes between quasiparticle modes which conserve the total quasiparticle density.
For the closed system, as number and energy are concentrated at the same end of the distribution in momentum space and since both need to be conserved, the scaling evolution takes the form of a critically accelerating wave front, Eqs.~\eq{WaveFrontScaling}, \eq{scalingParam1}, cf.~also our discussion in \Sect{WWTSolutions}.

The effective many-body coupling equals again the bare coupling, $g_\mathrm{eff}(p)\equiv g$, such that $\gamma_{\kappa}=0$, $m=m_{\kappa}=-2$. 
Analogously to the free case one obtains, from \Tab{ScalingRelConservationLaws} and Eqs.~
\eq{kappaQWT}, \eq{alphabetakappaQWT}, and \eq{app:betaQ},  the exponents 
\begin{align}
 \alpha_{Q}= 1-3d/5\,,
 \qquad
 \beta_{Q}&=-3/5\,,
 \qquad
 \kappa_{Q}=d-5/3\,,
 \label{eq:kappaQBogpert}
\end{align}
for the quasiparticle cascade, and, for the energy cascade,
\begin{align}
 \alpha_{P}= (4-3d)/7\,,
 \quad
 \beta_{P}&=-3/7\,,
 \quad
 \kappa_{P}=d-4/3\,.
 \label{eq:kappaPBogpert}
\end{align}

As above, we have numerically evaluated the wave-Boltzmann scattering integral \eq{KinScattIntCWL}, with bare $T$-matrix \eq{TitogbareQP} and a scaling ansatz of the type \eq{scalingfgL}  for the quasiparticle number distribution $n_{Q}(p,t)$.
Specifically, we choose the ansatz \eq{fmychoice}.
\Fig{IofkappaPertBog} shows the dependence of the scattering integral $I[n_{Q}](p)$ on $\kappa$, at the momentum $p=0.1p_{\Xi}$, for three different values of the infrared cutoff $p_{\Lambda}$.
The results indicate the way how the zeroes of $I[n_{Q}](\vec p,t)$ approach the predicted values $\kappa_{Q}=4/3$ and $\kappa_{P}=5/3$ as the IR cutoff is lowered.
Also the thermal zero at $\kappa_{T}=z=1$ is seen, where the number distribution exhibits Rayleigh-Jeans scaling $n_{Q}(p)= T/\omega_{p}\sim p^{-1}$.
Again, because $\kappa_{P}<d+z=4$, the sign of the slope $\partial I[n_{Q}]/\partial\kappa$ at the zeroes in $\kappa$ shows that only the inverse quasiparticle cascade should play a role in the perturbative dynamics of the closed system considered here.

The right panel of \Fig{IofkappaPertBog} shows the dependence of the wave-turbulent zeroes of the scattering integral on the momentum where the integral is evaluated, relative to the infrared cutoff. 
Again, the red dots correspond to the particle cascade  while green dots mark the zeroes corresponding to the energy cascade, confirming the analytically predicted cascade exponents $\kappa_{Q}$ and $\kappa_{P}$ are reached in the scaling limit.
As for the free-particle case, the figure shows that this limit requires a rather large quotient $p/p_{\Lambda}$.

In the \emph{collective-scattering regime} of high occupation numbers, i.e., for momenta $p\ll p_{\Xi}=(2mg\rho_\mathrm{nc})^{1/2}$,  the scaling evolution of a closed system with $z=1$ still represents a cascade, with a stationary distribution in the inertial range where the distribution scales as $n_{Q}(p)\sim p^{-\kappa}$.
However, the rescaling in time occurs in a self-similar manner and is critically slowed at large times.
Hence, the occupation number scales according to \Eq{NTFPscaling0f3}.
In this regime, for momenta larger than the IR cutoff, $p\gg p_{\Lambda}$, the effective many-body coupling takes the universal scaling form \eq{app:geffBogUniversal} with scaling exponent $\gamma_{\kappa}=2$.

Note that, in general, a rescaling of the quasiparticle distribution does not leave the total particle content invariant and thus  the density of non-condensed particles $\rho_\mathrm{nc}$  changes in time.
This is because the two quantities scale differently if $z\not=2$.
Hence, exchange of particles with, e.g., a thermal bath or a condensate mode is required.

The scaling exponents, in the collective-scattering regime ($\gamma_{\kappa}=2$, $m_{\kappa}=0$), read
\begin{align}
 \alpha_{Q}= d-1/3\,,
 \qquad
 \beta_{Q}&=1\,,
 \qquad
 \kappa_{Q}=d-1/3\,,
 \label{eq:kappaQBogcoll}
\end{align}
for the quasiparticle cascade, and 
\begin{align}
 \alpha_{P}= d\,,
 \qquad
 \beta_{P}&=1\,,
 \qquad
 \kappa_{P}=d\,,
 \label{eq:kappaPBogcoll}
\end{align}
for the energy cascade.
In all cases, $\alpha'_{Q}=\beta'_{Q}=0$.

As in \Sect{NumericalFree}, we have evaluated the dimensionally reduced scattering integral numerically and show, in \Fig{IofkappaCollBog}, its dependence on $\kappa$.
Note that only the exponent of the inverse quasiparticle cascade is safely within the region $\kappa<d=3$ where inverse-cascade dynamics is expected.
Hence, only this non-perturbative inverse quasiparticle cascade is expected to play a role, recall the discussion in \Sect{Turbulence}.
In analogy to the perturbative wave-turbulent cases, \Fig{IofkappaCollBog} (right) shows the zeroes of the scattering integral in the $p$--$\kappa$-plane.

We finally recall that the above analysis applies to a condensate with strongly occupied sound-wave modes ($z=1$), in the low-energy limit.
It gives a self-similar, critically slowed transport towards lower energies, with $\beta=1/z$, as for the free case with $z=2$.
The concomitant momentum exponent is $\kappa=d+z-(4-2\eta)/3=d-(1+2\eta)/3$. 
Note that, for $\eta=0$, this exponent has been found in related contexts, describing sound-wave excitations (rarefaction pulses) induced by vortex-antivortex annihilations \cite{Nowak:2011sk} as well as scaling solutions of the Kardar-Parisi-Zhang equation \cite{Mathey2014a.PhysRevA.92.023635}.

\end{appendix}


\begin{thebibliography}{128}%
\makeatletter
\providecommand \@ifxundefined [1]{%
 \@ifx{#1\undefined}
}%
\providecommand \@ifnum [1]{%
 \ifnum #1\expandafter \@firstoftwo
 \else \expandafter \@secondoftwo
 \fi
}%
\providecommand \@ifx [1]{%
 \ifx #1\expandafter \@firstoftwo
 \else \expandafter \@secondoftwo
 \fi
}%
\providecommand \natexlab [1]{#1}%
\providecommand \enquote  [1]{``#1''}%
\providecommand \bibnamefont  [1]{#1}%
\providecommand \bibfnamefont [1]{#1}%
\providecommand \citenamefont [1]{#1}%
\providecommand \href@noop [0]{\@secondoftwo}%
\providecommand \href [0]{\begingroup \@sanitize@url \@href}%
\providecommand \@href[1]{\@@startlink{#1}\@@href}%
\providecommand \@@href[1]{\endgroup#1\@@endlink}%
\providecommand \@sanitize@url [0]{\catcode `\\12\catcode `\$12\catcode
  `\&12\catcode `\#12\catcode `\^12\catcode `\_12\catcode `\%12\relax}%
\providecommand \@@startlink[1]{}%
\providecommand \@@endlink[0]{}%
\providecommand \url  [0]{\begingroup\@sanitize@url \@url }%
\providecommand \@url [1]{\endgroup\@href {#1}{\urlprefix }}%
\providecommand \urlprefix  [0]{URL }%
\providecommand \Eprint [0]{\href }%
\providecommand \doibase [0]{http://dx.doi.org/}%
\providecommand \selectlanguage [0]{\@gobble}%
\providecommand \bibinfo  [0]{\@secondoftwo}%
\providecommand \bibfield  [0]{\@secondoftwo}%
\providecommand \translation [1]{[#1]}%
\providecommand \BibitemOpen [0]{}%
\providecommand \bibitemStop [0]{}%
\providecommand \bibitemNoStop [0]{.\EOS\space}%
\providecommand \EOS [0]{\spacefactor3000\relax}%
\providecommand \BibitemShut  [1]{\csname bibitem#1\endcsname}%
\let\auto@bib@innerbib\@empty
\bibitem [{\citenamefont {Goldenfeld}(1992)}]{Goldenfeld1992a}%
  \BibitemOpen
  \bibfield  {author} {\bibinfo {author} {\bibfnamefont {N.}~\bibnamefont
  {Goldenfeld}},\ }\href {https://books.google.de/books?id=DdB1\_\_nl7CYC}
  {\emph {\bibinfo {title} {Lectures on phase transitions and the
  renormalization group}}},\ Frontiers in physics\ (\bibinfo  {publisher}
  {Addison-Wesley},\ \bibinfo {year} {1992})\BibitemShut {NoStop}%
\bibitem [{\citenamefont {Cardy}(1996)}]{Cardy1996a}%
  \BibitemOpen
  \bibfield  {author} {\bibinfo {author} {\bibfnamefont {J.}~\bibnamefont
  {Cardy}},\ }\href {https://books.google.de/books?id=Wt804S9FjyAC} {\emph
  {\bibinfo {title} {Scaling and Renormalization in Statistical Physics}}},\
  Cambridge Lecture Notes in Physics\ (\bibinfo  {publisher} {CUP, Cambridge,
  UK},\ \bibinfo {year} {1996})\BibitemShut {NoStop}%
\bibitem [{\citenamefont {Hohenberg}\ and\ \citenamefont
  {Halperin}(1977)}]{Hohenberg1977a}%
  \BibitemOpen
  \bibfield  {author} {\bibinfo {author} {\bibfnamefont {P.~C.}\ \bibnamefont
  {Hohenberg}}\ and\ \bibinfo {author} {\bibfnamefont {B.~I.}\ \bibnamefont
  {Halperin}},\ }\href {\doibase 10.1103/RevModPhys.49.435} {\bibfield
  {journal} {\bibinfo  {journal} {Rev. Mod. Phys.}\ }\textbf {\bibinfo {volume}
  {49}},\ \bibinfo {pages} {435} (\bibinfo {year} {1977})}\BibitemShut
  {NoStop}%
\bibitem [{\citenamefont {Janssen}(1979)}]{Janssen1979a}%
  \BibitemOpen
  \bibfield  {author} {\bibinfo {author} {\bibfnamefont {H.}~\bibnamefont
  {Janssen}},\ }in\ \href@noop {} {\emph {\bibinfo {booktitle} {Dynamical
  critical phenomena and related topics, Lecture Notes in Physics, vol. 104}}}\
  (\bibinfo  {publisher} {Springer, Heidelberg 1979},\ \bibinfo {year} {1979})\
  p.~\bibinfo {pages} {26}\BibitemShut {NoStop}%
\bibitem [{\citenamefont {R{\'{a}}cz}(1975)}]{Racz1975a.PLA.53.433}%
  \BibitemOpen
  \bibfield  {author} {\bibinfo {author} {\bibfnamefont {Z.}~\bibnamefont
  {R{\'{a}}cz}},\ }\href {\doibase
  http://dx.doi.org/10.1016/0375-9601(75)90656-8} {\bibfield  {journal}
  {\bibinfo  {journal} {Phys. Lett. A}\ }\textbf {\bibinfo {volume} {53}},\
  \bibinfo {pages} {433 } (\bibinfo {year} {1975})}\BibitemShut {NoStop}%
\bibitem [{\citenamefont {Fisher}\ and\ \citenamefont
  {R\'acz}(1976)}]{Fisher1976a.PhysRevB.13.5039}%
  \BibitemOpen
  \bibfield  {author} {\bibinfo {author} {\bibfnamefont {M.~E.}\ \bibnamefont
  {Fisher}}\ and\ \bibinfo {author} {\bibfnamefont {Z.}~\bibnamefont
  {R\'acz}},\ }\href {\doibase 10.1103/PhysRevB.13.5039} {\bibfield  {journal}
  {\bibinfo  {journal} {Phys. Rev. B}\ }\textbf {\bibinfo {volume} {13}},\
  \bibinfo {pages} {5039} (\bibinfo {year} {1976})}\BibitemShut {NoStop}%
\bibitem [{\citenamefont {Bausch}\ and\ \citenamefont
  {Janssen}(1976)}]{Bausch1976a}%
  \BibitemOpen
  \bibfield  {author} {\bibinfo {author} {\bibfnamefont {R.}~\bibnamefont
  {Bausch}}\ and\ \bibinfo {author} {\bibfnamefont {H.~K.}\ \bibnamefont
  {Janssen}},\ }\href {\doibase 10.1007/BF01420890} {\bibfield  {journal}
  {\bibinfo  {journal} {Z. Phys. B}\ }\textbf {\bibinfo {volume} {25}},\
  \bibinfo {pages} {275} (\bibinfo {year} {1976})}\BibitemShut {NoStop}%
\bibitem [{\citenamefont {Bausch}\ \emph {et~al.}(1979)\citenamefont {Bausch},
  \citenamefont {Eisenriegler},\ and\ \citenamefont {Janssen}}]{Bausch1979a}%
  \BibitemOpen
  \bibfield  {author} {\bibinfo {author} {\bibfnamefont {R.}~\bibnamefont
  {Bausch}}, \bibinfo {author} {\bibfnamefont {E.}~\bibnamefont
  {Eisenriegler}}, \ and\ \bibinfo {author} {\bibfnamefont {H.~K.}\
  \bibnamefont {Janssen}},\ }\href {\doibase 10.1007/BF01320218} {\bibfield
  {journal} {\bibinfo  {journal} {Z. Phys. B}\ }\textbf {\bibinfo {volume}
  {36}},\ \bibinfo {pages} {179} (\bibinfo {year} {1979})}\BibitemShut
  {NoStop}%
\bibitem [{\citenamefont {Diehl}(1986)}]{Diehl1986a}%
  \BibitemOpen
  \bibfield  {author} {\bibinfo {author} {\bibfnamefont {H.~W.}\ \bibnamefont
  {Diehl}},\ }in\ \href@noop {} {\emph {\bibinfo {booktitle} {Phase Transitions
  and Critical Phenomena}}}\ (\bibinfo  {publisher} {Academic Press, London},\
  \bibinfo {year} {1986})\BibitemShut {NoStop}%
\bibitem [{\citenamefont {Janssen}\ \emph {et~al.}(1989)\citenamefont
  {Janssen}, \citenamefont {Schaub},\ and\ \citenamefont
  {Schmittmann}}]{Janssen1989a}%
  \BibitemOpen
  \bibfield  {author} {\bibinfo {author} {\bibfnamefont {H.~K.}\ \bibnamefont
  {Janssen}}, \bibinfo {author} {\bibfnamefont {B.}~\bibnamefont {Schaub}}, \
  and\ \bibinfo {author} {\bibfnamefont {B.}~\bibnamefont {Schmittmann}},\
  }\href {\doibase 10.1007/BF01319383} {\bibfield  {journal} {\bibinfo
  {journal} {Z. Phys. B Cond. Mat.}\ }\textbf {\bibinfo {volume} {73}},\
  \bibinfo {pages} {539} (\bibinfo {year} {1989})}\BibitemShut {NoStop}%
\bibitem [{\citenamefont {Janssen}(1992)}]{Janssen1992a}%
  \BibitemOpen
  \bibfield  {author} {\bibinfo {author} {\bibfnamefont {H.}~\bibnamefont
  {Janssen}},\ }in\ \href@noop {} {\emph {\bibinfo {booktitle} {From phase
  transitions to chaos}}}\ (\bibinfo  {publisher} {World Scientific, Singapore
  1992},\ \bibinfo {year} {1992})\ p.~\bibinfo {pages} {68}\BibitemShut
  {NoStop}%
\bibitem [{\citenamefont {Calabrese}\ and\ \citenamefont
  {Gambassi}(2002)}]{Calabrese2002a.PhysRevE.65.066120}%
  \BibitemOpen
  \bibfield  {author} {\bibinfo {author} {\bibfnamefont {P.}~\bibnamefont
  {Calabrese}}\ and\ \bibinfo {author} {\bibfnamefont {A.}~\bibnamefont
  {Gambassi}},\ }\href {\doibase 10.1103/PhysRevE.65.066120} {\bibfield
  {journal} {\bibinfo  {journal} {Phys. Rev. E}\ }\textbf {\bibinfo {volume}
  {65}},\ \bibinfo {pages} {066120} (\bibinfo {year} {2002})},\ \Eprint
  {http://arxiv.org/abs/cond-mat/0203096} {arXiv:cond-mat/0203096
  [cond-mat.stat-mech]} \BibitemShut {NoStop}%
\bibitem [{\citenamefont {Calabrese}\ and\ \citenamefont
  {Gambassi}(2005)}]{Calabrese2005a.JPA38.05.R133}%
  \BibitemOpen
  \bibfield  {author} {\bibinfo {author} {\bibfnamefont {P.}~\bibnamefont
  {Calabrese}}\ and\ \bibinfo {author} {\bibfnamefont {A.}~\bibnamefont
  {Gambassi}},\ }\href {http://stacks.iop.org/0305-4470/38/i=18/a=R01}
  {\bibfield  {journal} {\bibinfo  {journal} {J. Phys. A: Math. Gen.}\ }\textbf
  {\bibinfo {volume} {38}},\ \bibinfo {pages} {R133} (\bibinfo {year}
  {2005})},\ \Eprint {http://arxiv.org/abs/cond-mat/0410357}
  {arXiv:cond-mat/0410357} \BibitemShut {NoStop}%
\bibitem [{\citenamefont
  {Gambassi}(2006)}]{Gambassi2006a.JPAConfSer.40.2006.13}%
  \BibitemOpen
  \bibfield  {author} {\bibinfo {author} {\bibfnamefont {A.}~\bibnamefont
  {Gambassi}},\ }\href {http://stacks.iop.org/1742-6596/40/i=1/a=002}
  {\bibfield  {journal} {\bibinfo  {journal} {J. Phys. Conf. Ser.}\ }\textbf
  {\bibinfo {volume} {40}},\ \bibinfo {pages} {13} (\bibinfo {year}
  {2006})}\BibitemShut {NoStop}%
\bibitem [{\citenamefont {Bray}\ and\ \citenamefont
  {Puri}(1991)}]{Bray1991a.PhysRevLett.67.2670}%
  \BibitemOpen
  \bibfield  {author} {\bibinfo {author} {\bibfnamefont {A.~J.}\ \bibnamefont
  {Bray}}\ and\ \bibinfo {author} {\bibfnamefont {S.}~\bibnamefont {Puri}},\
  }\href {\doibase 10.1103/PhysRevLett.67.2670} {\bibfield  {journal} {\bibinfo
   {journal} {Phys. Rev. Lett.}\ }\textbf {\bibinfo {volume} {67}},\ \bibinfo
  {pages} {2670} (\bibinfo {year} {1991})}\BibitemShut {NoStop}%
\bibitem [{\citenamefont {Bray}(1994)}]{Bray1994a.AdvPhys.43.357}%
  \BibitemOpen
  \bibfield  {author} {\bibinfo {author} {\bibfnamefont {A.~J.}\ \bibnamefont
  {Bray}},\ }\href {\doibase 10.1080/00018739400101505} {\bibfield  {journal}
  {\bibinfo  {journal} {Adv. Phys.}\ }\textbf {\bibinfo {volume} {43}},\
  \bibinfo {pages} {357} (\bibinfo {year} {1994})},\ \Eprint
  {http://arxiv.org/abs/cond-mat/9501089} {arXiv:cond-mat/9501 089} \BibitemShut
  {NoStop}%
\bibitem [{\citenamefont {{Bray}}\ \emph {et~al.}(2000)\citenamefont {{Bray}},
  \citenamefont {{Briant}},\ and\ \citenamefont
  {{Jervis}}}]{Bray2000PhRvL..84.1503B}%
  \BibitemOpen
  \bibfield  {author} {\bibinfo {author} {\bibfnamefont {A.~J.}\ \bibnamefont
  {{Bray}}}, \bibinfo {author} {\bibfnamefont {A.~J.}\ \bibnamefont
  {{Briant}}}, \ and\ \bibinfo {author} {\bibfnamefont {D.~K.}\ \bibnamefont
  {{Jervis}}},\ }\href {\doibase 10.1103/PhysRevLett.84.1503} {\bibfield
  {journal} {\bibinfo  {journal} {Phys. Rev. Lett.}\ }\textbf {\bibinfo
  {volume} {84}},\ \bibinfo {pages} {1503} (\bibinfo {year} {2000})},\ \Eprint
  {http://arxiv.org/abs/cond-mat/9902362} {cond-mat/9902362} \BibitemShut
  {NoStop}%
\bibitem [{\citenamefont {Frisch}(2004)}]{Frisch2004a}%
  \BibitemOpen
  \bibfield  {author} {\bibinfo {author} {\bibfnamefont {U.}~\bibnamefont
  {Frisch}},\ }\href@noop {} {\emph {\bibinfo {title} {Turbulence: The Legacy
  of A. N. Kolmogorov}}}\ (\bibinfo  {publisher} {Cambridge University Press},\
  \bibinfo {year} {2004})\BibitemShut {NoStop}%
\bibitem [{\citenamefont {Zakharov}\ \emph {et~al.}(1992)\citenamefont
  {Zakharov}, \citenamefont {{L'vov}},\ and\ \citenamefont
  {Falkovich}}]{Zakharov1992a}%
  \BibitemOpen
  \bibfield  {author} {\bibinfo {author} {\bibfnamefont {V.~E.}\ \bibnamefont
  {Zakharov}}, \bibinfo {author} {\bibfnamefont {V.~S.}\ \bibnamefont
  {{L'vov}}}, \ and\ \bibinfo {author} {\bibfnamefont {G.}~\bibnamefont
  {Falkovich}},\ }\href@noop {} {\emph {\bibinfo {title} {Kolmogorov Spectra of
  Turbulence I: Wave Turbulence}}}\ (\bibinfo  {publisher} {Springer, Berlin},\
  \bibinfo {year} {1992})\BibitemShut {NoStop}%
\bibitem [{\citenamefont {Nazarenko}(2011)}]{Nazarenko2011a}%
  \BibitemOpen
  \bibfield  {author} {\bibinfo {author} {\bibfnamefont {S.}~\bibnamefont
  {Nazarenko}},\ }\href@noop {} {\emph {\bibinfo {title} {Wave turbulence}}},\
  \bibinfo {series} {Lecture Notes in Physics}\ No.\ \bibinfo {number} {825}\
  (\bibinfo  {publisher} {Springer},\ \bibinfo {address} {Heidelberg},\
  \bibinfo {year} {2011})\ pp.\ \bibinfo {pages} {XVI, 279 S.}\BibitemShut
  {Stop}%
\bibitem [{\citenamefont {Tsubota}(2008)}]{Tsubota2008a}%
  \BibitemOpen
  \bibfield  {author} {\bibinfo {author} {\bibfnamefont {M.}~\bibnamefont
  {Tsubota}},\ }\href {\doibase 10.1143/JPSJ.77.111006} {\bibfield  {journal}
  {\bibinfo  {journal} {J. Phys. Soc. Jpn.}\ }\textbf {\bibinfo {volume}
  {77}},\ \bibinfo {pages} {111006} (\bibinfo {year} {2008})},\ \Eprint
  {http://arxiv.org/abs/0806.2737} {arXiv: 0806.2737 [cond-mat.other]}
  \BibitemShut {NoStop}%
\bibitem [{\citenamefont {Vinen}(2006)}]{Vinen2006a}%
  \BibitemOpen
  \bibfield  {author} {\bibinfo {author} {\bibfnamefont {W.}~\bibnamefont
  {Vinen}},\ }\href {\doibase 10.1007/s10909-006-9240-6} {\bibfield  {journal}
  {\bibinfo  {journal} {J. Low Temp. Phys.}\ }\textbf {\bibinfo {volume}
  {145}},\ \bibinfo {pages} {7} (\bibinfo {year} {2006})}\BibitemShut {NoStop}%
\bibitem [{\citenamefont {Braun}\ \emph {et~al.}(2015)\citenamefont {Braun},
  \citenamefont {Friesdorf}, \citenamefont {Hodgman}, \citenamefont
  {Schreiber}, \citenamefont {Ronzheimer}, \citenamefont {Riera}, \citenamefont
  {del Rey}, \citenamefont {Bloch}, \citenamefont {Eisert},\ and\ \citenamefont
  {Schneider}}]{Braun2014a.arXiv1403.7199B}%
  \BibitemOpen
  \bibfield  {author} {\bibinfo {author} {\bibfnamefont {S.}~\bibnamefont
  {Braun}}, \bibinfo {author} {\bibfnamefont {M.}~\bibnamefont {Friesdorf}},
  \bibinfo {author} {\bibfnamefont {S.~S.}\ \bibnamefont {Hodgman}}, \bibinfo
  {author} {\bibfnamefont {M.}~\bibnamefont {Schreiber}}, \bibinfo {author}
  {\bibfnamefont {J.~P.}\ \bibnamefont {Ronzheimer}}, \bibinfo {author}
  {\bibfnamefont {A.}~\bibnamefont {Riera}}, \bibinfo {author} {\bibfnamefont
  {M.}~\bibnamefont {del Rey}}, \bibinfo {author} {\bibfnamefont
  {I.}~\bibnamefont {Bloch}}, \bibinfo {author} {\bibfnamefont
  {J.}~\bibnamefont {Eisert}}, \ and\ \bibinfo {author} {\bibfnamefont
  {U.}~\bibnamefont {Schneider}},\ }\href {\doibase 10.1073/pnas.1408861112}
  {\bibfield  {journal} {\bibinfo  {journal} {PNAS}\ }\textbf {\bibinfo
  {volume} {112}},\ \bibinfo {pages} {3641} (\bibinfo {year} {2015})},\ \Eprint
  {http://arxiv.org/abs/1403.7199} {arXiv:1403.7199 [cond-mat.quant-gas]}
  \BibitemShut {NoStop}%
\bibitem [{\citenamefont
  {Lamacraft}(2007)}]{Lamacraft2007.PhysRevLett.98.160404}%
  \BibitemOpen
  \bibfield  {author} {\bibinfo {author} {\bibfnamefont {A.}~\bibnamefont
  {Lamacraft}},\ }\href {\doibase 10.1103/PhysRevLett.98.160404} {\bibfield
  {journal} {\bibinfo  {journal} {Phys. Rev. Lett.}\ }\textbf {\bibinfo
  {volume} {98}},\ \bibinfo {pages} {160404} (\bibinfo {year} {2007})},\
  \Eprint {http://arxiv.org/abs/cond-mat/0611017} {arXiv:cond-mat/0611017
  [cond-mat.stat-mech]} \BibitemShut {NoStop}%
\bibitem [{\citenamefont {Rossini}\ \emph {et~al.}(2009)\citenamefont
  {Rossini}, \citenamefont {Silva}, \citenamefont {Mussardo},\ and\
  \citenamefont {Santoro}}]{Rossini2009a.PhysRevLett.102.127204}%
  \BibitemOpen
  \bibfield  {author} {\bibinfo {author} {\bibfnamefont {D.}~\bibnamefont
  {Rossini}}, \bibinfo {author} {\bibfnamefont {A.}~\bibnamefont {Silva}},
  \bibinfo {author} {\bibfnamefont {G.}~\bibnamefont {Mussardo}}, \ and\
  \bibinfo {author} {\bibfnamefont {G.~E.}\ \bibnamefont {Santoro}},\ }\href
  {\doibase 10.1103/PhysRevLett.102.127204} {\bibfield  {journal} {\bibinfo
  {journal} {Phys. Rev. Lett.}\ }\textbf {\bibinfo {volume} {102}},\ \bibinfo
  {pages} {127204} (\bibinfo {year} {2009})},\ \Eprint
  {http://arxiv.org/abs/0810.5508} {arXiv:0810.5508 [cond-mat. stat-mech]}
  \BibitemShut {NoStop}%
\bibitem [{\citenamefont {Dalla~Torre}\ \emph {et~al.}(2013)\citenamefont
  {Dalla~Torre}, \citenamefont {Demler},\ and\ \citenamefont
  {Polkovnikov}}]{DallaTorre2013.PhysRevLett.110.090404}%
  \BibitemOpen
  \bibfield  {author} {\bibinfo {author} {\bibfnamefont {E.~G.}\ \bibnamefont
  {Dalla~Torre}}, \bibinfo {author} {\bibfnamefont {E.}~\bibnamefont {Demler}},
  \ and\ \bibinfo {author} {\bibfnamefont {A.}~\bibnamefont {Polkovnikov}},\
  }\href {\doibase 10.1103/PhysRevLett.110.090404} {\bibfield  {journal}
  {\bibinfo  {journal} {Phys. Rev. Lett.}\ }\textbf {\bibinfo {volume} {110}},\
  \bibinfo {pages} {090404} (\bibinfo {year} {2013})},\ \Eprint
  {http://arxiv.org/abs/1211.5145} {arXiv:1211.5145 [cond-mat.quant-gas]}
  \BibitemShut {NoStop}%
\bibitem [{\citenamefont {Gambassi}\ and\ \citenamefont
  {Calabrese}(2011)}]{Gambassi2011a.EPL95.6}%
  \BibitemOpen
  \bibfield  {author} {\bibinfo {author} {\bibfnamefont {A.}~\bibnamefont
  {Gambassi}}\ and\ \bibinfo {author} {\bibfnamefont {P.}~\bibnamefont
  {Calabrese}},\ }\href {\doibase 10.1209/0295-5075/95/66007} {\bibfield
  {journal} {\bibinfo  {journal} {Europhys. Lett.}\ }\textbf {\bibinfo
  {volume} {95}},\ \bibinfo {pages} {66007} (\bibinfo {year} {2011})},\ \Eprint
  {http://arxiv.org/abs/1012.5294} {arXiv:1012.5294 [cond-mat.stat-mech]}
  \BibitemShut {NoStop}%
\bibitem [{\citenamefont {Sciolla}\ and\ \citenamefont
  {Biroli}(2013)}]{Sciolla2013a.PhysRevB.88.201110}%
  \BibitemOpen
  \bibfield  {author} {\bibinfo {author} {\bibfnamefont {B.}~\bibnamefont
  {Sciolla}}\ and\ \bibinfo {author} {\bibfnamefont {G.}~\bibnamefont
  {Biroli}},\ }\href {\doibase 10.1103/PhysRevB.88.201110} {\bibfield
  {journal} {\bibinfo  {journal} {Phys. Rev. B}\ }\textbf {\bibinfo {volume}
  {88}},\ \bibinfo {pages} {201110} (\bibinfo {year} {2013})},\ \Eprint
  {http://arxiv.org/abs/1211.2572} {arXiv:1211.2572 [cond-mat.stat-mech]}
  \BibitemShut {NoStop}%
\bibitem [{\citenamefont {Smacchia}\ \emph {et~al.}(2015)\citenamefont
  {Smacchia}, \citenamefont {Knap}, \citenamefont {Demler},\ and\ \citenamefont
  {Silva}}]{Smacchia2015a.PhysRevB.91.205136}%
  \BibitemOpen
  \bibfield  {author} {\bibinfo {author} {\bibfnamefont {P.}~\bibnamefont
  {Smacchia}}, \bibinfo {author} {\bibfnamefont {M.}~\bibnamefont {Knap}},
  \bibinfo {author} {\bibfnamefont {E.}~\bibnamefont {Demler}}, \ and\ \bibinfo
  {author} {\bibfnamefont {A.}~\bibnamefont {Silva}},\ }\href {\doibase
  10.1103/PhysRevB.91.205136} {\bibfield  {journal} {\bibinfo  {journal} {Phys.
  Rev. B}\ }\textbf {\bibinfo {volume} {91}},\ \bibinfo {pages} {205136}
  (\bibinfo {year} {2015})},\ \Eprint {http://arxiv.org/abs/1409.1883}
  {arXiv:1409.1883 [cond-mat.stat-mech]} \BibitemShut {NoStop}%
\bibitem [{\citenamefont {Maraga}\ \emph {et~al.}(2015)\citenamefont {Maraga},
  \citenamefont {Chiocchetta}, \citenamefont {Mitra},\ and\ \citenamefont
  {Gambassi}}]{Maraga2015a.PhysRevE.92.042151}%
  \BibitemOpen
  \bibfield  {author} {\bibinfo {author} {\bibfnamefont {A.}~\bibnamefont
  {Maraga}}, \bibinfo {author} {\bibfnamefont {A.}~\bibnamefont {Chiocchetta}},
  \bibinfo {author} {\bibfnamefont {A.}~\bibnamefont {Mitra}}, \ and\ \bibinfo
  {author} {\bibfnamefont {A.}~\bibnamefont {Gambassi}},\ }\href {\doibase
  10.1103/PhysRevE.92.042151} {\bibfield  {journal} {\bibinfo  {journal} {Phys.
  Rev. E}\ }\textbf {\bibinfo {volume} {92}},\ \bibinfo {pages} {042151}
  (\bibinfo {year} {2015})},\ \Eprint {http://arxiv.org/abs/1506.04528}
  {arXiv:1506.04528 [cond-mat.stat-mech]} \BibitemShut {NoStop}%
\bibitem [{\citenamefont {Maraga}\ \emph {et~al.}(2016)\citenamefont {Maraga},
  \citenamefont {Smacchia},\ and\ \citenamefont
  {Silva}}]{Maraga2016b.PhysRevB.94.245122}%
  \BibitemOpen
  \bibfield  {author} {\bibinfo {author} {\bibfnamefont {A.}~\bibnamefont
  {Maraga}}, \bibinfo {author} {\bibfnamefont {P.}~\bibnamefont {Smacchia}}, \
  and\ \bibinfo {author} {\bibfnamefont {A.}~\bibnamefont {Silva}},\ }\href
  {\doibase 10.1103/PhysRevB.94.245122} {\bibfield  {journal} {\bibinfo
  {journal} {Phys. Rev. B}\ }\textbf {\bibinfo {volume} {94}},\ \bibinfo
  {pages} {245122} (\bibinfo {year} {2016})},\ \Eprint
  {http://arxiv.org/abs/1602.01763} {arXiv:1602.01763 [cond-mat.stat-mech]}
  \BibitemShut {NoStop}%
\bibitem [{\citenamefont {Chiocchetta}\ \emph {et~al.}(2015)\citenamefont
  {Chiocchetta}, \citenamefont {Tavora}, \citenamefont {Gambassi},\ and\
  \citenamefont {Mitra}}]{Chiocchetta2015a.PhysRevB.91.220302}%
  \BibitemOpen
  \bibfield  {author} {\bibinfo {author} {\bibfnamefont {A.}~\bibnamefont
  {Chiocchetta}}, \bibinfo {author} {\bibfnamefont {M.}~\bibnamefont {Tavora}},
  \bibinfo {author} {\bibfnamefont {A.}~\bibnamefont {Gambassi}}, \ and\
  \bibinfo {author} {\bibfnamefont {A.}~\bibnamefont {Mitra}},\ }\href
  {\doibase 10.1103/PhysRevB.91.220302} {\bibfield  {journal} {\bibinfo
  {journal} {Phys. Rev. B}\ }\textbf {\bibinfo {volume} {91}},\ \bibinfo
  {pages} {220302} (\bibinfo {year} {2015})},\ \Eprint
  {http://arxiv.org/abs/1411.7939} {arXiv:1411.7939 [cond-mat.quant-gas]}
  \BibitemShut {NoStop}%
\bibitem [{\citenamefont {Chiocchetta}\ \emph
  {et~al.}(2016{\natexlab{a}})\citenamefont {Chiocchetta}, \citenamefont
  {Tavora}, \citenamefont {Gambassi},\ and\ \citenamefont
  {Mitra}}]{Chiocchetta2016a.PhysRevB.94.134311}%
  \BibitemOpen
  \bibfield  {author} {\bibinfo {author} {\bibfnamefont {A.}~\bibnamefont
  {Chiocchetta}}, \bibinfo {author} {\bibfnamefont {M.}~\bibnamefont {Tavora}},
  \bibinfo {author} {\bibfnamefont {A.}~\bibnamefont {Gambassi}}, \ and\
  \bibinfo {author} {\bibfnamefont {A.}~\bibnamefont {Mitra}},\ }\href
  {\doibase 10.1103/PhysRevB.94.134311} {\bibfield  {journal} {\bibinfo
  {journal} {Phys. Rev. B}\ }\textbf {\bibinfo {volume} {94}},\ \bibinfo
  {pages} {134311} (\bibinfo {year} {2016}{\natexlab{a}})},\ \Eprint
  {http://arxiv.org/abs/1604.04614} {arXiv:1604.04614 [cond-mat.stat-mech]}
  \BibitemShut {NoStop}%
\bibitem [{\citenamefont {Chiocchetta}\ \emph
  {et~al.}(2016{\natexlab{b}})\citenamefont {Chiocchetta}, \citenamefont
  {Gambassi}, \citenamefont {Diehl},\ and\ \citenamefont
  {Marino}}]{Chiocchetta:2016waa.PhysRevB.94.174301}%
  \BibitemOpen
  \bibfield  {author} {\bibinfo {author} {\bibfnamefont {A.}~\bibnamefont
  {Chiocchetta}}, \bibinfo {author} {\bibfnamefont {A.}~\bibnamefont
  {Gambassi}}, \bibinfo {author} {\bibfnamefont {S.}~\bibnamefont {Diehl}}, \
  and\ \bibinfo {author} {\bibfnamefont {J.}~\bibnamefont {Marino}},\ }\href
  {\doibase 10.1103/PhysRevB.94.174301} {\bibfield  {journal} {\bibinfo
  {journal} {Phys. Rev. B}\ }\textbf {\bibinfo {volume} {94}},\ \bibinfo
  {pages} {174301} (\bibinfo {year} {2016}{\natexlab{b}})},\ \Eprint
  {http://arxiv.org/abs/1606.06272} {arXiv:1606.06272 [cond-mat.stat-mech]}
  \BibitemShut {NoStop}%
\bibitem [{\citenamefont {Chiocchetta}\ \emph {et~al.}(2017)\citenamefont
  {Chiocchetta}, \citenamefont {Gambassi}, \citenamefont {Diehl},\ and\
  \citenamefont {Marino}}]{Chiocchetta2016b.161202419C.PhysRevLett.118.135701}%
  \BibitemOpen
  \bibfield  {author} {\bibinfo {author} {\bibfnamefont {A.}~\bibnamefont
  {Chiocchetta}}, \bibinfo {author} {\bibfnamefont {A.}~\bibnamefont
  {Gambassi}}, \bibinfo {author} {\bibfnamefont {S.}~\bibnamefont {Diehl}}, \
  and\ \bibinfo {author} {\bibfnamefont {J.}~\bibnamefont {Marino}},\ }\href
  {\doibase 10.1103/PhysRevLett.118.135701} {\bibfield  {journal} {\bibinfo
  {journal} {Phys. Rev. Lett.}\ }\textbf {\bibinfo {volume} {118}},\ \bibinfo
  {pages} {135701} (\bibinfo {year} {2017})},\ \Eprint
  {http://arxiv.org/abs/1612.02419} {arXiv:1612.02419 [cond-mat.stat-mech]}
  \BibitemShut {NoStop}%
\bibitem [{\citenamefont {Marino}\ and\ \citenamefont
  {Diehl}(2016)}]{Marino2016a.PhysRevLett.116.070407}%
  \BibitemOpen
  \bibfield  {author} {\bibinfo {author} {\bibfnamefont {J.}~\bibnamefont
  {Marino}}\ and\ \bibinfo {author} {\bibfnamefont {S.}~\bibnamefont {Diehl}},\
  }\href {\doibase 10.1103/PhysRevLett.116.070407} {\bibfield  {journal}
  {\bibinfo  {journal} {Phys. Rev. Lett.}\ }\textbf {\bibinfo {volume} {116}},\
  \bibinfo {pages} {070407} (\bibinfo {year} {2016})},\ \Eprint
  {http://arxiv.org/abs/1508.02723} {arXiv:1508.02723 [cond-mat.quant-gas]}
  \BibitemShut {NoStop}%
\bibitem [{\citenamefont {{Marino}}\ and\ \citenamefont
  {{Diehl}}(2016)}]{Marino2016PhRvB..94h5150M}%
  \BibitemOpen
  \bibfield  {author} {\bibinfo {author} {\bibfnamefont {J.}~\bibnamefont
  {{Marino}}}\ and\ \bibinfo {author} {\bibfnamefont {S.}~\bibnamefont
  {{Diehl}}},\ }\href {\doibase 10.1103/PhysRevB.94.085150} {\bibfield
  {journal} {\bibinfo  {journal} {Phys. Rev. B}\ }\textbf {\bibinfo {volume}
  {94}},\ \bibinfo {pages} {085150} (\bibinfo {year} {2016})},\ \Eprint
  {http://arxiv.org/abs/1606.00452} {arXiv:1606.00452 [cond-mat.quant-gas]}
  \BibitemShut {NoStop}%
\bibitem [{\citenamefont {Damle}\ \emph {et~al.}(1996)\citenamefont {Damle},
  \citenamefont {Majumdar},\ and\ \citenamefont
  {Sachdev}}]{Damle1996a.PhysRevA.54.5037}%
  \BibitemOpen
  \bibfield  {author} {\bibinfo {author} {\bibfnamefont {K.}~\bibnamefont
  {Damle}}, \bibinfo {author} {\bibfnamefont {S.~N.}\ \bibnamefont {Majumdar}},
  \ and\ \bibinfo {author} {\bibfnamefont {S.}~\bibnamefont {Sachdev}},\ }\href
  {\doibase 10.1103/PhysRevA.54.5037} {\bibfield  {journal} {\bibinfo
  {journal} {Phys. Rev. A}\ }\textbf {\bibinfo {volume} {54}},\ \bibinfo
  {pages} {5037} (\bibinfo {year} {1996})},\ \Eprint
  {http://arxiv.org/abs/cond-mat/9511058} {arXiv:cond-mat/9511058} \BibitemShut
  {NoStop}%
\bibitem [{\citenamefont {Mukerjee}\ \emph {et~al.}(2007)\citenamefont
  {Mukerjee}, \citenamefont {Xu},\ and\ \citenamefont
  {Moore}}]{Mukerjee2007a.PhysRevB.76.104519}%
  \BibitemOpen
  \bibfield  {author} {\bibinfo {author} {\bibfnamefont {S.}~\bibnamefont
  {Mukerjee}}, \bibinfo {author} {\bibfnamefont {C.}~\bibnamefont {Xu}}, \ and\
  \bibinfo {author} {\bibfnamefont {J.~E.}\ \bibnamefont {Moore}},\ }\href
  {\doibase 10.1103/PhysRevB.76.104519} {\bibfield  {journal} {\bibinfo
  {journal} {Phys. Rev. B}\ }\textbf {\bibinfo {volume} {76}},\ \bibinfo
  {pages} {104519} (\bibinfo {year} {2007})},\ \Eprint
  {http://arxiv.org/abs/0704.3440} {arXiv:0704.3440 [cond-mat.stat-mech]}
  \BibitemShut {NoStop}%
\bibitem [{\citenamefont {Williamson}\ and\ \citenamefont
  {Blakie}(2016{\natexlab{a}})}]{Williamson2016a.PhysRevLett.116.025301}%
  \BibitemOpen
  \bibfield  {author} {\bibinfo {author} {\bibfnamefont {L.~A.}\ \bibnamefont
  {Williamson}}\ and\ \bibinfo {author} {\bibfnamefont {P.~B.}\ \bibnamefont
  {Blakie}},\ }\href {\doibase 10.1103/PhysRevLett.116.025301} {\bibfield
  {journal} {\bibinfo  {journal} {Phys. Rev. Lett.}\ }\textbf {\bibinfo
  {volume} {116}},\ \bibinfo {pages} {025301} (\bibinfo {year}
  {2016}{\natexlab{a}})},\ \Eprint {http://arxiv.org/abs/1504.06404}
  {arXiv:1504.06404 [cond-mat.quant-gas]} \BibitemShut {NoStop}%
\bibitem [{\citenamefont {{Hofmann}}\ \emph {et~al.}(2014)\citenamefont
  {{Hofmann}}, \citenamefont {{Natu}},\ and\ \citenamefont {{Das
  Sarma}}}]{Hofmann2014PhRvL.113i5702H}%
  \BibitemOpen
  \bibfield  {author} {\bibinfo {author} {\bibfnamefont {J.}~\bibnamefont
  {{Hofmann}}}, \bibinfo {author} {\bibfnamefont {S.~S.}\ \bibnamefont
  {{Natu}}}, \ and\ \bibinfo {author} {\bibfnamefont {S.}~\bibnamefont {{Das
  Sarma}}},\ }\href {\doibase 10.1103/PhysRevLett.113.095702} {\bibfield
  {journal} {\bibinfo  {journal} {Phys. Rev. Lett.}\ }\textbf {\bibinfo
  {volume} {113}},\ \bibinfo {pages} {095702} (\bibinfo {year} {2014})},\
  \Eprint {http://arxiv.org/abs/1403.1284} {arXiv:1403.1284
  [cond-mat.quant-gas]} \BibitemShut {NoStop}%
\bibitem [{\citenamefont {Williamson}\ and\ \citenamefont
  {Blakie}(2016{\natexlab{b}})}]{Williamson2016a.PhysRevA.94.023608}%
  \BibitemOpen
  \bibfield  {author} {\bibinfo {author} {\bibfnamefont {L.~A.}\ \bibnamefont
  {Williamson}}\ and\ \bibinfo {author} {\bibfnamefont {P.~B.}\ \bibnamefont
  {Blakie}},\ }\href {\doibase 10.1103/PhysRevA.94.023608} {\bibfield
  {journal} {\bibinfo  {journal} {Phys. Rev. A}\ }\textbf {\bibinfo {volume}
  {94}},\ \bibinfo {pages} {023608} (\bibinfo {year} {2016}{\natexlab{b}})},\
  \Eprint {http://arxiv.org/abs/1605.04016} {arXiv:1605.04016
  [cond-mat.quant-gas]} \BibitemShut {NoStop}%
\bibitem [{\citenamefont {Bourges}\ and\ \citenamefont
  {Blakie}(2017)}]{Bourges2016a.arXiv161108922B.PhysRevA.95.023616}%
  \BibitemOpen
  \bibfield  {author} {\bibinfo {author} {\bibfnamefont {A.}~\bibnamefont
  {Bourges}}\ and\ \bibinfo {author} {\bibfnamefont {P.~B.}\ \bibnamefont
  {Blakie}},\ }\href {\doibase 10.1103/PhysRevA.95.023616} {\bibfield
  {journal} {\bibinfo  {journal} {Phys. Rev. A}\ }\textbf {\bibinfo {volume}
  {95}},\ \bibinfo {pages} {023616} (\bibinfo {year} {2017})},\ \Eprint
  {http://arxiv.org/abs/1611.08922} {arXiv:1611.08922 [cond-mat.quant-gas]}
  \BibitemShut {NoStop}%
\bibitem [{\citenamefont {Berges}\ \emph {et~al.}(2008)\citenamefont {Berges},
  \citenamefont {Rothkopf},\ and\ \citenamefont {Schmidt}}]{Berges:2008wm}%
  \BibitemOpen
  \bibfield  {author} {\bibinfo {author} {\bibfnamefont {J.}~\bibnamefont
  {Berges}}, \bibinfo {author} {\bibfnamefont {A.}~\bibnamefont {Rothkopf}}, \
  and\ \bibinfo {author} {\bibfnamefont {J.}~\bibnamefont {Schmidt}},\ }\href
  {\doibase 10.1103/PhysRevLett.101.041603} {\bibfield  {journal} {\bibinfo
  {journal} {Phys. Rev. Lett.}\ }\textbf {\bibinfo {volume} {101}},\ \bibinfo
  {pages} {041603} (\bibinfo {year} {2008})},\ \Eprint
  {http://arxiv.org/abs/0803.0131} {arXiv:0803.0131 [hep-ph]} \BibitemShut
  {NoStop}%
\bibitem [{\citenamefont {Berges}\ and\ \citenamefont
  {Hoffmeister}(2009)}]{Berges:2008sr}%
  \BibitemOpen
  \bibfield  {author} {\bibinfo {author} {\bibfnamefont {J.}~\bibnamefont
  {Berges}}\ and\ \bibinfo {author} {\bibfnamefont {G.}~\bibnamefont
  {Hoffmeister}},\ }\href {\doibase 10.1016/j.nuclphysb.2008.12.017} {\bibfield
   {journal} {\bibinfo  {journal} {Nucl. Phys.}\ }\textbf {\bibinfo {volume}
  {B813}},\ \bibinfo {pages} {383} (\bibinfo {year} {2009})},\ \Eprint
  {http://arxiv.org/abs/0809.5208} {arXiv:0809.5208 [hep-th]} \BibitemShut
  {NoStop}%
\bibitem [{\citenamefont {Scheppach}\ \emph {et~al.}(2010)\citenamefont
  {Scheppach}, \citenamefont {Berges},\ and\ \citenamefont
  {Gasenzer}}]{Scheppach:2009wu}%
  \BibitemOpen
  \bibfield  {author} {\bibinfo {author} {\bibfnamefont {C.}~\bibnamefont
  {Scheppach}}, \bibinfo {author} {\bibfnamefont {J.}~\bibnamefont {Berges}}, \
  and\ \bibinfo {author} {\bibfnamefont {T.}~\bibnamefont {Gasenzer}},\ }\href
  {\doibase 10.1103/PhysRevA.81.033611} {\bibfield  {journal} {\bibinfo
  {journal} {Phys. Rev. A}\ }\textbf {\bibinfo {volume} {81}},\ \bibinfo
  {pages} {033611} (\bibinfo {year} {2010})},\ \Eprint
  {http://arxiv.org/abs/0912.4183} {arXiv:0912.4183 [cond-mat.quant-gas]}
  \BibitemShut {NoStop}%
\bibitem [{\citenamefont {Berges}\ and\ \citenamefont
  {Sexty}(2011)}]{Berges:2010ez}%
  \BibitemOpen
  \bibfield  {author} {\bibinfo {author} {\bibfnamefont {J.}~\bibnamefont
  {Berges}}\ and\ \bibinfo {author} {\bibfnamefont {D.}~\bibnamefont {Sexty}},\
  }\href {\doibase 10.1103/PhysRevD.83.085004} {\bibfield  {journal} {\bibinfo
  {journal} {Phys. Rev. D}\ }\textbf {\bibinfo {volume} {83}},\ \bibinfo
  {pages} {085004} (\bibinfo {year} {2011})},\ \Eprint
  {http://arxiv.org/abs/1012.5944} {arXiv:1012.5944 [hep-ph]} \BibitemShut
  {NoStop}%
\bibitem [{\citenamefont {Pi{\~n}eiro~Orioli}\ \emph
  {et~al.}(2015)\citenamefont {Pi{\~n}eiro~Orioli}, \citenamefont
  {Boguslavski},\ and\ \citenamefont {Berges}}]{Orioli:2015dxa}%
  \BibitemOpen
  \bibfield  {author} {\bibinfo {author} {\bibfnamefont {A.}~\bibnamefont
  {Pi{\~n}eiro~Orioli}}, \bibinfo {author} {\bibfnamefont {K.}~\bibnamefont
  {Boguslavski}}, \ and\ \bibinfo {author} {\bibfnamefont {J.}~\bibnamefont
  {Berges}},\ }\href {\doibase 10.1103/PhysRevD.92.025041} {\bibfield
  {journal} {\bibinfo  {journal} {Phys. Rev. D}\ }\textbf {\bibinfo {volume}
  {92}},\ \bibinfo {pages} {025041} (\bibinfo {year} {2015})},\ \Eprint
  {http://arxiv.org/abs/1503.02498} {arXiv:1503.02498 [hep-ph]} \BibitemShut
  {NoStop}%
\bibitem [{\citenamefont {Berges}(2016)}]{Berges:2015kfa}%
  \BibitemOpen
  \bibfield  {author} {\bibinfo {author} {\bibfnamefont {J.}~\bibnamefont
  {Berges}},\ }in\ \href@noop {} {\emph {\bibinfo {booktitle} {Proc. Int.
  School on Strongly Interacting Quantum Systems Out of Equilibrium, Les
  Houches}}},\ \bibinfo {editor} {edited by\ \bibinfo {editor} {\bibfnamefont
  {T.}~\bibnamefont {{Giamarchi et al.}}}}\ (\bibinfo  {publisher} {OUP,
  Oxford},\ \bibinfo {year} {2016})\ \Eprint {http://arxiv.org/abs/1503.02907}
  {arXiv:1503.02907 [hep-ph]} \BibitemShut {NoStop}%
\bibitem [{\citenamefont {Pr{\"u}fer}\ \emph {et~al.}(2018)\citenamefont
  {Pr{\"u}fer}, \citenamefont {Kunkel}, \citenamefont {Strobel}, \citenamefont
  {Lannig}, \citenamefont {Linnemann}, \citenamefont {Schmied}, \citenamefont
  {Berges}, \citenamefont {Gasenzer},\ and\ \citenamefont
  {Oberthaler}}]{Prufer:2018hto}%
  \BibitemOpen
  \bibfield  {author} {\bibinfo {author} {\bibfnamefont {M.}~\bibnamefont
  {Pr{\"u}fer}}, \bibinfo {author} {\bibfnamefont {P.}~\bibnamefont {Kunkel}},
  \bibinfo {author} {\bibfnamefont {H.}~\bibnamefont {Strobel}}, \bibinfo
  {author} {\bibfnamefont {S.}~\bibnamefont {Lannig}}, \bibinfo {author}
  {\bibfnamefont {D.}~\bibnamefont {Linnemann}}, \bibinfo {author}
  {\bibfnamefont {C.-M.}\ \bibnamefont {Schmied}}, \bibinfo {author}
  {\bibfnamefont {J.}~\bibnamefont {Berges}}, \bibinfo {author} {\bibfnamefont
  {T.}~\bibnamefont {Gasenzer}}, \ and\ \bibinfo {author} {\bibfnamefont
  {M.~K.}\ \bibnamefont {Oberthaler}},\ }\href {\doibase
  10.1038/s41586-018-0659-0} {\bibfield  {journal} {\bibinfo  {journal}
  {Nature}\ }\textbf {\bibinfo {volume} {563}},\ \bibinfo {pages} {217}
  (\bibinfo {year} {2018})},\ \Eprint {http://arxiv.org/abs/1805.11881}
  {arXiv:1805.11881 [cond-mat.quant-gas]} \BibitemShut {NoStop}%
\bibitem [{\citenamefont {Nowak}\ \emph {et~al.}(2011)\citenamefont {Nowak},
  \citenamefont {Sexty},\ and\ \citenamefont {Gasenzer}}]{Nowak:2010tm}%
  \BibitemOpen
  \bibfield  {author} {\bibinfo {author} {\bibfnamefont {B.}~\bibnamefont
  {Nowak}}, \bibinfo {author} {\bibfnamefont {D.}~\bibnamefont {Sexty}}, \ and\
  \bibinfo {author} {\bibfnamefont {T.}~\bibnamefont {Gasenzer}},\ }\href
  {\doibase 10.1103/PhysRevB.84.020506} {\bibfield  {journal} {\bibinfo
  {journal} {Phys. Rev. B}\ }\textbf {\bibinfo {volume} {84}},\ \bibinfo
  {pages} {020506(R)} (\bibinfo {year} {2011})},\ \Eprint
  {http://arxiv.org/abs/1012.4437v2} {arXiv:1012.4437v2 [cond-mat.quant-gas]}
  \BibitemShut {NoStop}%
\bibitem [{\citenamefont {Nowak}\ \emph {et~al.}(2012)\citenamefont {Nowak},
  \citenamefont {Schole}, \citenamefont {Sexty},\ and\ \citenamefont
  {Gasenzer}}]{Nowak:2011sk}%
  \BibitemOpen
  \bibfield  {author} {\bibinfo {author} {\bibfnamefont {B.}~\bibnamefont
  {Nowak}}, \bibinfo {author} {\bibfnamefont {J.}~\bibnamefont {Schole}},
  \bibinfo {author} {\bibfnamefont {D.}~\bibnamefont {Sexty}}, \ and\ \bibinfo
  {author} {\bibfnamefont {T.}~\bibnamefont {Gasenzer}},\ }\href {\doibase
  10.1103/PhysRevA.85.043627} {\bibfield  {journal} {\bibinfo  {journal} {Phys.
  Rev. A}\ }\textbf {\bibinfo {volume} {85}},\ \bibinfo {pages} {043627}
  (\bibinfo {year} {2012})},\ \Eprint {http://arxiv.org/abs/1111.6127}
  {arXiv:1111.6127 [cond-mat.quant-gas]} \BibitemShut {NoStop}%
\bibitem [{\citenamefont {Schole}\ \emph {et~al.}(2012)\citenamefont {Schole},
  \citenamefont {Nowak},\ and\ \citenamefont {Gasenzer}}]{Schole:2012kt}%
  \BibitemOpen
  \bibfield  {author} {\bibinfo {author} {\bibfnamefont {J.}~\bibnamefont
  {Schole}}, \bibinfo {author} {\bibfnamefont {B.}~\bibnamefont {Nowak}}, \
  and\ \bibinfo {author} {\bibfnamefont {T.}~\bibnamefont {Gasenzer}},\ }\href
  {\doibase 10.1103/PhysRevA.86.013624} {\bibfield  {journal} {\bibinfo
  {journal} {Phys. Rev. A}\ }\textbf {\bibinfo {volume} {86}},\ \bibinfo
  {pages} {013624} (\bibinfo {year} {2012})},\ \Eprint
  {http://arxiv.org/abs/1204.2487} {arXiv:1204.2487 [cond-mat.quant-gas]}
  \BibitemShut {NoStop}%
\bibitem [{\citenamefont {Karl}\ and\ \citenamefont
  {Gasenzer}(2017)}]{Karl2017b.NJP19.093014}%
  \BibitemOpen
  \bibfield  {author} {\bibinfo {author} {\bibfnamefont {M.}~\bibnamefont
  {Karl}}\ and\ \bibinfo {author} {\bibfnamefont {T.}~\bibnamefont
  {Gasenzer}},\ }\href {http://stacks.iop.org/1367-2630/19/i=9/a=093014}
  {\bibfield  {journal} {\bibinfo  {journal} {New J. Phys.}\ }\textbf {\bibinfo
  {volume} {19}},\ \bibinfo {pages} {093014} (\bibinfo {year} {2017})},\
  \Eprint {http://arxiv.org/abs/1611.01163} {arXiv:1611.01163
  [cond-mat.quant-gas]} \BibitemShut {NoStop}%
\bibitem [{\citenamefont {Erne}\ \emph {et~al.}(2018)\citenamefont {Erne},
  \citenamefont {B{\"u}cker}, \citenamefont {Gasenzer}, \citenamefont
  {Berges},\ and\ \citenamefont {Schmiedmayer}}]{Erne:2018gmz}%
  \BibitemOpen
  \bibfield  {author} {\bibinfo {author} {\bibfnamefont {S.}~\bibnamefont
  {Erne}}, \bibinfo {author} {\bibfnamefont {R.}~\bibnamefont {B{\"u}cker}},
  \bibinfo {author} {\bibfnamefont {T.}~\bibnamefont {Gasenzer}}, \bibinfo
  {author} {\bibfnamefont {J.}~\bibnamefont {Berges}}, \ and\ \bibinfo {author}
  {\bibfnamefont {J.}~\bibnamefont {Schmiedmayer}},\ }\href {\doibase
  10.1038/s41586-018-0667-0} {\bibfield  {journal} {\bibinfo  {journal}
  {Nature}\ }\textbf {\bibinfo {volume} {563}},\ \bibinfo {pages} {225}
  (\bibinfo {year} {2018})},\ \Eprint {http://arxiv.org/abs/1805.12310}
  {arXiv:1805.12310 [cond-mat.quant-gas]} \BibitemShut {NoStop}%
\bibitem [{\citenamefont {Eyink}\ and\ \citenamefont
  {Goldenfeld}(1994)}]{Eyink1994a}%
  \BibitemOpen
  \bibfield  {author} {\bibinfo {author} {\bibfnamefont {G.}~\bibnamefont
  {Eyink}}\ and\ \bibinfo {author} {\bibfnamefont {N.}~\bibnamefont
  {Goldenfeld}},\ }\href {\doibase 10.1103/PhysRevE.50.4679} {\bibfield
  {journal} {\bibinfo  {journal} {Phys. Rev. E}\ }\textbf {\bibinfo {volume}
  {50}},\ \bibinfo {pages} {4679} (\bibinfo {year} {1994})},\
  \Eprint {http://arxiv.org/abs/cond-mat/9407021} {arXiv:cond-mat/9407021} 
  \BibitemShut
  {NoStop}%
\bibitem [{\citenamefont {Gurarie}(1995)}]{Gurarie1995a}%
  \BibitemOpen
  \bibfield  {author} {\bibinfo {author} {\bibfnamefont {V.}~\bibnamefont
  {Gurarie}},\ }\href {\doibase 10.1016/0550-3213(95)00108-5} {\bibfield
  {journal} {\bibinfo  {journal} {Nucl. Phys. B}\ }\textbf {\bibinfo {volume}
  {441}},\ \bibinfo {pages} {569} (\bibinfo {year} {1995})},\
  \Eprint {http://arxiv.org/abs/hep-th/9405077} {arXiv:hep-th/9405077} 
  \BibitemShut
  {NoStop}%
\bibitem [{\citenamefont {Gasenzer}\ and\ \citenamefont
  {Pawlowski}(2008)}]{Gasenzer:2008zz}%
  \BibitemOpen
  \bibfield  {author} {\bibinfo {author} {\bibfnamefont {T.}~\bibnamefont
  {Gasenzer}}\ and\ \bibinfo {author} {\bibfnamefont {J.~M.}\ \bibnamefont
  {Pawlowski}},\ }\href {\doibase 10.1016/j.physletb.2008.10.049} {\bibfield
  {journal} {\bibinfo  {journal} {Phys. Lett.}\ }\textbf {\bibinfo {volume}
  {B670}},\ \bibinfo {pages} {135} (\bibinfo {year} {2008})},\ \Eprint
  {http://arxiv.org/abs/arXiv:0710.4627 [cond-mat.other]} {arXiv:0710.4627
  [cond-mat.other]} \BibitemShut {NoStop}%
\bibitem [{\citenamefont {Berges}\ and\ \citenamefont
  {Mesterh{\'a}zy}(2012)}]{Berges:2012ty}%
  \BibitemOpen
  \bibfield  {author} {\bibinfo {author} {\bibfnamefont {J.}~\bibnamefont
  {Berges}}\ and\ \bibinfo {author} {\bibfnamefont {D.}~\bibnamefont
  {Mesterh{\'a}zy}},\ }\bibfield  {booktitle} {\emph {\bibinfo {booktitle}
  {{Physics at all scales: The Renormalization Group. Proceedings, 49.
  Internationale Universit\"atswochen f\"ur Theoretische Physik, Winter
  School}}},\ }\href {\doibase 10.1016/j.nuclphysbps.2012.06.003} {\bibfield
  {journal} {\bibinfo  {journal} {Nucl. Phys. B (Proc. Suppl.)}\ }\textbf
  {\bibinfo {volume} {228}},\ \bibinfo {pages} {37} (\bibinfo {year} {2012})},\
  \Eprint {http://arxiv.org/abs/1204.1489} {arXiv:1204.1489 [hep-ph]}
  \BibitemShut {NoStop}%
\bibitem [{\citenamefont {Mathey}\ \emph {et~al.}(2015)\citenamefont {Mathey},
  \citenamefont {Gasenzer},\ and\ \citenamefont
  {Pawlowski}}]{Mathey2014a.PhysRevA.92.023635}%
  \BibitemOpen
  \bibfield  {author} {\bibinfo {author} {\bibfnamefont {S.}~\bibnamefont
  {Mathey}}, \bibinfo {author} {\bibfnamefont {T.}~\bibnamefont {Gasenzer}}, \
  and\ \bibinfo {author} {\bibfnamefont {J.~M.}\ \bibnamefont {Pawlowski}},\
  }\href {\doibase 10.1103/PhysRevA.92.023635} {\bibfield  {journal} {\bibinfo
  {journal} {Phys. Rev. A}\ }\textbf {\bibinfo {volume} {92}},\ \bibinfo
  {pages} {023635} (\bibinfo {year} {2015})}\BibitemShut {NoStop}%
\bibitem [{\citenamefont {Gasenzer}\ \emph {et~al.}(2010)\citenamefont
  {Gasenzer}, \citenamefont {Kessler},\ and\ \citenamefont
  {Pawlowski}}]{Gasenzer:2010rq}%
  \BibitemOpen
  \bibfield  {author} {\bibinfo {author} {\bibfnamefont {T.}~\bibnamefont
  {Gasenzer}}, \bibinfo {author} {\bibfnamefont {S.}~\bibnamefont {Kessler}}, \
  and\ \bibinfo {author} {\bibfnamefont {J.~M.}\ \bibnamefont {Pawlowski}},\
  }\href {\doibase 10.1140/epjc/s10052-010-1430-3} {\bibfield  {journal}
  {\bibinfo  {journal} {Eur. Phys. J. C}\ }\textbf {\bibinfo {volume} {70}},\
  \bibinfo {pages} {423} (\bibinfo {year} {2010})},\ \Eprint
  {http://arxiv.org/abs/1003.4163} {arXiv:1003.4163 [cond-mat.quant-gas]}
  \BibitemShut {NoStop}%
\bibitem [{\citenamefont {Berges}\ \emph
  {et~al.}(2015{\natexlab{a}})\citenamefont {Berges}, \citenamefont
  {Boguslavski}, \citenamefont {Schlichting},\ and\ \citenamefont
  {Venugopalan}}]{Berges:2014bba}%
  \BibitemOpen
  \bibfield  {author} {\bibinfo {author} {\bibfnamefont {J.}~\bibnamefont
  {Berges}}, \bibinfo {author} {\bibfnamefont {K.}~\bibnamefont {Boguslavski}},
  \bibinfo {author} {\bibfnamefont {S.}~\bibnamefont {Schlichting}}, \ and\
  \bibinfo {author} {\bibfnamefont {R.}~\bibnamefont {Venugopalan}},\ }\href
  {\doibase 10.1103/PhysRevLett.114.061601} {\bibfield  {journal} {\bibinfo
  {journal} {Phys. Rev. Lett.}\ }\textbf {\bibinfo {volume} {114}},\ \bibinfo
  {pages} {061601} (\bibinfo {year} {2015}{\natexlab{a}})},\ \Eprint
  {http://arxiv.org/abs/1408.1670} {arXiv:1408. 1670 [hep-ph]} \BibitemShut
  {NoStop}%
\bibitem [{\citenamefont {Svistunov}(1991)}]{Svistunov1991a}%
  \BibitemOpen
  \bibfield  {author} {\bibinfo {author} {\bibfnamefont {B.}~\bibnamefont
  {Svistunov}},\ }\href@noop {} {\bibfield  {journal} {\bibinfo  {journal} {J.
  Mosc. Phys. Soc.}\ }\textbf {\bibinfo {volume} {1}},\ \bibinfo {pages} {373}
  (\bibinfo {year} {1991})}\BibitemShut {NoStop}%
\bibitem [{\citenamefont {{Navon}}\ \emph {et~al.}(2016)\citenamefont
  {{Navon}}, \citenamefont {{Gaunt}}, \citenamefont {{Smith}},\ and\
  \citenamefont {{Hadzibabic}}}]{Navon2016a.Nature.539.72}%
  \BibitemOpen
  \bibfield  {author} {\bibinfo {author} {\bibfnamefont {N.}~\bibnamefont
  {{Navon}}}, \bibinfo {author} {\bibfnamefont {A.~L.}\ \bibnamefont
  {{Gaunt}}}, \bibinfo {author} {\bibfnamefont {R.~P.}\ \bibnamefont
  {{Smith}}}, \ and\ \bibinfo {author} {\bibfnamefont {Z.}~\bibnamefont
  {{Hadzibabic}}},\ }\href {\doibase 10.1038/nature20114} {\bibfield  {journal}
  {\bibinfo  {journal} {\nat}\ }\textbf {\bibinfo {volume} {539}},\ \bibinfo
  {pages} {72} (\bibinfo {year} {2016})},\ \Eprint
  {http://arxiv.org/abs/1609.01271} {arXiv:1609.01271 [cond-mat. quant-gas]}
  \BibitemShut {NoStop}%
\bibitem [{\citenamefont {Kagan}\ \emph {et~al.}(1992)\citenamefont {Kagan},
  \citenamefont {Svistunov},\ and\ \citenamefont {Shlyapnikov}}]{Kagan1992a}%
  \BibitemOpen
  \bibfield  {author} {\bibinfo {author} {\bibfnamefont {Y.}~\bibnamefont
  {Kagan}}, \bibinfo {author} {\bibfnamefont {B.~V.}\ \bibnamefont
  {Svistunov}}, \ and\ \bibinfo {author} {\bibfnamefont {G.~V.}\ \bibnamefont
  {Shlyapnikov}},\ }\href
  {http://www.jetp.ac.ru/cgi-bin/e/index/e/74/2/p279?a=list} {\bibfield
  {journal} {\bibinfo  {journal} {[Zh. Eksp. Teor. Fiz. 101, 528 (1992)] Sov.
  Phys. JETP}\ }\textbf {\bibinfo {volume} {74}},\ \bibinfo {pages} {279}
  (\bibinfo {year} {1992})}\BibitemShut {NoStop}%
\bibitem [{\citenamefont {Kagan}\ and\ \citenamefont
  {Svistunov}(1994)}]{Kagan1994a}%
  \BibitemOpen
  \bibfield  {author} {\bibinfo {author} {\bibfnamefont {Y.}~\bibnamefont
  {Kagan}}\ and\ \bibinfo {author} {\bibfnamefont {B.~V.}\ \bibnamefont
  {Svistunov}},\ }\href
  {http://www.jetp.ac.ru/cgi-bin/e/index/e/78/2/p187?a=list} {\bibfield
  {journal} {\bibinfo  {journal} {[Zh. Eksp. Teor. Fiz. 105, 353 (1994)] Sov.
  Phys. JETP}\ }\textbf {\bibinfo {volume} {78}},\ \bibinfo {pages} {187}
  (\bibinfo {year} {1994})}\BibitemShut {NoStop}%
\bibitem [{\citenamefont {Kagan}(1995)}]{Kagan1995a}%
  \BibitemOpen
  \bibfield  {author} {\bibinfo {author} {\bibfnamefont {Y.}~\bibnamefont
  {Kagan}},\ }\href@noop {} {\emph {\bibinfo {title} {Bose-{E}instein
  Condensation}}}\ (\bibinfo  {publisher} {Cambridge University Press},\
  \bibinfo {year} {1995})\ p.\ \bibinfo {pages} {202}\BibitemShut {NoStop}%
\bibitem [{\citenamefont {Semikoz}\ and\ \citenamefont
  {Tkachev}(1995)}]{Semikoz1995a.PhysRevLett.74.3093}%
  \BibitemOpen
  \bibfield  {author} {\bibinfo {author} {\bibfnamefont {D.~V.}\ \bibnamefont
  {Semikoz}}\ and\ \bibinfo {author} {\bibfnamefont {I.~I.}\ \bibnamefont
  {Tkachev}},\ }\href {\doibase 10.1103/PhysRevLett.74.3093} {\bibfield
  {journal} {\bibinfo  {journal} {Phys. Rev. Lett.}\ }\textbf {\bibinfo
  {volume} {74}},\ \bibinfo {pages} {3093} (\bibinfo {year}
  {1995})},\
  \Eprint {http://arxiv.org/abs/hep-ph/9409202} {arXiv:hep-ph/9409202} 
  \BibitemShut {NoStop}%
\bibitem [{\citenamefont {Semikoz}\ and\ \citenamefont
  {Tkachev}(1997)}]{Semikoz1997a}%
  \BibitemOpen
  \bibfield  {author} {\bibinfo {author} {\bibfnamefont {D.}~\bibnamefont
  {Semikoz}}\ and\ \bibinfo {author} {\bibfnamefont {I.}~\bibnamefont
  {Tkachev}},\ }\href@noop {} {\bibfield  {journal} {\bibinfo  {journal} {Phys.
  Rev. D}\ }\textbf {\bibinfo {volume} {55}},\ \bibinfo {pages} {489} (\bibinfo
  {year} {1997})},\ \Eprint {http://arxiv.org/abs/hep-ph/9507306}
  {hep-ph/9507306} \BibitemShut {NoStop}%
\bibitem [{\citenamefont {Berloff}\ and\ \citenamefont
  {Svistunov}(2002)}]{Berloff2002a}%
  \BibitemOpen
  \bibfield  {author} {\bibinfo {author} {\bibfnamefont {N.~G.}\ \bibnamefont
  {Berloff}}\ and\ \bibinfo {author} {\bibfnamefont {B.~V.}\ \bibnamefont
  {Svistunov}},\ }\href {\doibase 10.1103/PhysRevA.66.013603} {\bibfield
  {journal} {\bibinfo  {journal} {Phys. Rev. A}\ }\textbf {\bibinfo {volume}
  {66}},\ \bibinfo {pages} {013603} (\bibinfo {year} {2002})},\ \Eprint
  {http://arxiv.org/abs/cond-mat/0107209} {cond-mat/0107209} \BibitemShut
  {NoStop}%
\bibitem [{\citenamefont {Kozik}\ and\ \citenamefont
  {Svistunov}(2004)}]{Kozik2004a.PhysRevLett.92.035301}%
  \BibitemOpen
  \bibfield  {author} {\bibinfo {author} {\bibfnamefont {E.}~\bibnamefont
  {Kozik}}\ and\ \bibinfo {author} {\bibfnamefont {B.}~\bibnamefont
  {Svistunov}},\ }\href {\doibase 10.1103/PhysRevLett.92.035301} {\bibfield
  {journal} {\bibinfo  {journal} {Phys. Rev. Lett.}\ }\textbf {\bibinfo
  {volume} {92}},\ \bibinfo {pages} {035301} (\bibinfo {year}
  {2004})},\ \Eprint
  {http://arxiv.org/abs/cond-mat/0308193} {cond-mat/0308193} \BibitemShut {NoStop}%
\bibitem [{\citenamefont {Kozik}\ and\ \citenamefont
  {Svistunov}(2005{\natexlab{a}})}]{Kozik2005a.PhysRevLett.94.025301}%
  \BibitemOpen
  \bibfield  {author} {\bibinfo {author} {\bibfnamefont {E.}~\bibnamefont
  {Kozik}}\ and\ \bibinfo {author} {\bibfnamefont {B.}~\bibnamefont
  {Svistunov}},\ }\href {\doibase 10.1103/PhysRevLett.94.025301} {\bibfield
  {journal} {\bibinfo  {journal} {Phys. Rev. Lett.}\ }\textbf {\bibinfo
  {volume} {94}},\ \bibinfo {pages} {025301} (\bibinfo {year}
  {2005}{\natexlab{a}})},\ \Eprint
  {http://arxiv.org/abs/cond-mat/0408241} {cond-mat/0408241} \BibitemShut {NoStop}%
\bibitem [{\citenamefont {Kozik}\ and\ \citenamefont
  {Svistunov}(2005{\natexlab{b}})}]{Kozik2005a.PhysRevB.72.172505}%
  \BibitemOpen
  \bibfield  {author} {\bibinfo {author} {\bibfnamefont {E.}~\bibnamefont
  {Kozik}}\ and\ \bibinfo {author} {\bibfnamefont {B.}~\bibnamefont
  {Svistunov}},\ }\href {\doibase 10.1103/PhysRevB.72.172505} {\bibfield
  {journal} {\bibinfo  {journal} {Phys. Rev. B}\ }\textbf {\bibinfo {volume}
  {72}},\ \bibinfo {pages} {172505} (\bibinfo {year}
  {2005}{\natexlab{b}})},\ \Eprint
  {http://arxiv.org/abs/cond-mat/0505020} {cond-mat/0505020} \BibitemShut {NoStop}%
\bibitem [{\citenamefont {Kozik}\ and\ \citenamefont
  {Svistunov}(2009)}]{Kozik2009a}%
  \BibitemOpen
  \bibfield  {author} {\bibinfo {author} {\bibfnamefont {E.~V.}\ \bibnamefont
  {Kozik}}\ and\ \bibinfo {author} {\bibfnamefont {B.~V.}\ \bibnamefont
  {Svistunov}},\ }\href {\doibase 10.1007/s10909-009-9914-y} {\bibfield
  {journal} {\bibinfo  {journal} {J. Low Temp. Phys.}\ }\textbf {\bibinfo
  {volume} {156}},\ \bibinfo {pages} {215} (\bibinfo {year} {2009})},\ \Eprint
  {http://arxiv.org/abs/0904.1379} {arXiv:0904.1379 [cond-mat.stat-mech]}
  \BibitemShut {NoStop}%
\bibitem [{\citenamefont {{Nowak}}\ \emph {et~al.}(2014)\citenamefont
  {{Nowak}}, \citenamefont {{Schole}},\ and\ \citenamefont
  {{Gasenzer}}}]{Nowak:2012gd}%
  \BibitemOpen
  \bibfield  {author} {\bibinfo {author} {\bibfnamefont {B.}~\bibnamefont
  {{Nowak}}}, \bibinfo {author} {\bibfnamefont {J.}~\bibnamefont {{Schole}}}, \
  and\ \bibinfo {author} {\bibfnamefont {T.}~\bibnamefont {{Gasenzer}}},\
  }\href {\doibase 10.1088/1367-2630/16/9/093052} {\bibfield  {journal}
  {\bibinfo  {journal} {New J. Phys.}\ }\textbf {\bibinfo {volume} {16}},\
  \bibinfo {pages} {093052} (\bibinfo {year} {2014})},\ \Eprint
  {http://arxiv.org/abs/1206.3181v2} {arXiv:1206.3181v2 [cond-mat.quant-gas]}
  \BibitemShut {NoStop}%
\bibitem [{\citenamefont {Berges}\ and\ \citenamefont
  {Sexty}(2012)}]{Berges:2012us}%
  \BibitemOpen
  \bibfield  {author} {\bibinfo {author} {\bibfnamefont {J.}~\bibnamefont
  {Berges}}\ and\ \bibinfo {author} {\bibfnamefont {D.}~\bibnamefont {Sexty}},\
  }\href {\doibase 10.1103/PhysRevLett.108.161601} {\bibfield  {journal}
  {\bibinfo  {journal} {Phys. Rev. Lett.}\ }\textbf {\bibinfo {volume} {108}},\
  \bibinfo {pages} {161601} (\bibinfo {year} {2012})},\ \Eprint
  {http://arxiv.org/abs/1201.0687} {arXiv:1201.0687 [hep-ph]} \BibitemShut
  {NoStop}%
\bibitem [{\citenamefont {Davis}\ \emph {et~al.}(2017)\citenamefont {Davis},
  \citenamefont {Wright}, \citenamefont {Gasenzer}, \citenamefont {Gardiner},\
  and\ \citenamefont {Proukakis}}]{Davis:2016hwt}%
  \BibitemOpen
  \bibfield  {author} {\bibinfo {author} {\bibfnamefont {M.~J.}\ \bibnamefont
  {Davis}}, \bibinfo {author} {\bibfnamefont {T.~M.}\ \bibnamefont {Wright}},
  \bibinfo {author} {\bibfnamefont {T.}~\bibnamefont {Gasenzer}}, \bibinfo
  {author} {\bibfnamefont {S.~A.}\ \bibnamefont {Gardiner}}, \ and\ \bibinfo
  {author} {\bibfnamefont {N.~P.}\ \bibnamefont {Proukakis}},\ }in\ \href@noop
  {} {\emph {\bibinfo {booktitle} {Universal Themes of Bose-Einstein
  Condensation}}},\ \bibinfo {editor} {edited by\ \bibinfo {editor}
  {\bibfnamefont {D.~W.}\ \bibnamefont {Snoke}}, \bibinfo {editor}
  {\bibfnamefont {N.~P.}\ \bibnamefont {Proukakis}}, \ and\ \bibinfo {editor}
  {\bibfnamefont {P.~B.}\ \bibnamefont {Littlewood}}}\ (\bibinfo  {publisher}
  {CUP, Cambridge},\ \bibinfo {year} {2017})\ \Eprint
  {http://arxiv.org/abs/1601.06197} {arXiv:1601.06197 [cond-mat.quant-gas]}
  \BibitemShut {NoStop}%
\bibitem [{\citenamefont {Popov}(1972)}]{Popov1972a}%
  \BibitemOpen
  \bibfield  {author} {\bibinfo {author} {\bibfnamefont {V.~N.}\ \bibnamefont
  {Popov}},\ }\href {\doibase 10.1007/BF01028373} {\bibfield  {journal}
  {\bibinfo  {journal} {Theor. Math. Phys.}\ }\textbf {\bibinfo {volume}
  {11}},\ \bibinfo {pages} {565} (\bibinfo {year} {1972})}\BibitemShut
  {NoStop}%
\bibitem [{\citenamefont {Popov}(1983)}]{Popov1983a}%
  \BibitemOpen
  \bibfield  {author} {\bibinfo {author} {\bibfnamefont {V.~N.}\ \bibnamefont
  {Popov}},\ }\href@noop {} {\emph {\bibinfo {title} {Functional Integrals in
  Quantum Field Theory and Statistical Physics}}}\ (\bibinfo  {publisher}
  {Reidel},\ \bibinfo {address} {Dordrecht},\ \bibinfo {year} {1983})\
  Chap.~\bibinfo {chapter} {6}\BibitemShut {NoStop}%
\bibitem [{\citenamefont {Stamper-Kurn}\ and\ \citenamefont
  {Ueda}(2013)}]{Stamper-Kurn2013a.RevModPhys.85.1191}%
  \BibitemOpen
  \bibfield  {author} {\bibinfo {author} {\bibfnamefont {D.~M.}\ \bibnamefont
  {Stamper-Kurn}}\ and\ \bibinfo {author} {\bibfnamefont {M.}~\bibnamefont
  {Ueda}},\ }\href {\doibase 10.1103/RevModPhys.85.1191} {\bibfield  {journal}
  {\bibinfo  {journal} {Rev. Mod. Phys.}\ }\textbf {\bibinfo {volume} {85}},\
  \bibinfo {pages} {1191} (\bibinfo {year} {2013})},\ \Eprint
  {http://arxiv.org/abs/1205.1888} {arXiv:1205.1888 [cond-mat.quant-gas]}
  \BibitemShut {NoStop}%
\bibitem [{\citenamefont {{Karl}}\ \emph {et~al.}(2013)\citenamefont {{Karl}},
  \citenamefont {{Nowak}},\ and\ \citenamefont {{Gasenzer}}}]{Karl:2013mn}%
  \BibitemOpen
  \bibfield  {author} {\bibinfo {author} {\bibfnamefont {M.}~\bibnamefont
  {{Karl}}}, \bibinfo {author} {\bibfnamefont {B.}~\bibnamefont {{Nowak}}}, \
  and\ \bibinfo {author} {\bibfnamefont {T.}~\bibnamefont {{Gasenzer}}},\
  }\href {\doibase 10.1038/srep02394} {\bibfield  {journal} {\bibinfo
  {journal} {Sci. Rep.}\ }\textbf {\bibinfo {volume} {3}},\ \bibinfo {eid}
  {2394} (\bibinfo {year} {2013}),\ 10.1038/srep02394},\ \Eprint
  {http://arxiv.org/abs/1302.1122} {arXiv:1302.1122 [cond-mat.quant-gas]}
  \BibitemShut {NoStop}%
\bibitem [{\citenamefont {Karl}\ \emph {et~al.}(2013)\citenamefont {Karl},
  \citenamefont {Nowak},\ and\ \citenamefont {Gasenzer}}]{Karl:2013kua}%
  \BibitemOpen
  \bibfield  {author} {\bibinfo {author} {\bibfnamefont {M.}~\bibnamefont
  {Karl}}, \bibinfo {author} {\bibfnamefont {B.}~\bibnamefont {Nowak}}, \ and\
  \bibinfo {author} {\bibfnamefont {T.}~\bibnamefont {Gasenzer}},\ }\href
  {\doibase 10.1103/PhysRevA.88.063615} {\bibfield  {journal} {\bibinfo
  {journal} {Phys. Rev. A}\ }\textbf {\bibinfo {volume} {88}},\ \bibinfo
  {pages} {063615} (\bibinfo {year} {2013})},\ \Eprint
  {http://arxiv.org/abs/1307.7368} {arXiv:1307.7368 [cond-mat.quant-gas]}
  \BibitemShut {NoStop}%
\bibitem [{\citenamefont {Zache}\ \emph {et~al.}(2017)\citenamefont {Zache},
  \citenamefont {Kasper},\ and\ \citenamefont {Berges}}]{Zache:2017dnz}%
  \BibitemOpen
  \bibfield  {author} {\bibinfo {author} {\bibfnamefont {T.~V.}\ \bibnamefont
  {Zache}}, \bibinfo {author} {\bibfnamefont {V.}~\bibnamefont {Kasper}}, \
  and\ \bibinfo {author} {\bibfnamefont {J.}~\bibnamefont {Berges}},\ }\href
  {\doibase 10.1103/PhysRevA.95.063629} {\bibfield  {journal} {\bibinfo
  {journal} {Phys. Rev. A}\ }\textbf {\bibinfo {volume} {95}},\ \bibinfo
  {pages} {063629} (\bibinfo {year} {2017})},\ \Eprint
  {http://arxiv.org/abs/1704.02271} {arXiv:1704.02271 [cond-mat.quant-gas]}
  \BibitemShut {NoStop}%
\bibitem [{\citenamefont {Deng}\ \emph {et~al.}(2018)\citenamefont {Deng},
  \citenamefont {Schlichting}, \citenamefont {Venugopalan},\ and\ \citenamefont
  {Wang}}]{Deng:2018xsk}%
  \BibitemOpen
  \bibfield  {author} {\bibinfo {author} {\bibfnamefont {J.}~\bibnamefont
  {Deng}}, \bibinfo {author} {\bibfnamefont {S.}~\bibnamefont {Schlichting}},
  \bibinfo {author} {\bibfnamefont {R.}~\bibnamefont {Venugopalan}}, \ and\
  \bibinfo {author} {\bibfnamefont {Q.}~\bibnamefont {Wang}},\ }\href {\doibase
  10.1103/PhysRevA.97.053606} {\bibfield  {journal} {\bibinfo  {journal} {Phys.
  Rev. A}\ }\textbf {\bibinfo {volume} {97}},\ \bibinfo {pages} {053606}
  (\bibinfo {year} {2018})},\ \Eprint {http://arxiv.org/abs/1801.06260}
  {arXiv:1801.06260 [hep-th]} \BibitemShut {NoStop}%
\bibitem [{\citenamefont {Schmied}\ \emph {et~al.}(2019)\citenamefont
  {Schmied}, \citenamefont {Mikheev},\ and\ \citenamefont
  {Gasenzer}}]{Schmied:2018upn.PhysRevLett.122.170404}%
  \BibitemOpen
  \bibfield  {author} {\bibinfo {author} {\bibfnamefont {C.-M.}\ \bibnamefont
  {Schmied}}, \bibinfo {author} {\bibfnamefont {A.~N.}\ \bibnamefont
  {Mikheev}}, \ and\ \bibinfo {author} {\bibfnamefont {T.}~\bibnamefont
  {Gasenzer}},\ }\href {\doibase 10.1103/PhysRevLett.122.170404} {\bibfield
  {journal} {\bibinfo  {journal} {Phys. Rev. Lett.}\ }\textbf {\bibinfo
  {volume} {122}},\ \bibinfo {pages} {170404} (\bibinfo {year}
  {2019})},\ \Eprint {https://arxiv.org/abs/1807.07514} {arXiv:1807.07514 
  [cond-mat.quant-gas]}\BibitemShut {NoStop}%
\bibitem [{\citenamefont {Berges}\ \emph
  {et~al.}(2015{\natexlab{b}})\citenamefont {Berges}, \citenamefont
  {Boguslavski}, \citenamefont {Schlichting},\ and\ \citenamefont
  {Venugopalan}}]{Berges:2015ixa}%
  \BibitemOpen
  \bibfield  {author} {\bibinfo {author} {\bibfnamefont {J.}~\bibnamefont
  {Berges}}, \bibinfo {author} {\bibfnamefont {K.}~\bibnamefont {Boguslavski}},
  \bibinfo {author} {\bibfnamefont {S.}~\bibnamefont {Schlichting}}, \ and\
  \bibinfo {author} {\bibfnamefont {R.}~\bibnamefont {Venugopalan}},\ }\href
  {\doibase 10.1103/PhysRevD.92.096006} {\bibfield  {journal} {\bibinfo
  {journal} {Phys. Rev. D}\ }\textbf {\bibinfo {volume} {92}},\ \bibinfo
  {pages} {096006} (\bibinfo {year} {2015}{\natexlab{b}})},\ \Eprint
  {http://arxiv.org/abs/1508.03073} {arXiv:1508.03073 [hep-ph]} \BibitemShut
  {NoStop}%
\bibitem [{\citenamefont {Schachner}\ \emph {et~al.}(2017)\citenamefont
  {Schachner}, \citenamefont {Pi{\~n}eiro~Orioli},\ and\ \citenamefont
  {Berges}}]{Schachner:2016frd}%
  \BibitemOpen
  \bibfield  {author} {\bibinfo {author} {\bibfnamefont {A.}~\bibnamefont
  {Schachner}}, \bibinfo {author} {\bibfnamefont {A.}~\bibnamefont
  {Pi{\~n}eiro~Orioli}}, \ and\ \bibinfo {author} {\bibfnamefont
  {J.}~\bibnamefont {Berges}},\ }\href {\doibase 10.1103/PhysRevA.95.053605}
  {\bibfield  {journal} {\bibinfo  {journal} {Phys. Rev. A}\ }\textbf {\bibinfo
  {volume} {95}},\ \bibinfo {pages} {053605} (\bibinfo {year} {2017})},\
  \Eprint {http://arxiv.org/abs/1612.03038} {arXiv:1612.03038
  [cond-mat.quant-gas]} \BibitemShut {NoStop}%
\bibitem [{\citenamefont {Berges}\ \emph {et~al.}(2017)\citenamefont {Berges},
  \citenamefont {Boguslavski}, \citenamefont {Chatrchyan},\ and\ \citenamefont
  {Jaeckel}}]{Berges:2017ldx}%
  \BibitemOpen
  \bibfield  {author} {\bibinfo {author} {\bibfnamefont {J.}~\bibnamefont
  {Berges}}, \bibinfo {author} {\bibfnamefont {K.}~\bibnamefont {Boguslavski}},
  \bibinfo {author} {\bibfnamefont {A.}~\bibnamefont {Chatrchyan}}, \ and\
  \bibinfo {author} {\bibfnamefont {J.}~\bibnamefont {Jaeckel}},\ }\href
  {\doibase 10.1103/PhysRevD.96.076020} {\bibfield  {journal} {\bibinfo
  {journal} {Phys. Rev. D}\ }\textbf {\bibinfo {volume} {96}},\ \bibinfo
  {pages} {076020} (\bibinfo {year} {2017})},\ \Eprint
  {http://arxiv.org/abs/1707.07696} {arXiv:1707.07696 [hep-ph]} \BibitemShut
  {NoStop}%
\bibitem [{\citenamefont {Walz}\ \emph {et~al.}(2018)\citenamefont {Walz},
  \citenamefont {Boguslavski},\ and\ \citenamefont {Berges}}]{Walz:2017ffj}%
  \BibitemOpen
  \bibfield  {author} {\bibinfo {author} {\bibfnamefont {R.}~\bibnamefont
  {Walz}}, \bibinfo {author} {\bibfnamefont {K.}~\bibnamefont {Boguslavski}}, \
  and\ \bibinfo {author} {\bibfnamefont {J.}~\bibnamefont {Berges}},\ }\href
  {\doibase 10.1103/PhysRevD.97.116011} {\bibfield  {journal} {\bibinfo
  {journal} {Phys. Rev. D}\ }\textbf {\bibinfo {volume} {97}},\ \bibinfo
  {pages} {116011} (\bibinfo {year} {2018})},\ \Eprint
  {http://arxiv.org/abs/1710.11146} {arXiv:1710.11146 [hep-ph]} \BibitemShut
  {NoStop}%
\bibitem [{\citenamefont {{Mikheev}}\ \emph {et~al.}(2018)\citenamefont
  {{Mikheev}}, \citenamefont {{Schmied}},\ and\ \citenamefont
  {{Gasenzer}}}]{Mikheev2018a.arXiv180710228M}%
  \BibitemOpen
  \bibfield  {author} {\bibinfo {author} {\bibfnamefont {A.~N.}\ \bibnamefont
  {{Mikheev}}}, \bibinfo {author} {\bibfnamefont {C.-M.}\ \bibnamefont
  {{Schmied}}}, \ and\ \bibinfo {author} {\bibfnamefont {T.}~\bibnamefont
  {{Gasenzer}}},\ }\href@noop {} {\bibfield  {journal} {\bibinfo  {journal}
  {ArXiv e-prints}\ } (\bibinfo {year} {2018})},\ \Eprint
  {http://arxiv.org/abs/1807.10228} {arXiv:1807.10228 [cond-mat.quant-gas]}
  \BibitemShut {NoStop}%
\bibitem [{\citenamefont {Gasenzer}\ \emph {et~al.}(2012)\citenamefont
  {Gasenzer}, \citenamefont {Nowak},\ and\ \citenamefont
  {Sexty}}]{Gasenzer:2011by}%
  \BibitemOpen
  \bibfield  {author} {\bibinfo {author} {\bibfnamefont {T.}~\bibnamefont
  {Gasenzer}}, \bibinfo {author} {\bibfnamefont {B.}~\bibnamefont {Nowak}}, \
  and\ \bibinfo {author} {\bibfnamefont {D.}~\bibnamefont {Sexty}},\ }\href
  {\doibase 10.1016/j.physletb.2012.03.031} {\bibfield  {journal} {\bibinfo
  {journal} {Phys. Lett.}\ }\textbf {\bibinfo {volume} {B710}},\ \bibinfo
  {pages} {500} (\bibinfo {year} {2012})},\ \Eprint
  {http://arxiv.org/abs/1108.0541} {arXiv:1108.0541 [hep-ph]} \BibitemShut
  {NoStop}%
\bibitem [{\citenamefont {Schmidt}\ \emph {et~al.}(2012)\citenamefont
  {Schmidt}, \citenamefont {Erne}, \citenamefont {Nowak}, \citenamefont
  {Sexty},\ and\ \citenamefont {Gasenzer}}]{Schmidt:2012kw}%
  \BibitemOpen
  \bibfield  {author} {\bibinfo {author} {\bibfnamefont {M.}~\bibnamefont
  {Schmidt}}, \bibinfo {author} {\bibfnamefont {S.}~\bibnamefont {Erne}},
  \bibinfo {author} {\bibfnamefont {B.}~\bibnamefont {Nowak}}, \bibinfo
  {author} {\bibfnamefont {D.}~\bibnamefont {Sexty}}, \ and\ \bibinfo {author}
  {\bibfnamefont {T.}~\bibnamefont {Gasenzer}},\ }\href {\doibase
  10.1088/1367-2630/14/7/075005} {\bibfield  {journal} {\bibinfo  {journal}
  {New J. Phys.}\ }\textbf {\bibinfo {volume} {14}},\ \bibinfo {pages} {075005}
  (\bibinfo {year} {2012})},\ \Eprint {http://arxiv.org/abs/1203.3651}
  {arXiv:1203.3651 [cond-mat.quant-gas]} \BibitemShut {NoStop}%
\bibitem [{\citenamefont {Gasenzer}\ \emph {et~al.}(2014)\citenamefont
  {Gasenzer}, \citenamefont {McLerran}, \citenamefont {Pawlowski},\ and\
  \citenamefont {Sexty}}]{Gasenzer:2013era}%
  \BibitemOpen
  \bibfield  {author} {\bibinfo {author} {\bibfnamefont {T.}~\bibnamefont
  {Gasenzer}}, \bibinfo {author} {\bibfnamefont {L.}~\bibnamefont {McLerran}},
  \bibinfo {author} {\bibfnamefont {J.~M.}\ \bibnamefont {Pawlowski}}, \ and\
  \bibinfo {author} {\bibfnamefont {D.}~\bibnamefont {Sexty}},\ }\href
  {\doibase 10.1016/j.nuclphysa.2014.07.030} {\bibfield  {journal} {\bibinfo
  {journal} {Nucl. Phys. A}\ }\textbf {\bibinfo {volume} {930}},\ \bibinfo
  {pages} {163} (\bibinfo {year} {2014})},\ \Eprint
  {http://arxiv.org/abs/1307.5301} {arXiv:1307.5301 [hep-ph]} \BibitemShut
  {NoStop}%
\bibitem [{\citenamefont {Moore}(2016)}]{Moore:2015adu}%
  \BibitemOpen
  \bibfield  {author} {\bibinfo {author} {\bibfnamefont {G.~D.}\ \bibnamefont
  {Moore}},\ }\href {\doibase 10.1103/PhysRevD.93.065043} {\bibfield  {journal}
  {\bibinfo  {journal} {Phys. Rev. D}\ }\textbf {\bibinfo {volume} {93}},\
  \bibinfo {pages} {065043} (\bibinfo {year} {2016})},\ \Eprint
  {http://arxiv.org/abs/1511.00697} {arXiv: 1511.00697 [hep-ph]} \BibitemShut
  {NoStop}%
\bibitem [{Note1()}]{Note1}%
  \BibitemOpen
  \bibinfo {note} {The U($N$) symmetric model, at low energies and a total
  density set by a chemical potential, in fact has $N-1$ Goldstone modes with
  free dispersion (\ref {eq:freedispersion}) and one density mode,
  characterized by strongly suppressed density fluctuations and enhanced phase
  fluctuations, with linear, Bogoliubov-sound dispersion (\ref
  {eq:BogDispersion}), cf.~Ref.~\cite {Mikheev2018a.arXiv180710228M}.}\BibitemShut {Stop}%
\bibitem [{\citenamefont {Berges}\ and\ \citenamefont
  {Jaeckel}(2015)}]{Berges:2014xea}%
  \BibitemOpen
  \bibfield  {author} {\bibinfo {author} {\bibfnamefont {J.}~\bibnamefont
  {Berges}}\ and\ \bibinfo {author} {\bibfnamefont {J.}~\bibnamefont
  {Jaeckel}},\ }\href {\doibase 10.1103/PhysRevD.91.025020} {\bibfield
  {journal} {\bibinfo  {journal} {Phys. Rev. D}\ }\textbf {\bibinfo {volume}
  {91}},\ \bibinfo {pages} {025020} (\bibinfo {year} {2015})},\ \Eprint
  {http://arxiv.org/abs/1402.4776} {arXiv:1402.4776 [hep-ph]} \BibitemShut
  {NoStop}%
\bibitem [{Note2()}]{Note2}%
  \BibitemOpen
  \bibinfo {note} {And of $\alpha $, in principle, which however follows from
  $\beta $ by means of conservation laws, cf.~Sect.~\ref
  {sec:UniversalDynamics}.}\BibitemShut {Stop}%
\bibitem [{Note3()}]{Note3}%
  \BibitemOpen
  \bibinfo {note} {Locality of the integral in momentum space is to be
  understood on a logarithmic scale. It ensures that the scaling remains
  largely unaffected by the cutoffs, cf., e.g.~Ref.~\cite
  {Zakharov1992a}.}\BibitemShut {Stop}%
\bibitem [{\citenamefont {Newton}(1982)}]{Newton1982a}%
  \BibitemOpen
  \bibfield  {author} {\bibinfo {author} {\bibfnamefont {R.~G.}\ \bibnamefont
  {Newton}},\ }\href@noop {} {\emph {\bibinfo {title} {Scattering Theory of
  Waves and Particles}}}\ (\bibinfo  {publisher} {Springer, New York},\
  \bibinfo {year} {1982})\BibitemShut {NoStop}%
\bibitem [{Note4()}]{Note4}%
  \BibitemOpen
  \bibinfo {note} {Taking also into account three-wave scattering of Bogoliubov
  modes involving the condensate is beyond the scope of this article and will
  be done elsewhere. This implies that exchange of particles with the
  condensate mode is not captured here yet.}\BibitemShut {Stop}%
\bibitem [{Note5()}]{Note5}%
  \BibitemOpen
  \bibinfo {note} {We point out that we choose a notation different from that
  in earlier work, e.g.~Refs.~\cite
  {Berges:2008wm,Scheppach:2009wu,Orioli:2015dxa}, effectively replacing
  $\kappa \to \kappa -\eta $, to take into account that in equilibrium, both,
  $\rho $ and $F$ scale with the same anomalous dimension $\eta $ while out of
  equilibrium the scaling dimension of $F$ gets modified by $\kappa \protect
  \not =z$, cf.~App.~\ref {app:RelEqCrPh}. Note furthermore that the scaling
  (\ref {eq:mainT:ScalingF}) along the center-time axis, $x_{0}=y_{0}=t$, is
  different in character from scaling known in the context of initial-slip and
  ageing dynamics \cite
  {Janssen1989a,Janssen1992a,Calabrese2002a.PhysRevE.65.066120,Calabrese2005a.JPA38.05.R133,Gambassi2006a.JPAConfSer.40.2006.13}.
  See App.~\ref {app:RelEqCrPh}.}\BibitemShut {Stop}%
\bibitem [{Note6()}]{Note6}%
  \BibitemOpen
  \bibinfo {note} {$\Pi ^{R}$ according to Eq.\protect \tmspace +\thinmuskip
  {.1667em}\protect \textup {\hbox {\mathsurround \z@ \protect \normalfont
  (\ignorespaces \ref {eq:app:Pif_free_withf}\unskip \@@italiccorr )}} scales
  as $p_{\Lambda }^{2-\kappa }$, and the integral over $y$ in (\ref
  {eq:app:Pif_free_withf}) scales, by Eqs.~(\ref {eq:sol_pi_free}) and (\ref
  {eq:pitilde_xgge}), as $\sim x^{-1}\sim p_{\Lambda }/p$, giving in total $\Pi
  ^{R}\sim p_{\Lambda }^{3-\kappa }$, as in (\ref {eq:PiRscaling-in-pL}) for
  $d=3$, $\eta =0$, $z=2$.}\BibitemShut {Stop}%
\bibitem [{Note7()}]{Note7}%
  \BibitemOpen
  \bibinfo {note} {The integral (\ref {eq:sol_pi_free}) in the case $z=2$, $d=3$, $\eta=0$ is UV divergent already for $\kappa<2$. However, for $\kappa>d+z-2(2-\eta)$, the leading divergence is an imaginary constant,
  the last term in Eq.\protect \tmspace +\thinmuskip {.1667em}\protect \textup
  {\hbox {\mathsurround \z@ \protect \normalfont (\ignorespaces \ref
  {eq:pikappa-one-x-pl}\unskip \@@italiccorr )}} which drops out when inserting
  $\protect \mathaccentV {tilde}07E\pi _{\kappa }$ into $\Pi
  ^{R}$.}\BibitemShut {Stop}%
\bibitem [{\citenamefont {Bransch{\"a}del}\ and\ \citenamefont
  {Gasenzer}(2008)}]{Branschadel:2008sk}%
  \BibitemOpen
  \bibfield  {author} {\bibinfo {author} {\bibfnamefont {A.}~\bibnamefont
  {Bransch{\"a}del}}\ and\ \bibinfo {author} {\bibfnamefont {T.}~\bibnamefont
  {Gasenzer}},\ }\href {\doibase 10.1088/0953-4075/41/13/135302'} {\bibfield
  {journal} {\bibinfo  {journal} {J. Phys. B}\ }\textbf {\bibinfo {volume}
  {41}},\ \bibinfo {pages} {135302} (\bibinfo {year} {2008})},\ \Eprint
  {http://arxiv.org/abs/0801.4466} {arXiv:0801.4466 [cond-mat.other]}
  \BibitemShut {NoStop}%
\bibitem [{Note8()}]{Note8}%
  \BibitemOpen
  \bibinfo {note} {The exponent $\alpha $ in Ref.~\cite {Svistunov1991a}
  translates to $d/z-1$ in our work, $\gamma $ to $2(d+m_{\kappa })/z-3$.
  Hence, $\varepsilon ^{\gamma /2+1}\sim p^{z(\gamma /2+1)}\sim p^{d+m_{\kappa
  }-z/2}$.}\BibitemShut {Stop}%
\bibitem [{\citenamefont {Svistunov}\ \emph {et~al.}(2015)\citenamefont
  {Svistunov}, \citenamefont {Babaev},\ and\ \citenamefont
  {Prokof'ev}}]{Svistunov2015a.SuperfluidStatesofMatter}%
  \BibitemOpen
  \bibfield  {author} {\bibinfo {author} {\bibfnamefont {B.}~\bibnamefont
  {Svistunov}}, \bibinfo {author} {\bibfnamefont {E.}~\bibnamefont {Babaev}}, \
  and\ \bibinfo {author} {\bibfnamefont {N.}~\bibnamefont {Prokof'ev}},\ }\href
  {https://www.crcpress.com/Superfluid-States-of-Matter/Svistunov-Babaev-Prokofev/p/book/9781439802755}
  {\emph {\bibinfo {title} {Superfluid States of Matter}}}\ (\bibinfo
  {publisher} {CRC Press},\ \bibinfo {year} {2015})\BibitemShut {NoStop}%
\bibitem [{Note9()}]{Note9}%
  \BibitemOpen
  \bibinfo {note} {Note that we choose the scaling form (\ref
  {eq:fmychoice_free}) as it is easier to integrate analytically than the form
  (\ref {eq:scalingf}). Sufficiently far away from the cutoff scale we do not
  expect this to affect our results.}\BibitemShut {Stop}%
\bibitem [{\citenamefont {Falkovich}\ and\ \citenamefont
  {Ryzhenkova}(1992)}]{Falkovich1991a:PhysFlB4.92.594}%
  \BibitemOpen
  \bibfield  {author} {\bibinfo {author} {\bibfnamefont {G.~E.}\ \bibnamefont
  {Falkovich}}\ and\ \bibinfo {author} {\bibfnamefont {I.~V.}\ \bibnamefont
  {Ryzhenkova}},\ }\href {\doibase 10.1063/1.860257} {\bibfield  {journal}
  {\bibinfo  {journal} {Phys. Fl. B: Plasma Phys.}\ }\textbf {\bibinfo {volume}
  {4}},\ \bibinfo {pages} {594} (\bibinfo {year} {1992})}\BibitemShut {NoStop}%
\bibitem [{Note10()}]{Note10}%
  \BibitemOpen
  \bibinfo {note} {Note that in Ref.~\cite {Svistunov1991a}, the formulation
  was given in terms of $\varepsilon _{1}(t)\sim p_{\Lambda }(t)^{2}$, i.e.,
  $n(\varepsilon ,t)\sim \varepsilon _{1}(t)^{-7/6}f(\varepsilon /\varepsilon
  _{1}(t))$, with $f(x)\propto x^{-\alpha }$ where $\alpha =7/6$.}\BibitemShut
  {Stop}%
\bibitem [{\citenamefont {Luttinger}\ and\ \citenamefont
  {Ward}(1960)}]{Luttinger1960a}%
  \BibitemOpen
  \bibfield  {author} {\bibinfo {author} {\bibfnamefont {J.~M.}\ \bibnamefont
  {Luttinger}}\ and\ \bibinfo {author} {\bibfnamefont {J.~C.}\ \bibnamefont
  {Ward}},\ }\href {\doibase 10.1103/PhysRev.118.1417} {\bibfield  {journal}
  {\bibinfo  {journal} {Phys. Rev.}\ }\textbf {\bibinfo {volume} {118}},\
  \bibinfo {pages} {1417} (\bibinfo {year} {1960})}\BibitemShut {NoStop}%
\bibitem [{\citenamefont {Baym}(1962)}]{Baym1962a}%
  \BibitemOpen
  \bibfield  {author} {\bibinfo {author} {\bibfnamefont {G.}~\bibnamefont
  {Baym}},\ }\href {\doibase 10.1103/PhysRev.127.1391} {\bibfield  {journal}
  {\bibinfo  {journal} {Phys. Rev.}\ }\textbf {\bibinfo {volume} {127}},\
  \bibinfo {pages} {1391} (\bibinfo {year} {1962})}\BibitemShut {NoStop}%
\bibitem [{\citenamefont {Cornwall}\ \emph {et~al.}(1974)\citenamefont
  {Cornwall}, \citenamefont {Jackiw},\ and\ \citenamefont
  {Tomboulis}}]{Cornwall1974a}%
  \BibitemOpen
  \bibfield  {author} {\bibinfo {author} {\bibfnamefont {J.~M.}\ \bibnamefont
  {Cornwall}}, \bibinfo {author} {\bibfnamefont {R.}~\bibnamefont {Jackiw}}, \
  and\ \bibinfo {author} {\bibfnamefont {E.}~\bibnamefont {Tomboulis}},\ }\href
  {\doibase 10.1103/PhysRevD.10.2428} {\bibfield  {journal} {\bibinfo
  {journal} {Phys. Rev. D}\ }\textbf {\bibinfo {volume} {10}},\ \bibinfo
  {pages} {2428} (\bibinfo {year} {1974})}\BibitemShut {NoStop}%
\bibitem [{\citenamefont {Berges}(2002)}]{Berges:2001fi}%
  \BibitemOpen
  \bibfield  {author} {\bibinfo {author} {\bibfnamefont {J.}~\bibnamefont
  {Berges}},\ }\href {\doibase 10.1016/S0375-9474(01)01295-7} {\bibfield
  {journal} {\bibinfo  {journal} {Nucl. Phys.}\ }\textbf {\bibinfo {volume}
  {A699}},\ \bibinfo {pages} {847} (\bibinfo {year} {2002})},\ \Eprint
  {http://arxiv.org/abs/hep-ph/0105311} {hep-ph/0105311} \BibitemShut {NoStop}%
\bibitem [{\citenamefont {Aarts}\ \emph {et~al.}(2002)\citenamefont {Aarts},
  \citenamefont {Ahrensmeier}, \citenamefont {Baier}, \citenamefont {Berges},\
  and\ \citenamefont {Serreau}}]{Aarts:2002dj}%
  \BibitemOpen
  \bibfield  {author} {\bibinfo {author} {\bibfnamefont {G.}~\bibnamefont
  {Aarts}}, \bibinfo {author} {\bibfnamefont {D.}~\bibnamefont {Ahrensmeier}},
  \bibinfo {author} {\bibfnamefont {R.}~\bibnamefont {Baier}}, \bibinfo
  {author} {\bibfnamefont {J.}~\bibnamefont {Berges}}, \ and\ \bibinfo {author}
  {\bibfnamefont {J.}~\bibnamefont {Serreau}},\ }\href {\doibase
  10.1103/PhysRevD.66.045008} {\bibfield  {journal} {\bibinfo  {journal} {Phys.
  Rev. D}\ }\textbf {\bibinfo {volume} {66}},\ \bibinfo {pages} {045008}
  (\bibinfo {year} {2002})},\ \Eprint {http://arxiv.org/abs/hep-ph/0201308}
  {hep-ph/0201308} \BibitemShut {NoStop}%
\bibitem [{\citenamefont {Berges}(2005)}]{Berges:2004yj}%
  \BibitemOpen
  \bibfield  {author} {\bibinfo {author} {\bibfnamefont {J.}~\bibnamefont
  {Berges}},\ }\href {\doibase 10.1063/1.1843591} {\bibfield  {journal}
  {\bibinfo  {journal} {AIP Conf. Proc.}\ }\textbf {\bibinfo {volume} {739}},\
  \bibinfo {pages} {3} (\bibinfo {year} {2005})},\ \Eprint
  {http://arxiv.org/abs/hep-ph/0409233} {hep-ph/0409233} \BibitemShut {NoStop}%
\bibitem [{\citenamefont {Gasenzer}(2009)}]{Gasenzer2009a}%
  \BibitemOpen
  \bibfield  {author} {\bibinfo {author} {\bibfnamefont {T.}~\bibnamefont
  {Gasenzer}},\ }\href {\doibase 10.1140/epjst/e2009-00960-5} {\bibfield
  {journal} {\bibinfo  {journal} {Eur. Phys. J. ST}\ }\textbf {\bibinfo
  {volume} {168}},\ \bibinfo {pages} {89} (\bibinfo {year} {2009})},\ \Eprint
  {http://arxiv.org/abs/0812.0004} {arXiv:0812.0004 [cond-mat.other]}
  \BibitemShut {NoStop}%
\bibitem [{\citenamefont {Berges}\ and\ \citenamefont
  {Gasenzer}(2007)}]{Berges:2007ym}%
  \BibitemOpen
  \bibfield  {author} {\bibinfo {author} {\bibfnamefont {J.}~\bibnamefont
  {Berges}}\ and\ \bibinfo {author} {\bibfnamefont {T.}~\bibnamefont
  {Gasenzer}},\ }\href {\doibase 10.1103/PhysRevA.76.033604} {\bibfield
  {journal} {\bibinfo  {journal} {Phys. Rev. A}\ }\textbf {\bibinfo {volume}
  {76}},\ \bibinfo {pages} {033604} (\bibinfo {year} {2007})},\ \Eprint
  {http://arxiv.org/abs/cond-mat/0703163} {cond-mat/0703163} \BibitemShut
  {NoStop}%
\bibitem [{\citenamefont {Keldysh}(1965)}]{Keldysh1964a}%
  \BibitemOpen
  \bibfield  {author} {\bibinfo {author} {\bibfnamefont {L.~V.}\ \bibnamefont
  {Keldysh}},\ }\href
  {http://www.jetp.ac.ru/cgi-bin/e/index/e/20/4/p1018?a=list} {\bibfield
  {journal} {\bibinfo  {journal} {[Zh. Eksp. Teor. Fiz. {\bfseries 47}, 1515
  (1964)] Sov. Phys. JETP}\ }\textbf {\bibinfo {volume} {20}},\ \bibinfo
  {pages} {1018} (\bibinfo {year} {1965})}\BibitemShut {NoStop}%
\bibitem [{\citenamefont {Kadanoff}\ and\ \citenamefont
  {Baym}(1995)}]{KadanoffBaym1995a}%
  \BibitemOpen
  \bibfield  {author} {\bibinfo {author} {\bibfnamefont {L.~P.}\ \bibnamefont
  {Kadanoff}}\ and\ \bibinfo {author} {\bibfnamefont {G.}~\bibnamefont
  {Baym}},\ }\href@noop {} {\emph {\bibinfo {title} {Quantum {S}tatistical
  {M}echanics}}},\ \bibinfo {edition} {2nd}\ ed.\ (\bibinfo  {publisher}
  {Addison-Wesley},\ \bibinfo {year} {1995})\BibitemShut {NoStop}%
\bibitem [{\citenamefont {Rammer}(2007)}]{Rammer2007a}%
  \BibitemOpen
  \bibfield  {author} {\bibinfo {author} {\bibfnamefont {J.}~\bibnamefont
  {Rammer}},\ }\href@noop {} {\emph {\bibinfo {title} {Quantum {F}ield {T}heory
  of {N}on-equilibrium {S}tates}}},\ \bibinfo {edition} {1st}\ ed.\ (\bibinfo
  {publisher} {CUP, Cambridge, UK},\ \bibinfo {year} {2007})\BibitemShut
  {NoStop}%
\bibitem [{\citenamefont {Arrizabalaga}\ \emph {et~al.}(2005)\citenamefont
  {Arrizabalaga}, \citenamefont {Smit},\ and\ \citenamefont
  {Tranberg}}]{Arrizabalaga:2005tf}%
  \BibitemOpen
  \bibfield  {author} {\bibinfo {author} {\bibfnamefont {A.}~\bibnamefont
  {Arrizabalaga}}, \bibinfo {author} {\bibfnamefont {J.}~\bibnamefont {Smit}},
  \ and\ \bibinfo {author} {\bibfnamefont {A.}~\bibnamefont {Tranberg}},\
  }\href {\doibase 10.1103/PhysRevD.72.025014} {\bibfield  {journal} {\bibinfo
  {journal} {Phys. Rev. D}\ }\textbf {\bibinfo {volume} {72}},\ \bibinfo
  {pages} {025014} (\bibinfo {year} {2005})},\ \Eprint
  {http://arxiv.org/abs/hep-ph/0503287} {hep-ph/0503287} \BibitemShut {NoStop}%
\bibitem [{\citenamefont {Gasenzer}\ \emph {et~al.}(2005)\citenamefont
  {Gasenzer}, \citenamefont {Berges}, \citenamefont {Schmidt},\ and\
  \citenamefont {Seco}}]{Gasenzer:2005ze}%
  \BibitemOpen
  \bibfield  {author} {\bibinfo {author} {\bibfnamefont {T.}~\bibnamefont
  {Gasenzer}}, \bibinfo {author} {\bibfnamefont {J.}~\bibnamefont {Berges}},
  \bibinfo {author} {\bibfnamefont {M.~G.}\ \bibnamefont {Schmidt}}, \ and\
  \bibinfo {author} {\bibfnamefont {M.}~\bibnamefont {Seco}},\ }\href {\doibase
  10.1103/PhysRevA.72.063604} {\bibfield  {journal} {\bibinfo  {journal} {Phys.
  Rev. A}\ }\textbf {\bibinfo {volume} {72}},\ \bibinfo {pages} {063604}
  (\bibinfo {year} {2005})},\ \Eprint {http://arxiv.org/abs/cond-mat/0507480}
  {cond-mat/0507480} \BibitemShut {NoStop}%
\bibitem [{Note11()}]{Note11}%
  \BibitemOpen
  \bibinfo {note} {Alternatively, one could look at a $T=0$ quantum critical
  point where the equilibrium scaling reads $g^{(1)}(\protect \mathbf
  {r})=s^{d+z-2+\eta }g^{(1)}(s\protect \mathbf {r})$ and $\eta _\protect
  \mathrm {neq}=\eta -\kappa $. In this case, the equilibrium momentum exponent
  is $\kappa =0$, consistent with a vacuum distribution.}\BibitemShut {Stop}%
\bibitem [{\citenamefont {Bateman}\ and\ \citenamefont
  {Project}(1955)}]{bateman1955higher}%
  \BibitemOpen
  \bibfield  {author} {\bibinfo {author} {\bibfnamefont {H.}~\bibnamefont
  {Bateman}}\ and\ \bibinfo {author} {\bibfnamefont {B.~M.}\ \bibnamefont
  {Project}},\ }\href {https://books.google.de/books?id=IfaJAQAACAAJ} {\emph
  {\bibinfo {title} {{Higher Transcendental Functions}}}},\ \bibinfo {series}
  {{Bateman Manuscript Project California Institute of Technology}}\ No.\
  \bibinfo {number} {v. 1}\ (\bibinfo  {publisher} {McGraw-Hill},\ \bibinfo
  {year} {1955})\BibitemShut {NoStop}%
\bibitem [{\citenamefont {Chantesana}(2017)}]{Chantesana2017a}%
  \BibitemOpen
  \bibfield  {author} {\bibinfo {author} {\bibfnamefont {I.}~\bibnamefont
  {Chantesana}},\ }\href@noop {} {\bibinfo {type} {Ph{D} thesis}},\ \bibinfo
  {school} {Ruprecht-Karls-Universit{\"a}t Heidelberg} (\bibinfo {year}
  {2017})\BibitemShut {NoStop}%
\bibitem [{Note12()}]{Note12}%
  \BibitemOpen
  \bibinfo {note} {The case of an integer $\kappa $ requires further discussion
  due to the non-simple pole structure of the integral representation of the
  hypergeometric function in the form of a Mellin-Barnes integral. As our
  numerical results presented in Sect.~\ref {sec:SummaryNumerical} demonstrate,
  there is no discontinuity in the transition from non-integer to integer
  values of $\kappa $. We therefore use the expressions for non-integer $\kappa
  $ and take the limit to an integer value.}\BibitemShut {Stop}%
\bibitem [{\citenamefont {Huang}(1987)}]{Huang1987a}%
  \BibitemOpen
  \bibfield  {author} {\bibinfo {author} {\bibfnamefont {K.}~\bibnamefont
  {Huang}},\ }\href@noop {} {\emph {\bibinfo {title} {Statistical Mechanics}}}\
  (\bibinfo  {publisher} {Wiley, New York},\ \bibinfo {year}
  {1987})\BibitemShut {NoStop}%
\bibitem [{\citenamefont {Kraichnan}(1967)}]{Kraichnan1967a}%
  \BibitemOpen
  \bibfield  {author} {\bibinfo {author} {\bibfnamefont {R.}~\bibnamefont
  {Kraichnan}},\ }\href {\doibase 10.1063/1.1762301} {\bibfield  {journal}
  {\bibinfo  {journal} {Phys. Fl.}\ }\textbf {\bibinfo {volume} {10}},\
  \bibinfo {pages} {1417} (\bibinfo {year} {1967})}\BibitemShut {NoStop}%
\end{thebibliography}


%

\end{document}